\documentclass[12pt]{report}

\usepackage{fancyhdr}
\usepackage{rotating}
\usepackage{a4wide}
\usepackage[utf8]{inputenc}
\usepackage{multirow}
\usepackage{pbox}
\usepackage{tabularx}
\usepackage{lscape}
\usepackage{graphicx}
\usepackage{array,graphicx}
\usepackage{booktabs}
\usepackage{pifont}
\usepackage{xcolor}
\usepackage{fancyvrb}
\usepackage{enumerate}
\usepackage{pifont}
\usepackage{listings}
\usepackage{amsmath}
\usepackage{amssymb}
\usepackage{algorithm}
\usepackage{algpseudocode}
\usepackage[backend=bibtex]{biblatex}

\newcommand{\cmark}{\textcolor{green}{\ding{51}}}%
\newcommand{\xmark}{\textcolor{red}{\ding{55}}}%

\newcommand{\Break}{\State \textbf{break} }

\DeclareMathOperator{\y}{y}
\DeclareMathOperator{\K}{K}
\DeclareMathOperator{\p}{p}
\DeclareMathOperator{\Z}{Z}
\DeclareMathOperator{\PP}{P}
\DeclareMathOperator{\dd}{d}
\DeclareMathOperator{\T}{T}
\DeclareMathOperator{\C}{C}
\DeclareMathOperator{\B}{B}
\DeclareMathOperator{\prt}{par}
\DeclareMathOperator{\anc}{anc}
\DeclareMathOperator{\area}{area}
\DeclareMathOperator{\class}{class}
\DeclareMathOperator{\aut}{aut}
\DeclareMathOperator{\aff}{aff}
\DeclareMathOperator{\aaf}{aa}
\DeclareMathOperator{\inst}{inst}
\DeclareMathOperator{\addr}{addr}
\DeclareMathOperator{\coun}{coun}
\DeclareMathOperator{\rf}{ref}
\DeclareMathOperator{\sm}{sim}
\DeclareMathOperator{\score}{score}
\DeclareMathOperator{\cor}{cor}
\DeclareMathOperator{\sw}{sw}
\DeclareMathOperator{\lnt}{l}

\numberwithin{equation}{chapter}
\numberwithin{algorithm}{chapter}

\def\thetitle{New Methods for Metadata Extraction\\from Scientific Literature}

\def\theauthor{Dominika Tkaczyk}

\def\themonth{November}
\def\theyear{2015}
\def\thesupervisor{Professor Marek Niezg\'{o}dka, PhD, DSc}
\def\thesupervisors{\L{}ukasz Bolikowski, PhD}

\def\themonthyear{\themonth{} \theyear}

\author{\theauthor}
\title{\thetitle}

\def\titlepages{\newpage
\thispagestyle{empty}
\begin{centering}
\large
Systems Research Institute, Polish Academy of Sciences\\
\vspace{4cm}
\Large
\theauthor\\
\large
ICM, University of Warsaw\\
\vspace{0.85cm}
\LARGE
\thetitle\\
\vspace{3cm}
\normalsize
\textit{Ph.D. Thesis}\\
\vspace{3cm}
Supervisor:
\vspace{0.25cm}
\thesupervisor\\
\vspace{-7px}
Auxiliary supervisor:
\thesupervisors\\
\vspace{0.5cm}
ICM, University of Warsaw\\
\vfill
\themonthyear\\
\end{centering}

\newpage
}

\bibliography{bibliography}

\begin{document}

\newcolumntype{L}[1]{>{\raggedright\arraybackslash}p{#1}}
\newcolumntype{C}[1]{>{\centering\arraybackslash}p{#1}}
\newcolumntype{R}[1]{>{\raggedleft\arraybackslash}p{#1}}

\titlepages

\clearpage

\thispagestyle{empty}
\cleardoublepage

\pagenumbering{roman}

\begin{abstract}

Spreading the ideas and announcing new discoveries and findings in the scientific world is typically
realized by publishing and reading scientific literature. Within the past few decades we have
witnessed digital revolution, which moved scholarly communication to electronic media and also
resulted in a substantial increase in its volume. Nowadays keeping track with the latest scientific
achievements poses a major challenge for the researchers. Scientific information overload is a
severe problem that slows down scholarly communication and knowledge propagation across the
academia.

Modern research infrastructures facilitate studying scientific literature by providing intelligent
search tools, proposing similar and related documents, building and visualizing interactive citation
and author networks, assessing the quality and impact of the articles using citation-based
statistics, and so on. In order to provide such high quality services the system requires the access
not only to the text content of stored documents, but also to their machine-readable metadata. Since
in practice good quality metadata is not always available, there is a strong demand for a reliable
automatic method of extracting machine-readable metadata directly from source documents.

Our research addresses these problems by proposing an automatic, accurate and flexible algorithm for
extracting wide range of metadata directly from scientific articles in born-digital form. Extracted
information includes basic document metadata, structured full text and bibliography section.

Designed as a universal solution, proposed algorithm is able to handle a vast variety of publication
layouts with high precision and thus is well-suited for analyzing heterogeneous document 
collections. This was achieved by employing supervised and unsupervised machine-learning algorithms
trained on large, diverse datasets. The evaluation we conducted showed good performance of proposed
metadata extraction algorithm. The comparison with other similar solutions also proved our algorithm
performs better than competition for most metadata types.

Proposed method is a reliable and accurate solution to the problem of extracting the metadata from 
documents. It allows modern research infrastructures to provide intelligent tools and services
supporting the process of consuming the growing volume of scientific literature by the readers,
which results in facilitating the communication among the scientists and the overall improvement of
the knowledge propagation and the quality of the research in the scientific world. 

\end{abstract}

\clearpage
\thispagestyle{empty}
\cleardoublepage

\setcounter{tocdepth}{1}
\tableofcontents
\clearpage

\listoffigures
\clearpage
\listoftables
\clearpage
\renewcommand\lstlistlistingname{List of Listings}
\lstlistoflistings
\clearpage
\listofalgorithms
\clearpage

\thispagestyle{empty}
\cleardoublepage

\pagenumbering{arabic}
\setcounter{page}{1}


\chapter{Overview}
This chapter is an introduction to the research described in the thesis. First we sketch the 
background of the work and explain our motivations. Then we formulate the problem we focus on and
outline the proposed solution and its features. Finally, we state the key contributions and briefly
describe the thesis structure.

\section{Background and Motivation}
The background of the research described in the thesis is related to scholarly communication in
digital era and the problems it encounters. Our main objective is to facilitate the communication
among the scientists and improve the knowledge propagation in the scientific world. These goals are
accomplished by equipping digital libraries and research infrastructures with means allowing them to
support the process of consuming the growing volume of scientific literature by researchers and
scientists.

\subsubsection{Scholarly Communication}
In the scientific world communicating the ideas, describing the planned, ongoing and completed
research and finally reporting discoveries and project results is typically realized by publishing
and reading scientific literature, mostly in the form of articles published in journals or
conference proceedings. Originally scientific literature was distributed in the form of printed
paper, but within the last 30 years we have witnessed the digital revolution which has moved this 
aspect of scientific communication to electronic media.

Along with the media change we have also observed a huge increase in the volume of available
scientific literature. The exact total number of existing scientific articles is not known, but the
statistics gathered from popular electronic databases show the scale we are dealing with. For 
example DBLP database\footnote{http://dblp.uni-trier.de/}, which provides bibliographic information
on scientific literature from computer science discipline only, currently contains approximately 3
million records. PubMed Central\footnote{http://www.ncbi.nlm.nih.gov/pmc/} is a full text free
archive of 3.6 million biomedical and life sciences journal articles. 
PubMed\footnote{http://www.ncbi.nlm.nih.gov/pubmed}, a freely available index of biomedical
abstracts, including the entire MEDLINE database, contains 25 million references. Finally, Scopus
database\footnote{http://www.scopus.com/}, which collects publications from a much wider range of
disciplines that DBLP or PubMed, currently contains 57 million records.

There also have been a number of attempts of estimating the total number of scientific articles or a
specific subset of them. For example Bjork {\it et al.}~\cite{BjorkRL09} estimated the number of
peer-reviewed journal articles published by 2006 to be about 1,350,000 using data from the ISI 
citation database\footnote{http://ip-science.thomsonreuters.com}. Jinha~\cite{Jinha2010258} used
this result and a number of assumptions related to a steady increase in the number of researchers,
journals and articles, and arrived at the estimation of the total number of journal articles ever
published to be more than 50 million as of 2009. Finally, Khabsa and Giles~\cite{Khabsa2014} studied
the volume of scholarly documents written in English and available on the web by analysing the
coverage of two popular academic search engines: Google Scholar\footnote{https://scholar.google.pl/}
and Microsoft Academic Search\footnote{http://academic.research.microsoft.com/}. Their estimates
show that at least 114 million documents are accessible on the web, with at least 27 million 
available without any subscription or payment.

\begin{figure}[h]
  \centering
  \includegraphics[width=0.7\textwidth]{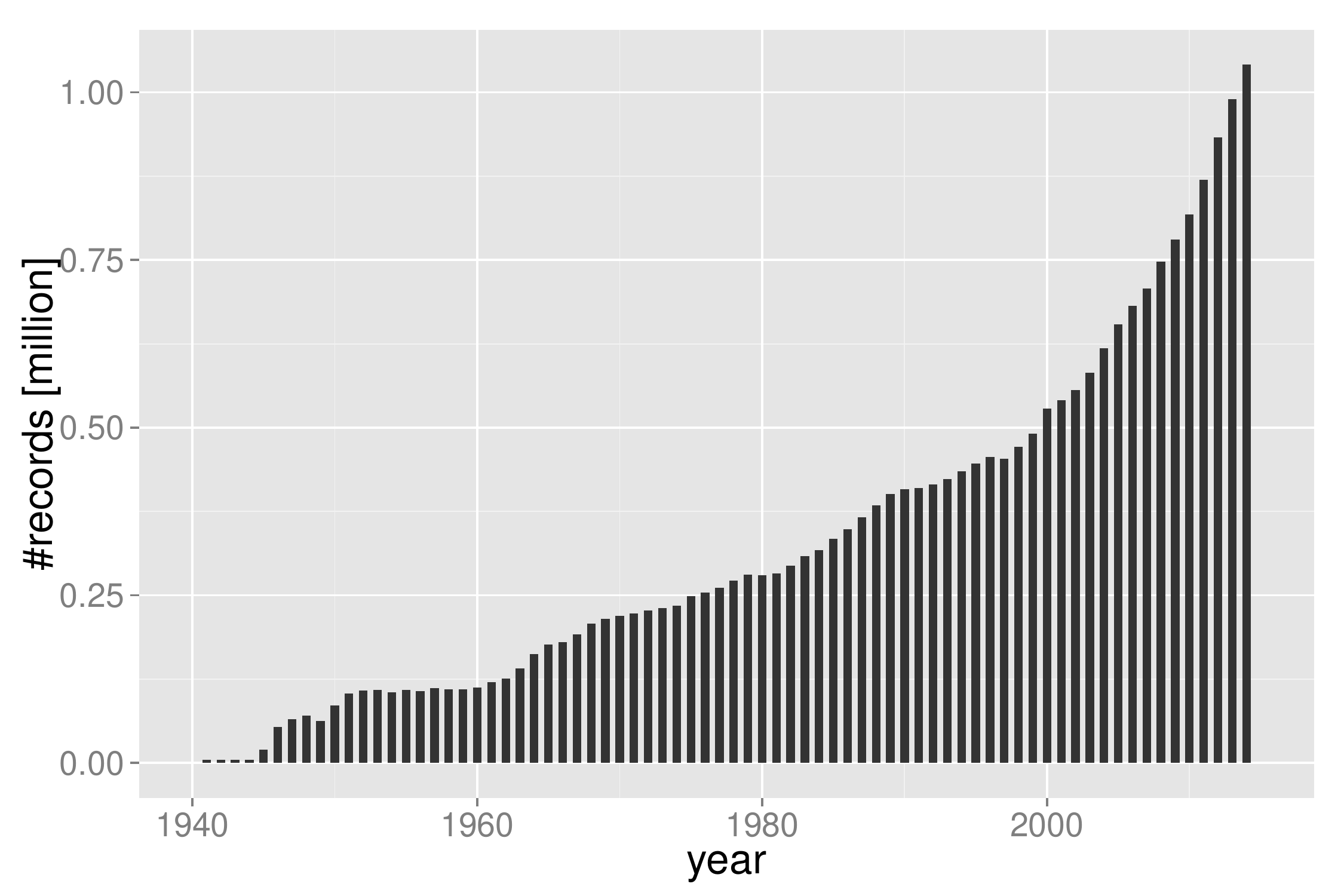}
  \caption[The number of records by year of publication in PubMed]{The number of records by year of
  publication in PubMed database. The data shows clear growing trend.}
  \label{fig:stats-pubmed}
\end{figure}

\begin{figure}[h]
  \centering
  \includegraphics[width=0.7\textwidth]{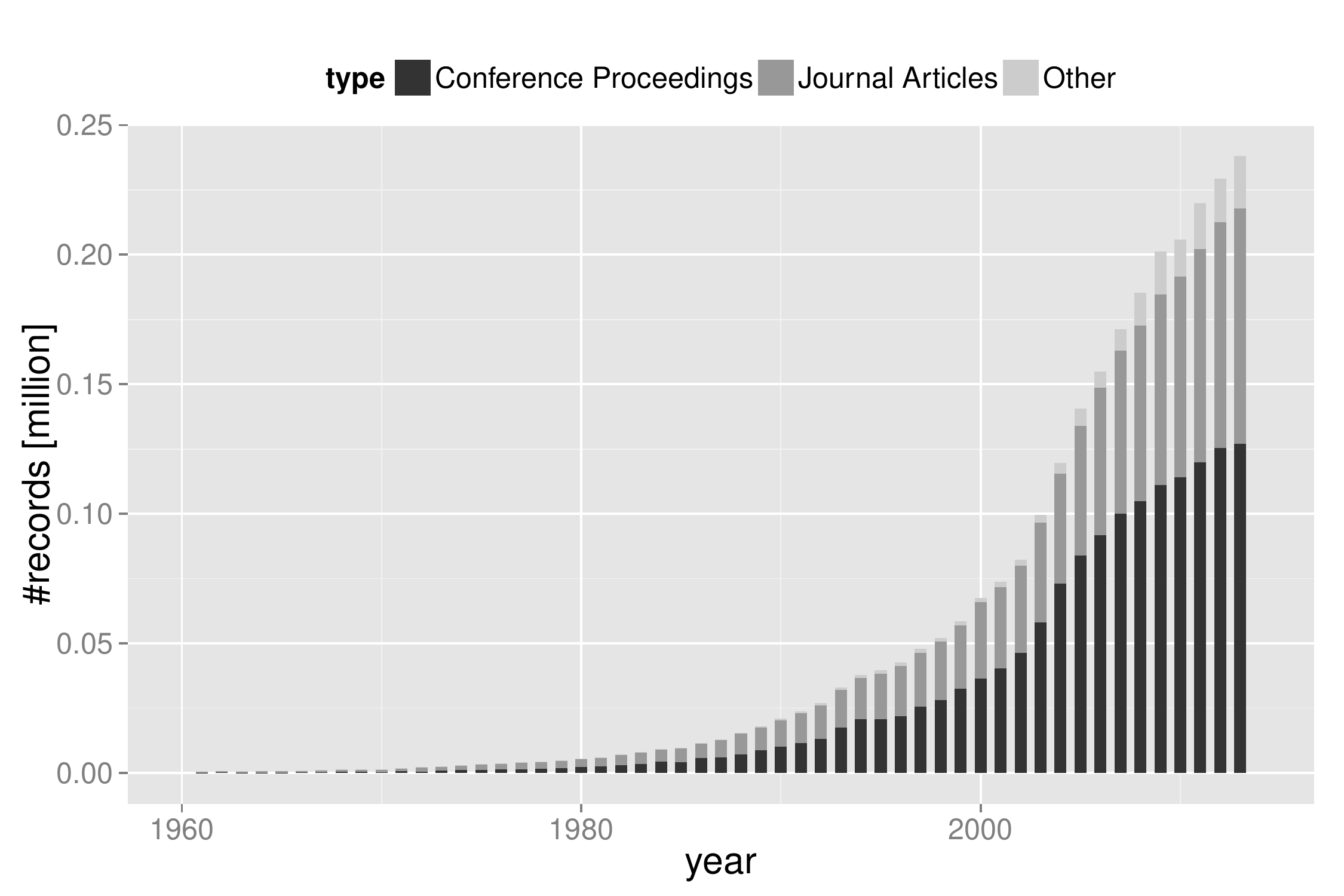}
  \caption[The number of records by year of publication and type in DBLP]{The number of records by
  year of publication and type in DBLP database. The data shows clear growing trend, in particular
  in both journal articles and conference proceedings categories.}
  \label{fig:stats-dblp}
\end{figure}

In addition to the total volume of already published scientific literature being huge, we are also 
observing a substantial increase in the number of new articles published every year. According to
Larsen and von Ins~\cite{LarsenI10}, there are no indications that the growth rate in the number of
published peer-reviewed journal articles has decreased within the last 50 years, and at the same
time, the publication using new channels, such as conference proceedings, open archives and web 
pages, is growing fast. The statistical data obtained from DBLP and PubMed databases show similar
trends (Figures~\ref{fig:stats-pubmed} and~\ref{fig:stats-dblp}).

Writing and publishing articles is only one side of scholarly communication. At the other end there
are the consumers of the literature, usually also scientists and researchers, interested in new
ideas and discoveries in their own field, or trying to get familiar with the state of the art in new
fields. Keeping track of the latest scientific findings and achievements published in journals or
conference proceedings is a crucial aspect of their work. Ignoring this task results in deficiencies
in the knowledge related to the latest discoveries and trends, which in turn can lower the quality
of their own research, make results assessment much harder and significantly limit the possibility
to find new interesting research areas and challenges.

Unfortunately, due to the huge and still growing volume of scientific literature, keeping track with 
the latest achievements is a major challenge for the researchers. Scientific information overload is
a severe problem that slows down the scholarly communication and knowledge propagation across the
academia.

\subsubsection{Digital Libraries}
The digital era resulted not only in moving the literature from paper to digital media, but in fact
changed the way modern research is conducted. Research infrastructures equip the users with the
resources and services supporting all stages of the research in many disciplines. Digital libraries
provide means for storing, organizing and accessing digital collections of research-related data of
all kinds, such as documents, datasets or tools.

\begin{figure}[h]
  \centering
  \includegraphics[width=1\textwidth]{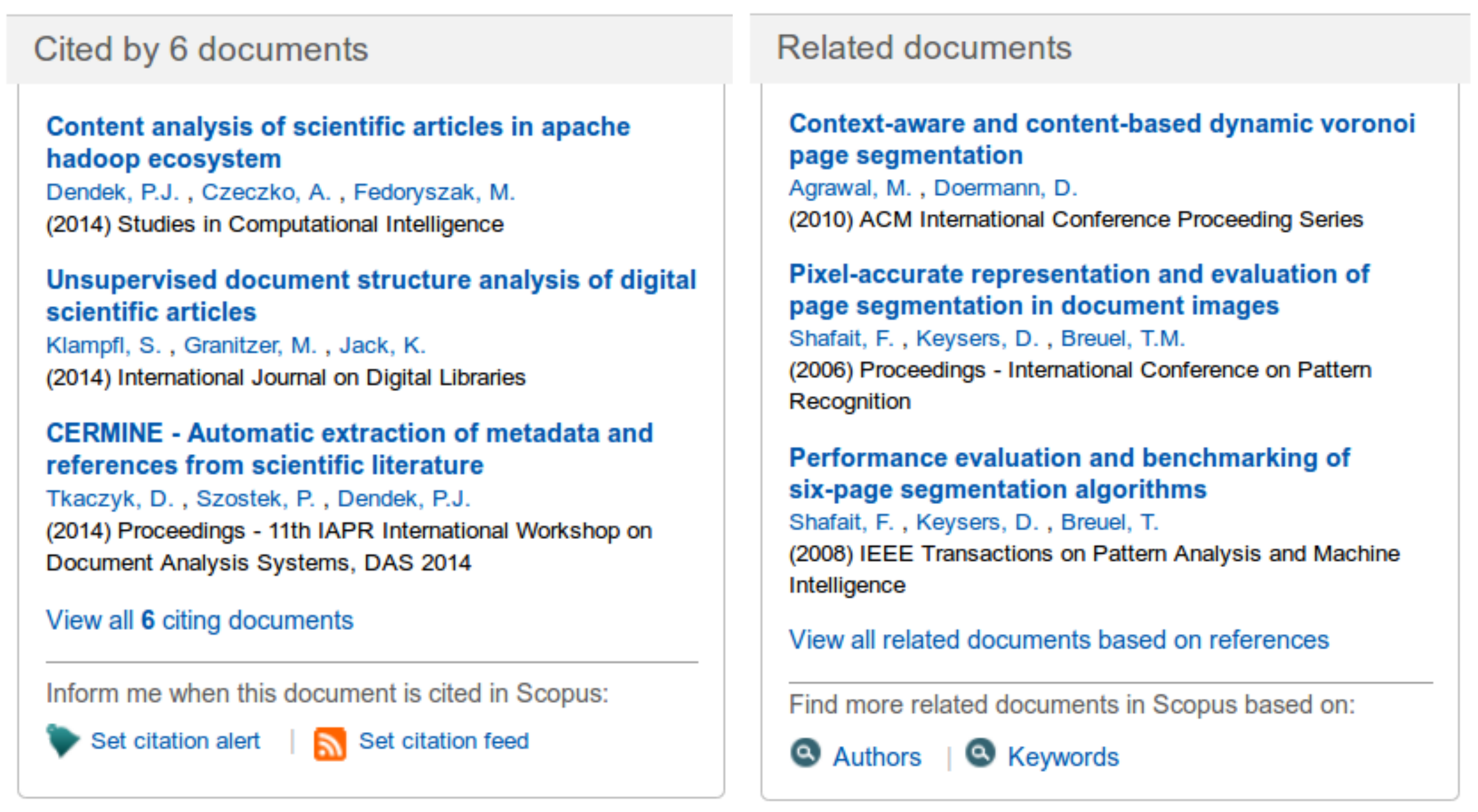}
  \caption[Citation and related documents networks in Scopus]{Example screenshots from Scopus
  system. Scopus enables the users to navigate through citation network using "cited by" list. The
  system also proposes documents related to the current one based on references, authors or
  keywords.}
  \label{fig:scopus}
\end{figure}

\begin{figure}[h]
  \centering
  \includegraphics[width=1\textwidth]{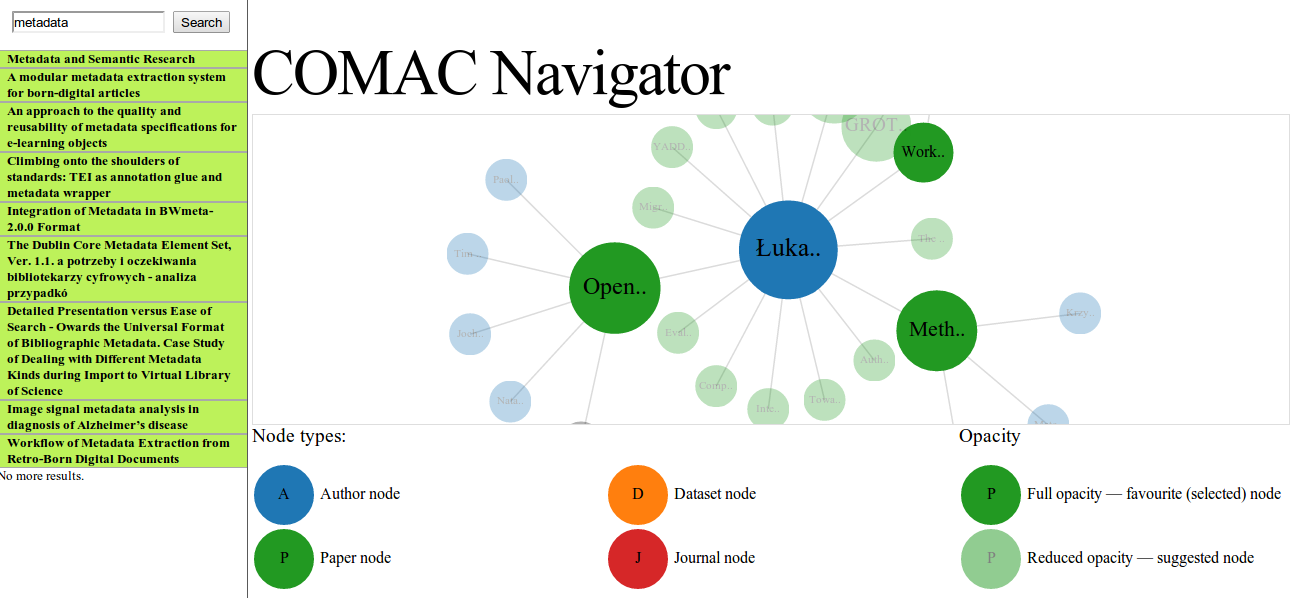}
  \caption[Documents and authors network in COMAC]{The visualization of a fragment of documents and
  authors network generated by COMAC (Common Map of Academia)~\cite{JurkiewiczWWFD14}.}
  \label{fig:comac}
\end{figure}

These modern infrastructures support the process of studying scientific literature by providing
intelligent search tools, proposing similar and related documents (Figure~\ref{fig:scopus}),
building and visualizing interactive citation and author networks (Figure~\ref{fig:comac}), 
providing various citation-based statistics, and so on. This enables the users to effectively 
explore the map of science, quickly get familiar with the current state of the art of a given 
problem and reduce the volume of articles to read by retrieving only the most relevant and
interesting positions.

Unfortunately, building the services supporting the readers is not a trivial task. Such intelligent,
high-quality services and tools require reliable, machine-readable metadata of the digital library
resources. Unfortunately, in practice a large portion of the resources is typically available to a
great extent as unstructured text, intended for human readers, but poorly understood by machines.
Good quality metadata is not always available, sometimes it is missing, full of errors or
fragmentary, even for fairly recently published articles.

\begin{figure}[h]
  \centering
  \includegraphics[width=0.9\textwidth]{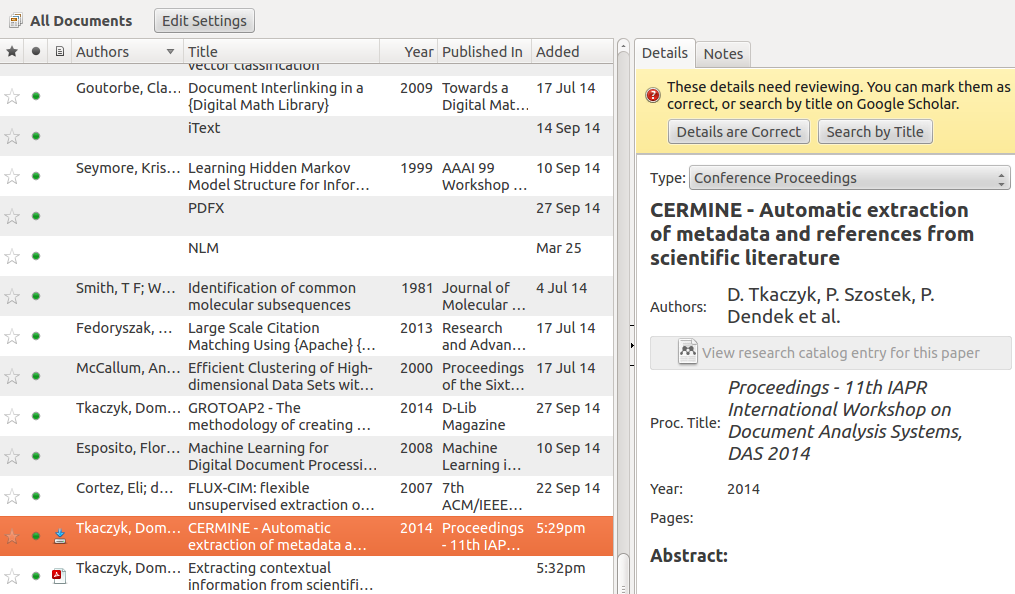}
  \caption[Intelligent metadata input interface from Mendeley]{An example of an intelligent metadata
  input interface from Mendeley. The system automatically extracts basic metadata from uploaded
  source document and asks the user to verify and correct the results, if needed.}
  \label{fig:mendeley}
\end{figure}

There are two complementary solutions to this problem. The easiest way to provide high quality 
metadata for scientific documents is to gather this information directly from the author when a
document is submitted to the system for the first time. Since we are interested in a wide range of
metadata, possibly including the metadata of all the references placed in the document and its full
text, inputting the metadata even for a single document can be tedious and time-consuming, and thus
error-prone. Therefore it would be very helpful to assist the user by providing the metadata
extracted from the document automatically, which can be then verified and corrected manually. Such
solutions result in a substantial time saving and much better metadata quality. An example of such 
an intelligent interface from Mendeley is shown in Figure~\ref{fig:mendeley}.

On the other hand, digital libraries already have to deal with huge number of existing documents 
with missing or fragmentary metadata records. Since processing this huge volume by human experts
would be extremely ineffective, we have to rely on automatic tools able to process large collections
and provide reliable metadata for the documents in an unsupervised manner. Unfortunately, already 
existing metadata extraction tools are not accurate, flexible or comprehensive enough.

\section{Problem Statement}
The main goal of our research is to solve the problem of missing metadata information by providing
an automatic, accurate and flexible algorithm for extracting wide range of metadata directly from
scientific articles.

Even limited to analysing scientific literature only, the problem of extracting the document's
metadata remains difficult and challenging, mainly due to the vast diversity of possible layouts and
styles used in articles. In different documents the same type of information can be displayed in
different places using a variety of formatting styles and fonts. For instance, a random subset of
125,000 documents from PubMed Central contains publications from nearly 500 different publishers,
many of which use original layouts and styles in their articles.

In general solving the metadata extraction problem requires addressing the two major tasks: the
analysis of the layout of the document, the difficulty of which varies with the input document 
formats, and understanding the roles played by all the fragments in the document.

The result of the research is an accurate automatic algorithm for extracting rich metadata directly
from a scientific publication. Proposed algorithm takes a single publication in PDF format on the
input, performs a thorough analysis of the document and outputs a structured machine-readable
metadata record containing:
\begin{itemize}
\item a rich set of document's basic metadata, such as title, abstract, keywords, authors' full
names, their affiliations and email addresses, journal name, volume, issue, year of publication,
etc.,
\item a list of references to other documents given in the article along with their metadata such as
the document's authors, title, journal name or year,
\item structured full text with sections and subsections hierarchy.
\end{itemize}

Designed as a universal solution, the algorithm is able to handle a vast variety of scientific 
articles reasonably well, instead of being perfect in processing a limited number of document
layouts only. We achieved this by employing supervised and unsupervised machine-learning algorithms
trained on large, diverse datasets. This decision made the method well-suited for analysing 
heterogeneous document collections, and also resulted in increased maintainability of the system, as
well as its ability to adapt to new, previously unseen document layouts.

Since our main objective was to provide a useful, accurate solution to a practical problem, machine
learning-based solutions are accompanied with a number of rules and heuristics. This approach proved
to work very well in practice, although perhaps lacks the simplicity and elegance of algorithms 
based purely on machine learning.

The evaluation we conducted showed good performance of the proposed metadata extraction algorithm.
The comparison to other similar systems also proved our algorithm performs better than competition
for most metadata types.

Proposed algorithm is very useful in the context of digital libraries for both automatic extraction
of reliable metadata from large heterogeneous document collections and assisting the users in the
process of providing metadata for the submitted documents.

The implementation of the algorithm is available as an open-source Java 
library\footnote{https://github.com/CeON/CERMINE} and a web 
service\footnote{http://cermine.ceon.pl}.

\section{Key Contributions}
The extraction algorithm we developed is based to a great extent on well-known supervised and
unsupervised machine-learning techniques accompanied with heuristics. Nevertheless, the research
contains the following innovatory ideas and extensions:
\begin{enumerate}
\item One of the key contributions is the architecture of the entire extraction workflow and the
decomposition of the problem into smaller, well-defined tasks.
\item The page segmentation algorithm was enhanced with a few modifications increasing its accuracy.
\item We developed a large set of numeric features for text fragments of the document capturing all
aspects of the content and appearance of the text and allowing to classify fragments with high
accuracy.
\item We also developed a set of features for citation and affiliation tokens, which allow to parse
affiliations and citations with high accuracy.
\item A clustering-based approach was proposed for extracting reference strings from the document.
\item We also proposed an algorithm based on normal scores of various statistics for selecting 
section header lines from the text content of the document.
\item Finally, we developed an efficient, scalable method of building gold standard publication
datasets.
\end{enumerate}

\section{Thesis Structure}
The thesis is structured as follows. In Chapter~\ref{chap:lit-review} we describe current state of
the art with respect to scientific document analysis and automatic metadata extraction. 
Chapter~\ref{chap:algorithm} provides all the details related to the overall algorithm architecture,
its internal decomposition into individual tasks and approaches employed for solving all of them. In
Chapter~\ref{chap:evaluation} we thoroughly describe the datasets and methodology used to assess the
quality of the algorithm and report the evaluation results, including the comparison with other 
similar systems. Chapter~\ref{chap:summary} summarizes the research. Appendix~\ref{chap:app-results}
provides the detailed results of the evaluation and all the tests performed, and finally
Appendix~\ref{chap:app-arch} covers the practical aspects of the available algorithm implementation.

\clearpage
\thispagestyle{empty}
\cleardoublepage

\chapter{State of the Art}
\label{chap:lit-review}
In this chapter we present the current state of the art with respect to extracting metadata and 
content from scientific literature. We start with a brief description of document formats used for
creating and storing academic articles (Section~\ref{sec:formats}). In Section~\ref{sec:ml} we
present machine learning techniques relevant to our work. Finally, in Section~\ref{sec:sota} we
discuss various algorithms and approaches to the problem of metadata extraction and list existing
systems and tools that can be used for analysing scientific articles.

\section{Metadata and Content Formats}
\label{sec:formats}
In this section we present a number of document formats useful for creating and storing academic
articles, focusing on the most popular ones. Described formats are optimized for many different 
purposes, and as a result they differ a lot in the type of information they are able to store and
the stage of the document's life they are mostly useful in. In the context of automatic document
analysis, the most important feature of a format is its machine-readability, which determines the
ability of extracting the information from the documents by automatic tools.

In general we deal with three types of formats:
\begin{itemize}
\item formats optimized for creating and editing the documents, such as MS Word formats or LaTeX,
\item formats optimized for presentation, mostly used for exchanging and storing the documents, but
not for manipulating them, such as PDF,
\item modern, machine-readable formats storing various aspects of documents, such as the content and 
physical and/or logical structure.
\end{itemize}

One of the most popular formats used for creating and editing documents are of course those related
to Microsoft Word\footnote{https://products.office.com/en-us/word}, a widely used word processor. 
Microsoft Word uses several file formats, and the default one varies with the version of the 
software.

In the 1990s and early 2000s the default format was 
.DOC\footnote{https://msdn.microsoft.com/en-us/library/office/cc313153\%28v=office.12\%29.aspx}. It
is a very complex binary format, where a document is in fact a hierarchical file system within a 
file. The format was optimized for the software performance during editing and viewing the files,
and not for machine understanding. For many years .DOC format specification was closed. Some
specifications for Microsoft Office 97 were first published in 1997 under a restrictive license, and
remained available until 1999. Since 2006 the specification was available under a restrictive 
license on request. In 2008 Microsoft released a .DOC format specification under the Microsoft Open 
Specification Promise. Unfortunately, due to the format complexity and missing descriptions of some
features, automatic analysis of .DOC files still requires some amount of reverse engineering.

Starting from Microsoft Office 2007, the default format is Office Open
XML\footnote{http://officeopenxml.com/}, which comprises formats for word processing documents,
spreadsheets and presentations as well as specific formats for mathematical formulae, graphics,
bibliographies etc. The format uses WordprocessingML as the markup language for word processing
documents. In comparison to .DOC, OOXML is much more machine-readable thanks to the usage of XML and
open specifications.

Another format used for creating and editing documents, popular especially in academia, is
LaTeX\footnote{https://www.latex-project.org/}. As opposed to Microsoft Word, writers using LaTeX
write in plain text and use markup tagging to define styles, the document structure, mathematical
formulae, citations, and so on. LaTeX uses the TeX typesetting program for formatting its output,
and is itself written in the TeX macro language. LaTeX documents can be processed by machines,
although it is often used as an intermediate format only.

Portable Document Format (PDF)~\cite{pdfref} is currently the most popular format for exchanging and
storing the documents, including the contents of scientific publications. The format is optimized
for presentation, PDF documents look the same no matter what application software, operating system
or hardware is used for creating or viewing them.

A PDF document is in fact a collection of objects that together specify the appearance of a list of
pages and their content. A single page contains a PDF content stream which is a sequence of text,
graphics and image objects printed on the page, along with all the information related to the
position and appearance of all the objects.

A text object in a PDF stream specifies the text to be painted on the page, as well as the font,
size, position, and other geometric features used to print the text. Listing~\ref{lst:pdftextobj}
shows an example text object, which results in writing a string "PDF" using 10-point font identified
by F13 font source (typically Helvetica), 360 typographic points from the bottom of the page and 288
typographic points from its left edge.

A text object can contain three types of operators:
\begin{itemize}
\item text state operators, used to set and modify text state parameters, such as character spacing,
word spacing, horizontal scaling, text font and text font size,
\item text positioning operators, which control the placement of chunks that are subsequently
painted, for example they can be used to move the current position to the next line with or without 
an offset,
\item text showing operators, used to paint the text accordingly to the current state and position
parameters.
\end{itemize}

\begin{lstlisting}[caption=Example PDF text object, label=lst:pdftextobj]
BT
  /F13 10 Tf
  288 360 Td
  ( PDF ) Tj
ET
\end{lstlisting}

Depending on the software and method used to create a PDF file, a single text-showing operator can
be used to print a single character, word, line, or any other chunk of continuous text without line
breaks. Spaces may be included in the text strings painted on the pages, or may be a result of
moving the current cursor to a different position. Some text decorations, such as underline or
strikethrough, can be produced using specialized fonts or printed independently on top of the text
as geometric objects.

What is more, PDF format does not preserve any information related to the logical structure of the
text, such as words, lines, paragraphs, enumerations, column layout, sections, section titles or
even the reading order of text chunks. This information has to be deduced from the geometric 
features of the text chunks. The text in a PDF file may be also present not in the form of text
operators, but as images of scanned pages. In such cases only optical character recognition can be
used to extract the text content from a file. All these issues make PDF format very difficult to
understand by machines.

Another format specifying the precise positions and the appearance of the text in a document is
TrueViz~\cite{LeeK03}, an XML-based, machine-readable format. TrueViz stores the geometric structure
of the document containing pages, zones, lines, words and characters, along with their positions,
dimensions, font information and the reading order.

Modern XML-based machine-readable formats can be used for storing both structured metadata and the
content of the documents, preserving various characteristics related to the appearance and meaning
of the text. For example NLM JATS\footnote{http://jats.nlm.nih.gov/} (Journal Article Tag Suite)
defines a rich set of XML elements and attributes for describing scientific publications. Documents
in JATS format can store a wide range of structured metadata of the document (title, authors,
affiliations, abstract, journal name, identifiers, etc.), the full text (the hierarchy of sections,
headers and paragraphs, structured tables, equations, etc.), the document's bibliography in the
form of a list of references along with their identifiers and metadata, and also the information
related to the text formatting.

Other similar XML-based formats are: the format developed by Text Encoding Initiative 
(TEI)\footnote{http://www.tei-c.org} which is semantic rather than presentational, and Dublin Core
Schema\footnote{http://dublincore.org/schemas/}, a small set of vocabulary terms that can be used to
describe documents.

In our algorithm we use three formats described above. PDF, as the most popular format for storing 
the documents in digital libraries, is the input format to the entire algorithm. TrueViz is used as
an intermediate format to serialize the geometric model of the input document inferred from the PDF 
file. The output format is NLM JATS, as a widely used machine-readable format able to store both the 
metadata of the document as well as structured full text in hierarchical form.

\section{Relevant Machine Learning Techniques}
\label{sec:ml}
This section describes briefly machine learning tasks and techniques related to extracting metadata
from documents. We focus mainly on the algorithms used in our work.

\subsection{General Classification}
Classification is one of the most useful technique in the context of extracting information from
documents. Classification can be used to determine the roles played in the document by its fragments
of various granularity.

Classification refers to the problem of assigning a category (a label from a known label set) to an 
instance. In supervised machine learning this is achieved by learning a model (a classification 
function) from a set of instances with known labels, called the training set, and applying the 
learned function to new instances with unknown labels. Instances are typically represented by
features of various types (binary, numerical, categorical).

There are many known classification algorithms, for example: linear classifiers (including LDA, 
naive Bayes, logistic regression, Support Vector Machines), which make classification decisions
based on a linear combination of instance features, k-Nearest Neighbors algorithm, in which the
decision is based on the labels of instances close to the input instance according to some metric,
or decision trees, which make a decision based on a sequence of "questions" related to the values of
individual features.

Our extraction algorithm makes extensive use of Support Vector Machines. 
SVM~\cite{BoserGV92,Vapnik98,Cristianini10} is a powerful classification technique able to handle a
large variety of input and work effectively even with training data of a small size. SVM is a binary
classifier (able to handle label sets containing exactly two elements) based on finding the optimal
separation hyperplane between the observations of two classes. It is little prone to overfitting,
generalizes well, does not require a lot of parameters and can deal with highly dimensional data.
SVM is widely used for content classification and achieves very good results in practice.

Let the classification instances be represented by vectors of real-valued features. Let us also 
assume that the label set contains exactly two elements. SVM is a linear model of the form

\[ \y(\mathbf{x}) = \mathbf{w}^T \phi(\mathbf{x}) + b \]

where
\begin{itemize}
\item $\mathbf{x}$ is a feature vector representing the classification instance,
\item $\phi(\mathbf{x})$ denotes a fixed feature-space transformation,
\item $\mathbf{w}$ and $b$ are parameters determined during the training based on the training 
instances,
\item new instances are classified according to the sign of $\y(\mathbf{x})$.
\end{itemize}

The training set contains $N$ vectors $\mathbf{x}_1, ..., \mathbf{x}_N$ with corresponding target
values (labels) $t_1, ..., t_N$, where $t_n \in \{-1,1\}$. If we assume the training set is 
linearly separable, then there exists at least one choice of the parameters $\mathbf{w}$ and $b$
such that the function $y$ satisfies $\y(\mathbf{x_n}) > 0$ for points having $t_n = 1$ and
similarly, $\y(\mathbf{x_n}) < 0$ for points having $t_n = -1$. In short, we have $t_n\y(\mathbf{x})
> 0$ for all training data points. There of course might exist many choices of the parameters
separating the classes entirely. The objective of the training phase is to find the parameters
resulting in the best separation.

In SVM we are interested in finding the parameters which maximize the margin in the training set,
which is the smallest distance between the decision boundary and any of the points of a given class.
Formally, the task of the training is to find

\begin{equation*}
\begin{aligned}
& \underset{\mathbf{w},b}{\text{arg max}}
& & \frac{1}{||\mathbf{w}||} \min_n [t_n (\mathbf{w}^T \phi(x_n) + b)]
\end{aligned}
\end{equation*}

Since the direct solution of this problem would be very complex, we often convert this into an
equivalent problem easier to solve by rescaling $\mathbf{w}$ and $b$, so that we have $t_n 
(\mathbf{w}^T \phi(x_n) + b) = 1$ for the point that is the closest to the decision boundary. For
all data points we then have $t_n (\mathbf{w}^T \phi(x_n) + b) \geq 1$ and the optimization problem
now becomes the equivalent 

\begin{equation*}
\begin{aligned}
& \underset{\mathbf{w},b}{\text{arg min}}
& & \frac{1}{2} ||\mathbf{w}||^2 \\
& \text{subject to}
& & t_n (\mathbf{w}^T \phi(x_n) + b) \geq 1
\end{aligned}
\end{equation*}

If the feature space $\phi(\mathbf{x})$ is not linearly separable, then there is no hyperplane
separating the training data points of two classes. To deal with this, we allow some data points to
be on the wrong side of the hyperplane, but we use a penalty which increases with the distance to
the decision boundary. We introduce slack variables, $\xi_n \geq 0$, one per training instance, and
the condition $t_n (\mathbf{w}^T \phi(x_n) + b) \geq 1$ becomes $t_n (\mathbf{w}^T \phi(x_n) + b)
\geq 1 - \xi_n$. Our new optimization problem now becomes the following:

\begin{equation*}
\begin{aligned}
& \underset{\mathbf{w},b,\xi}{\text{arg min}}
& & \frac{1}{2} ||\mathbf{w}||^2 + C \sum_{n=1}^N \xi_n \\
& \text{subject to}
& & t_n (\mathbf{w}^T \phi(x_n) + b) \geq 1 - \xi_n \\
& & & \xi_n \geq 0
\end{aligned}
\end{equation*}

where $C > 0$ is the regularization parameter (the penalty parameter of the error term).

For practical reasons, we usually operate not on $\phi(\mathbf{x})$ function directly, but on a
kernel function $\K(\mathbf{x}_i,\mathbf{x}_j) = \phi(\mathbf{x}_i)^T\phi(\mathbf{x}_j)$. The most
popular kernel functions are:
\begin{itemize}
\item linear: $\K(\mathbf{x}_i,\mathbf{x}_j) = \mathbf{x}_i^T\mathbf{x}_j$
\item polynomial: $\K(\mathbf{x}_i,\mathbf{x}_j) = (\gamma\mathbf{x}_i^T\mathbf{x}_j + r)^d$, $\gamma 
> 0$
\item radial basis function (RBF): $\K(\mathbf{x}_i,\mathbf{x}_j) = \exp(-\gamma||\mathbf{x}_i - 
\mathbf{x}_j||^2)$, $\gamma > 0$
\item sigmoid: $\K(\mathbf{x}_i,\mathbf{x}_j) = \tanh(\gamma\mathbf{x}_i^T\mathbf{x}_j + r)$
\end{itemize}

The kernel function, as well as its parameters $\gamma$, $r$ and/or $d$ are typically set prior to
the training. Usually some procedure is adopted in order to determine the best kernel function and
its parameters. One of the most widely used is a grid search, in which various combinations of
parameters are used to assess the classifier performance on a validation set and the parameters 
giving the best scores are chosen.

Since SVM is a binary classifier, usually we need a strategy of dealing with multiple target 
classes. Two most popular strategies are one-vs.-all and one-vs.-one. In one-vs.-all strategy we
train a single classifier per class, with the instances of the given class as positive samples and
all other samples as negatives. In one-against-one approach~\cite{Knerr90} we train a single 
classifier per each pair of classes using only the samples of these classes, resulting in $k(k - 
1)/2$ classifiers for $k$ classes. During the classification a voting strategy might be used. Early
works of applying this strategy to SVM-based classification include, for example~\cite{Kreel99}.

\subsection{Sequence Classification}
A special case of classification, sequence classification, is also often encountered in document
analysis domain. In sequence classification we are interested in analysing a sequence of instances
rather than independent instances.

More formally, the input is a sequence of instances and we are interested in finding a sequence of
corresponding class labels from a known label set. We would like to predict a vector $\mathbf{y} = 
{y_0, y_1, \dots, y_T}$ of class labels given an observed feature vector $\mathbf{x}$, which is
typically divided into feature vectors ${\mathbf{x}_0, \mathbf{x}_1, \dots, \mathbf{x}_T}$. Each
$\mathbf{x}_s$ contains various information about the instance at position $s$.

Sequence classification can be approached as any other classification problem by simply treating
sequence elements as independent classification instances, and the successor and/or predecessor 
relations might be reflected in the instances features. On the other hand, sequences can also be
seen as special cases of graphs and analysed with graphical modelling tools. In graphical modelling
we model probabilistically arbitrary graphs, which represent the conditional dependence structure 
between random variables (labels and features).

A lot of effort in learning with graphical models has focused on generative models that explicitly
model a joint probability distribution $\p(y, x)$ over both features and output labels and usually 
have the form $\p(y)\p(x|y)$. One very popular approach from this family are Hidden Markov Models
(HMM). 

HMM models a chain of variables, where every variable can be in a certain state (states correspond
to the labels) and emit observations (observations correspond to the features). In HMM we assume
that each state depends only on its immediate predecessor, and each observation variable depends
only on the current variable's state. The model comprises the initial probability distribution for
the state of the first variable in a sequence, the transition probability distribution from one
variable's state to the next variable and emission probability distributions. The classification is
performed with the use of Viterbi algorithm, which infers the most probable label sequence based on
observed features. 

Apart from generative models, another family of approaches are discriminative models, which instead
of modelling the joint probability focus only on the conditional distribution $\p(y|x)$. This
approach does not include modelling $\p(x)$, which is not needed for classification and often hard 
to model because it may contain many highly dependent features. Modelling the dependencies among
inputs can lead to complex and unmanageable models, but ignoring them can result in reduced
performance. Because dependencies that involve only variables in $x$ play no role in the 
discriminative models, they are better suited to including rich, overlapping features.

A very popular model in the discriminative family is Conditional Random Fields
(CRF)~\cite{LaffertyMP01}. CRF combines the advantages of classification and graphical modeling,
bringing together the ability to model multivariate, highly dependent data with the ability to
leverage a large number of input features for prediction. CRF can be seen as a discriminative
variant of HMM.

In general CRF can be used to model arbitrary graphs. A special case, a linear-chain CRF models
sequences of variables and is a distribution $\p(\mathbf{y}|\mathbf{x})$ that takes the form

\[ \p(\mathbf{y}|\mathbf{x}) = \frac{1}{\Z(\mathbf{x})} \exp \Big(\sum_{t=1}^T \sum_{k=1}^k
\lambda_k f_k(y_{t-1},y_t,\mathbf{x},t)\Big) \]

where
\begin{itemize}
\item $\{f_k(y,y',\mathbf{x},i)\}_{k=1}^K$ is a set of real-valued feature functions, which 
typically are based on two consecutive class labels and the entire observations sequence,
\item $\lambda = \{\lambda_k\} \in R^K$ is a vector of feature weights which are learned during the
training phase,
\item $\Z(\mathbf{x})$ is a normalization function to make $\p(\mathbf{y}|\mathbf{x})$ a valid
probability:

\[ \Z(\mathbf{x}) = \sum_{\mathbf{y}} \sum_{t=1}^T \sum_{k=1}^k \lambda_k f_k(y_{t-1},y_t,
\mathbf{x},t) \]

\end{itemize}

After we have trained the model, we can predict the labels of a new input $\mathbf{x}$ using the
most likely labeling by calculating

\[ \underset{y}{\text{arg max}} \quad \p(y|x) \]

In the case of a linear-chain CRF, finding the most probable label sequence can be performed
efficiently and exactly by variants of the standard dynamic programming algorithm for HMM, Viterbi
algorithm.

CRF is trained by a maximum likelihood estimation, that is, the parameters are chosen such that the
training data has the highest probability under the given model.

\subsection{Clustering}
Clustering is another very useful technique in document analysis. It can be employed whenever we
wish to group a set of objects into disjoint subsets called clusters, such that the objects in the
same cluster have similar characteristics. Two widely used clustering techniques are: hierarchical
clustering and k-means clustering.

Hierarchical clustering not only groups objects into clusters, but also results in a hierarchy of 
clusters. In a bottom-up approach each object starts as a single-element cluster, and the clusters
are iteratively merged accordingly to a certain strategy. In a top-down approach we start with a
single cluster containing the entire set and the clusters are then iteratively split. Both 
approaches result in a tree-like hierarchy, where the root is a cluster containing the entire set,
the leaves represent the individual elements and the remaining nodes are clusters of various
granularity.

In order to decide which clusters should be combined or where a cluster should be split we need a
measure of distance between sets of observations. This is typically based on a distance metric 
between individual points. Some commonly used metrics are:
\begin{itemize}
\item minimum distance between pairs of observations (single linkage clustering):
\[ \dd(A,B) = \min\{\dd(a,b) : a \in A, b \in B\} \]
\item maximum distance between pairs of observations (complete linkage clustering): 
\[ \dd(A,B) = \max\{\dd(a,b) : a \in A, b \in B\} \]
\item average distance between pairs of observations (average linkage clustering): 
\[ \dd(A,B) = \frac{1}{|A||B|} \sum_{a \in A}\sum_{b \in B} \dd(a,b) \]
\end{itemize}

In k-means clustering the number of target clusters has to be known in advance. The clusters are 
represented by the means of the observations and each data point belongs to the closest mean 
according to a given distance metric.

More formally, given a set of observations $(\mathbf{x}_1, \mathbf{x}_2, \dots, \mathbf{x}_n)$,
where each observation is a real vector, k-means clustering aims to partition the observations into
$k \leq n$ sets $S = {S_1, S_2, \dots, S_k}$ so that the within-cluster sum of squares is minimized.
In other words, its objective is to find:

\begin{equation*}
\begin{aligned}
& \underset{S}{\text{arg min}}
& & \sum_{i=1}^k \sum_{\mathbf{x} \in S_i} ||\mathbf{x} - \mathbf{\mu}_i||^2
\end{aligned}
\end{equation*}

where ${\mu}_i$ is the mean of vectors in $S_i$.

K-means algorithm works in iterations. At the beginning we choose $k$ vectors as the initial
centroids of the clusters. Then every point in the data set is assigned to the nearest cluster 
centroid and the centroids are recalculated as the means of their assigned points. This is repeated
until there are any changes in the points assignments. The algorithm converges to a local minimum,
but there is no guarantee a global minimum will be found. To obtain better results the algorithm can
be repeated several times with different initial centroids.

\section{Document Analysis}
\label{sec:sota}
In this section we describe the state of the art in the area of scientific literature analysis. The 
section covers a number of tasks related to the problem, including layout analysis and information
extraction.

Extracting metadata and content from scientific articles and other documents is a well-studied
problem. Older approaches expected scanned documents on the input and were prepared for executing
full digitization from bitmap images. Nowadays we have to deal with growing amount of born-digital
documents, which do not require individual character recognition.

Extracting information from documents is a complex problem and usually has to be divided into 
smaller subtasks. Typical tasks related to the extraction problem include:
\begin{itemize}
\item Preprocessing, which can be understood as parsing the input document and preparing a model of
it for further analysis. The difficulty depends heavily on the input format. In the case of scanned
documents, an OCR has to be performed. For PDFs the input text objects need to be parsed. Highly
machine-readable formats, such as NLM JATS or TrueViz are comparatively easy to process.
\item Page segmentation, in which we detect basic objects on the pages of the document, for example
text lines or blocks (zones). Similarly as before, depending on the format it might be sufficient to 
parse the input XML-based file, or a more complicated analysis of mutual positions of chunks or 
characters has to be performed.
\item Reading order resolving, in which we determine the order, in which all the text elements 
should be read. In western culture the text is usually read from the top to the bottom and from left
to right, but the resolver has to take into account the column layout, various decorations such as 
page numbers, headers or footers, text elements floating around images and tables, etc.
\item Region classification, in which we detect the roles played by different regions in the
document. The classification may be performed on the instances of various kinds (such as zones,
lines or text chunks, images, other graphical objects) and can be based on many different features
related to both text content and the way the objects are displayed on the document's pages.
\item Parsing, which assigns sequences of labels to text string. Parsing is typically used to detect 
metadata in shorter fragments of the text, such as citation or affiliation strings, author full
names, etc.
\end{itemize}

In the following subsections we discuss the state of the art in the following tasks: page 
segmentation (Section~\ref{sec:ssegm}), reading order resolving (Section~\ref{sec:rores}), document 
content classification  (Section~\ref{sec:class}), sequence parsing (Section~\ref{sec:pars}). 
Finally, in Section~\ref{sec:systems} we describe available systems and tools for processing 
scientific publications and extracting useful metadata and content from them.

\subsection{Page Segmentation}
\label{sec:ssegm}
Page segmentation refers to the task of detecting objects of various kinds in a document's pages. 
The objects we are interested in can be zones (regions separated geometrically from other parts,
such as blocks of text or images), text lines and/or words. Most approaches assume an image of the
page is present on the input, and require additional OCR phase, as well as noise removal. Some of
the algorithms can be adapted to analyzing born-digital documents, where the input is rather a bag
of characters or chunks appearing on the page.

One of the most famous and widely used page segmentation algorithm is XY-cut proposed by Nagy {\it 
et al.}~\cite{NagySV92}. XY-cut is a top-down algorithm which recursively divides the input page
into blocks. The result of the algorithm is a tree, in which the root represents the entire document
page and the leaf nodes are the final blocks. The tree is built from the top, by recursively 
dividing the current rectangular region into two rectangular parts by cutting it horizontally or
vertically. The place of a cut is determined by detecting valleys (empty horizontal or vertical 
space touching both up and down, or left and right fragment edge). By default the widest valley is
chosen as the cut and the entire process stops when there are no more valleys wider than a 
predefined threshold. XY-cut is a simple and efficient algorithm, though sensitive to the skew of
the page.

Run-length smearing algorithm (RLSA) proposed by Wong {\it et al.}~\cite{WongCW82} expects an image
of the page on the input and analyses the bitmap of white and black pixels. It is based on a simple
observation, that zones typically are dense and contain a lot of black pixels separated only by a
small number of white pixels. The first phase of the algorithm is called smearing. During smearing
the sequences of pixels (rows or columns of the bitmap) are analysed and black pixels separated by
only a low number (less than some predefined threshold) of white pixels are joined together by
transforming the separating white pixels into black ones. Smearing is performed vertically and
horizontally separately with different thresholds and the resulting bitmaps are then combined in a
logical AND operation. Then, one additional horizontal smearing is performed using a smaller 
threshold, which results in a smoothed final bitmap. Next, connected component analysis is performed
on the pixels to obtain document zones. Finally, each block's mean height and mean run-length of
black pixels is compared to the mean values calculated over all blocks on the page. Based on this
each block is classified into one of four classes: text, horizontal black line, vertical black line
or image.

The whitespace analysis algorithm proposed by Baird~\cite{Baird94} is based on analysing the
structure of the white background in document images. First, the algorithm finds a set of maximal
white rectangles called covers, the union of which completely covers the background. Then, the
covers are sorted using a sorting key based on the rectangle area combined with a weighting
function, which assigns higher weight to tall and long rectangles (as meaningful separators of text
blocks). Next, we gradually construct the union of the covers in the resulting order, covering more
and more of the white background. At each step, connected components within the remaining uncovered
parts are considered candidates for text blocks. This process stops at some point, determined by a
stopping rule, which results in the final segmentation. The stopping rule is defined as a predicate
function of two numerical properties of segmentations: the sorting key of the covers and the
fraction of the cover set used so far.

Breuel~\cite{breuel02} describes two geometric algorithms for solving layout analysis-related
problems: finding a set of maximal empty rectangles covering the background whitespace of a document
page image and finding constrained maximum likelihood matches of geometric text lines in the 
presence of obstacles. The combination of these algorithms can be used to find text lines in a
document in the following manner: after finding the background rectangles, they are evaluated as
candidates for column separators (called gutters or obstacles) based on their aspect ratio, width,
text columns width and proximity to text-sized connected components, and finally, the whitespace
rectangles representing the gutters are used as obstacles in a robust least square text-line
detection algorithm. This approach is not sensitive to font size, font style, or scan resolution.

The Docstrum algorithm proposed by O'Gorman~\cite{OGorman93} is a bottom-up page segmentation 
approach based on the analysis of the nearest-neighbor pairs of connected components extracted from
the document image. After noise removal, $K$ nearest neighbors are found for each connected
component. Then, the histograms of the distances and angles between nearest-neighbor pairs are
constructed. The peak of the angle histogram gives the dominant skew (the text line orientation
angle) in the document image. This skew estimate is used to compute within-line nearest neighbor
pairs. Next, text lines are found by computing the transitive closure on within-line nearest
neighbor pairs using a threshold. Finally, text-lines are merged to form text blocks using a
parallel distance threshold and a perpendicular distance threshold. The algorithm uses a significant
number of threshold values and behaves the best if they are tuned for a particular document
collection.

The Voronoi-diagram based segmentation algorithm proposed by Kise {\it et al.}~\cite{KiseSI98} is
also a bottom-up approach. It is based on a generalization of Voronoi diagram called area Voronoi
diagram, where the regions are generated by a set of non-overlapping figures of any shape rather
than individual points, and the distance between a point and a figure is defined as a minimal
distance between the point and any point belonging to a figure. At the beginning the algorithm
computes the connected components and samples the points from the boundaries of them. Then, an area
Voronoi diagram is generated using sample points obtained from the borders of the connected 
components. The Voronoi edges that pass through any connected component are deleted to obtain an
area Voronoi diagram. Finally, superfluous Voronoi edges are deleted to obtain boundaries of
document components. An edge is considered superfluous if the minimum distance between its
associated connected components is small, the area ratio of the two connected components is above a
certain threshold or at least one of its terminals is neither shared by another Voronoi edge nor
lies on the edge of the document image. The algorithm works well even for non-Manhattan layouts and
is not sensitive to line skew or text orientation.

The above six algorithms were evaluated and compared by Shafait {\it et al.}~\cite{ShafaitKB08}. 
They propose a pixel-accurate representation of a document's page along with several performance
measures to identify and analyze different classes of segmentation errors made by page segmentation
algorithms. The algorithms were evaluated using a well-known University of Washington III (UW-III)
database~\cite{Guyon97}, which consists of 1,600 English document images with Manhattan layouts
scanned from different archival journals with manually edited ground-truth of entity bounding boxes,
including text and non-text zones, text lines and words. On average, Docstrum along with the
Voronoi-based algorithm achieved the lowest error rates in most categories. Docstrum is also the
only algorithm, which by default detects both text lines and zones.

In addition, International Conference on Document Analysis and Recognition (ICDAR) hosted a number
of page segmentation competitions starting from 2001. The last competition for tools and systems of
general segmentation purpose took place in 2009~\cite{AntonacopoulosPBP09}. Its aim was to evaluate
new and existing page segmentation methods using a realistic dataset and objective performance
measures. The dataset used comprised both technical articles and magazine pages and was selected
from the expanded PRImA dataset~\cite{AntonacopoulosBPP09}.

In 2009 four systems were submitted to the competition. DICE (Document Image Content Extraction)
system is based on classifying individual pixels into machine-print text, handwriting text and 
photograph~\cite{BairdMAC07}, followed by a post-classification methodology~\cite{AnBX07} which 
enforces local uniformity without imposing a restricted class of region shapes. The Fraunhofer 
Newspaper Segmenter system is based on white~\cite{breuel02} and black~\cite{ZhengLDP01} separator 
detection followed by a hybrid approach for page segmentation~\cite{JainY98}. The REGIM-ENIS method 
is designed primarily for degraded multi-script multi-lingual complex official documents, containing 
also tabular structures, logos, stamps, handwritten text and images. Finally, 
Tesseract~\cite{Smith09}, an extension of the Tesseract OCR system, in which the page layout 
analysis method uses bottom-up methods, including binary morphology and connected component 
analysis, to estimate the type (text, image, separator, or unknown) of connected components.

According to the competition results, the Fraunhofer Newspaper Segmenter method performed the best,
improving on both the state-of-the-art methods (ABBYY FineReader and OCRopus) and the best methods
of the ICDAR2007 page segmentation competition.

The segmentation algorithms can be adapted to process born-digital documents, although in some cases
it might be non-trivial. In the case of born-digital document we often deal with characters, their
dimensions and positions, but we lack the pixel-accurate representation of the pages, and thus the
algorithms analysing individual pixels, such as RLSA or Voronoi would require preprocessing.

\subsection{Reading Order Resolving}
\label{sec:rores}
Another task related to document layout analysis is reading order resolving. Reading order resolving 
aims at determining the order, in which all the elements on a given page should be read. One might
be interested in the order of elements of various types, such as zones, lines, or words. An accurate
solution has to take into account many different aspects of the document layout, such as: column 
layout, presence of images and other text fragments not belonging to the main text flow, various
language scripts, etc.

XY-cut algorithm, described in the previous section, can be naturally extended to output extracted
zones in their natural reading order. In XY-cut algorithm every cut divides the current page 
fragment into two blocks positioned left-right or up-down to each other. If we assume that the text
should be read from left to right and from top to bottom, then it is enough to always assign left
or top part to the left child in the constructed tree, and similarly, right and bottom part should
be always assigned to the right child. After the tree is complete, an in-order leaves traversal 
gives the resulting reading order of the extracted zones.

There are, however, a few serious problems with this approach. First of all, the algorithm is not
able to process non-manhattan layouts (such as pages containing L-shaped zones). This is not a big 
problem in the case of scientific publications, since most of them use manhattan layout. There is
also an issue of choosing the right threshold for the minimum width of the valley, which might vary
from one document to another. The most problematic is the issue of choosing the best cut when there
are a number of possible valleys to choose from. The default decision in XY-cut is to cut the
regions along the widest valley, which works well for page segmentation, but often results in
incorrect reading order. For example a multi-column page might get cut horizontally in the middle
dividing all the columns, because the gaps between paragraphs or sections happen to occur in one
horizontal line creating a valley wider than the gap between the columns.

Various approaches addressing these issues have been proposed. For example Ishitani {\it et 
al.}~\cite{Ishitani03} describe a bottom up approach, in which some objects on the page are merged
prior to applying XY-cut, using three heuristics based on local geometric features, text orientation
and distance among vertically adjacent layout objects. As observed by Meunier~\cite{Meunier05}, this
aims at reducing the probability of dealing with multiple cutting alternatives, but it does not
entirely prevent them from occurring. Meunier proposes a different approach, in which an optimal
sequence of XY cuts is determined using dynamic programming and a score function, which prefers
column-based reading order. This results in a cutting strategy which favors vertical cuts against
horizontals ones, based on the heights of blocks.

Another example of a reading order algorithm is the approach proposed by Breuel~\cite{Breuel03}. It
is based on topological sorting and can be used for determining the reading order of text lines. At
the beginning, four simple rules are used to determine the order between a subset of line pairs on a
page, giving a partial order. The rules are based on mutual coordinate positions and overlap, for 
example: line segment $a$ comes before line segment $b$ if their ranges of x-coordinates overlap and
if line segment $a$ is above line segment $b$ on the page. Finally, a topological sorting algorithm
is applied to find a global order consistent with previously determined partial order.

As opposed to previous methods, Aiello {\it et al.}~\cite{AielloMT02} propose to employ linguistic 
features in addition to the geometric hints. In their approach the input is first divided into two
parts: metadata and body. The reading order is determined separately for these subsets, and finally
the two orders are combined using rules. Each reading order is determined in two steps performed by
the following modules: a spatial reasoning module, based on spatial relations, and a natural 
language processing module, based on lexical analysis. The modules are applied in order: first, the
spatial reasoning module identifies a number of possible reading orders by solving a 
constraint-satisfaction problem, where constraints correspond to rules such as "documents are 
usually read from top to bottom and from left to right". The natural language processing module
identifies the linguistically most probable reading orders among those returned by the first module
using part-of-speech tagging and assessing the probabilities of POS sequences of possible reading
orders.

Finally, Malerba {\it et al.}~\cite{MalerbaCB08} propose a learning-based method for reading order
detection. In their approach the domain specific knowledge required for this task is automatically
acquired from a set of training examples by applying logic programming techniques. The input of the
learning algorithm is the description of chains of layout components defined by the user, and the
output is a logical theory which defines two predicates: "first to read" and "successor". The
algorithm uses only spatial information of the page elements. In the recognition phase learned rules
are used to reconstruct the reading order, which in this case contains reading chains and may not
define a total ordering.

\subsection{Content Classification}
\label{sec:class}
Content classification, the purpose of which is to find the roles played by different objects in the
document, is a crucial task in analysing the documents. The problem has been addressed by numerous
researchers. Proposed solutions differ a lot in the approach used (usually rule- or machine
learning-based), classified objects (zones, lines or text chunks), used features and characteristics
(geometrical, formatting, textual, etc.) and target labels. Some examples of the target labels
include: {\it title}, {\it authors}, {\it affiliation}, {\it address}, {\it email}, {\it abstract},
{\it keywords}, but also {\it header}, {\it footer}, {\it page number}, {\it body text}, {\it 
citation}.

Rule-based systems were more popular among the older algorithms. Such approach does not require 
building a training set and performing the training, but since the rules are usually constructed 
manually, it does not generalize well and is not easily maintainable. Rule-based approaches are 
well-suited for homogeneous and stable document sets, with only few different documents layouts.

An example of a rule-based classification is PDFX system described by Constantin {\it et 
al.}~\cite{ConstantinPV13}. In this approach page elements are converted to geometric and textual
features and hand-made rules are used to label them. The target label set contains: front matter
labels ({\it title}, {\it abstract}, {\it author}, {\it author footnote}), body matter labels ({\it 
body text}, {\it h1 title}, {\it h2 title}, {\it h3 title}, {\it image}, {\it table}, {\it figure
caption}, {\it table caption}, {\it figure reference}, {\it table reference}, {\it bibliographic
item}, {\it citation}) and other ({\it header}, {\it footer}, {\it side note}, {\it page number},
{\it email}, {\it URI}).

Flynn {\it et al.}~\cite{FlynnZMZZ07} describe a system which also can be seen as a variant of the 
rule-based approach. Their algorithm uses a set of templates associated with document layouts. A
template can be understood as a set of rules for labelling page elements. A document is first 
assigned to a group of documents of similar layout and then corresponding template is used to assign
labels to elements. The target label set depends on the layout, some examples include: {\it title}, 
{\it author}, {\it date}.

Also in the algorithm proposed by Giuffrida {\it et al.}~\cite{GiuffridaSY00} hand-made rules are 
used to label text chunks. In this approach, text strings annotated with spatial and visual 
properties, such as positions, page number and font metrics, are used as "facts" in a knowledge
base. Basic document metadata, including title, authors, affiliations, relations author-affiliation,
is extracted by a set of hand-made rules that reason upon those facts. Some examples of rules
include: "the title is written in the top half of the first page with the biggest font", or
"authors' list follows the title immediately".

Mao {\it et al.}~\cite{MaoKT04} also propose a rule-based system for extracting basic metadata,
including the title, authors, affiliations and abstract, from scanned medical journals. The system
is used for MEDLINE database. Its iterative process includes human intervention, which corrects the 
zone labelling obtained from the previous rules. Corrected results are then used to develop 
specialized geometric and contextual features and new rules from a set of issues of each journal.

Rule-based approaches are especially popular for locating the regions containing references. This is
related to the fact that fairly common and clear differences between these sections and other 
document parts can be easily expressed by hand-made rules and heuristics.

For example Pdf-extract\footnote{http://labs.crossref.org/pdfextract/} system uses a combination of
visual cues and content traits to detect references sections. The rules are to a great extent based
on the observation that the reference section of a scientific document tends to have a significantly
higher ratio of proper names, initials, years and punctuation with comparison to other regions.

In the approach proposed by Gao {\it et al.}~\cite{GaoTL09} a rule-based method is used to locate
citation regions in electronic books. The rules are based on the percentage of text lines in the
page containing certain Chinese words such as "reference", "bibliography", years and family names.

Also in the system described by Gupta {\it et al.}~\cite{GuptaMCS09} the reference blocks are found
by estimating the probability that each paragraphs belongs to references using parameters based on
paragraph length, presence of keywords, author names, years and other text clues.

Kern and Klampfl~\cite{KernK13} also propose a heuristics-based approach for locating the references
sections. Their algorithm first iterates over all blocks in the reading order and uses regular
expressions and a dictionary of references section titles to find the reference headers. Then all
the lines are collected until another section heading, for example "Acknowledgement", 
"Autobiographical", "Table", "Appendix", "Exhibit", "Annex", "Fig", "Notes", or the end of the 
document, is found. Headers and footers lines are recognized based on the comparison of the blocks 
across neighboring pages based on their content and geometrical position and they are not added to
the references content.

Supervised machine learning-based approaches are far more popular for classifying the document
fragments. They are more flexible and generalize better, in particular when we have to deal with
diverse document collections. Proposed methods differ in classification algorithms, document
fragments that undergo the classification (text chunks, lines or blocks) and extracted features. 
Some examples of the classification algorithms used for this task include: Hidden Markov Models,
Support Vector Machines, neural classifiers, Maximum Entropy, Conditional Random Fields.

For example, Cui and Chen~\cite{CuiC10} propose a classification approach in which text blocks 
(small pieces of text, often smaller than one logical line) are classified with an HMM classifier
using features based on location and the font information. The target labels include: {\it title},
{\it author}, {\it affiliation}, {\it address}, {\it email} and {\it abstract}. A straightforward
HMM-based approach would just label the stream of text blocks, but the authors modified it to take
into account the structure of the lines containing the classified blocks. Based on the location of
the text chunks, the HMM state transition matrix is divided into two separate matrices: one for the
state transition probability within the same line and the other for the state transition probability
between lines. A modified Viterbi algorithm uses these new matrices to find the most probable label
sequence.

Han {\it et al.}~\cite{HanGMZZF03} perform a two-stage classification of document header text lines
with the use of Support Vector Machines and only text-related features. They use a rich set of 
labels: {\it title}, {\it author}, {\it affiliation}, {\it address}, {\it note}, {\it email}, {\it
date}, {\it abstract}, {\it introduction}, {\it phone}, {\it keyword}, {\it web}, {\it degree}, {\it
pubnum} and {\it page}. In the first step the lines are classified independently of each other using
features related to text and dictionaries. The second step makes use of the sequential information
among lines by extending the feature vectors with the classes of a number of preceding and following
lines. Iteratively a new classifier is trained using extended feature vectors and lines are
reclassified, until the process converges (there are only few changes in the class assignments).

Another example of an SVM-based approach is described by Kovacevic {\it et 
al.}~\cite{Kovacevic2011}. In their method the lines of text on the first page of documents are 
classified into the following classes: {\it title}, {\it authors}, {\it affiliation}, {\it address},
{\it email}, {\it abstract}, {\it keywords} and {\it publication note} using both geometric
(formatting, positions) and text-related (lexical, NER) features. The authors experimented with
different models (decision trees, Naive Bayes, k-Nearest Neighbours and Support Vector Machines) and
different strategies for multi-class classification. Based on the results obtained during the
classification experiments, an SVM model with a one-vs.-all strategy was chosen, as giving the best
performance on a manually produced test set.

Lu {\it et al.}~\cite{LuKWG08} also use SVM to classify the lines of the text in scanned scientific 
journals. They use the following classes: {\it title}, {\it author}, {\it volume}, {\it issue}, {\it
start page}, {\it end page}, {\it start page index} and {\it start page image} and geometric,
formatting and textual features of the text lines. The approach is tested on scanned historical
documents nearly two centuries old. 

Marinai~\cite{Marinai09} proposes a Multi-Layer Perceptron (MLP) classifier to identify regions that
could contain the title and the authors of the paper by classifying the text blocks. The features 
include: graphical features (related to the position on the page, the width and height of the 
region, which page it is on), textual features (the number of characters, bold or italics 
characters), and neighbor features (such as the number of neighboring regions and their distance).

Team-Beam algorithm proposed by Kern {\it et al.}~\cite{KernJHG12} uses an enhanced Maximum Entropy
classifier for assigning labels to document fragments. The approach works in two stages: first the
blocks are classified as {\it title}, {\it subtitle}, {\it journal}, {\it abstract}, {\it author},
{\it email}, {\it affiliation}, {\it author-mixed} or {\it other}, and then the tokens within blocks
related to author metadata are classified as {\it given name}, {\it middle name}, {\it surname}, 
{\it index}, {\it separator}, {\it email}, {\it affiliation-start}, {\it affiliation}, {\it other}.
The algorithm uses and enhanced version of Maximum Entropy classifier, which takes the 
classification decision of preceding instances into account to improve the performance and to 
eliminate unlikely label sequences. The features used for classification are derived from the 
layout, the formatting, the words within and around a text block, and common name lists.

In the approach proposed by Lopez~\cite{Lopez09} the regions of the document are classified using 11
different CRF models cooperating together at various levels of a document's structure. Each
specialized model aims at solving a concrete classification task. The main model classifies the
fragments of the entire document into {\it header}, {\it body}, {\it references}, etc. Other models
are used for classifying the header fragments, parsing affiliation, author and dates strings,
classifying body parts into titles, paragraphs and figures, parsing references and so on. Each model
has its own set of features and training sets. The features are based on position-, lexical- and
layout-related information.

Cuong {\it et al.}~\cite{CuongCKL15} also use CRF for labelling the fragments of the documents. In
their approach the input is a document in plain text, and therefore they do not use geometric hints
present for example in the PDF files. They describe methods for solving three tasks: reference 
parsing (where the reference tokens are labelled as {\it title}, {\it author}, {\it year}, etc.), 
section labelling (where the sequence of document's sections are given the functional label, such 
{\it abstract}, {\it acknowledgement}, {\it background}, {\it categories}, {\it conclusions}, {\it 
discussions}, {\it evaluation}, {\it general terms}, {\it introduction}, {\it methodology}, {\it 
references}, {\it related works}) and finally assigning labels such as {\it author} and {\it 
affiliation} to the lines of the document's header. The instances are classified using a higher
order semi-Markov Conditional Random Fields to model long-distance label sequences, improving upon
the performance of the linear-chain CRF model.

Finally, Zou {\it et al.}~\cite{ZouLT10} propose a binary SVM classifier for locating the references
sections in the document. The text zones are classified using both geometric and textual features.

\subsection{Sequence Parsing}
\label{sec:pars}
Parsing refers to extracting metadata from strings by annotating their fragments with labels from a
particular label set. In the context of scientific documents analysis parsing can be used for 
example to extract metadata such as title, authors, source or date from citation strings, dividing
authors' fullnames into given names and surnames, recognizing days, months and years in date strings
or extracting institution name, address and country name from an affiliation string.

Similarly as in the case of content classification, there are two widely used families of approaches
used for parsing. One popular family of methods is based on regular expressions or knowledge bases.
The advantage of these techniques is that they usually can be implemented in a straightforward 
manner without gathering any training data or performing the training.

For example Gupta {\it et al.}~\cite{GuptaMCS09} propose a simple regexp-based approach for
classifying fragments of citation strings as particular metadata classes: {\it authors}, {\it 
title}, {\it publication} and {\it year of publication}. The regular expressions are hand-made and
the algorithm is also enhanced by using a publication database of a domain of interest (zoology) to
lookup the title in case the default approach failed.

Jonnalagadda and Topham~\cite{Jonnalagadda11} describe NEMO system, which is able to parse 
affiliations using rules and a number of dictionaries. The parsing includes extracting fragments
related to {\it country}, {\it email address}, {\it URL}, {\it state}, {\it city}, {\it street
address} and {\it organization name}. The fragments are extracted in consecutive steps using 30
different manually verified dictionaries, such as the dictionaries in Geoworldmap 
database\footnote{http://www.geobytes.com/geoworldmap/}, the mapping between internet domains and
countries, stop words list, organization-related keywords, address-related keywords or zip code
dictionary.

Day {\it et al.}~\cite{DayTSHLWWOH07} propose a knowledge-based approach for parsing citations in
order to extract the following metadata: {\it author}, {\it title}, {\it journal}, {\it volume}, 
{\it issue}, {\it year}, and {\it page}. The method is based on a hierarchical knowledge 
representation framework called INFOMAP. First the data from the Journal Citation Reports (JCR)
indexed by the ISI and digital libraries is collected and fed into the knowledge base. The format of
INFOMAP is a tree-like knowledge representation scheme that organizes knowledge of reference
concepts in a hierarchical fashion, which contains characteristic patterns occurring in citations.
To extract metadata from a citation, the template matching engine uses dynamic programming to match
it with the syntax templates.

Vilarinho {\it et al.}~\cite{VilarinhoSGMM07} also propose a knowledge-based approach for citation 
parsing. In their method, the knowledge base stores typical words for each citation metadata type,
which is then used to assign labels to the citation tokens. After that, tokens left unassociated in
the previous step are further analyzed and labels are assigned to them based on rules related to
their neighbourhood and relative position in the citation string.

Unfortunately, similarly as in the case of general classification, rule-based approaches are poorly
adaptable and do not generalize well. For this reason, machine learning-based approaches, which are
much more flexible, are far more popular for sequence parsing. These methods typically leverage the
sequence-related information in addition to the tokens themselves, either by decoding them in the
features or using dedicated algorithms for sequence labelling.

For example Zhang {\it et al.}~\cite{ZhangZLT11} propose SVM for classifying the reference's tokens 
into the following classes: {\it citation number}, {\it author names}, {\it article title}, {\it 
journal title}, {\it volume}, {\it pagination} and {\it publication year}. Their method uses
structural SVM~\cite{TsochantaridisHJA04}, an extension of SVM designed for predicting complex
structured outputs, such as sequences, trees and graphs. The features are related to dictionaries of
author names, article titles and journal titles, patterns for name initials or years, the presence
of digits and letters and the position of the token. Additionally, two kinds of contextual features
are used: the features of the neighboring tokens and the labels assigned to those tokens.

Hetzner~\cite{Hetzner08} proposes to parse citation strings using HMM in order to extract: {\it 
author}, {\it booktitle}, {\it date}, {\it editor}, {\it institution}, {\it journal title}, {\it 
location}, {\it note}, {\it number}, {\it pages}, {\it publisher}, {\it techtitle}, {\it title} and
{\it volume}. The model includes two HMM states per each metadata class: a "first" state for the
first token in the subsequence and a "rest" state, along with a set of separator states (which
represent words and punctuation that are not part of metadata fields) for every metadata class pair,
and a terminating "end" state. The tokens are mapped to a small alphabet of emission symbols, which
is composed of symbols representing punctuation, particular words, classes of words and word 
features.

Yin {\it et al.}~\cite{YinZDY04} propose to parse citations using a bigram HMM, where emission 
symbols are token words. Different from the traditional HMM, which typically uses word frequency, 
this model also considers the words' bigram sequential relation and position information in text
fields. In particular, a modified model is used for computing the emission probability, while 
keeping the structure of HMM unchanged. In the new model, the probability of emitting symbol at
given state composes of beginning emission probability (the probability that the state emits word as
the first word) and inner emission probability (the probability that the state emits the word as the
inner word).

Ojokoh {\it et al.}~\cite{OjokohZT11} propose even more advanced approach based on a trigram HMM,
where a state of a current token depends on the states of two preceding tokens, instead of one. A
modified Viterbi algorithm is used to infer the most probable sequence of token labels. Only 20
symbols are used for the emission alphabet, these are based on specific characters (for example a
comma, a dot, a hyphen), regular expressions (for example checking whether the token is a number),
also a dictionary of state names and a list of common words found in specific metadata fields.

Definitely the most popular technique for citation parsing is linear-chain CRF. In practice, it
achieves better results than HMM and is more flexible as able to handle a lot of overlapping 
features of the tokens, whereas in HMM the tokens have to be mapped to a dictionary of emission
symbols.

ParsCit described by Councill {\it et al.}~\cite{CouncillGK08} is an open-source library for 
citation parsing based on CRF. The labels assigned to the citation tokens include: {\it author},
{\it booktitle}, {\it date}, {\it editor}, {\it institution}, {\it journal}, {\it location}, {\it 
note}, {\it pages}, {\it publisher}, {\it tech}, {\it title} and {\it volume}. The features are
related to the token identity, punctuation, numbers, letters, cases, dictionaries (for example
dictionaries of publisher names, place names, surnames, female and male names, and months).

Also Gao {\it et al.}~\cite{GaoTL09} use CRF to parse citations in Chinese electronic books in order
to extract: {\it author}, {\it editor}, {\it title}, {\it publisher}, {\it date}, {\it page number},
{\it issue}, {\it volume}, {\it journal}, {\it conference}, {\it book}, {\it note}, {\it location}
and {\it URL}. The parsing is supported by a knowledge base storing the most common words in
citation strings, the punctuation marks used to separate fields, Chinese family names, English 
names, publishing houses in China, journal names, conference names, places and dates, and so on.
Apart from textual features, layout-related features are also used. Finally, the tool takes 
advantage of document layout consistency to enhance citation parsing through clustering techniques.
The main citation format used in the book is detected and used to correct minor mistakes occurred 
during parsing.

Kern and Klampfl~\cite{KernK13} also propose a citation parsing algorithm based on CRF. The model 
uses the following labels of the tokens: {\it author given name}, {\it author surname}, {\it author
other}, {\it editor}, {\it title}, {\it date}, {\it publisher}, {\it issue}, {\it book}, {\it 
pages}, {\it location}, {\it conference}, {\it source}, {\it volume}, {\it edition}, {\it issue},
{\it url}, {\it note}, and {\it other}. In order to integrate sequence-related information, the 
algorithm takes the classification decision of four preceding instances into account. In addition to 
the typical text-related features, the model also incorporates layout and formatting information
using a set of binary features specifying whether the font of the tokens inside a sliding window
from -2 to +2 tokens is equal to the font of the current token.

Another example of a CRF-based approach is the citation parser proposed by Zhang {\it et
al.}~\cite{ZhangCY11}. The algorithm extracts: {\it author} (further separated into surname and
given name), {\it title}, {\it source} (for example journal, conference, or other source of
publication), {\it volume}, {\it pages} (further separated into first page and last page), and {\it 
year}. The features are based on traits such as whether the token contains a capital letter, all
capital letters, a digit, all digits, and other symbols (such as Roman and Greek characters, 
hyphens, etc.), as well as the length of the token.

In Enlil system described by Do {\it et al.}~\cite{DoCCK13} linear-chain CRF classifier is used to
parse both author names and affiliation strings in order to recognize the {\it name} (author or
organization name), {\it symbol} (characters marking the relations between authors and affiliations)
and {\it separator}. The features are both text- (such as token identity, punctuation, number) and
layout-related (fonts, subscript, superscript).

CRF is also used extensively in GROBID system described by Lopez~\cite{Lopez09} to parse various
entities, for example citations, affiliations, author names or date strings. Each task has its own
CRF model, training set and a set of features, which are based on position, lexical and layout
information.

Cuong {\it et al.}~\cite{CuongCKL15} is another example of a CRF-based citation parser. In their
approach the citation tokens are labelled as: {\it author}, {\it booktitle}, {\it date}, {\it 
editor}, {\it institution}, {\it journal}, {\it location}, {\it note}, {\it pages}, {\it publisher},
{\it tech}, {\it title} or {\it volume}. The tokens are classified using a higher order semi-Markov
Conditional Random Fields to model long-distance label sequences, improving upon the performance of
the linear-chain CRF.

Finally, Zou {\it et al.}~\cite{ZouLT10} compared two algorithms for citation parsing. One relies on
sequence statistics and trains a CRF. The other focuses on local feature statistics and trains an 
SVM to classify each individual word, which is followed by a search algorithm that systematically
corrects low confidence labels if the label sequence violates a set of predefined rules. The 
approaches achieved very similar high accuracies.

\subsection{Extraction Systems}
\label{sec:systems}
This section describes tools and systems able to extract various types of metadata and content from
scientific literature. The approaches differ in the scope of extracted information, methods used,
input and output formats, availability and licenses.

Typically at the beginning of the document processing some kind of layout analysis is performed, and
then the regions of the document are classified using various algorithms. The metadata extracted 
from documents usually contains the title, authors, affiliations, emails, abstract, keywords, and so
on. These fragments are usually located in the document using rules or machine learning. Extracting
bibliography-related information typically includes locating the references sections in the document
using rules or machine learning, splitting their content into individual references and parsing 
them. The analysis of the middle part of the document might require locating the paragraphs, tables,
figures, section titles, sometimes determining the hierarchy of sections or the roles of the 
sections as well.

For example Flynn {\it et al.}~\cite{FlynnZMZZ07} propose a metadata extraction approach which can
be seen as a variant of a rule-based approach. First input PDF documents are OCRed using a 
commercial tool ScanSoft's OmniPage Pro, which results in a XML-based representation containing the
layout and the text organized in pages, regions, paragraphs, lines and words, accompanied by
information such as font face, font size and font style, alignment and spacings. The metadata is 
then extracted using independent templates, each of which is a set of simple rules associated with a
particular document layout. Processed document is first assigned to a group of documents of similar
layout, and then the corresponding template is used to extract the document's metadata.

Mao {\it et al.}~\cite{MaoKT04} propose a rule-based system for extracting {\it title}, {\it 
author}, {\it affiliation}, and {\it abstract} from scanned medical journals. The system is used
for MEDLINE database. The documents are first OCRed, and then undergo an iterative process which 
includes human intervention for correction the zone labelling resulting from applied rules.
Corrected results are then used to develop geometric and contextual features and rules optimized 
for the set of issues of a given journal.

Hu {\it et al.}~\cite{HuLCMZ05} describe a machine learning-based approach for extracting titles 
from general documents, including presentations, book chapters, technical papers, brochures, reports
and letters. As a case study, Word and PowerPoint document are used. During pre-processing, the 
units (text chunks with uniform format) are extracted from the top region of the first page of a
document. These units are then transformed to features and classified as {\it title\_begin}, {\it 
title\_end} or {\it other}. Two types of features were used: format features (font size, alignment,
boldface, the presence of blank lines) and linguistic features (keywords specific for titles and
other document parts, number of words). Four models were employed (Maximum Entropy Model, Perceptron
with Uneven Margins, Maximum Entropy Markov Model, and Voted Perceptron). The authors have observed
that Perceptron-based models perform better in terms of extraction accuracy.

Cui and Chen~\cite{CuiC10} describe a system for extracting {\it title}, {\it author}, {\it 
affiliation}, {\it address}, {\it email} and {\it abstract} from PDF documents. In this approach, 
text extraction and page segmentation is done with the use of {\tt pdftohtml}, a third-party 
open-source tool. The resulting HTML document contains a set of text blocks, which are small pieces
of text, often less than one logical line, along with their location and font information. These 
blocks are labelled with the target metadata classes with the use of an enhanced HMM classifier.

Han {\it et al.}~\cite{HanGMZZF03} extract metadata ({\it title}, {\it author}, {\it affiliation},
{\it address}, {\it note}, {\it email}, {\it date}, {\it abstract}, {\it introduction}, {\it phone},
{\it keyword}, {\it web}, {\it degree} and {\it page}) from the headers of scientific papers in 
plain text format. The metadata is extracted by classifying the text lines with the use of a 
two-stage SVM classification based on text-related features.

Another example of an SVM-based approach is the metadata extractor used in CRIS systems described by
Kovacevic {\it et al.}~\cite{Kovacevic2011}. In this approach PDF articles are first converted to
HTML, which preserves the formatting and layout-related information. Then, the lines of text in the
first page of the document are classified using both geometric and text-related features. Extracted
metadata contains: {\it title}, {\it authors}, {\it affiliation}, {\it address}, {\it email}, {\it 
abstract}, {\it keywords} and {\it publication note}.

Lu {\it et al.}~\cite{LuKWG08} analyse scanned scientific journals in order to obtain volume 
metadata (such as {\it name}, {\it number}), issue metadata ({\it volume number}, {\it issue 
number}, etc.) and article metadata ({\it title}, {\it authors}, {\it volume}, {\it issue} and {\it
pages range}). In their approach scanned pages are first converted to text using OCR techniques.
Then, rule-based pattern matching on the feature vectors of the text lines is used to recognize and
analyze volume and issue title pages, while article metadata is extracted using SVM and geometric,
formatting, distance, layout and textual features of text lines. The approach is tested on scanned
historical documents nearly two centuries old.

Marinai~\cite{Marinai09} first extracts characters from PDF documents using JPedal package, which 
results in a set of basic objects on each page accompanied with additional information such as their
position and font size. Then, the blocks are merged in the horizontal and vertical directions, 
avoiding to join separate columns or paragraphs, using simple rule-based heuristics. Each region is
then transformed to feature vectors and a Multi-Layer Perceptron (MLP) classifier is used to 
identify regions that could contain the title and the authors of the paper. The classifier uses 
features related to both the layout and the text. Additionally, the information gathered from DBLP
citation database is used to assist the tool by checking the extracted metadata.

Enlil, described by Do {\it et al.}~\cite{DoCCK13}, is a tool able to extract authors, affiliations
and relations between them from scientific publications in PDF format. In this approach a PDF file
is first OCRed with the use of OmniPage, which results in an XML version of the document that stores
both the textual and spatial information for each word that appears on each page. The system is 
built on top of SectLabel module from ParsCit~\cite{LuongNK10}, which is used to detect authors and
affiliations blocks in the text by classifying text lines using CRF. The lines classified as author
or affiliation are then tokenized into chunks and the tokens are labelled using a linear-chain CRF
classifier with the following classes: {\it name}, {\it symbol} and {\it separator}. The model uses 
both text- and layout-related features. Finally, a binary SVM classifier is applied to 
author-affiliation pairs to extract the relations between them. The features used in this model are
related to the information provided by the parsing module and the distances between the author and
the affiliation fragments.

In the citation extraction system described by Gupta {\it et al.}~\cite{GuptaMCS09} the documents
are first scanned, OCRed and converted into PDF format. The PDF documents are then converted into
HTML using Abby PDF Reader. Then reference block is found by estimating the probability that each
paragraph belongs to references using parameters based on paragraph length, presence of keywords, 
the author names, the presence of the year and other text clues. Regular expressions are then used 
to extract metadata from citation strings. The algorithm also uses external publication database to 
correct the extraction results.

Zou {\it et al.}~\cite{ZouLT10} propose a two-step process using statistical machine learning
algorithms for extracting bibliography data from medical articles in HTML format. The algorithm 
first locates the references with a binary SVM classifier using geometric and text features for text 
zones. For reference parsing two algorithms were used: CRF and SVM followed by a search algorithm
that systematically corrects low confidence labels if the label sequence violates a set of 
predefined rules.

Gao {\it et al.}~\cite{GaoTLLQW11} describe CEBBIP, a tool able to extract the chapter and section
hierarchy from Chinese books. The overall approach is based on the observation that within a book,
some features related to formatting and fonts are shared among elements of the same type, such as
the headings, footnotes or citations. At the beginning the tool performs page layout analysis by
merging small page objects (eg. characters, words, lines) into bigger ones in a bottom-up manner
using position and font-related heuristics. Then the global typographies characteristics, such as
columns, header and footer, page body area, text line directions, line spacing of body text, and 
fonts used in various components (headings, paragraphs, etc.) are extracted. For example, to detect
headers and footers, similarities of the text and geometric position between the top/bottom lines on
neighboring pages is exploited. Columns are identified by detecting the recurring white spaces in
multiple pages. Then the page objects are clustered based on general typesseting, and the output
clusters serve as the prototypes of similar blocks. After that, the system uses a learning-based
classification method to label the blocks in each cluster as headings, figure/table captions,
footnotes. The table of contents hierarchy is extracted from the "Table of contents" section with
the use of heuristics and associated with the headings extracted from the text.

CEBBIP is also able to extract the bibliographic data~\cite{GaoTL09}. In this approach a rule-based
method is used to locate citation data in a book and the data is segmented into citation strings of
individual references with the use of heuristics based on the citation markers and spaces. A
learning-based approach (CRF) is employed to parse citation strings. Finally, the tool takes 
advantage of document layout consistency to enhance citation data segmentation and parsing through
clustering techniques. The main citation format used in the book is detected and used to correct the
parsing results.

Giuffrida {\it et al.}~\cite{GiuffridaSY00} extract the content from PostScript files using a tool
based on {\tt pstotext}, which results in a set of strings of text annotated with spatial/visual
properties, such as positions, page number and font metrics. These strings become "facts" in a 
knowledge base. Basic document metadata, including title, authors, affiliations, relations 
author-affiliation, is extracted by a set of hand-made rules that reason upon those facts.

The same system also uses rules to extract the titles of the sections from the text of the document.
The algorithm first determines whether the section titles are numbered. If the sections are 
numbered, various schemes of numbering are examined. If the sections are not numbered, heuristics
based on the text size and line space are used. Additionally the algorithm looks for titles commonly
appearing in the documents, such as "Introduction", "Overview", "Motivation" or "References" to find
hints for the font size typical for the titles in a given document.

PDFX, described by Constantin {\it et al.}~\cite{ConstantinPV13}, is a rule-based system able to 
extract basic metadata, structured full text and unparsed reference strings from scientific 
publications. PDFX can be used for converting scholarly articles in PDF format to their XML 
representation by annotating fragments of the input documents. The analysis comprises two main 
stages. During the first one the geometric model of the article's content is constructed to 
determine the organization of textual and graphical units on every page using a library from the 
Utopia Documents PDF reader. The model comprises pages, words and bitmap images, along with their
features such as bounding box, orientation, textual content or font information. During the second
stage, different logical units are identified by rules based on their discriminative features. The
following block types are used: {\it title}, {\it author}, {\it abstract}, {\it author footnote},
{\it reference}, {\it body}, {\it (sub)section}, {\it (sub)section heading}, {\it figure}, {\it 
table}, {\it caption}, {\it figure/table reference}. PDFX is a closed source system, available only
as a web service\footnote{http://pdfx.cs.man.ac.uk/}.

Pdf-extract\footnote{http://labs.crossref.org/pdfextract/} is an open-source tool for identifying
and extracting semantically significant regions of scholarly articles in PDF format. Pdf-extract can 
be used to extract the title and a list of unparsed bibliographic references of the document. The
tool uses a combination of visual cues and content traits to perform structural analysis in order to
determine columns, headers, footers and sections, detect references sections and finally extract
individual references. Locating the references section is based on the observation, that it tends to
have a significantly higher ratio of proper names, initials, years and punctuation than other 
sections. The reference section is divided into individual references also based on heuristics.

Team-Beam algorithm proposed by Kern {\it et al.}~\cite{KernJHG12} is able to extract a basic set of
metadata (title, subtitle, the name of the journal, conference or venue, abstract, the names of the
authors, the e-mail addresses of the authors and their affiliations) from PDF documents using an
enhanced Maximum Entropy classifier. In their approach first the PDF file is processed by PDFBox 
library\footnote{https://pdfbox.apache.org/}. Then, clustering techniques are used in order to build
words, lines and text blocks from the output obtained from PDFBox library. The structure is built
from the bottom, and each type is built by two steps: merge (done by hierarchical clustering) and
split (k-means clustering for splitting incorrectly merged objects). Then, reading order is 
determined using the approach described in~\cite{AielloMT02}. Next, a machine learning approach is
employed to extract the metadata in two stages: first the blocks of text are classified into 
metadata types and then the tokens within blocks related to author metadata are classified in order
to extract given names, middle names, surnames, etc.

Team-Beam also contains a bibliography extraction component described by Kern and 
Klampfl~\cite{KernK13}. In this approach, the references section is located using heuristics related
to a list of typical section titles. The individual references are extracted based on a simple
version of k-means clustering algorithm applied to the text lines. The only feature used is the
minimal x-coordinate of a line's bounding box. The algorithm clusters the line set into two subsets,
representing the first lines of the references and the rest. Finally, the references are parsed 
using a CRF token classifier.

Team-Beam provides also the functionality of extracting the body text of the article along with the 
table of contents hierarchy~\cite{KlampflGJK14} based on an unsupervised method. After performing 
the segmentation and detecting the reading order, the text blocks are categorized by a sequential
pipeline of detectors, each of which labels a specific type of block: decorations, such as page
numbers, headers, and footers (done by analysing the similarity between blocks on the neighbouring
pages), figure and table captions (based on heuristics), main text (hierarchical clustering applied
to blocks based on alignment, font, and width-related features), section headings (heuristics based
on fonts, distances and regular expressions), sparse blocks and tables (again heuristics). Each of
these detectors is completely model-free and unsupervised.

Lopez~\cite{Lopez09} describes GROBID, a system useful for analysing scientific publication in PDF
format. GROBID uses pdf2xml/Xpdf for processing PDF files and CRF in order to extract a rich set of 
document's metadata, full text with section titles, paragraphs and figures, and a list of parsed
bibliographic references with their metadata. The extraction is performed by a cascade of 11 
specialized CRF models, which start by labelling the parts of the document as header, body, 
references, etc., and then focuses on each part. There are separate models specializing in 
classifying the header parts, parsing affiliations, author data, dates, selecting the header title
lines, paragraphs, figures, extracting individual references and parsing them. Each model has its
own set of features and training data. The features are based on position, lexical and layout
information. The system is available as open-source\footnote{https://github.com/kermitt2/grobid}.

ParsCit system, described by Councill {\it et al.}~\cite{CouncillGK08}, is an open-source system for
extracting parsed references from plain text files. Reference sections are identified using 
heuristics, which search for particular section titles such as "References" or "Bibliography". If
such a label is found too early in the document, the process continues. The references sections are
then split into individual strings also using heuristics. Regular expressions are used for detecting 
characteristic markers indicating the beginning of the reference, and if no such markers are found,
the system splits the lines based on their length and other indicators related to whether the line
ends with a dot and whether it contains tokens characteristic for author names. Reference parsing is
realized by a CRF model labelling token sequence in the reference string. The tokens' features are
related to punctuation, numbers, letters, cases and dictionaries (such as publisher names, place
names, surnames, female and male names, months).

ParsHed, described by Cuong {\it et al.}~\cite{CuongCKL15}, is a ParsCit's module able to extract
the basic metadata (title, authors, abstract, etc.) from the document's plain text. The extraction 
is done by classifying the header's text lines using higher order semi-Markov Conditional Random
Fields, which model long-distance label sequences, improving upon the performance of the 
linear-chain CRF model.

SectLabel, described by Luong {\it et al.}~\cite{LuongNK10} is also a ParsCit's module, useful for
extracting a logical structure of a document, both in PDF or plain text formats. In this approach, 
PDFs are first converted to XML using Nuance OmniPage, and the output XML includes the coordinates
of paragraphs, lines and words within a page, alignment, font size, font face or format. SectLabel
performs then two tasks: logical structure classification and generic section classification. For
both tasks CRF with lexical and formatting features is used. In the first task the ordered sequence
of document's text lines is labeled with the following categories: {\it address}, {\it affiliation}, 
{\it author}, {\it bodyText}, {\it categories}, {\it construct}, {\it copyright}, {\it email}, {\it  
equation}, {\it figure}, {\it figureCaption}, {\it footnote}, {\it keywords}, {\it listItem}, {\it 
note}, {\it page}, {\it reference}, {\it sectionHeader}, {\it subsectionHeader}, {\it 
subsubsectionHeader}, {\it table}, {\it tableCaption} and {\it title}. In the second task, the 
sequence of the headers of the sections is labelled with a class denoting the purpose of the
section, including: {\it abstract}, {\it categories}, {\it general terms}, {\it keywords}, {\it 
introduction}, {\it background}, {\it related  work}, {\it methodology}, {\it evaluation}, {\it 
discussion}, {\it conclusions}, {\it acknowledgments}, and {\it references}. ParsCit is also an open
source system\footnote{http://aye.comp.nus.edu.sg/parsCit/}.

Unfortunately, as of now, there has not been any competition for evaluating the tools extracting
rich metadata from scientific publications. Semantic Publishing Challenge\footnote{http://2015.eswc-conferences.org/program/semwebeval}\textsuperscript{,}\footnote{https://github.com/ceurws/lod/wiki/SemPub2015}, 
hosted in 2015 by European Semantic Web Conference\footnote{http://2015.eswc-conferences.org/},
contained some tasks related to analysing scientific articles. Our algorithm~\cite{TkaczykB15} was
the winner in Task 2, which included the problems of extracting authors, affiliations and citations
from PDF documents~\cite{DiIorioLDV15}.

The extraction systems available online, namely GROBID, ParsCit, PDFX and Pdf-extract, are the most
similar to our work in terms of potential applications in the context of digital libraries and 
research infrastructures. Unfortunately, they all have major drawbacks. Pdf-extract is rule-based
and focuses almost only on extracting the references without parsing them. PDFX is also rule-based
and closed-source, available only as a web service. The open-source version of ParsCit processes
only plain text documents, ignoring the layout, and also does not output structured document
metadata record, but rather the annotated input text along with the confidence scores. Finally,
GROBID does not extract the section hierarchy and uses the same machine-learning method for all
tasks, without taking into account the specifics of particular problems.

Our research brings together various advantages of the previous works, resulting in one accurate
solution. The algorithm is comprehensive and analyses the entire document, including the document's
header, body and references. It processes born-digital PDF documents and focuses not only on the
textual features, but also on the layout and appearance of the text, which carries a lot of valuable
hints for classification. Extensive use of machine learning-based algorithms makes the system 
well-suited for heterogeneous document collections, increases its flexibility and the ability to
adapt to new, previously unseen document layouts. Careful decomposition of the extraction problem
allows for treating each task independently of others, with each solution selected and tuned for the
task's specific needs. The system is open-source and available as a Java 
library\footnote{https://github.com/CeON/CERMINE} and a web 
service\footnote{http://cermine.ceon.pl}. In Section~\ref{chap:evaluation} we report the results of 
the comparison of our method to four similar systems available online: GROBID, ParsCit, PDFX and 
Pdf-extract.

\clearpage
\thispagestyle{empty}
\cleardoublepage

\chapter{Document Content Extraction}
\label{chap:algorithm}
In this chapter we describe the algorithm for extracting basic metadata, structured body content and
the bibliography from academic articles in born-digital form. The chapter provides the details 
related to the algorithm's architecture, all the individual tasks it performs during the analysis of
a document, and the approaches employed for solving them.

The extraction algorithm~\cite{TkaczykBCR12,TkaczykSDFB14,TkaczykSFDB15,TkaczykTB15} is based on a 
modular workflow composed of a number of paths and steps. The purpose of each step is to address a
single task with clearly defined input and output. Thanks to such modular architecture individual
steps can be maintained separately, which allows for evaluation, testing, adjustment and replacement
of individual components without touching other parts of the workflow.

The algorithm is designed as a universal solution, able to handle a vast variety of scientific
publications reasonably well, instead of being perfect in processing a limited number of document
layouts only. This was achieved by employing supervised and unsupervised machine-learning algorithms
trained on large and diverse datasets. The result is also increased maintainability of the system,
as well as its ability to adapt to new, previously unseen document layouts.

The chapter is organized as follows. Section~\ref{sec:overview} describes the overall architecture
of the extraction algorithm, as well as its input and output. The remaining sections contain
detailed descriptions of all algorithm parts, including: layout analysis (Section~\ref{sec:layout}),
document content classification (Section~\ref{sec:classification}), metadata extraction 
(Section~\ref{sec:metadata}), bibliography extraction (Section~\ref{sec:bibliography}) and
structured body extraction (Section~\ref{sec:body}).

\section{Algorithm Overview}
\label{sec:overview}
The extraction algorithm accepts a single scientific publication on the input, inspects the entire
content of the document and outputs a single record containing: the document's metadata information,
parsed bibliography and structured body content.

The input document format is PDF~\cite{pdfref}. The algorithm is optimized for processing 
born-digital documents and does not perform optical character recognition. As a result, PDF 
documents containing scanned pages in the form of images are not properly processed.

The output format is NLM JATS\footnote{http://jats.nlm.nih.gov/}. The output contains all the
information the algorithm is able to extract from a given document structured into three parts: {\it 
front} (the document's metadata), {\it body} (the middle part of the document, its proper text in a
structured form containing the hierarchy of sections) and {\it back} (the bibliography section).

\begin{figure}
  \centering
  \includegraphics[width=0.6\textwidth]{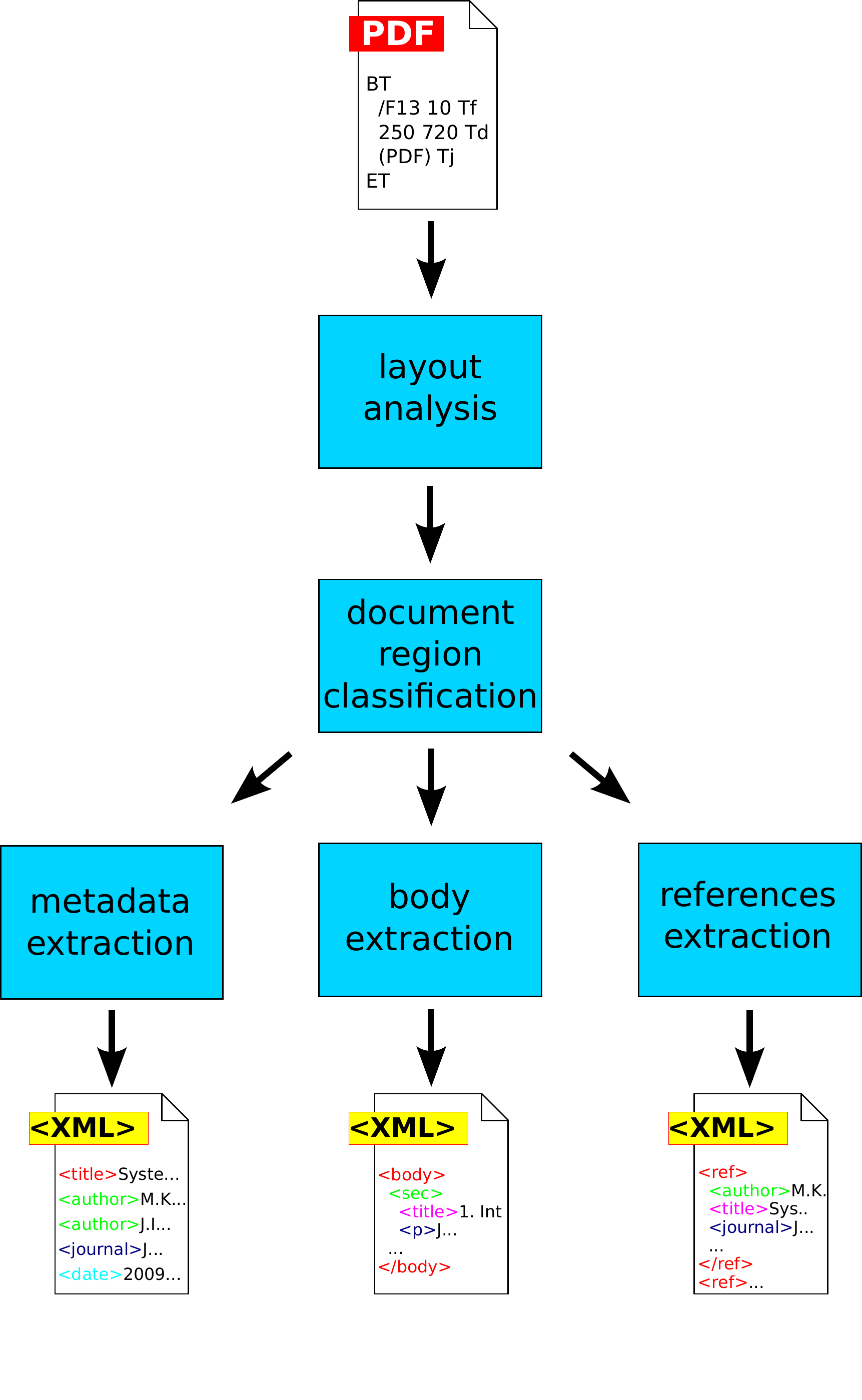}
  \caption[The architecture of the extraction algorithm]{The architecture of the extraction
  algorithm. The algorithm accepts a single PDF file on the input. The first stage of the process is
  layout analysis, followed by assigning functional classes to all the text fragments of the
  document during document region classification. Finally, in three parallel stages the following
  information is extracted: document's metadata, structured body content and parsed bibliographic
  references.}
  \label{fig:overview}
\end{figure}

The algorithm is composed of the following stages (Figure~\ref{fig:overview}):
\begin{itemize}
\item Layout analysis (Section~\ref{sec:layout}) --- The initial part of the entire algorithm. 
During layout analysis the input PDF file is analysed in order to extract all text fragments and 
their geometric characteristics.
\item Document region classification (Section~\ref{sec:classification}) --- The goal of the
classification is to assign a single label to each text fragment of the document. The labels denote
the function a given fragment plays in the document.
\item Metadata extraction (Section~\ref{sec:metadata}) --- In this stage structured metadata is
extracted from previously labelled document.
\item Bibliography extraction (Section~\ref{sec:bibliography}) --- The purpose of this stage is to
extract parsed bibliography in a structured form from previously labelled document.
\item Body extraction (Section~\ref{sec:body}) --- The goal of this stage is to extract full text
and section hierarchy from labelled document.
\end{itemize}

\section{Document Layout Analysis}
\label{sec:layout}
Document layout analysis is the initial phase of the extraction algorithm. Its goal is to detect all
the text fragments in the input PDF document, compute their geometric characteristics and produce a
geometric hierarchical model of the document.

The input is a single file in PDF format and the output is a geometric hierarchical model of the
document. The model holds the entire text content of the article, while also preserving the 
information related to various aspects of the way elements are displayed in the input PDF file.

Intuitively, the output model represents the document as a list of pages, each page contains a list
of text zones (blocks), each zone contains a list of lines, each line contains a list of words, and
finally each word contains a list of characters. Each element in this hierarchical structure can be
described by its text content and the position on its page. The order of the elements in the lists
corresponds to the natural reading order of the text, that is the order, in which the fragments 
should be read. In this tree structure every text element belongs to exactly one element of higher
level.

A single page of a given document is a rectangle-shaped area, where the text elements are placed.
The position of any point $p$ on the page is defined by two coordinates: $p_X$ (the horizontal
distance to the left edge of the page) and $p_Y$ (the vertical distance to the top edge of the 
page). The origin of the coordinate system is the left upper corner of the page, and the coordinates
are given in typographic points (1 typographic point equals to 1/72 of an inch). The positions of
all the text elements are defined with respect to this coordinate system.

The model stores text elements of various granularity: characters, words, lines and zones. Every 
text element belongs to exactly one document page and represents a fragment of a text written on the
page. The position of the element on its page is defined by two points: left upper and right lower
corner of its bounding box, which is a rectangle with edges parallel to the page's edges enclosing a
given text element. 

Formally the levels in the model can be defined in terms of sets. For any set $S$ let's denote its
partition as $\PP(S)$. In other words, $\PP(S)$ is any set meeting the following conditions:
\begin{itemize}
\item $\forall_{p \in \PP(S)} \quad p \subset S$
\item $\emptyset \notin \PP(S)$
\item $\bigcup_{p \in \PP(S)} p = S$
\item $\forall_{p,r \in \PP(S)} \quad (p \neq r \enskip \Rightarrow \enskip p \cap r = \emptyset)$
\end{itemize}

For a given document $D$ let's define the following sets:
\begin{itemize}
\item Characters. Let $C_D$ be the set of all characters visible in the document. For every 
character $c \in C_D$ we define its text $\T(c) \in \Sigma_D$, where $\Sigma_D = \{'a', 'b', ..., 
'A', 'B', ...\}$ is the alphabet used within document $D$, and its bounding box given by two points:
left upper corner $\B_1(c) \in R^2$ and right lower corner $\B_2(c) \in R^2$.
\item Words. Let $W_D = \PP(C_D)$ be the set of all words in the document. Intuitively, a word is a
continuous sequence of characters placed in one line with no spaces between them. Punctuation marks
and typographical symbols can be separate words or parts of adjacent words, depending on the 
presence of spaces. 
\item Lines. Let $L_D = \PP(W_D)$ be the set of all lines in the document. Intuitively, a line is a
sequence of words that forms a consistent fragment of the document's text. Words placed 
geometrically in the same line of the page, which are parts of neighbouring columns, should not
belong to the same line in the model. Hyphenated words that are divided into two lines should 
appear in the model as two separate words that belong to different lines.
\item Zones. Let $Z_D = \PP(L_D)$ be the set of all zones in the document. Intuitively, A zone is a
consistent fragment of the document's text, geometrically separated from surrounding fragments and
not divided into columns.
\item Pages. Finally, let $P_D = \PP(Z_D)$ be the set of all pages in the document.
\end{itemize}

We can also define a parent function, which for any character, word, line or zone returns the
element's parent in the structure:

\[ \prt : C_D \cup W_D \cup L_D \cup Z_D \to W_D \cup L_D \cup Z_D \cup P_D \]

such that
\begin{itemize}
\item $\forall_{c \in C_D} \quad (\prt(c) \in W_D \enskip \wedge \enskip c \in \prt(c))$
\item $\forall_{w \in W_D} \quad (\prt(w) \in L_D \enskip \wedge \enskip w \in \prt(w))$
\item $\forall_{l \in L_D} \quad (\prt(l) \in Z_D \enskip \wedge \enskip l \in \prt(l))$
\item $\forall_{z \in Z_D} \quad (\prt(z) \in P_D \enskip \wedge \enskip z \in \prt(z))$
\end{itemize}

The sets $C_D$, $W_D$, $L_D$, $Z_D$ and $P_D$ are totally ordered sets. The order corresponds to the
natural reading order of the elements in the document, that is the order in which the text should
be read. The order of the elements respects the set hierarchy, in particular

\[ \forall_{S \in \{C_D, W_D, L_D, Z_D\}} \quad \forall_{e_1, e_2 \in S} \quad (e_1 \leq e_2 
\enskip \Rightarrow \enskip \prt(e_1) \leq \prt(e_2)) \]

For every word, line and zone we also define a bounding box as a minimal rectangle enclosing all
contained elements:
\begin{itemize}
\item $\forall_{w \in W_D} \quad (\B_1(w) = \B_{min}(w) \enskip \wedge \enskip \B_2(w) = \B_{max}
(w))$
\item $\forall_{l \in L_D} \quad (\B_1(l) = \B_{min}(l) \enskip \wedge \enskip \B_2(l) = \B_{max}
(l))$
\item $\forall_{z \in Z_D} \quad (\B_1(z) = \B_{min}(z) \enskip \wedge \enskip \B_2(z) = \B_{max}
(z))$
\end{itemize}

where
\begin{itemize}
\item $\B_{min}(x) = (\min\{\B_{1,X}(y) \enskip | \enskip y \in x\}, \min\{\B_{1,Y}(y) \enskip | 
\enskip y \in x\})$
\item $\B_{max}(x) = (\max\{\B_{2,X}(y) \enskip | \enskip y \in x\}, \max\{\B_{2,Y}(y) \enskip | 
\enskip y \in x\})$
\end{itemize}

The model of a document described in this section is built incrementally by three steps executed in
a sequence: character extraction (Section~\ref{sec:charextr}), page segmentation 
(Section~\ref{sec:segm}) and reading order resolving (Section~\ref{sec:rores}). Each steps updates
the structure with new information. Table~\ref{tab:layout} summarizes the basic information about
the steps.

\begin{table}[h]
\renewcommand{\arraystretch}{1.4}
\renewcommand{\tabcolsep}{3pt}
\centering
  \begin{tabular}{ | m{155px} | m{185px} | m{65px} | }
	\hline
    \multicolumn{1}{|c|}{Step} & \multicolumn{1}{c|}{Goal} & \multicolumn{1}{c|}{Implementation}
    \\ \hline\hline
    
    1. {\bf Character extraction} & Extracting individual characters along with their page
    coordinates and dimensions from the input PDF file. & {\bf iText \mbox{library}} \\ \hline
     
	2. {\bf Page segmentation} & Constructing the document's geometric hierarchical structure
	 containing (from the top level) pages, zones, lines, words and characters, along with their
	 page coordinates and dimensions. & {\bf enhanced Docstrum}
	 \\\hline

	3. {\bf Reading order resolving} & Determining the reading order for all structure elements. &
	 {\bf bottom-up heuristics}\\\hline

  \end{tabular}
\caption[The decomposition of layout analysis]{The decomposition of layout analysis stage into 
individual steps.}
\label{tab:layout}
\end{table}

\subsection{Character Extraction}
\label{sec:charextr}
Character extraction is the first step of the entire extraction process. Its purpose is to parse the
input PDF file and build initial simple geometric model of the input document, which stores only the
pages and individual characters.

Let $D$ be the given input document. The purpose of character extraction is to:
\begin{itemize}
\item determine $P_D$ --- a set of document's pages along with their order,
\item determine $C_D$ --- a set of characters visible in the document,
\item assign characters to pages, that is find a function $\anc_{cp}: C_D \to P_D$, which for given
character $c \in C_D$ returns the page the character is displayed on.
\end{itemize}

Character extraction does not find other elements of the model, and does not determine the order of
the characters. The output of character extraction is a list of pages, each of which contains a set
of characters.

The implementation of character extraction is based on open-source PDF parsing library 
iText\footnote{http://itextpdf.com/}. The document's pages and their order are explicitly given in
the source of the input file. To extract characters, we iterate over all text-related PDF operators,
keeping track of the current text state and text positioning parameters. During the iteration we
extract text strings from text-showing operators along with their bounding boxes. The strings are
then split into individual characters and their individual widths and positions are calculated.
Finally, all the coordinates are translated from the PDF coordinate system to the system used in our
geometric model. The mapping between characters and pages is determined directly by the position of
the text-showing operators in the input PDF file.

\begin{figure}[h]
  \centering
  \includegraphics[width=0.6\textwidth]{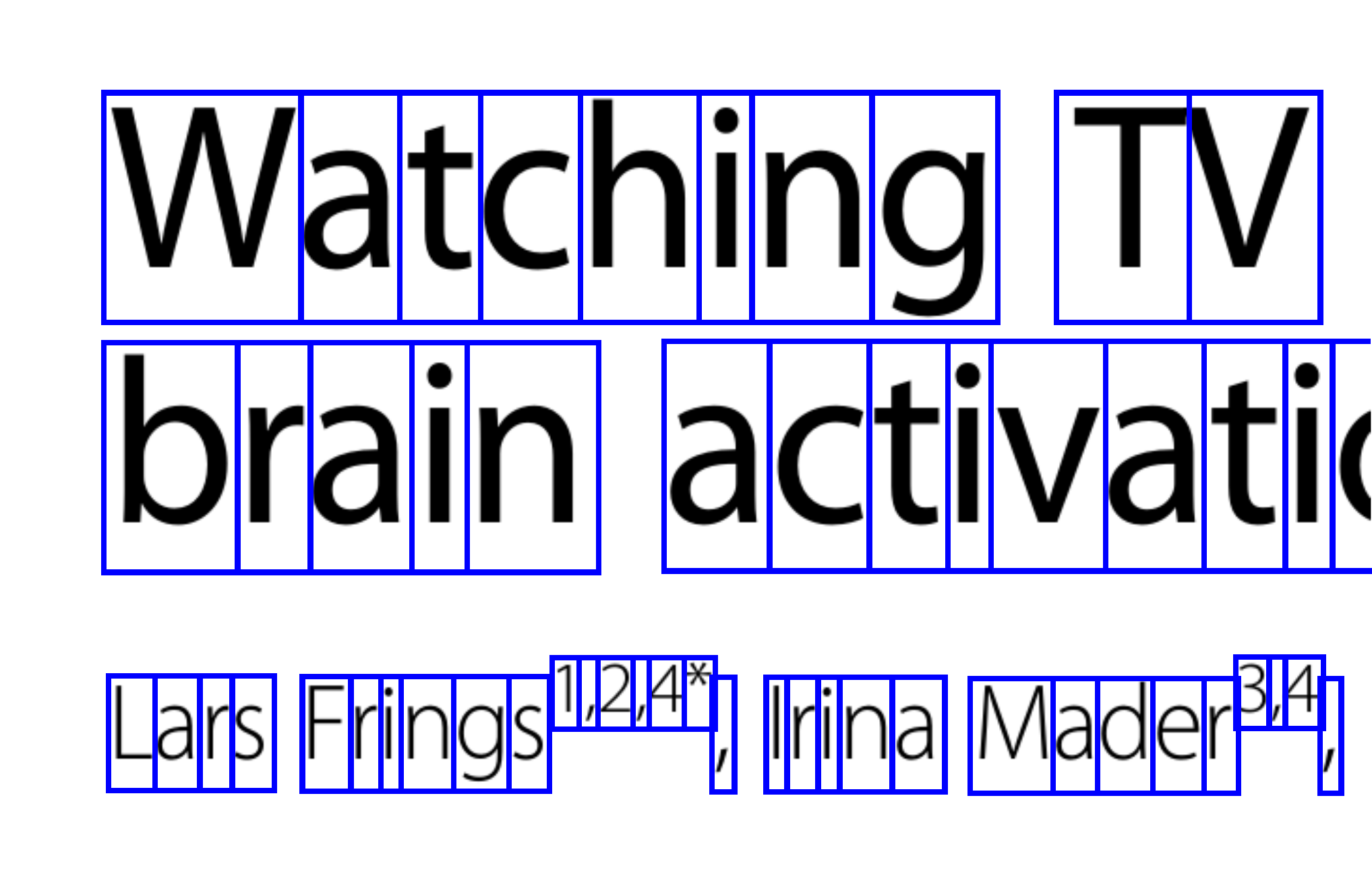}
  \caption[The bounding boxes of characters]{An example fragment of a page from a scientific
  publication. The rectangles are the bounding boxes of individual characters.}
  \label{fig:ex-chars}
\end{figure}

Due to the features of the PDF format and iText library, the resulting bounding boxes are in fact 
not the smallest rectangles enclosing the characters, but often are slightly bigger, depending on
the font and size used. In particular the bounding boxes of the characters printed in the same line
using the same font have usually the same vertical position and height. Figure~\ref{fig:ex-chars}
shows an example fragment of a page from a scientific publication with characters' bounding boxes, 
as returned by iText library. Fortunately, these approximate coordinates are sufficient for further
steps of the algorithm.

For performance reasons we enhanced character extraction with an additional cleaning phase, which in 
some rare cases reduces the number of extracted characters. In general the PDF text stream can 
contain text-showing operators, which do not result in any text visible to the document's reader. 
For example a text string might be printed in a position outside of the current page, or text
fragments can be printed in the same place, causing one fragment to cover the other. We also
encountered PDF files, in which text-showing operators were used for printing image fragments, which
resulted in tens of thousands tiny characters on one page, that do not contribute to the proper text
content of the document. In such rare cases it is very difficult to extract a logical, useful text
from the PDF stream. What is more, the number of characters is a significant factor of the algorithm
performance (more details are given in Section~\ref{sec:time}). The algorithm attempts to detect
such problems during character extraction step and reduce the number of characters by removing
suspicious characters, if needed.

The cleaning phase comprises the following steps. First, we remove those characters, that would not
be visible on the page, because their coordinates are outside of the pages limits. Then, we detect
and remove duplicates, that is characters with the same text and bounding boxes. Finally, we check
whether the density of the characters on each page is within a predefined threshold. If the overall
density exceeds the limit, we use a small sliding window to detect highly dense regions and all the
characters from these regions are removed.

Individual characters extracted in this step are the input for the page segmentation step.

\subsection{Page Segmentation}
\label{sec:segm}
The goal of page segmentation is to extract the remaining levels of the model described previously:
words, lines and zones. It is achieved by grouping characters into larger objects in a bottom-up 
manner.

Let $D$ be the given document, $P_D$ --- a list of document's pages and $C_D$ --- a set of extracted
characters. From character extraction step we also have the function $\anc_{cp}: C_D \to P_D$, which
assigns a parent page to every character.

The purpose of page segmentation is to find:
\begin{itemize}
\item $W_D$: a partition of the set $C_D$ corresponding to the words of the document,
\item $L_D$: a partition of the set $W_D$ corresponding to the text lines of the document,
\item $Z_D$: a partition of the set $L_D$ corresponding to the text zones of the document.
\end{itemize}

The function $\prt$, which is defined by the partitions, should satisfy the following condition:

$$\anc_{cp} = \prt \circ \prt \circ \prt \circ \prt$$

Page segmentation does not determine the order of the elements. The result of page segmentation is
a partial model, in which the analysed document is represented by a list of pages, each of which
contains a set of zones, each of which contains a set of lines, each of which contains a set of 
words, each of which contains a set of characters.

\begin{figure}[h]
  \centering
  \includegraphics[width=0.7\textwidth]{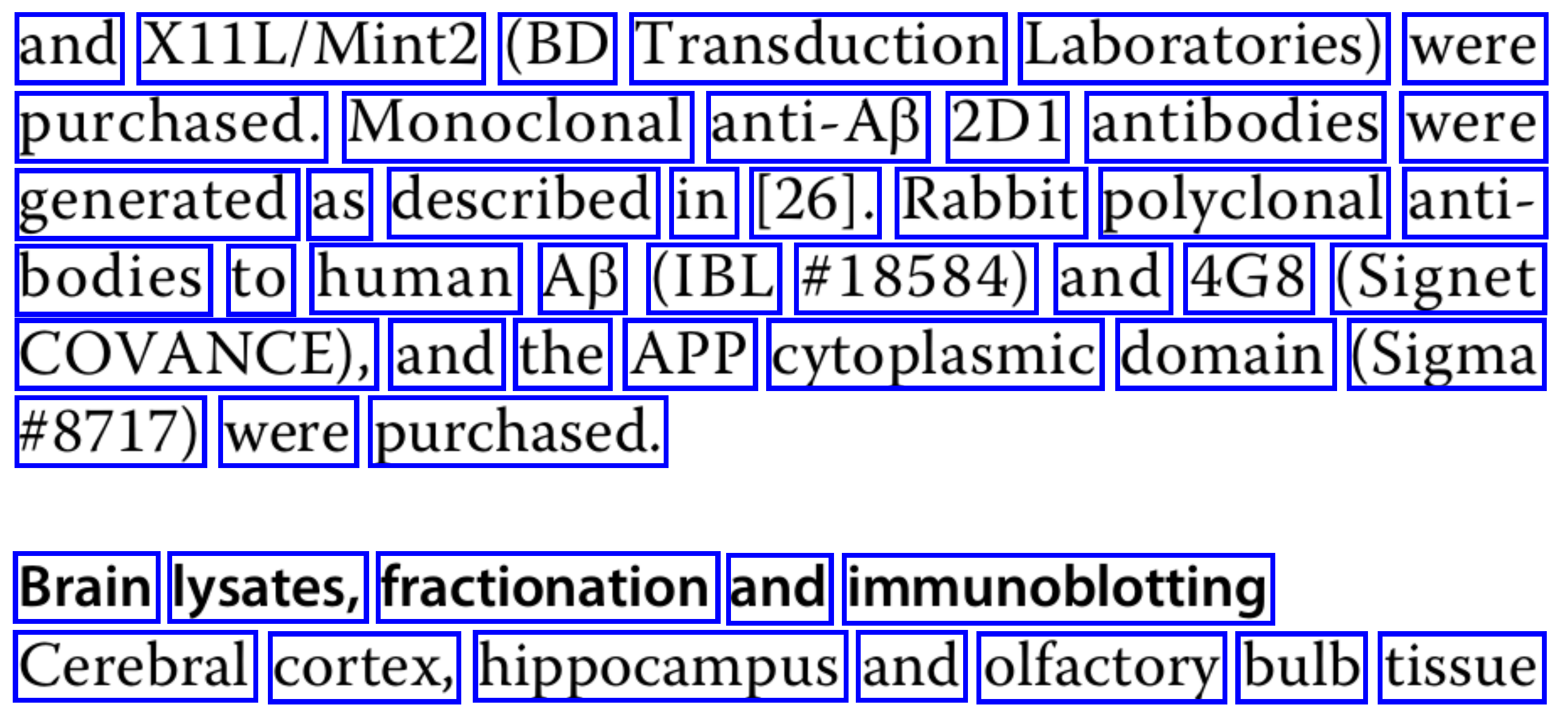}
  \caption[The bounding boxes of words]{An example fragment of a page from a scientific publication.
  The rectangles mark the bounding boxes of words.}
  \label{fig:ex-words}
\end{figure}

Figure~\ref{fig:ex-words} shows a group of words with their bounding boxes printed on a page of a
scientific publication. As the picture shows, punctuation marks as well as hyphenation characters 
usually belong to the proper word preceding them. In general words in the model should be understood 
geometrically rather than logically --- as a continuous sequence of characters without a space or
other white character.

\begin{figure}[h]
  \centering
  \includegraphics[width=0.9\textwidth]{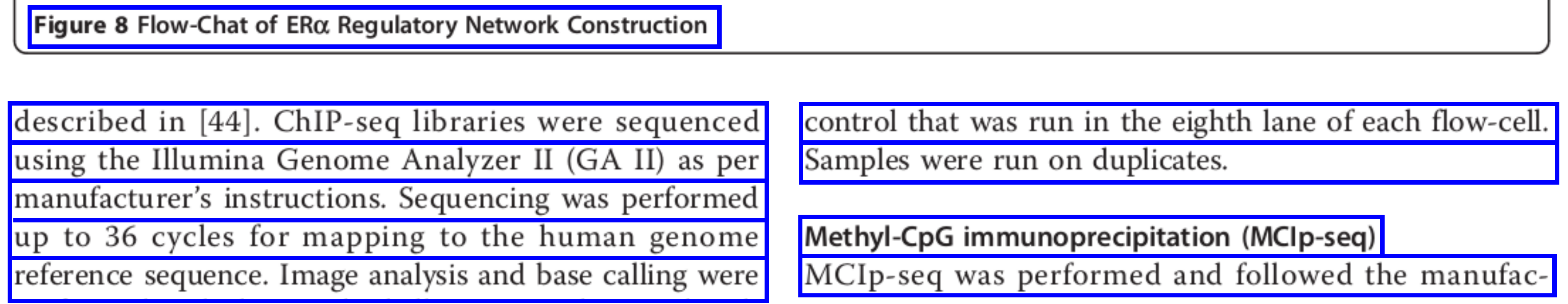}
  \caption[The bounding boxes of lines]{An example fragment of a page from a scientific publication.
  The rectangles mark the bounding boxes of lines.}
  \label{fig:ex-lines}
\end{figure}

Figure~\ref{fig:ex-lines} shows a group of lines with their bounding boxes. As shown in the picture,
the lines respect the multi-column document layout.

\begin{figure}[h]
  \centering
  \includegraphics[width=0.9\textwidth]{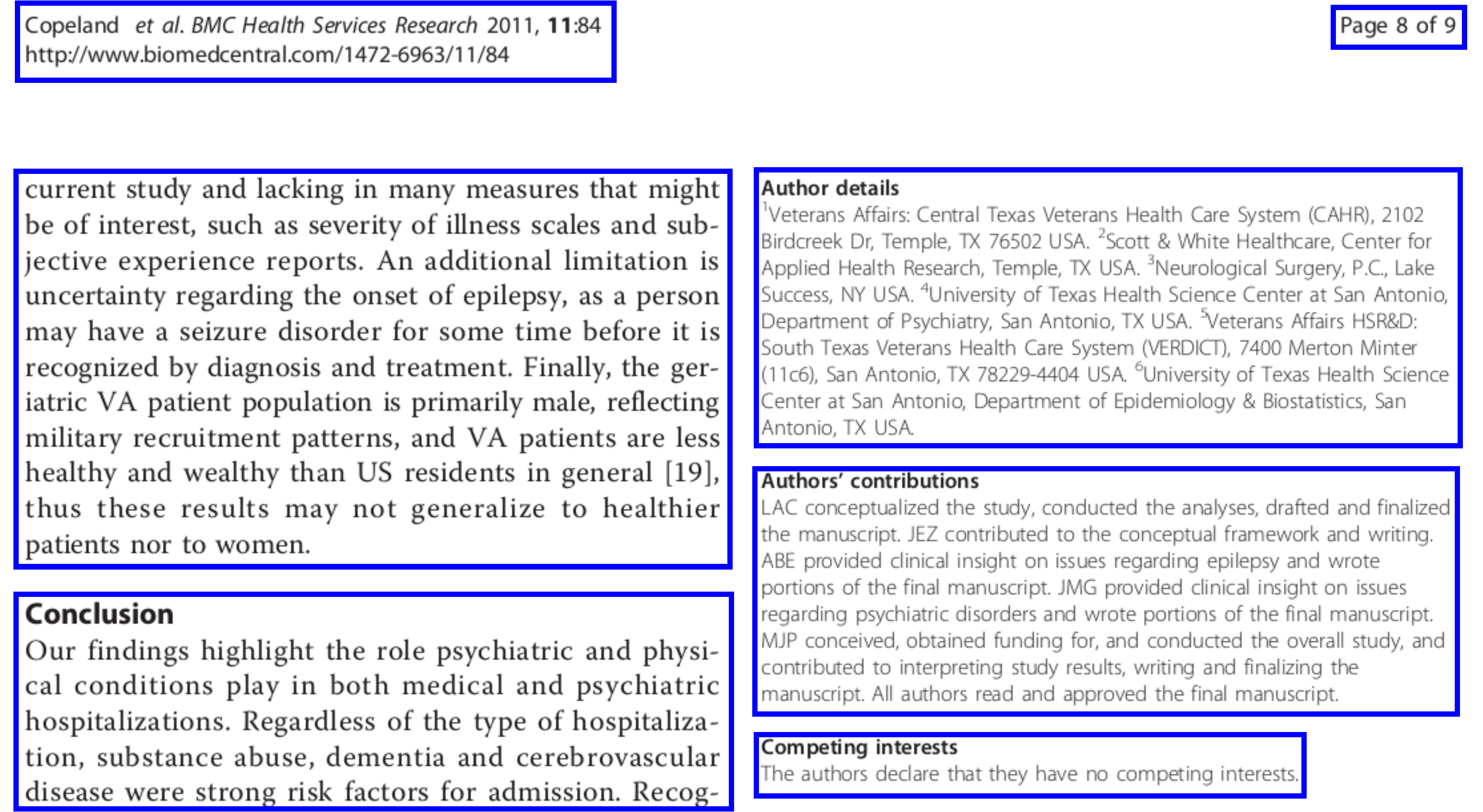}
  \caption[The bounding boxes of zones]{An example fragment of a page from a scientific publication.
  The rectangles mark the bounding boxes of zones.}
  \label{fig:ex-zones}
\end{figure}

Figure~\ref{fig:ex-zones} shows a fragment of a scientific publication with example zones and their
bounding boxes. In general a zone contains lines that are close to each other, even if they play a
different role in the document (for example section title and paragraph).

\begin{figure}[h]
  \centering
  \includegraphics[width=0.7\textwidth]{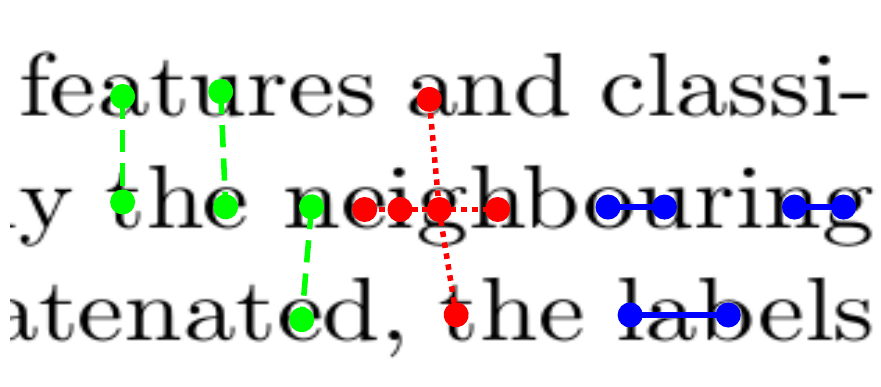}
  \caption[Nearest neighbours in Docstrum algorithm]{An example fragment of a text zone in a
  scientific article. The figure shows five nearest neighbours of a given character (red dotted
  lines), neighbours placed in the same line used to determine in-line spacing (blue solid lines),
  and neighbours placed approximately in the line perpendicular to the text line orientation used to
  determine between-line spacing (green dashed lines).}
  \label{fig:segm}
\end{figure}

Page segmentation is implemented with the use of a bottom-up Docstrum algorithm~\cite{OGorman93}.
Docstrum is an accurate algorithm able to recognize both text lines and zones. The algorithm can be 
fairly easily adapted to process born-digital documents: it is sufficient to treat individual
characters as connected components, which in the original algorithm are calculated from a page 
image.

\begin{figure}[h]
  \centering
  \includegraphics[width=0.9\textwidth]{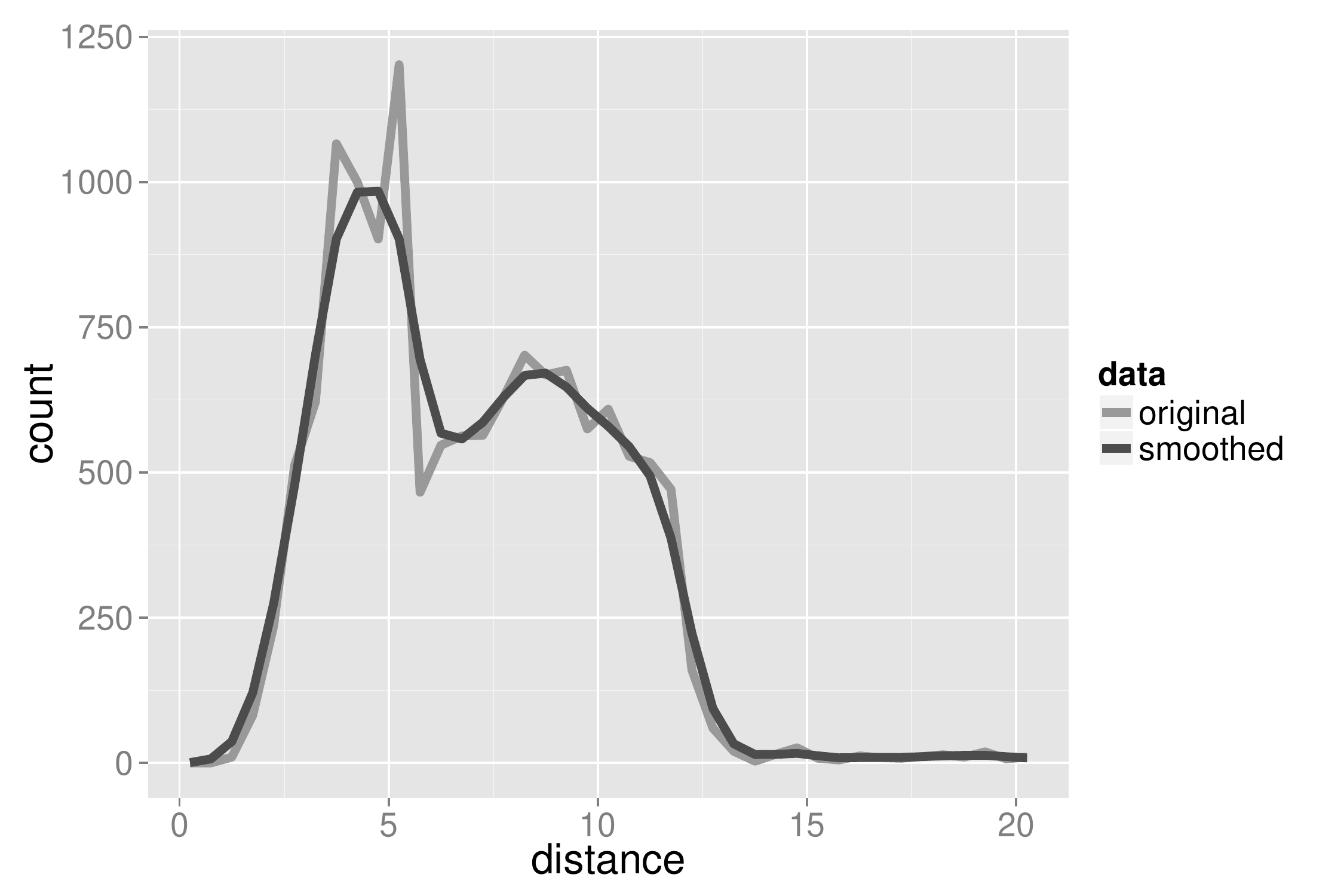}
  \caption[Nearest-neighbour distance histogram in Docstrum algorithm]{An example of a 
  nearest-neighbour distance histogram. The figure shows both original and smoothed versions of the
  histogram. The peak distance chosen based on the original data would be the global maximum, even
  though the histogram contains two close peaks of similarly high frequency. Thanks to smoothing
  both local peaks are taken into account, shifting the resulting peak slightly to the left and
  yielding more reliable results.}
  \label{fig:smoothing}
\end{figure}

In our case the algorithm's input is a single page containing a set of characters, which are 
clustered hierarchically based on geometric traits. The algorithm is based to a great extent on the
analysis of the nearest-neighbor pairs of individual characters:
\begin{enumerate}
\item First, five nearest components for every character on the page are identified (red dotted 
lines in Figure~\ref{fig:segm}). The distance between two characters is the Euclidean distance
between the centers of their bounding boxes.
\item In order to calculate the text orientation (the skew angle) we analyze the histogram of the
angles between the elements of all nearest-neighbor pairs. The peak value is assumed to be the angle
of the text. Since in the case of born-digital documents the skew is almost always horizontal, this
step would be more useful for documents in the form of scanned pages. All the histograms used in
Docstrum are smoothed to avoid detecting local abnormalities. An example of a smoothed histogram is
shown in Figure~\ref{fig:smoothing}.
\item Next, within-line spacing is estimated by detecting the peak of the histogram of distances
between the nearest neighbors. For this histogram we use only those pairs, in which the angle 
between components is similar to the estimated text orientation angle (blue solid lines in 
Figure~\ref{fig:segm}).
\item Similarly, between-line spacing is also estimated with the use of a histogram of the distances
between the nearest-neighbor pairs. In this case we include only those pairs, that are placed
approximately in the line perpendicular to the text line orientation (green dashed lines in
Figure~\ref{fig:segm}).
\item Next, line segments are found by performing a transitive closure on within-line 
nearest-neighbor pairs. To prevent joining line segments belonging to different columns, the 
components are connected only if the distance between them is sufficiently small.
\item The zones are then constructed by grouping the line segments on the basis of heuristics 
related to spatial and geometric characteristics. Each line segment pair is examined and the 
decision is made whether they should be in the same zone. If both horizontal and vertical distance
are within predefined limits, the current zones of the line segments are merged.
\item Finally, line segments belonging to the same zone and placed in one line horizontally are
merged into final text lines.
\end{enumerate}

All the threshold values used in the algorithm have been obtained by manual experiments performed on
a validation dataset. The experiments also resulted in adding a few improvements to the 
Docstrum-based implementation of page segmentation:
\begin{itemize}
\item the distance between connected components, which is used for grouping components into line
segments, has been split into horizontal and vertical distance (based on estimated text orientation
angle),
\item fixed maximum distance between lines that belong to the same zone has been replaced with a
value scaled relatively to the line height,
\item merging of lines belonging to the same zone has been added,
\item rectangular smoothing window has been replaced with Gaussian smoothing window,
\item merging of highly overlapping zones has been added,
\item words determination based on within-line spacing has been added.
\end{itemize}

Section~\ref{sec:segm} reports the results of the comparison of the performance of the original 
Docstrum and the enhanced version used in our algorithm.

The resulting hierarchical structure is the input for the next step, reading order resolving.

\subsection{Reading Order Resolving}
\label{sec:rores}
The purpose of reading order resolving is to determine the right sequence, in which all the 
structure elements should be read. More formally, its task is to find a total order for the sets of
zones, lines, words and characters. The order of the pages is explicitly given in the input PDF 
file.

\begin{figure}[h]
  \centering
  \includegraphics[width=0.6\textwidth]{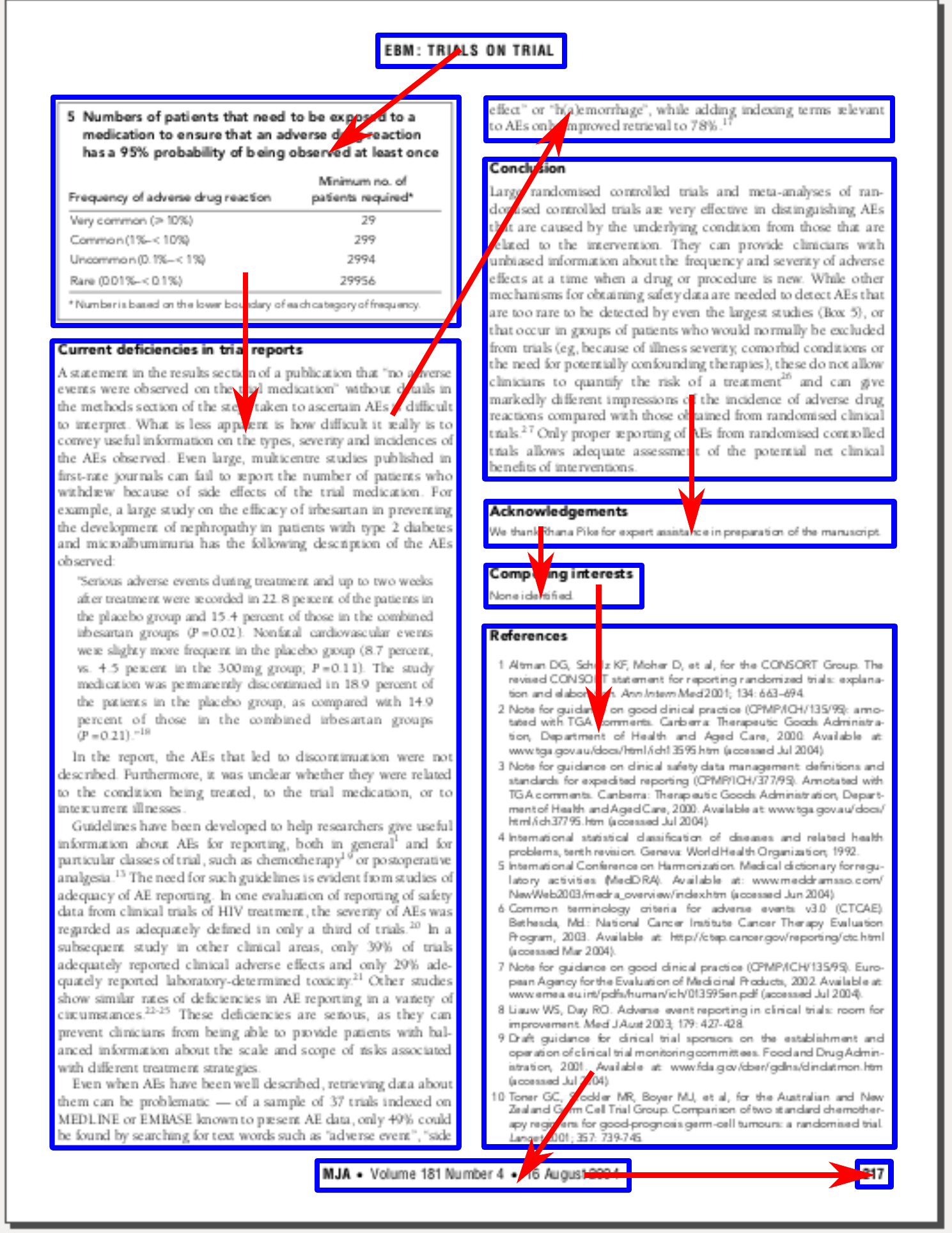}
  \caption[The reading order of zones]{An example page from a scientific publication. The image
  shows the zones and their reading order.}
  \label{fig:ro}
\end{figure}

An example document page with a reading order of the zones is shown in Figure~\ref{fig:ro}. The
reading order is very important in the context of the body of the document and bibliography 
sections, but much less meaningful for the areas of the document containing metadata.

Algorithm~\ref{alg:rores} shows the pseudocode of reading order resolving step. The algorithm is
based on a bottom-up strategy:
\begin{enumerate}
\item At the beginning the characters are sorted within words horizontally, from left to right (line 
6 in Algorithm~\ref{alg:rores}).
\item Similarly, the words are sorted within lines also horizontally, from left to right (line 8 in
Algorithm~\ref{alg:rores}).
\item Next, the lines are sorted vertically within zones, from top to bottom (line 10 in
Algorithm~\ref{alg:rores}). 
\item In the final step we sort zones. Sorting zones is done with the use of simple heuristics
similar to those used in PDFMiner
tool\footnote{http://www.unixuser.org/$\sim$euske/python/pdfminer/}. We make use of an observation
that the natural reading order in most modern languages descends from top to bottom, if successive
zones are aligned vertically, otherwise it traverses from left to right. There are few exceptions to
this rule, for example Arabic script, and such cases would currently not be handled properly by the
algorithm. 
\end{enumerate}

\begin{figure}[h]
  \centering
  \includegraphics[width=0.8\textwidth]{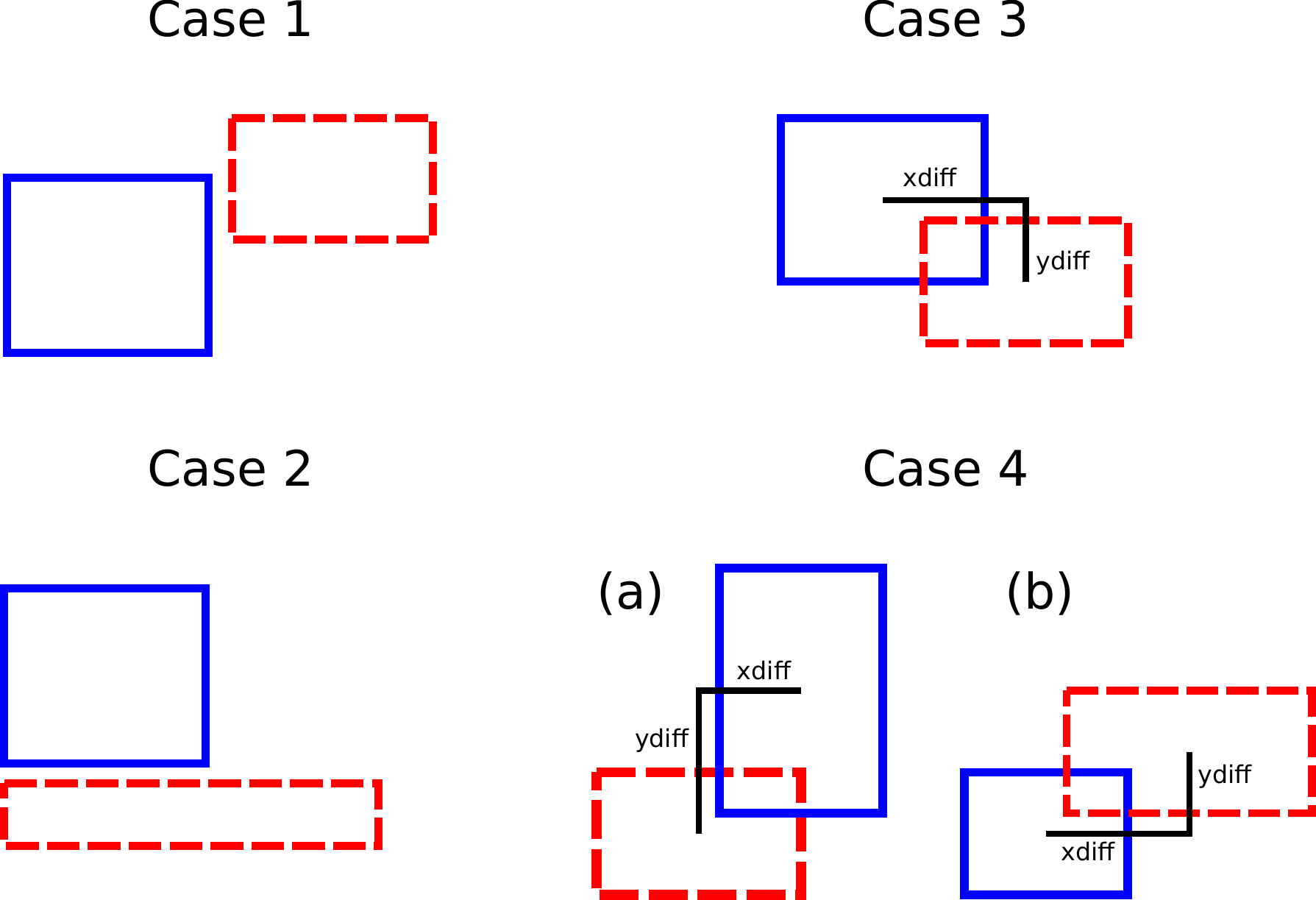}
  \caption[The mutual position of two zone groups]{Different cases of the mutual position of two
  zone groups. In each case the blue solid group will precede the red dashed one in the resulting
  reading order. Case 1: if the groups can be separated by a vertical line, the first is the group
  closer to the left. Case 2: if the groups can be separated by a horizontal line, the first is the
  group closer to the top of the page. Case 3: if the groups cannot be fully separated by any line,
  and {\it xdiff} and {\it ydiff} are both positive or both negative, the first is the group closer
  to the left upper corner of the page. Case 4: if the groups cannot be fully separated by any line,
  and {\it xdiff} and {\it ydiff} have different signs, the decision is based on the sign of $xDiff
  + yDiff$.}
  \label{fig:rores}
\end{figure}

The zones are sorted in the following steps:
\begin{enumerate}
\item We use the following formula to calculate the distance between all pairs of zones on a given 
page:
\begin{align*}
\dd(z_1, z_2) =  (\area(\{z_1, z_2\}) - \area(z_1) - \area(z_2))\\
* (0.5 + \min(\cos_L(z_1, z_2), \cos_M(z_1, z_2)))
\end{align*}
where 
\begin{itemize}
\item for any zone $z \in Z_D$ $\area(z)$ is the area of the zone's bounding box,
\item for any zone set $S \subset Z_D$ placed in the same page, $\area(S)$ is the area of the
smallest rectangle containing all the zones in $S$,
\item $\cos_L(z_1, z_2)$ is the cosine of the slope connecting the centers of left edges of the
zones,
\item $\cos_M(z_1, z_2)$ is the cosine of the slope connecting the centers of the zones.
\end{itemize}
We use the angle of the slope of the vector connecting zones to make sure that in general zones
aligned vertically are closer than those aligned horizontally.
\item Using this distance we apply a hierarchical clustering algorithm, repeatedly joining the
closest zones and zone sets. This results in a binary tree, where the leaves represent individual
zones, other nodes can be understood as groups of zones and the root represents the set of all zones
on the page (line 12 in Algorithm~\ref{alg:rores}).
\item Next, we visit every node in the tree and swap the children if needed (lines 13-17 in
Algorithm~\ref{alg:rores}). The decision process for every node is based on a sequence of rules. The
first matched rule determines the decision result:
\begin{enumerate}
\item if two groups can be separated by a vertical line, their order is determined by the 
x-coordinate (case 1 in Figure~\ref{fig:rores}),
\item if two groups can be separated by a horizontal line, their order is determined by the
y-coordinate (case 2 in Figure~\ref{fig:rores}),
\item if the groups overlap, we calculate $xDiff$ and $yDiff$, which are horizontal and vertical
distance between the centers of the right and left child of the node. The children are swapped if
$xDiff + yDiff < 0$ (case 3 and 4 in Figure~\ref{fig:rores}).
\end{enumerate}
\item Finally, an in-order tree traversal gives the desired zones order (line 18 in 
Algorithm~\ref{alg:rores}).
\end{enumerate}

Reading order resolving concludes the layout extraction stage of the extraction algorithm. The 
result is a fully featured geometric model of the document, containing the entire text content of
the input file as well as the geometric characteristics related to the way the text is displayed in
the input PDF file.

\begin{algorithm}[H]
\caption[Reading order resolving]{Reading order resolving algorithm}
\label{alg:rores}
\begin{algorithmic}[1]
\Function{SortDocument}{document}
\For{page $\in$ {\sc Pages}(document)}
\For{zone $\in$ {\sc Zones}(page)}
\For{line $\in$ {\sc Lines}(zone)}
\For{word $\in$ {\sc Words}(line)}
\State {\sc SortHorizontally}(word) \Comment{sort characters from left to right}
\EndFor
\State {\sc SortHorizontally}(line) \Comment{sort words from left to right}
\EndFor
\State {\sc SortVertically}(zone)  \Comment{sort lines from top to bottom}
\EndFor
\State tree $\gets$ {\sc BuildTree}({\sc Zones}(page)) \Comment{build hierarchical tree by clustering}
\For{node $\in$ {\sc Nodes}(tree)}
\If{{\sc ShouldBeSwapped}({\sc LeftChild}(node), {\sc RightChild}(node))}
\State {\sc SwapChildren}(node) \Comment{swap children if needed}
\EndIf
\EndFor
\State {\sc sort}(page, tree) \Comment{sort zones accordingly to in-order tree traversal}
\EndFor
\EndFunction
\State
\Function{ShouldBeSwapped}{leftNode, rightNode}
\State leftGroup $\gets$ {\sc Zones}(leftNode) \Comment{group of zones represented by left node}
\State rightGroup $\gets$ {\sc Zones}(rightNode) \Comment{group of zones represented by right node}
\If{{\sc MaxX}(leftGroup) $\leq$ {\sc MinX}(rightGroup)}
\State\Return false \Comment{groups can be separated by a vertical line}
\EndIf
\If{{\sc MaxX}(rightGroup) $\leq$ {\sc MinX}(leftGroup)}
\State\Return true \Comment{groups can be separated by a vertical line}
\EndIf
\If{{\sc MaxY}(leftGroup) $\leq$ {\sc MinY}(rightGroup)}
\State\Return false \Comment{groups can be separated by a horizontal line}
\EndIf
\If{{\sc MaxY}(rightGroup) $\leq$ {\sc MinY}(leftGroup)}
\State\Return true \Comment{groups can be separated by a horizontal line}
\EndIf
\State \Return {\sc DiffX}(rightGroup, leftGroup) $+$ {\sc DiffY}(rightGroup, leftGroup) $< 0$
\EndFunction
\end{algorithmic}
\end{algorithm}

\section{Document Region Classification}
\label{sec:classification}
The goal of content classification is to determine the role played by every zone in the document by
assigning a general category to it. We use the following classes: {\it metadata} (document's 
metadata, containing title, authors, abstract, keywords, and so on), {\it references} (the 
bibliography section), {\it body} (publication's text, sections, section titles, equations, figures
and tables, captions) and {\it other} (acknowledgments, conflicts of interests statements, page
numbers, etc.).

Formally, the goal of document region classification is to find a function

\[ \class: Z_D \to \{metadata, references, body, other\} \]

The classification is performed by a Support Vector Machine classifier using a large set of zone 
features of various nature. SVM is a very powerful classification technique able to handle a large
variety of input and work effectively even with training data of a small size. The algorithm is
little prone to overfitting. It does not require a lot of parameters and can deal with highly
dimensional data. SVM is widely used for content classification and achieves very good results in
practice.

The features we developed capture various aspects of the content and surroundings of the zones and
can be divided into the following categories:
\begin{itemize}
\item geometric --- based on geometric attributes, some examples include: zone's height and width,
height to width ratio, zone's horizontal and vertical position, the distance to the nearest zone,
empty space below and above the zone, mean line height, whether the zone is placed at the top, 
bottom, left or right side of the page;
\item sequential --- based on sequence-related information, some examples include: the label of the
previous zone (according to the reading order), the presence of the same text blocks on the 
surrounding pages, whether the zone is placed in the first/last page of the document;
\item formatting --- related to text formatting in the zone, examples include: font size in the
current and adjacent zones, the amount of blank space inside zones, mean indentation of text lines 
in the zone;
\item lexical --- based upon keywords characteristic for different parts of narration, such as: 
affiliations, acknowledgments, abstract, keywords, dates, referen\-ces, or article type; these
features typically check, whether the text of the zone contains any of the characteristic keywords;
\item heuristics --- based on heuristics of various nature, such as the count and percentage of 
lines, words, uppercase words, characters, letters, upper/lowercase letters, digits, whitespaces,
punctuation, brackets, commas, dots, etc; also whether each line starts with enumeration-like
tokens, or whether the zone contains only digits.
\end{itemize}

The features used by the classifier were selected semi-automatically from a set of 103 features with
the use of the zone validation dataset. The final version of the classifier uses 54 features. More
details about the selection procedure and results can be found in 
Section~\ref{sec:cont-class-features}.

The best SVM parameters were also estimated automatically using the zone validation dataset. More
detailed results can be found in Section~\ref{sec:cont-class-pars}.

Since our problem is a multiclass classification problem, it is reduced to a number of binary
classifiers with the use of "one vs. one" strategy.

Document region classification allows to split the content of the input file into three areas of 
interest: metadata, body and references, which are later on analysed in three parallel specialized
extraction paths.

\section{Metadata Extraction}
\label{sec:metadata}
The geometric model of the input document enhanced with zone categories is the input to metadata
extraction stage, which is a part of the algorithm specializing in extracting the proper metadata of
the document. During metadata extraction only zones labelled as {\it metadata} are analysed.

The algorithm is able to extract the following information:
\begin{itemize}
\item title (string): the title of the document,
\item authors (a list of strings): the full names of all the authors, in the order given in the
document,
\item affiliations (a list of tuples): a list of parsed affiliations of the authors of the document,
in the order given in the document; a single affiliation contains:
\begin{itemize}
\item raw text of the affiliation (string),
\item organization name (string),
\item address (string),
\item country (string and two-character country ISO code).
\end{itemize}
\item relations author-affiliation,
\item emails (a list of strings): a list of emails of the authors of the document,
\item relations author-email,
\item abstract (string): the abstract provided by the authors,
\item keywords (a list of strings): the article's keywords listed in the document,
\item journal (string): the name of the journal in which the article was published,
\item volume (string): the volume in which the article was published,
\item issue (string): the issue in which the article was published,
\item year (string): the year of publication,
\item pages (string): the pages range of the published article,
\item DOI (string): DOI identifier of the document.
\end{itemize}

The algorithm analyses only the content of the input document, and only the information explicitly
given in the document are extracted. No information is acquired from external sources or inferred
based on the text of the document. All information listed above is optional, and there is no
guarantee that it will appear in the resulting metadata record.

The default output format is NLM JATS. Listing~\ref{lst:metadata} shows an example metadata record.

\begin{lstlisting}[caption=Example document metadata record in NLM JATS format,label=lst:metadata]
<article>
  <front>
    <journal-meta>
      <journal-title-group>
        <journal-title>Dhaka Univ. J. Pharm. Sci.</journal-title>
      </journal-title-group>
    </journal-meta>
    <article-meta>
      <title-group>
        <article-title>Phytochemical and Biological investigations of Phoenix paludosa Roxb.
        </article-title>
      </title-group>
      <contrib-group>
        <contrib contrib-type="author">
          <string-name>Farzana Alam</string-name>
          <xref ref-type="aff" rid="2">2</xref>
        </contrib>
        <contrib contrib-type="author">
          <string-name>Mohammad S. Rahman</string-name>
          <xref ref-type="aff" rid="1">1</xref>
        </contrib>
        <contrib contrib-type="author">
          <string-name>Md. Shahanur Alam</string-name>
          <xref ref-type="aff" rid="2">2</xref>
        </contrib>
        <contrib contrib-type="author">
          <string-name>Md. Khalid Hossain</string-name>
          <xref ref-type="aff" rid="1">1</xref>
        </contrib>
        <contrib contrib-type="author">
          <string-name>Md. Aslam Hossain</string-name>
          <xref ref-type="aff" rid="1">1</xref>
        </contrib>
        <contrib contrib-type="author">
          <string-name>Mohammad A. Rashid</string-name>
          <email>rashidma@univdhaka.edu</email>
          <xref ref-type="aff" rid="0">0</xref>
          <xref ref-type="aff" rid="1">1</xref>
        </contrib>
        <aff id="0">
          <label>0</label>
          <institution>Centre for Biomedical Research, University of Dhaka</institution>
          ,
          <addr-line>Dhaka-1000</addr-line>
          ,
          <country country="BD">Bangladesh</country>
        </aff>
        <aff id="1">
          <label>1</label>
          <institution>Department of Pharmaceutical Chemistry, Faculty of Pharmacy, University of Dhaka</institution>
          ,
          <addr-line>Dhaka-1000</addr-line>
          ,
          <country country="BD">Bangladesh</country>
        </aff>
        <aff id="2">
          <label>2</label>
          <institution>Department of Pharmacy, The University of Asia Pacific</institution>
          ,
          <addr-line>Dhaka-1000</addr-line>
          ,
          <country country="BD">Bangladesh</country>
        </aff>
      </contrib-group>
      <abstract>
        <p>Lupeol (1), epilupeol (2) and B-sitosterol (3) were isolated from the n-hexane and the carbon tetrachloride soluble fraction of a methanol extract of the leaves of Phoenix paludosa Roxb. The n-hexane, carbon tetrachloride and chloroform soluble materials from the concentrated methanol extract were subjected to antimicrobial screening and brine shrimp lethality bioassay. All of the partitionates showed insensitivity to microbial growth, while the n-hexane, chloroform and methanol soluble fractions showed significant cytotoxicity having LC50 2.17 ug/ml, 2.77 ug/ml and 2.46 ug/ml, respectively. This is the first report of isolation of the compounds 1-3 and bioactivities of P. paludosa.</p>
      </abstract>
      <volume>8</volume>
      <issue>1</issue>
      <fpage>7</fpage>
      <lpage>10</lpage>
      <pub-date>
        <year>2009</year>
      </pub-date>
    </article-meta>
  </front>
</article>
\end{lstlisting}

\begin{table}[h]
\renewcommand{\arraystretch}{1.4}
\renewcommand{\tabcolsep}{3pt}
\centering
\begin{tabular}{ | m{150px} | m{210px} | m{70px} | }
	\hline
    \multicolumn{1}{|c|}{Step} & \multicolumn{1}{c|}{Goal} &
    \multicolumn{1}{c|}{Implementation} \\ \hline\hline
    
    1. {\bf Metadata zone classification} & Classifying the zones labelled previously as {\it 
    metadata} into specific metadata classes. & {\bf SVM}\\ \hline
	 
    2. {\bf Authors and affiliations extraction} & Extracting individual author names, affiliation
    strings and determining the relations between them. & {\bf heuristics}\\ \hline

	3. {\bf Affiliation parsing} & Extracting {\it organization}, {\it address} and {\it country}
	from affiliation strings. & {\bf CRF}\\ \hline
	 
	4. {\bf Metadata cleaning} & Extracting atomic metadata information from labelled zones, 
	cleaning and forming the final record. & {\bf simple rules}\\\hline
	 
  \end{tabular}
\caption[The decomposition of metadata extraction]{The decomposition of metadata extraction stage
into independent steps.}
\label{tab:metadata}
\end{table}

Table~\ref{tab:metadata} lists the steps executed during the metadata extraction stage. The details
of the implementations are provided in the following sections: metadata zone classification
(Section~\ref{sec:metadata-classification}), authors and affiliations extraction 
(Section~\ref{sec:relations}), affiliation parsing (Section~\ref{sec:aff-parsing}) and metadata 
cleaning (Section~\ref{sec:cleaning}).

\subsection{Metadata Classification}
\label{sec:metadata-classification}
Metadata classification is the first step in the metadata extraction stage. Its goal is to classify
all zones labelled previously as {\it metadata} into specific metadata classes: {\it title} (the
title of the document), {\it author} (the names of the authors), {\it affiliation} (authors'
affiliations), {\it editor} (the names of the editors), {\it correspondence} (addresses and emails),
{\it type} (the type specified in the document, such as "research article", "editorial" or "case
study"), {\it abstract} (document's abstract), {\it keywords} (keywords listed in the document), 
{\it bib\_info} (for zones containing various bibliographic information, such as journal name, 
volume, issue, DOI, etc.), {\it dates} (the dates related to the process of publishing the article).

Formally, the goal of metadata zone classification is to find a function

\[ \class_M: Z_{DM} \to L_M \]

where

\[ Z_{DM} = \{z \in Z_D \enskip | \enskip \class(z) = metadata\} \]

\begin{align*}
L_M = \{title, author, affiliation, editor, correspondence,\\
type, abstract, keywords, bib\_info, dates\}
\end{align*}

Metadata classifier is based on Support Vector Machines and is implemented in a similar way as 
category classification. The classifiers differ in target zone labels, the features and SVM
parameters used. The features, as well as SVM parameters were selected using the same procedure,
described in Sections~\ref{sec:cont-class-features} and~\ref{sec:cont-class-pars}. The final 
classifier contains 53 features.

The decision of splitting zone classification into two separate classification steps, as opposed
to implementing only one classification step, was based mostly on aspects related to the workflow
architecture and maintenance. In fact both tasks have different characteristics and needs. The goal
of the category classifier is to divide the article's content into three general areas of interest,
which can be then analysed independently in parallel, while metadata classifier focuses on far more
detailed analysis of only a small subset of all zones.

The implementation of the category classifier is more stable: the target label set does not change,
and once trained on a reasonably large and diverse dataset, the classifier performs well on other
layouts as well. On the other hand, metadata zones have much more variable characteristics across
different layouts, and from time to time there is a need to tune the classifier or retrain it using
a wider document set. What is more, in the future the classifier might be extended to be able to
capture new labels, not considered before (for example a special label for zones containing both
author and affiliation, a separate label for categories or general terms).

For these reasons we decided to implement content classification in two separate steps. As a result
the two tasks can be maintained independently, and for example adding another metadata label to the
algorithm does not change the performance of recognizing the bibliography sections. It is also
possible that in the future the metadata classifier will be reimplemented using a different 
technique, allowing to add new training cases incrementally, for example using a form of online
machine learning.

As a result of metadata classification the zones labelled previously as {\it metadata} have specific 
metadata labels assigned, which gives the algorithm valuable hints where different metadata types
are located in the document.

\subsection{Affiliation-Author Relation Determination}
\label{sec:relations}
As a result of classifying the document's fragments, we usually obtain a few regions labelled as 
{\it author} or {\it affiliation}. In this step individual author names and affiliation strings are 
extracted and the relations between them are determined.

More formally, the goal of author-affiliation relation extraction is to determine for a given
document $D$:
\begin{itemize}
\item $\aut(D)$ --- a list of document's author full names,
\item $\aff(D)$ --- a set of document's affiliation strings,
\item $\aaf(D)$ --- a relation author-affiliation, where $(aut, aff) \in \aaf(D)$ if and only if the
affiliation string represents the author's affiliation.
\end{itemize}

\begin{figure}[h]
  \centering
  \includegraphics[width=0.7\textwidth]{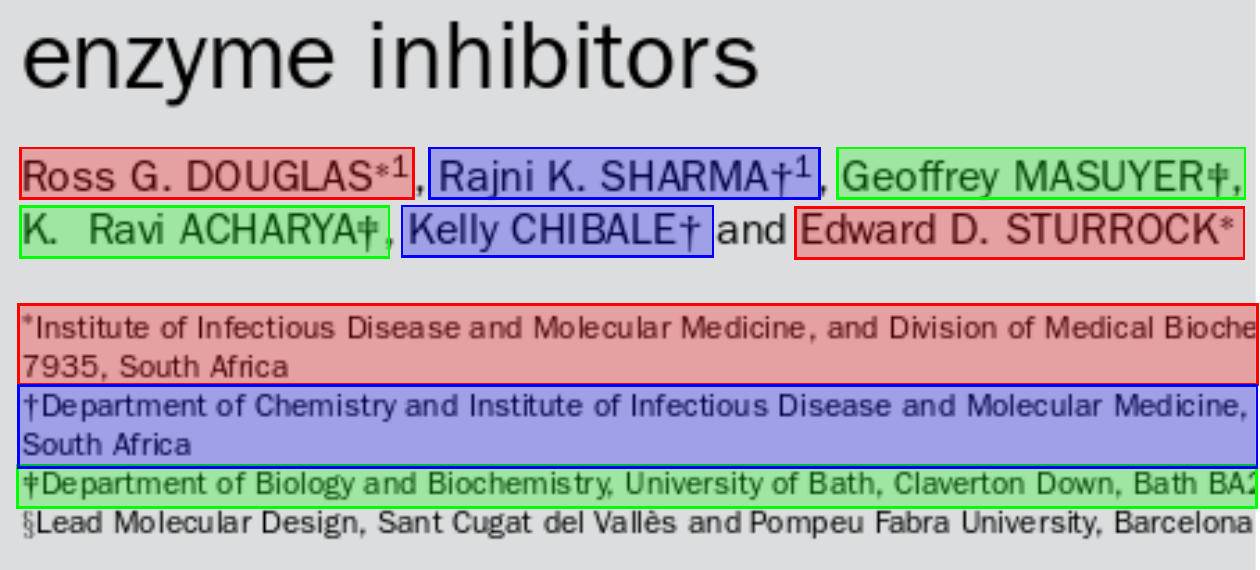}
  \caption[Affiliations associated with authors using indexes]{An example fragment of a page from a
  scientific publication with authors and affiliations zones. In this case the relations author-
  affiliation (coded with colors) can be determined with the use of upper indexes.}
  \label{fig:aff-example}
\end{figure}

\begin{figure}[h]
  \centering
  \includegraphics[width=0.7\textwidth]{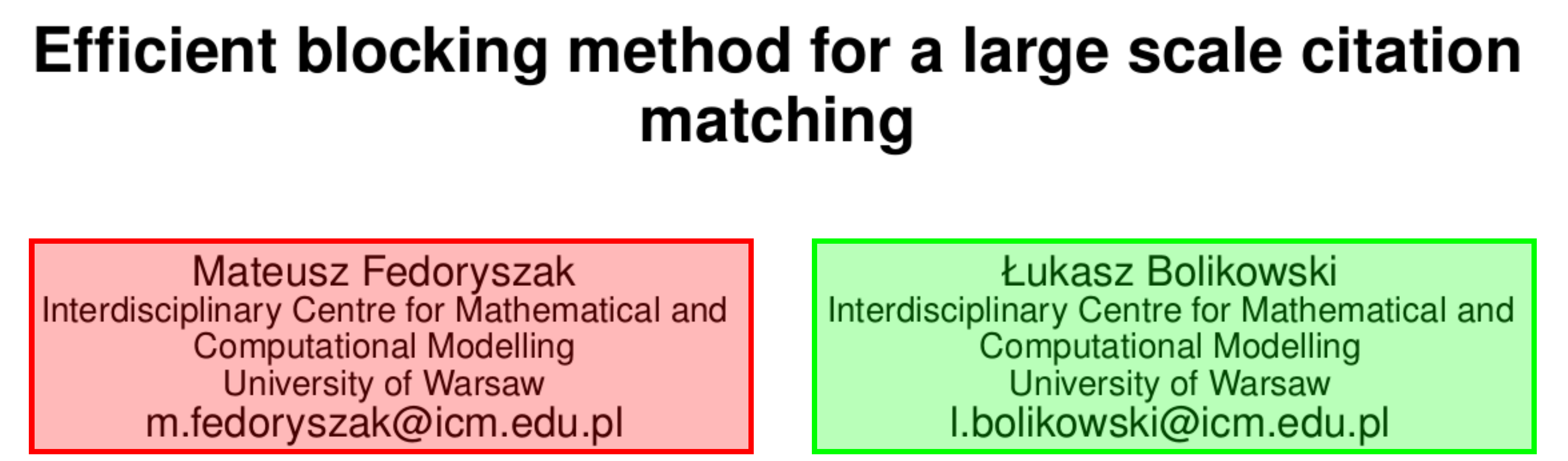}
  \caption[Affiliations associated with authors using distance]{An example fragment of a page from a
  scientific publication with authors and affiliations zones. In this case the relations 
  author-affiliation can be determined using the distance and belonging to the same zone. The
  content of such a zone is split with the use of regular expressions.}
  \label{fig:aff-example-2}
\end{figure}

In general the implementation is based on heuristics and regular expressions, and the details depend
on article's layout. There are two main styles used: (1) author names are grouped together in a form
of a list, and affiliations are also placed together below the author's list, at the bottom of the
first page or even just before the bibliography section (an example is shown in 
Figure~\ref{fig:aff-example}), and (2) each author is placed in a separate zone along with its
affiliation and email address (an example is shown in Figure~\ref{fig:aff-example-2}).

First step is to recognize the type of layout of a given document. If the document contains at least 
two zones labelled as affiliation placed approximately in the same horizontal line, the algorithm
treats it as type (2), otherwise --- as type (1).

In the case of a layout of the first type (Figure~\ref{fig:aff-example}), at the beginning authors'
lists are split using a predefined lists of separators. Then we detect affiliation indexes based on 
predefined lists of symbols and also geometric features, in particular y-position of the characters.
Detected indexes are then used to split affiliation lists and assign affiliations to authors.

In the case of a layout of the second type (Figure~\ref{fig:aff-example-2}), each author is already
assigned to its affiliation by being placed in the same zone. It is therefore enough to detect 
author name, affiliation and email address. We assume the first line of such a zone is the author
name, email is detected based on regular expressions, and the rest is treated as the affiliation
string.

\subsection{Affiliation Parsing}
\label{sec:aff-parsing}
\begin{figure}[ht]
  \centering
  \includegraphics[width=0.8\textwidth]{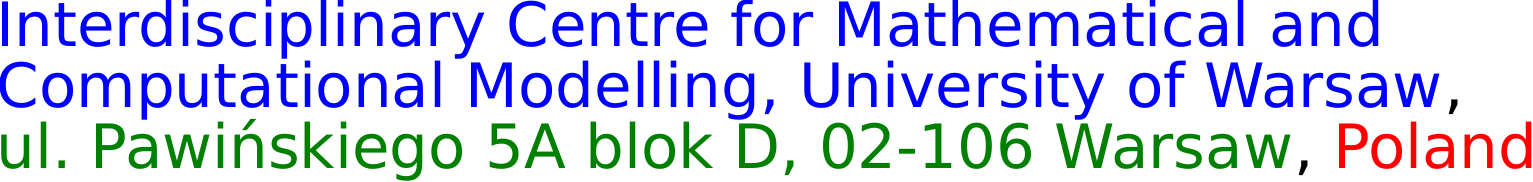}
  \caption[An example of a parsed affiliation string]{An example of a parsed affiliation string.
  Colors mark fragments related to {\it institution}, {\it address} and {\it country}.}
  \label{fig:parsed-example}
\end{figure}

Extracted affiliation strings are the input to affiliation parsing step~\cite{TkaczykTB15}, the goal
of which is to recognize affiliation fragments related to {\it institution}, {\it address} and {\it 
country}. Additionally, country names are decorated with their ISO codes. 
Figure~\ref{fig:parsed-example} shows an example of a parsed affiliation string.

More formally, let $\Sigma_D$ be the alphabet used in the document $D$ and $s \in \Sigma_D^*$ --- 
the non-empty affiliation string. Let's also denote as $S$ a set of all (possibly empty) substrings
of $s$:

\[ S = \{w \in \Sigma_D^* \enskip | \enskip \exists_{w_1, w_2 \in \Sigma_D^*} \enskip w_1ww_2 = s\}
\]

The goal of affiliation parsing is to find:
\begin{itemize}
\item $\inst(s) \in S$ --- the name of the institution,
\item $\addr(s) \in S$ --- the address of the institution,
\item $\coun(s) \in S$ --- the name of the country,
\end{itemize}

such that $\inst(s)$, $\addr(s)$ and $\coun(s)$ are pairwise non-overlapping substrings.

The first step of affiliation parsing is tokenization. The input string $s$ is divided into a list
of tokens $t_1, ..., t_k$, such that $s = t_1t_2...t_{k-1}t_k$, and each $t_i \in S$ is a maximum
continuous substring containing only letters, only digits or a single other character.

After tokenization each token is classified as {\it institution}, {\it address}, {\it country} or
{\it other}. The classification is done by a linear-chain Conditional Random Fields classifier, 
which is a state-of-the-art technique for sequence classification able to model sequential 
relationships and handle a lot of overlapping features.

The classifier uses the following binary features:
\begin{itemize}
\item WORD --- every word (the token itself) corresponds to a feature.
\item RARE --- whether the word is rare, that is whether the training set contains less than a
predefined threshold occurrences or it.
\item NUMBER --- whether the token is a number.
\item ALLUPPER --- whether it is all uppercase word.
\item ALLLOWER --- whether it is all lowercase word.
\item STARTUPPER --- whether it is a lowercase word that starts with an uppercase letter.
\item COUNTRY --- whether the token is contained in the dictionary of country words.
\item INSTITUTION --- whether the token is contained in the dictionary of institution words.
\item ADDRESS --- whether the token is contained in the dictionary of address words.
\end{itemize}

The dictionaries were compiled by hand using the resources from~\cite{Jonnalagadda11}. All the 
features exists in five versions: for the current token, for the two preceding tokens, and for the 
two following tokens.

After the classification the neighbouring tokens with the same label are concatenated. The resulting  
$\inst(s)$, $\addr(s)$ and $\coun(s)$ are the first occurrences of substrings labelled accordingly. 
Theoretically, the affiliation can contain multiple fragments of a certain label; in practice,
however, as a result of the training data we used, one affiliation contains usually at most one
substring of each kind: {\it institution}, {\it address} and {\it country}.

\subsection{Metadata Cleaning}
\label{sec:cleaning}
The purpose of the final step of metadata extraction stage is to gather the information from 
labelled zones, extracted author names, parsed affiliations and relations between them, clean the 
metadata and export the final record.

The cleaning is done with a set of heuristic-based rules. The algorithm performs the following
operations:
\begin{itemize}
\item removing the ligatures from the text,
\item concatenating zones labelled as {\it abstract},
\item removing hyphenation from the abstract based on regular expressions,
\item as type is often placed just above the title, it is removed from the {\it title} zone if 
needed (based on a small dictionary of types),
\item extracting email addresses from {\it correspondence} and {\it affiliation} zones using regular
expressions,
\item associating email addresses with authors based on author names,
\item pages ranges placed directly in {\it bib\_info} zones are parsed using regular expressions, 
\item if there is no pages range given explicitly in the document, we also try to retrieve it from
the pages numbers on each page,
\item parsing dates using regular expressions, 
\item journal, volume, issue and DOI are extracted from {\it bib\_info} zones based on regular
expressions.
\end{itemize}

Metadata cleaning is the final step of the metadata extraction stage. It results in the final 
metadata record of the document, containing the proper document metadata and exported as {\it front}  
section of the resulting NLM JATS file.

\section{Bibliography Extraction}
\label{sec:bibliography}
Bibliography extraction is next to metadata extraction another specialized extraction stage of the 
algorithm. During bibliography extraction, zones labelled previously as {\it references} are 
analyzed in order to extract parsed bibliographic references listed in the document.

The result of bibliography extraction is a list of bibliographic references, each of which is a 
tuple that can contain the following information:
\begin{itemize}
\item raw reference (string): raw text of the reference, as it was given in the input document,
\item type (string): type of the referenced document; possible values are: {\it journal paper}, {\it 
conference paper}, {\it technical report},
\item title (string): the title of the referenced document,
\item authors (a list of pairs of given name and surname): the full names of all the authors,
\item source (string): the name of the journal in which the article was published or the name of the 
conference,
\item volume (string): the volume in which the article was published,
\item issue (string): the issue in which the article was published,
\item year (string): the year of publication,
\item pages (a pair of first and last page): the range of pages of the article,
\item DOI (string): DOI identifier of the referenced document.
\end{itemize}

Each reference on the output contains the raw text and type; other information is optional. The 
output of bibliography extraction corresponds to the {\it back} section of the resulting NLM JATS
record. Listing~\ref{lst:metadata} shows an example of such a section.

\begin{lstlisting}[caption=Example document bibliography in NLM JATS format,label=lst:metadata]
<article>
  <back>
    <ref-list>
      <ref>
        <mixed-citation>[1]
          <string-name>
            <given-names>E.</given-names>
            <surname>Braunwald</surname>
          </string-name>, 
          <article-title>Shattuck lecture: cardiovascular medicine at the turn of the millennium: triumphs, concerns, and opportunities</article-title>,
          <source>New England Journal of Medicine</source>,
          vol. <volume>337</volume>, 
          no. <issue>19</issue>, 
          pp. <fpage>1360</fpage>-<lpage>1369</lpage>,
          <year>1997</year>.
        </mixed-citation>
      </ref>
      <ref>
        <mixed-citation>[2]
          <string-name>
            <given-names>R. L.</given-names>
            <surname>Campbell</surname>
          </string-name>,
          <string-name>
            <given-names>R.</given-names>
            <surname>Banner</surname>
          </string-name>,
          <string-name>
            <given-names>J.</given-names>
            <surname>Konick-McMahan</surname>
          </string-name>, and
          <string-name>
            <given-names>M. D.</given-names>
            <surname>Naylor</surname>
          </string-name>,
          <article-title>Discharge planning and home follow-up of the elderly patient with heart failure</article-title>, 
          <source>The Nursing Clinics of North America</source>,
          vol. <volume>33</volume>,
          no. <issue>3</issue>,
          pp. <fpage>497</fpage>-<lpage>513</lpage>,
          <year>1998</year>.
        </mixed-citation>
      </ref>
      <ref>
        <mixed-citation>[3]
          <string-name>
            <given-names>S. A.</given-names>
            <surname>Murray</surname>
          </string-name>,
          <string-name>
            <given-names>K.</given-names>
            <surname>Boyd</surname>
          </string-name>,
          <string-name>
            <given-names>M.</given-names>
            <surname>Kendall</surname>
          </string-name>,
          <string-name>
            <given-names>A.</given-names>
            <surname>Worth</surname>
          </string-name>,
          <string-name>
            <given-names>T. F.</given-names>
            <surname>Benton</surname>
          </string-name>, and
          <string-name>
            <given-names>H.</given-names>
            <surname>Clausen</surname>
          </string-name>,
          <article-title>Dying of lung cancer or cardiac failure: prospective qualitative interview study of patients and their carers in the community</article-title>,
          <source>British Medical Journal</source>,
          vol. <volume>325</volume>,
          no. <issue>7370</issue>,
          pp. <fpage>929</fpage>-<lpage>932</lpage>,
          <year>2002</year>.
        </mixed-citation>
      </ref>
    </ref-list>
  </back>
</article>
\end{lstlisting}

\begin{table}[h]
\renewcommand{\arraystretch}{1.4}
\renewcommand{\tabcolsep}{3pt}
\centering
\begin{tabular}{ | m{180px} | m{185px} | m{65px} | }
	\hline
    \multicolumn{1}{|c|}{Step} & \multicolumn{1}{c|}{Goal} &
    \multicolumn{1}{c|}{Implementation} \\ \hline\hline
    	
	1. {\bf Reference strings extraction} & Dividing the content of {\it references} zones into
	 individual reference strings. & {\bf k-means clustering}\\ \hline
	 
	2. {\bf Reference parsing} & Extracting metadata information from references strings. & 
	{\bf CRF}\\\hline
	 
	3. {\bf Reference cleaning} & Cleaning and exporting the final record. & {\bf 
	heuristics}\\\hline  
	 
  \end{tabular}
\caption[The decomposition of bibliography extraction]{The decomposition of bibliography extraction
stage into independent steps.}
\label{tab:bibliography}
\end{table}

Table~\ref{tab:bibliography} lists the steps executed during bibliography extraction stage. The
detailed descriptions are provided in the following sections: reference extraction 
(Section~\ref{sec:refs-extr}), reference parsing (Section~\ref{sec:refs-parsing}) and reference
cleaning (Section~\ref{sec:refs-cleaning}).

\subsection{References Extraction}
\label{sec:refs-extr}
Zones labelled as {\it references} by category classifier contain a list of reference strings, each
of which can span over one or more text lines. The goal of reference strings extraction is to split
the content of those zones into individual reference strings.

Let's denote as $Z_{DR}$ the set of all zones in the document labelled as {\it references}:

\[ Z_{DR} = \{z \in Z_D \enskip | \enskip \class(z) = references\} \]

Let also $L_{ZR}$ be the set of all lines from the references zones:

\[ L_{ZR} = \bigcup Z_{DR} \]

The goal of reference extraction is to find a partition $R = P(L_{ZR})$ such that each set $r \in 
R$ along with the order inherited from the set $L_{ZR}$ represents a single reference string. Let's 
denote as $\rf(l) \in R$ the reference of a given line $l \in L_{ZR}$, that is $r = \rf(l) 
\Leftrightarrow l \in r$. The partition $R$ should respect the reading order in the line set, that
is

\[ \forall_{l_1,l_2,l_3 \in L_{ZR}} \quad \big(l_1 \leq l_2 \leq l_3 \enskip \wedge \enskip \rf(l_1) 
= \rf(l_3) \enskip \Rightarrow \enskip \rf(l_1) = \rf(l_2)\big) \]

\begin{figure*}[h]
  \centering
  \includegraphics[width=.95\textwidth]{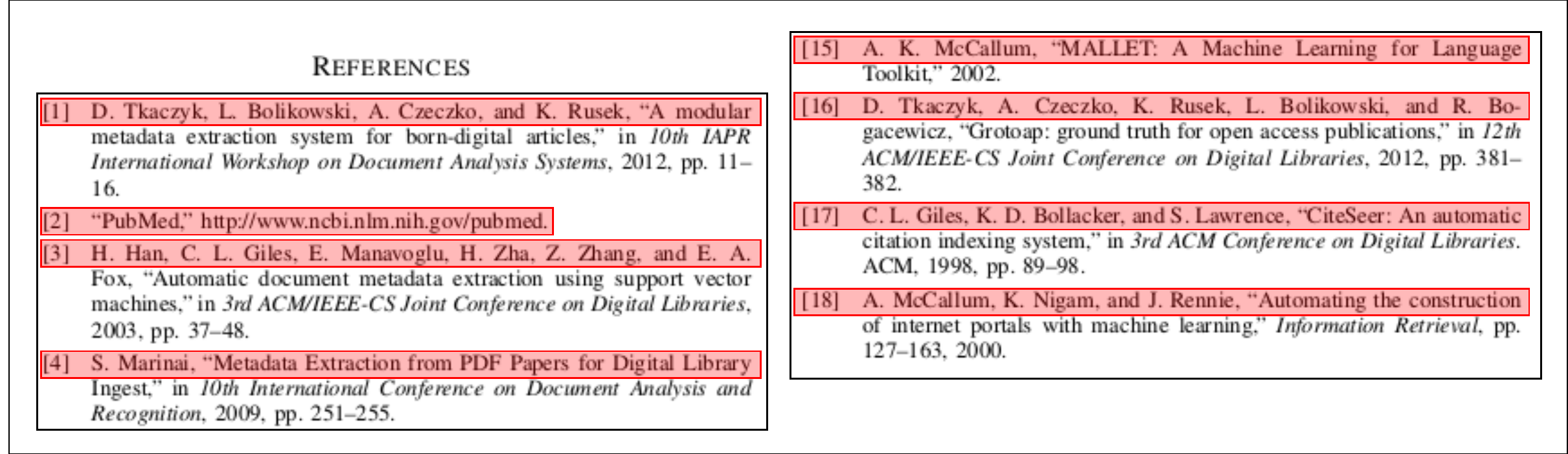}
  \caption[The references section with marked first lines of the references]{A fragment of the
  references section of an article. Marked lines are the first lines of their references. After
  detecting these lines, the references section content can be easily split to form consecutive
  references strings.}
  \label{fig:ref-extr}
\end{figure*}

Each line $l \in L_{ZR}$ belongs to exactly one reference string, some of them are first lines of
their reference, others are inner or last ones. The sequence of all text lines belonging to
bibliography section can be represented by the following regular expression:
\begin{Verbatim}[samepage=true]
(
  <first line of a reference>
  (
    <inner line of a reference>*
    <last line of a reference>
  )?
)*
\end{Verbatim}

The task of grouping text lines into consecutive references can be solved by determining which lines
are the first lines of their references. A set of such lines is shown in Figure~\ref{fig:ref-extr}.
More formally, we are interested in finding a set $FL_{ZR} \subset L_{ZR}$, such that

\[ |FL_{ZR}| = |R| \quad \wedge \quad \bigg( l \in FL_{ZR} \quad \Leftrightarrow \quad \Big( 
\forall_{p \in L_{ZR}} \quad \big( p < l \enskip \Rightarrow \enskip \rf(p) \neq \rf(l) \big) \Big) 
\bigg) \]

Finding the set $FL_{ZR}$ is equivalent to finding the partition $R$, since every set $r \in R$ can 
be constructed by taking a first line $l \in FL_{ZR}$ and adding all the following lines until the 
next first line or the end of the line sequence is reached.

The pseudocode of the algorithm is presented in Algorithm~\ref{alg:refextr}. To find the set
$FL_{ZR}$, we transform all lines to feature vectors and cluster them into two disjoint subsets.
Ideally one of them is the set of all first lines ($FL_{ZR}$) and the other is equal to $L_{ZR} -
FL_{ZR}$. The cluster containing the first line in $L_{ZR}$ (the smallest with respect to the order)
is assumed to be equal to $FL_{ZR}$.

\begin{algorithm}[H]
\caption[Reference strings extraction]{Reference strings extraction algorithm}
\label{alg:refextr}
\begin{algorithmic}[1]
\Function{ExtractReferenceStrings}{refLines}
\State refLinesFV $\gets$ {\sc ExtractFeatures}(refLines) \Comment{extract features for the lines}
\State clusters $\gets$ {\sc KMeans}(refLinesFV, 2) \Comment{cluster lines into two subsets}
\State firstCluster $\gets$ clusters[0]
\State refStrings $\gets ()$ \Comment{resulting reference strings list}
\State actString $\gets ()$ \Comment{current reference}
\State {\sc Add}(actString, refLines[0])
\For{i: 1 to {\sc Length}(refLines)-1} 
\If{firstCluster = clusters[i]} \Comment{current reference is completed}
\State {\sc Add}(refStrings, {\sc Concatenate}(actString))
\State actString $\gets ()$
\EndIf
\State {\sc Add}(actString, refLines[i])
\EndFor
\State {\sc Add}(refStrings, {\sc Concatenate}(actString))
\State \Return refStrings
\EndFunction
\end{algorithmic}
\end{algorithm}

For clustering lines we use k-means algorithm with Euclidean distance metric. In this case $k = 2$, 
since the line set is clustered into two subsets. As initial centroids we set the first line's 
feature vector and the vector with the largest distance to the first one. We use the following 
features:
\begin{itemize}
\item whether the line starts with an enumeration pattern --- this feature activates only if there 
exists a preceding line with the same pattern, but labelled with the previous number, and if there
exists a following line with the same pattern, but labelled with the next number,
\item whether the previous line ends with a dot,
\item the ratio of the length of the previous line to the width of the previous line's zone,
\item whether the indentation of the current line within its zone is above a certain threshold,
\item whether the vertical distance between the line and the previous one is above a certain 
threshold (calculated based on the minimum distance between references lines in the document).
\end{itemize}

The result of the references extraction step is a list of bibliographic references in the form of 
raw strings, that undergo parsing in the next step.

\subsection{References Parsing}
\label{sec:refs-parsing}
\begin{figure}[h]
  \centering
  \includegraphics[width=0.7\textwidth]{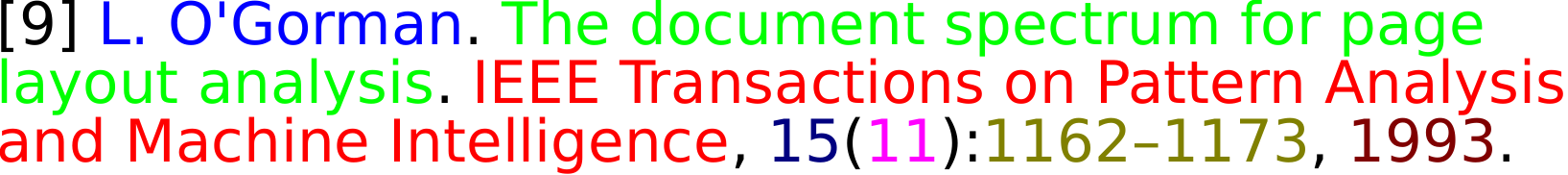}
  \caption[An example of a parsed bibliographic reference]{An example of a bibliographic reference
  with various metadata information highlighted using different colors, these are in order: {\it 
  author}, {\it title}, {\it journal}, {\it volume}, {\it issue}, {\it pages} and {\it year}.}
  \label{fig:reference}
\end{figure}

Reference strings extracted previously contain important reference metadata. During parsing metadata
is extracted from reference strings and the result is the list of document's parsed bibliographic
references. The information we extract from the strings include: {\it author} (including given name
and surname), {\it title}, {\it source}, {\it volume}, {\it issue}, {\it pages} (including the first
and last page number from a range) and {\it year}. An example of a parsed reference is shown in
Figure~\ref{fig:reference}.

Formally, the task can be defined similarly as the task of affiliation parsing described in 
Section~\ref{sec:aff-parsing}. The implementation is also similar: first a reference string is
tokenized into a sequence of tokens. The tokens are then transformed into vectors of features and
classified by a linear-chain CRF classifier. The classifiers differ in target labels and used 
features.

The token classifier uses the following token labels: {\it first\_name} (author's first name or
initial), {\it surname} (author's surname), {\it title}, {\it source} (journal or conference name),
{\it volume}, {\it issue}, {\it page\_first} (the lower bound of pages range), {\it page\_last} (the
upper bound of pages range), {\it year} and {\it text} (for separators and other tokens without a
specific label).

The main feature is the token (the word) itself. This feature activates only if the number of its
occurrences in the validation dataset exceeds a certain threshold. We also developed 34 additional 
binary features:
\begin{itemize}
\item features checking whether all characters in the token are: digits, letters, letters or digits,
lowercase letters, uppercase letters, Roman numerals;
\item whether the token starts with an uppercase letter;
\item whether the token is: a single digit, a lowercase letter, an uppercase letter;
\item whether the token is present in the dictionaries of: cities, publisher words, series words,
source words, number/issue words, pages words, volume words;
\item whether the token is: an opening/closing parenthesis, an opening/closing square bracket, a
comma, a dash, a dot, a quotation mark, a slash;
\item whether the token is equal to "and" or "\&";
\item whether the token is a dash placed between words;
\item whether the token is a single quote placed between words;
\item whether the token is a year.
\end{itemize}

It is worth to notice that the token's label depends not only on its feature vector, but also on the 
features of the surrounding tokens. To reflect this in the classifier, the token's feature vector
contains not only features of the token itself, but also features of two preceding and two following
tokens, similarly as in the case of the affiliation parser.

After token classification fragments labelled as {\it first\_\-name} and {\it surname} are joined
together based on their order to form consecutive authors, and similarly fragments labelled as {\it 
page\_first} and {\it page\_last} are joined together to form pages range. Additionally, in the case
of {\it title} or {\it source} labels, the neighbouring tokens with the same label are concatenated.

As a result of reference parsing step, we have a list of the document's bibliographic references,
each of which is a tuple containing the raw reference strings as well as the metadata extracted from
it.

\subsection{References Cleaning}
\label{sec:refs-cleaning}
Similarly to metadata cleaning, references cleaning is the last step of the bibliography extraction 
stage. Its purpose is to clean previously extracted data and export the final record.

During references cleaning the following operations are performed:
\begin{itemize}
\item The ligatures are removed from the text.
\item Hyphenation is removed from the strings based on regular expressions.
\item DOI is recognized in the reference strings by a regular expression. The reference parser is
not responsible for extracting this information, because the dataset used for training the token
classifier does not contain enough references with DOI.
\item Finally, the type of the reference ({\it journal paper}, {\it conference proceedings} or {\it 
technical report}) is detected by searching for specific keywords in the reference string.
\end{itemize}

Reference cleaning is the last step of bibliography extraction. The entire stage results in a list
of parsed bibliographic references, corresponding to the {\it back} section of the output NLM JATS
record.

\section{Structured Body Extraction}
\label{sec:body}
Structured body extraction is, next to metadata extraction and bibliography extraction, another
specialized extraction stage of the algorithm. The purpose of structured body extraction is to 
obtain the main text of the document in the hierarchical form composed of sections, subsections and
subsubsections by the analysis of the middle region of the document labelled previously as {\it 
body}.

Intuitively, the result of structured body extraction is the full text of the document represented
by a list of sections, each of which might contain a list of subsections, each of which might 
contain a list of subsubsections. Each structure part (section, subsection and subsubsection) has 
the title and the text content.

More formally, for a given document $D$ we denote as $S_D$ the set of all structure parts. We have
$S_D = S^1_D \cup S^2_D \cup S^3_D$, where $S^1_D \neq \emptyset$ is a set of the sections of the
document, $S^2_D$ is a (possibly empty) set of subsections and $S^3_D$ is a (possibly empty) set of
subsubsections. The following statements are also true for the structured parts: 

\[ \forall_{S_1,S_2 \in \{S^1_D,S^2_D,S^3_D\}} \quad (S_1 \neq S_2 \enskip \Rightarrow \enskip S_1
\cap S_2 = \emptyset ) \]

\[ S^2_D = \emptyset \enskip \Rightarrow \enskip S^3_D = \emptyset \]

The hierarchical structure of the document parts is defined by a parent function $\prt : S^2_D \cup
S^3_D \to S^1_D \cup S^2_D$, which maps the elements to their parents in the structure, in 
particular:

\begin{itemize}
\item $\forall_{s \in S^2_D} \quad \prt(s) \in S^1_D$
\item $\forall_{s \in S^3_D} \quad \prt(s) \in S^2_D$
\end{itemize}

All the sets $S_D, S^1_D$, $S^2_D$ and $S^3_D$ are totally ordered sets, where the order corresponds 
to the natural reading order of the parts of the document. The order of the elements also respects
the section hierarchy, in particular:

\[ \forall_{S \in \{S^2_D,S^3_D\}} \quad \forall_{s_1, s_2 \in S} \quad \big( s_1 \leq s_2 \enskip
\Rightarrow \enskip \prt(s_1) \leq \prt(s_2) \big) \]

Every structure part $s \in S_D$ has its title $\T(s) \in \Sigma_D^*$ and the text content $\C(s)
\in \Sigma_D^*$. The text content is understood as the text associated directly with the given
element, in particular the text contents of the children of a given element are not part of its text
content; in order to obtain the full content of a given element one has to recursively iterate over
its descendants and concatenate their contents. The text content of every element precedes the text
content of its descendants with respect to the document's reading order.

The output of body extraction corresponds to the {\it body} section of the resulting NLM JATS
record. Listing~\ref{lst:body} shows an example of such a section. The paragraphs are shortened for
conciseness.

\begin{lstlisting}[caption=Example document body in NLM JATS format,label=lst:body]
<article>
  <body>
    <sec>
      <title>Introduction</title>
      <sec>
        <title>Challenges</title>
        <p>In the West, carers of children with disabilities...</p>
        <p>Existing research has shown that carers of children...</p>
        <p>...</p>
      </sec>
      <sec>
        <title>Coping strategies</title>
        <p>Evidence suggests that...</p>
      </sec>
    </sec>
    <sec>
      <title>Design and methodology</title>
      <sec>
        <title>Study design</title>
        <p>We employed a qualitative phenomenological approach...</p>
      </sec>
      <sec>
        <title>Sample size determination and sampling procedure</title>
        <p>Children for the study were selected from 104 children...</p>
      </sec>
      <sec>
        <title>Development of research tools</title>
        <p>A checklist of questions was developed...</p>
      </sec>
      <sec>
        <title>Methods of data collection</title>
        <sec>
          <title>In-depth interviews</title>
          <p>In-depth interviews were conducted...</p>
        </sec>
      </sec>
      <sec>
        <title>Observations</title>
        <p>Passive observations consisted of systematic watching...</p>
      </sec>
      <sec>
        <title>Data analysis</title>
        <p>Inductive analysis as described by...</p>
        <p>Data triangulation from the interviews...</p>
      </sec>
      <sec>
        <title>Ethical consideration</title>
        <p>The study was approved by...</p>
      </sec>
    </sec>
  </body>
</article>
\end{lstlisting}

\begin{table}[h]
\renewcommand{\arraystretch}{1.4}
\renewcommand{\tabcolsep}{3pt}
\centering
\begin{tabular}{ | m{170px} | m{190px} | m{65px} | }
	\hline
    \multicolumn{1}{|c|}{Step} & \multicolumn{1}{c|}{Goal} &
    \multicolumn{1}{c|}{Implementation} \\ \hline\hline
    	
	1. {\bf Text content filtering} & Filtering out fragments related to the tables, images and
	equations from {\it 	body} parts of the document. & {\bf SVM}\\ \hline
	 
	2. {\bf Section headers detection} & Detecting the {\it body} lines containing the titles of
	sections, subsections and subsubsections. &
	 {\bf heuristics}\\\hline
	 
	3. {\bf Section hierarchy determination} & Dividing the section headers into levels and 
	building the section hierarchy. & {\bf heuristic clustering}\\\hline  
	
	4. {\bf Structured body cleaning} & Cleaning and exporting the final structured body content. & 
	{\bf heuristics}\\\hline  
	 
  \end{tabular}
\caption[The decomposition of body extraction]{The decomposition of body extraction stage into
independent steps.}
\label{tab:body}
\end{table}

Table~\ref{tab:body} lists the steps executed during body extraction stage. The detailed 
descriptions are provided in the following sections: text content filtering 
(Section~\ref{sec:text-filtering}), section headers detection (Section~\ref{sec:toc-headers}), 
section hierarchy determination (Section~\ref{sec:toc-hierarchy}) and structured body cleaning 
(Section~\ref{sec:toc-cleaning}).

\subsection{Text Content Filtering}
\label{sec:text-filtering}
Text content filtering is the first step in the body extraction stage. The purpose of this step is 
to locate all the relevant (containing section titles and paragraphs) parts in the body of the
document. The task is accomplished by classifying the {\it body} zones into one of the two classes:
{\it body\_content} (the parts we are interested in) and {\it body\_other} (all non-relevant
fragments, such as tables, table captions, the text belonging to images, image captions, equations,
etc).

More formally, the goal of text filtering is to find a function

\[ \class_B: Z_{DB} \to L_B \]

where

\[ Z_{DB} = \{z \in Z_D \enskip | \enskip \class(z) = body\} \]
\begin{align*}
L_B = \{body\_content, body\_other\}
\end{align*}

The classifier is based on Support Vector Machines and is implemented in a similar way as category
and metadata classifiers. It differs from them in target zone labels, the features and SVM
parameters used. The features, as well as SVM parameters were selected using the same procedure as
before, described in Sections~\ref{sec:cont-class-features} and~\ref{sec:cont-class-pars}. The final
body classifier contains 63 features capturing both the geometric and textual characteristics of 
classified zones.

As a result of text content filtering the set of zones labelled as {\it body} is split into two 
groups: the fragments containing section titles and paragraphs, which are further analyzed in the
following steps, and the non-relevant fragments like tables and figures, which are ignored.

\subsection{Section Headers Detection}
\label{sec:toc-headers}
The goal of section headers detection is to find all the headers in the {\it body\_content} zones,
that is lines containing the titles of the sections, subsections and subsubsections of the document.

Let's denote as $Z_{DC}$ the set of all zones in the document $D$ labelled as {\it body\_content}:

\[ Z_{DC} = \{z \in Z_D \enskip | \enskip \class(z) = body \land \class_B(z) = body\_content\} \]

Let also $L_{ZC}$ be the set of all lines from the {\it body\_content} zones:

\[ L_{ZC} = \bigcup Z_{DC} \]

Every title of a section, subsection or subsubsection spans over one or more subsequent body content
lines. In this step we are interested in finding a set of all headers $H_D = \{L_1, L_2, ..., 
L_{|S_D|}\}$ such that
\begin{itemize}
\item every header $L_i \subset L_{ZC}$, $L_i \neq \emptyset$ is a non-empty set of {\it 
body\_content} lines,
\item the headers are pairwise disjoint: 

\[ \forall_{L_i, L_j \in H_D} \quad ( L_i \neq L_j \enskip \Rightarrow \enskip L_i \cap L_j =
\emptyset ) \]

\item every header is a continuous sequence of lines:

\[ \forall_{L_i \in H_D} \quad \forall_{l_1,l_2,l_3 \in L_{ZC}} \quad ( l_1 \in L_i \enskip \land 
\enskip l_3 \in L_i \enskip \land \enskip l_1 \leq l_2 \leq l_3 \enskip \Rightarrow \enskip l_2 \in 
L_i )\]
\end{itemize}

\begin{figure}
  \centering
  \includegraphics[width=0.9\textwidth]{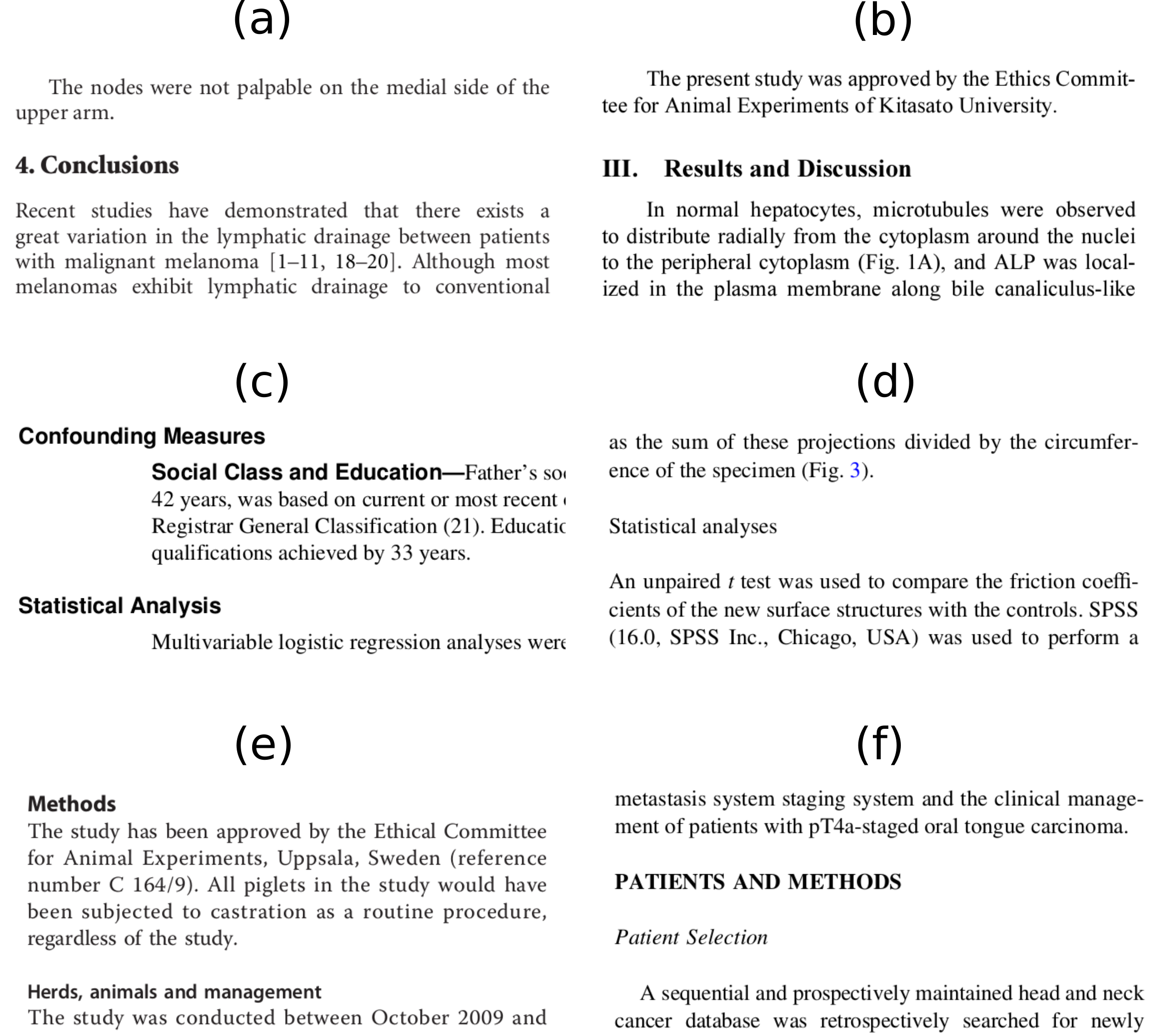}
  \caption[Examples of headers in different layouts]{The examples of headers in different layouts.
  The header lines are always different from the paragraph lines in some way: (a) with the use of
  enumerations, (b) another type of enumeration, (c) the indentation, (d) the distance between
  lines, (e) and (f) the fonts.}
  \label{fig:head-ex}
\end{figure}

Regardless of the document layout the header lines are always different in some way from much more
numerous paragraph lines. In different layouts different ways to make header lines stand out are
used; for example they can differ in the fonts, the size of the text, the enumeration patterns, the
indentation or distance between lines, or any combination of these features. Some examples of header
lines in different layouts are shown in Figure~\ref{fig:head-ex}.

Since the header lines are always different in some way from the text used in the document's
paragraphs, and also far less common among the {\it body\_content} lines, we are in fact looking
for outliers in the set $L_{ZC}$. To make use of this observation, we analyze the entire populations
of various feature values of the lines in $L_{ZC}$ for a given input document. For a single line and
a specific feature calculated for the line we can obtain the standard score of this observation with
respect to the entire population we are dealing with:

\[ z = \frac{x-\mu}{\sigma} \]

where:
\begin{itemize}
\item $x$ is the value of the feature of interest calculated for the current line,
\item $\mu$ is the mean of the feature values over the population of $L_{ZC}$,
\item $\sigma$ is the standard deviation of the feature values in the population.
\end{itemize}

The closer $z$ is to $0$, the more typical the line is with respect to the given feature. We use
standard scores for the following features to detect the outliers in the populations:
\begin{itemize}
\item $z_h$ --- line height,
\item $z_l$ --- line length,
\item $z_x$ --- the x-coordinate (the distance to the left edge of the page),
\item $z_d$ --- the distance from the previous line,
\item $z_f$ --- the font (fonts are encoded as subsequent natural numbers).
\end{itemize}

Unfortunately, although the standard scores proved to be very useful, they are not enough to detect
the header lines with high precision. In practise we often have to deal with errors caused by the 
classification in the previous step, which results in small fragments of tables, images or captions
still present in the set $L_{ZC}$. Since these fragments also visually differ from the paragraphs,
they often appear as outliers as well. To counteract this we employed additional heuristics to
filter out such lines.

In general the algorithm consists of two phases:
\begin{enumerate}
\item In the first phase we detect the first line of every header. More formally, we are interested
in finding a set $FH_D = \{l_1, l_2, ..., l_{|S_D|}\}$, such that

\[ \forall_{1 \leq i \leq |S_D|} \quad ( l_i \in L_i \quad \land \quad \forall_{s \in L_i} \enskip
l_i \leq s ) \]

\item In the second phase we find the remaining header lines, that is the set $\bigcup H_D - FH_D$.
\end{enumerate}

The pseudocode of the first phase is presented in Algorithm~\ref{alg:fheader}. In the first phase we
perform the following operations:
\begin{enumerate}
\item First, we iterate over all lines from the set $L_{ZC}$ and select the candidates for the first 
header lines. A line becomes a candidate if it is the first line in its zone and if it starts with
an uppercase letter or a typical enumeration pattern (lines 3-7 in Algorithm~\ref{alg:fheader}).
\item Next, we remove candidates that do not meet certain criteria (lines 8-12 in 
Algorithm~\ref{alg:fheader}). A line is removed from the candidate set if at least one of the
following conditions is true:
\begin{itemize}
\item the line is too long (large $z_l$)
\item the text size is too small (small $z_h$)
\item the line is printed using a typical font, typical distance from the previous line and X
coordinate ($z_f$, $z_d$ and $z_x$ all close to $0$),
\item the text of the line contains a pattern characteristic for table or figure caption,
\item the text of the line contains characters typical for equations,
\item less than half of the characters in the line are letters,
\item the text of the line does not contain any 4-character continuous sequence of letters,
\item none of the five lines following the examined line starts with an uppercase letter.
\end{itemize}
\item Next, using the candidate set we identify a set of fonts typical for headers in a given
document (lines 13-19 in Algorithm~\ref{alg:fheader}). A font is considered typical for headers if
it appears in at least 3 candidate lines and its $|z_f|$ exceeds a certain threshold, which means it
is an outlier with respect to the fonts. If headers are printed using the same font as the
paragraphs, the font information cannot be safely used to extend the candidate set.
\item Next, we use the set of header fonts to extend the candidate set (lines 20-24 in
Algorithm~\ref{alg:fheader}). We iterate again over all relevant lines and mark them as candidates
if they use one of the header fonts and if they start with an uppercase letter or a typical
enumeration pattern.
\item Finally, we delete candidate lines using similar criteria as in step 3, only this time the
thresholds for normal scores are more tolerant (lines 25-29 in Algorithm~\ref{alg:fheader}).
\end{enumerate}

\begin{algorithm}[H]
\caption[First header lines extraction]{First header lines extraction algorithm}
\label{alg:fheader}
\begin{algorithmic}[1]
\Function{ExtractFirstHeaderLines}{bodyLines}
\State candidates $\gets \emptyset$
\For{line $\in$ bodyLines}
\If{{\sc IsFirstInZone}(line) and {\sc MatchesHeaderRegexp}(line)}
\State {\sc Add}(candidates, line)
\EndIf
\EndFor
\For{candidate $\in$ candidates}
\If{{\sc ViolatesHeaderCriteriaStrict}(candidate)}
\State {\sc Delete}(candidates, candidate)
\EndIf
\EndFor
\State headerFonts $\gets \emptyset$
\For{candidate $\in$ candidates}
\State font $\gets$ {\sc Font}(candidate)
\If{{\sc NumberOfLinesWithFont}(font, candidates) $\geq 3$ and {\sc IsOutlier}(font)}
\State {\sc Add}(headerFonts, font)
\EndIf
\EndFor
\For{line $\in$ bodyLines}
\If{{\sc Contains}(headerFonts, {\sc Font}(line)) and {\sc MatchesHeaderRegexp}(line)}
\State {\sc Add}(candidates, line)
\EndIf
\EndFor
\For{candidate $\in$ candidates}
\If{{\sc ViolatesHeaderCriteriaSoft}(candidate)}
\State {\sc Delete}(candidates, candidate)
\EndIf
\EndFor
\State\Return candidates
\EndFunction
\end{algorithmic}
\end{algorithm}

\begin{figure}
  \centering
  \includegraphics[width=0.6\textwidth]{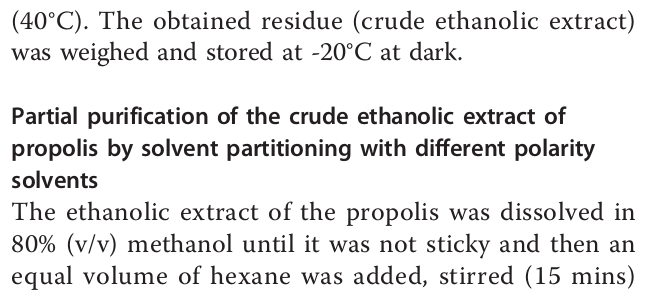}
  \caption[An example of a multi-line header]{An example of a multi-line header consisting of three
  subsequent lines.}
  \label{fig:head-multi}
\end{figure}

This approach allows to identify only the first line of each header, represented by the set $FH_D$,
which corresponds to the target set $H_D$. The sets might not be equal, as a header can span over
more lines (an example is presented in Figure~\ref{fig:head-multi}). In the last phase we examine
each first line and select additional lines following it, if needed.

To detect additional header lines, we inspect the lines directly following every first header line
found in the previous phase. The pseudocode of the algorithm is presented in 
Algorithm~\ref{alg:addheader}. For every first header line we execute the following steps:
\begin{enumerate}
\item First we select the candidates for additional lines. We iterate over the lines directly 
following the main header line. A line is consider a candidate if its height is similar to the 
height of the main header line and if it is printed using the same font. We stop the iteration if we
come across a line that does not meet these criteria, if we come across another first header line 
or if we already added three lines to the candidates list.
\item Selected candidates are added to the proper element in the set $H_D$ as additional lines if at
least two of the following conditions are met:
\begin{itemize}
\item the last candidate line is noticeably shorter than the main header line,
\item the last candidate is printed with a different indentation than the line following it,
\item the line following the last candidate starts with an uppercase,
\item the last candidate is written using a different font than the line following it.
\end{itemize}
\end{enumerate}

\begin{algorithm}[h]
\caption[Header lines extraction]{Header lines extraction algorithm}
\label{alg:addheader}
\begin{algorithmic}[1]
\Function{ExtractHeaders}{firstHeaderLines}
  \State headerLines $\gets \emptyset$
  \For{line $\in$ firstHeaderLines}
    \State header $\gets \emptyset$
    \State {\sc Add}(header, line)
    \State candidates $\gets \emptyset$
    \State actLine $\gets$ {\sc Next}(line)
    \While{actLine $\neq$ null and {\sc Size}(candidates) $\leq$ 3}
      \If{{\sc Font}(actLine) $\neq$ {\sc Font}(line)}
        \Break
      \EndIf
      \If{!{\sc AreSimilar}({\sc Height}(actLine), {\sc Height}(line)}
        \Break
      \EndIf
      \State {\sc Add}(candidates, actLine)
      \State actLine $\gets$ {\sc Next}(actLine)
    \EndWhile
    \If{{\sc EvaluateCandidates}(candidates)}
      \State {\sc AddAll}(header, candidates)
    \EndIf
    \State {\sc Add}(headersLines, header)
  \EndFor
  \State\Return headerLines
\EndFunction
\end{algorithmic}
\end{algorithm}

The resulting set of headers $H_D$ corresponds to $S_D$ --- the set of the sections, subsections and 
subsubsections of the document.

\subsection{Section Hierarchy Determination}
\label{sec:toc-hierarchy}
The purpose of section hierarchy determination is to reconstruct the hierarchy of sections from the
set $H_D$. More precisely, our goal is to divide the set $H_D$ into disjoint subsets $H^1_D$, 
$H^2_D$ and $H^3_D$ corresponding to the sets of sections, subsections and subsubsections of the
document, respectively. We are also interested in finding a parent function $\prt_H : H^2_D \cup
H^3_D \rightarrow H^1_D \cup H^2_D$, which maps subsections and subsubsections to their parents in
the hierarchy and corresponds to the function $\prt$.

To find the sets $H^1_D$, $H^2_D$ and $H^3_D$, we make use of an observation that the headers of
sections on the same hierarchy level are usually printed using a similar style. In the first step we
cluster the set of headers $H_D$ with the use of the following simple clustering algorithm:
\begin{enumerate}
\item We maintain a set of headers with no cluster assignment. At the beginning the set consists of
all headers from $H_D$.
\item We randomly choose a header from the set. Then we compare the first line of selected header to
the first lines of all the headers from the set and mark those which have the same font and similar
height. All the selected headers form a new cluster and are removed from the set.
\item This step is repeated until the set is empty.
\end{enumerate}

Let $CL_D = (cl_1, cl_2, ... cl_{|S_D|})$ be the sequence of clusters of the elements of the set 
$H_D = \{L_1, L_2, ..., L_{|S_D|}\}$ sorted with respect to the reading order of the headers. Since
we are interested only in three section levels (scientific publications rarely contain more levels),
we keep only the first three clusters appearing in the sequence $CL_D$. All the headers belonging to
the remaining clusters are deleted from the header set.

Next we divide $H_D$ into the sets $H^1_D$, $H^2_D$ and $H^3_D$ and determine the final hierarchy:
\begin{enumerate}
\item The set $H^1_D$ is equal to $cl_1$, the cluster of the first header $L_1$.
\item We then divide the sequence of the headers into continuous subsequences, such that each
subsequence starts with a header from $H^1_D$ and contains all the following headers until the next 
header of the first level (or the end of the sequence) is reached. These sequences represent the
sections with their descendants and are then processed independently.
\item If the header sequence contains more than one header, the cluster of the second element in the 
sequence is assumed to represent the second hierarchy level in this section. Thus we select all the
headers in the sequence belonging to this cluster and form subsections.
\item All the remaining lines in the sequence form subsubsections.
\item For every subsection and subsubsection we set the parent section to be the last preceding 
section of the higher level.
\end{enumerate}

The algorithm results in the division of the set $H_D$ into sets $H^1_D$, $H^2_D$ and $H^3_D$, which
along with the reading order and parent function corresponds directly to the desired sections,
subsections and subsubsections hierarchy of the document.

\subsection{Structured Body Cleaning}
\label{sec:toc-cleaning}
Body cleaning is the final step in the structured body extraction path of the algorithm. Its purpose
is to form the final section hierarchy along with the section titles and text content and clean the 
data.

The sections, subsections and subsubsections, and their hierarchy are formed based on the sets 
$H^1_D$, $H^2_D$ and $H^3_D$ and the parent function. The title of a content part is the 
concatenated text of all its header lines, in their reading order. Similarly, the text content of a
content part is the concatenated text of the paragraph lines placed between its header and the 
header of the next content part.

The cleaning phase comprises removing the ligatures and end-of-line hyphenation based on regular
expressions.

The result of the structured body extraction is a hierarchical structure of sections, subsections
and subsubsections of the document along with all the content part titles and the text content. The 
extracted structured content corresponds to the {\it body} section of the resulting NLM JATS record.

\section{Our Contributions}
The proposed extraction algorithm is based to a great extend on a well-known supervised and 
unsupervised machine-learning techniques accompanied with heuristics. This section summarizes all
the innovatory ideas and extensions we proposed.

One of the key contributions is the architecture of the entire extraction workflow and the
decomposition of the problem into smaller, well-defined tasks. We designed four processing paths:
an initial path containing preprocessing, layout analysis and initial classification, and three
specialized extraction paths analysing three separate document regions: metadata extraction,
structured body extraction and bibliography extraction path. Each path consists of several steps
executed in a sequence. Each step solves a single, well defined problem and its implementation is
independent of other workflow parts.

For the first workflow step, character extraction, iText library was used. Based on the observations
related to various rare cases and problems with PDF files, we added an additional cleaning step
(Section~\ref{sec:charextr}), the goal of which is to reduce the number of extracted characters by
removing duplicates, characters not visible in the resulting file, or characters used for objects
other than the text of the document.

The page segmentation algorithm, Docstrum, was enhanced with a few modifications related to
identifying words, merging lines, scaling the thresholds relatively to line heights, etc.
(Section~\ref{sec:segm}). The evaluation we performed showed our modifications increased the 
accuracy of recognizing lines and zones in the document (Section~\ref{sec:segm}).

We developed a large set of 103 numeric features for document's text zones capturing all aspects of
the content and appearance of the text and allowing to classify fragments with high accuracy
(Sections~\ref{sec:classification}, \ref{sec:metadata-classification} and~\ref{sec:text-filtering}).
The features were subjected to semi-automatic selection process separately for every classifier,
which reduced the dimensionality of the feature vectors and resulted in only the most useful 
features being included in the models.

We also developed a set of features for citation and affiliation tokens, which allow to parse
affiliations and citations with high accuracy (Sections~\ref{sec:aff-parsing} 
and~\ref{sec:refs-parsing}). These features are based purely on the text of the tokens and are
accompanied with dictionaries of words commonly appearing as particular metadata fields.

A clustering-based approach was proposed for extracting reference strings from the document. This 
method does not require expensive training set preparation and parameter learning, while still 
achieving very good results (Section~\ref{sec:refs-extr}).

We also proposed an algorithm based on normal scores of various statistics for selecting section
header lines from the text content of the document (Section~\ref{sec:toc-hierarchy}). This is based
on a simple observation that within one document numerous text paragraph lines have similar values 
of certain features, and thus the farther the line's feature value is from the document's mean, the
higher probability that as an outlier it is not a paragraph line.

Finally, we designed a clustering-based algorithm for retrieving the hierarchy of sections 
(Section~\ref{sec:toc-hierarchy}). The approach is based on an observation, that section titles of
the same hierarchy level are printed using similar style. As an unsupervised algorithm, it does not
require gathering training data and performing the training.

\section{Limitations}
The extraction algorithm described in this chapter has a number of limitations.

Currently the algorithm does not include any optical character recognition phase, it analyses only
the PDF text stream found in the input document. As a result, PDF documents containing scanned pages
in the form of images are not properly processed.

There is also another problem related to analysing the underlying PDF text stream rather than the
visible shapes printed on the document's pages. In some cases the underlying text might not be the
same as the visible one, for example due to objects covering other objects, or custom mappings 
between the characters and printed glyphs used in a given PDF. In such rare cases the information 
extracted by the algorithm might not match the text visible to the document readers.

The extraction algorithm analyses only the content of a single PDF file, and only the information
explicitly given in the document has a chance to be extracted. The algorithm does not infer any new
information based on the content, in particular it does not extract the language the document is
written in based on n-grams, compile the document summary by selecting the most important sentences
and phrases or extract the keywords from the text using statistical methods.

Finally, the extraction algorithm does not analyse some regions typically present in scientific 
articles, such as: tables, acknowledgments, conflict statements or authors' contributions.

\clearpage
\thispagestyle{empty}
\cleardoublepage

\chapter{Experiments and Evaluation}
\label{chap:evaluation}
This chapter contains all the details related to the experiments we performed on the algorithm and 
its individual steps, as well as the methodology and results of the evaluation used to assess the
quality of the extraction process and to compare it to other existing methods.

The data used for the experiments is based on the following resources: PubMed Central Open Access
Subset\footnote{http://www.ncbi.nlm.nih.gov/pubmed}, Directory of Open Access 
Journals\footnote{https://doaj.org/}, Elsevier\footnote{http://www.elsevier.com/} and Cora Citation
Matching dataset\footnote{http://people.cs.umass.edu/$\sim$mccallum/data.html}. All the datasets
used in the experiments were prepared semi-automatically.

The experiments we performed comprise the adjustment of the Support Vector Machine-based
classifiers, including semi-automatic feature selection and automatic best SVM parameters choosing
with the use of a validation dataset.

The most important part of the chapter is the methodology and the results of the evaluation of the
entire extraction algorithm. We evaluated the performance of the algorithm with respect to its 
entire functionality: extracting the basic document metadata, bibliography and structured text
content. The scores achieved by our algorithm were also compared to the performance of four similar
solutions.

We also performed the evaluation of the following individual steps of the algorithm: page 
segmentation, category classification, metadata classification, body classification, affiliation
parsing and reference parsing. Other steps were not directly evaluated, mainly due to the fact, that
creating ground truth datasets for them would be very time-consuming. Since all the steps affect the
final extraction results, they were all evaluated indirectly by the assessment of the performance of
the entire algorithm.

The chapter is organized as follows. Section~\ref{sec:data-sets} contains all the details related to
the datasets used for experiments as well as the procedures used for creating them. The following 
sections provide the information related to the methodology and the results of the evaluation of the
individual steps of the algorithm: page segmentation (Section~\ref{sec:segm}), content, metadata and
body classification (Section~\ref{sec:cont-class}), affiliation parsing 
(Section~\ref{sec:aff-parsing}) and reference parsing (Section~\ref{sec:ref-parsing}). In
Section~\ref{sec:extr-eval} we report the results of the evaluation of the entire extraction 
algorithm and the comparison with other similar tools. Finally Section~\ref{sec:time} covers the 
time performance issues of the extraction algorithm.

\section{Datasets Overview}
\label{sec:data-sets}
The following resources were used as a basis for the datasets used in all the experiments: PubMed
Central Open Access Subset, Directory of Open Access Journals, Elsevier and Cora Citation Matching.

Directory of Open Access Journals (DOAJ) is an online directory providing access to open access,
peer-reviewed journals. DOAJ contains nearly 2 million articles in the form of PDF files and the
corresponding metadata in Dublin Core format.

Elsevier is a publisher publishing over 250,000 articles a year in over 3,000 journals. 
Unfortunately, its resources are not open access and were available for our experiments based on a
dedicated license.

PubMed Central Open Access Subset (PMC) contains over 1 million life sciences publications in PDF
format, and their corresponding metadata in the form of NLM JATS files. NLM JATS files contain a 
rich set of document's metadata (title, authors, affiliations, abstract, journal name, etc.),
structured full text (sections, section titles, paragraphs, tables, equations), and also document's
parsed bibliography.

Cora Citation Matching dataset, commonly referred to as Cora-ref dataset, is a widely used set of
parsed bibliographic citations useful for evaluating citation parsing and matching solutions.

In case of most experiments these datasets could not be used directly, due to missing or fragmentary
information and the differences in the formats or labelling. We employed these resources as a basis
to construct our own datasets, which were then used in the experiments. 

\begin{table}[h]
\renewcommand{\arraystretch}{1.1}
\renewcommand{\tabcolsep}{3pt}
\centering
  \begin{tabular}{ | p{75px} | >{\raggedright}p{95px} | >{\raggedright}p{60px} | p{75px} | p{115px} 
  | }
	\hline
    \multicolumn{1}{|c|}{Type} & \multicolumn{1}{c|}{Contents} & \multicolumn{1}{c|}{Format} &
    \multicolumn{1}{c|}{Sources} & \multicolumn{1}{c|}{Purposes} \\ \hline\hline

    {\bf structure set} & geometric models of the documents & TrueViz & DOAJ, PMC & the evaluation 
    of the page segmentation, the training and the evaluation of the zone classifiers \\ \hline

	{\bf metadata set} & pairs of PDF files and corresponding metadata records & PDF + NLM JATS &
	PMC, Elsevier & the evaluation of the entire algorithm, the comparison with other similar tools
	\\\hline
    
    {\bf affiliation set} & affiliation strings with tagged metadata & NLM JATS & PMC & the 
    evaluation of the affiliation parser \\ \hline
	
	{\bf citation set} & citation strings with tagged metadata & NLM JATS & Cora-ref, PMC & the 
    evaluation of the citation parser \\ \hline
	
  \end{tabular}
\caption[The dataset types used for the experiments]{The summary of all the dataset types used for
the experiments.}
\label{tab:set-types}
\end{table}

Table~\ref{tab:set-types} summarizes all the dataset types used for the experiments. Structure 
datasets contain the documents in the form of serialized geometric models of their content along 
with the zone labels. Metadata datasets contain the pairs of the source PDF files and their
corresponding metadata records. The affiliation dataset contains the affiliation strings with tagged
metadata and similarly, the citation dataset contains the citation strings with tagged metadata. The
following sections provide more details related to the contents of the datasets, their purpose and
the process of creating them.

\subsection{Structure Sets}
\label{sec:str-sets}
A structure set contains the documents in the form of serialized geometric models of their content 
along with the zone labels. We use TrueViz~\cite{LeeK03} as a serialization format. Structure sets
were used for the evaluation of the page segmenter, and also for training and evaluation of the zone
classifiers. We use two structure datasets: GROTOAP~\cite{grotoap,TkaczykCRBB12} and
GROTOAP2~\cite{grotoap2,TkaczykSB14}, which were built in a different way and are used for slightly
different purposes.

\subsubsection{GROTOAP}
GROTOAP (GROund Truth for Open Access Publications)~\cite{grotoap,TkaczykCRBB12} is a structure set
built semi-automatically upon a group of scientific articles from DOAJ database. GROTOAP consists
of:
\begin{itemize}
\item 113 scientific publications in PDF format,
\item corresponding ground truth files in TrueViz format containing the geometric hierarchical
structure of the input documents along with their zone labels.
\end{itemize}

The process of creating GROTOAP dataset was semi-automatic and consisted of the following phases:
\begin{enumerate}
\item selecting and downloading a set of publications based on metadata from DOAJ database,
\item automatic extraction of hierarchical structure using our extraction tools,
\item manual correction of the results of automatic structure extraction with the use of a visual
editing tool.
\end{enumerate}

All the publications in the dataset were taken from journals distinguished with SPARC Europe Seal
for Open Access Journals. We harvested Directory of Open Access Journals for basic metadata of all
articles from these journals, including links to full texts. Next, we randomly selected articles to
be downloaded. The selected group contains one article from each four journals published by the same
publisher, 113 articles in total. Articles published by the same publisher have usually very similar
layout, and as a result the layout distribution in the group is similar to the layout distribution
in the entire DOAJ database.

To minimize the time needed for manual correction, first we processed selected publications using
tools implemented in our metadata extraction process in order to automatically extract their 
structure. First, individual characters and their bounding boxes were extracted, then we segmented
pages. Finally, zone classifiers assigned labels to all zones in the documents.

During the last phase TrueViz files created automatically were corrected by a group of people with 
the use of a visualization and editing tool. The correction phase included verifying words, lines 
and zones generated by metadata extraction tools, splitting incorrectly merged objects and merging 
incorrectly split ones, verifying and correcting labels assigned to zones. To minimize human errors
we performed an additional checking phase, which included an inspection and approval by an
independent judge.

The last two phases were executed twice. In the first round we used the default version of the
extraction tools on a very small subset of GROTOAP. This resulted in a substantial number of errors,
since the tools were not trained or adjusted in any way to perform well on the layouts present in 
the dataset. After the manual correction of these files, they were used to retrain and adjust the
classification algorithms. The new versions were then used in the second round on the remaining
documents, which resulted in much less errors and less time spent on the manual correction.

Since GROTOAP's creation process required a manual correction of every document, it does not scale
well. The resulting dataset is relatively small and every attempt to expand it is time-consuming and
expensive. The layout distribution in the dataset resembles the distribution in the DOAJ database, 
but its small size resulted in a lot of journals and layouts missing. Due to the small size and lack
of diversity, the algorithms trained on GROTOAP did not generalize well enough and performed worse 
on diverse sets. To solve this problem, we decided to use a different, scalable approach to create
much larger dataset, which resulted in GROTOAP2.

\subsubsection{GROTOAP2}
GROTOAP2~\cite{grotoap2,TkaczykSB14} is a successor of GROTOAP. GROTOAP2 is also a structure set, 
and can be used for the same purposes. The dataset is based on born-digital PDF documents from
PubMed Central Open Access Subset. In contrast to GROTOAP, the method used to construct GROTOAP2 is
scalable and efficient, which allowed to construct much larger and diverse dataset. 
Table~\ref{tab:grotoap-comparison} compares the basic size-related parameters of GROTOAP and 
GROTOAP2.

\begin{table}[h]
\begin{center}
\renewcommand{\tabcolsep}{3pt}
\renewcommand{\arraystretch}{1.2}
\begin{tabular}{| l | r | r |}
	\hline
	{\bf Number of} & {\bf GROTOAP} & {\bf GROTOAP2}\\ \hline
	\hspace{10pt}publishers & 12 & 208 \\ \hline
	\hspace{10pt}journals & 113 & 1,170 \\ \hline
	\hspace{10pt}documents & 113 & 13,210 \\ \hline
	\hspace{10pt}pages & 1,031 & 119,334 \\ \hline
	\hspace{10pt}zones & 20,121 & 1,640,973\\ \hline
	\hspace{10pt}zone labels & 20 & 22\\ \hline
\end{tabular}
\end{center}
\caption[The comparison of the parameters of GROTOAP and GROTOAP2 datasets]{The comparison of the
parameters of GROTOAP and GROTOAP2 datasets. The table shows the numbers of different publishers and
journals included in both datasets, as well as the numbers of documents, pages, zones and zone
labels.}
\label{tab:grotoap-comparison}
\end{table}

GROTOAP2 contains 13,210 scientific publications in the form of the geometric model of their content
serialized using TrueViz format. The final list of zone labels in GROTOAP2 includes: {\it abstract},
{\it acknowledgment}, {\it affiliation}, {\it author}, {\it bib\_info}, {\it body}, {\it 
conflict\_statement}, {\it copyright}, {\it correspondence}, {\it dates}, {\it editor}, {\it 
equation}, {\it figure}, {\it glossary}, {\it keywords}, {\it page\_number}, {\it references}, {\it 
table}, {\it title}, {\it title\_author}, {\it type}, {\it unknown}.

\begin{figure}[h]
  \centering
  \includegraphics[width=0.6\textwidth]{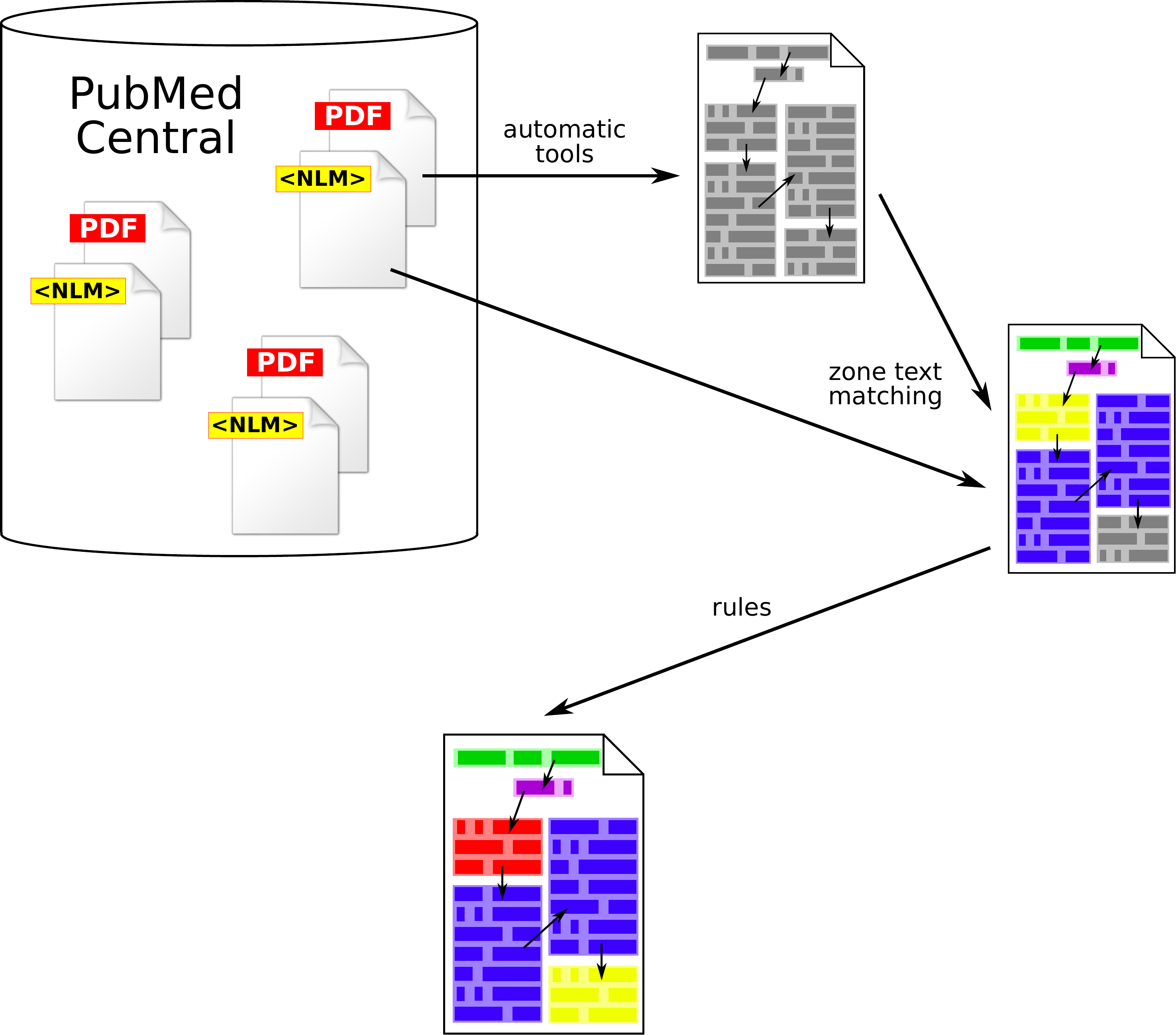}
  \caption[Semi-automatic method of creating GROTOAP2 dataset]{Semi-automatic method of creating
  GROTOAP2 dataset. First, automatic tools extracted the hierarchical geometric structure and the
  reading order of a document. Next, we automatically assigned labels to zones by matching their
  text to labelled fragments from NLM JATS files. Finally, additional rules were developed manually
  and applied to the dataset in order to increase the labelling accuracy.}
  \label{fig:grotoap2}
\end{figure}

GROTOAP2 was created semi-automatically from PMC resources (Figure~\ref{fig:grotoap2}). Since the
ground truth files in PMC (the metadata records in NLM JATS format) contain only the annotated text 
of documents, and do not preserve any geometric features related to the way the text is displayed in
the PDF files, they could not be directly used as a structure set. Instead we used the text 
annotations from the ground truth files to assign labels to zones automatically, while the zones, 
lines and words themselves were constructed using our extraction and segmentation tools. More 
precisely, GROTOAP2 was created in the following steps:
\begin{enumerate}
\item First, a large group of files were randomly selected from PMC resources. We obtained both the
input PDF files and the corresponding metadata records.
\item The PDF files were processed automatically by our algorithms in order to extract the 
hierarchical geometric structure and the reading order. More precisely, character extraction, page
segmentation and reading order resolving steps were executed.
\item The text content of every zone was then compared to labelled text from NLM JATS files with the
use of Smith-Watermann sequence alignment algorithm~\cite{Smith1981}. By selecting the annotated 
text fragment the most similar to the zone's text we were able to assign labels to the zones in the
structure.
\item Files with a lot of zones labelled as "unknown", that is zones, for which the labelling 
process was unable to find a label, were filtered out.
\item A small random sample of the remaining files was inspected manually. This resulted in 
identifying a number of repeated problems and errors in the dataset. 
\item Based on the results of the analysis, we developed a set of heuristic-based rules and applied
them to the entire dataset in order to increase the labelling accuracy.
\end{enumerate}

In the first phase of ground truth generation process we used our automatic algorithms. First, the
characters were extracted from PDF files. Then, the characters were grouped into words, words into
lines and finally lines into zones by the page segmenter. After that reading order analysis was
performed resulting in elements at each hierarchy level being stored in the order reflecting how
people read manuscripts.

In the second phase the text content of each zone extracted previously was matched against the 
labelled text fragments extracted from corresponding NLM JATS files. We used Smith-Watermann 
sequence alignment algorithm~\cite{Smith1981} to measure the similarity of two text strings. For
every zone, a string with the highest similarity score above a certain threshold was chosen from all
the strings extracted from the ground truth file. The label of the chosen string was then used to
assign a functional label to the zone. If this approach failed, the process tried to use
"accumulated" distance, which makes it possible to assign a label to these zones that form together
one ground truth entry, but in the geometric structure were segmented into several parts. If again
none of the similarity scores exceeded the threshold value, the zone was labelled as {\it unknown}.
After processing the entire page, an additional attempt to assign a label to every {\it unknown}
zone based on the labels of its neighbouring zones was made.

\begin{figure}[ht]
  \centering
  \includegraphics[width=0.8\textwidth]{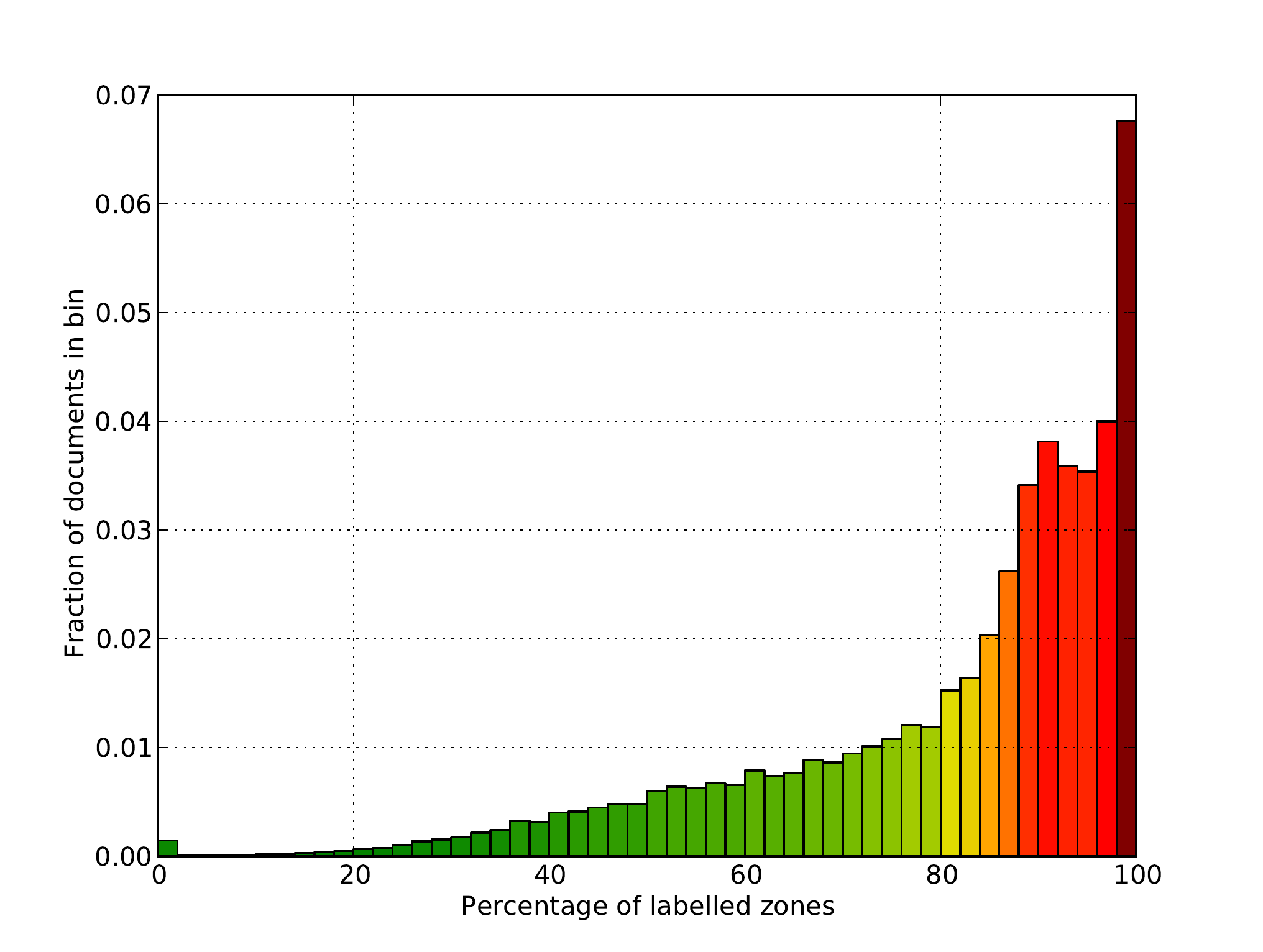}
  \caption[The histogram of the zone label coverage of the documents]{Histogram of documents in the
  dataset having given percentage of zones with an assigned specific (not {\it unknown}) class
  value.}
  \label{fig:dataset-histogram}
\end{figure}

Data in NLM JATS files in PMC vary greatly in quality from perfectly labelled down to containing no
valuable information. Poor quality NLMs result in sparsely labelled zones in generated TrueViz 
files, as the labelling process has no data to compare the zone text content to. Hence, it was
necessary to filter documents whose zones are labelled in satisfying measure, that is the documents
containing only a small number of zones labelled as {\it unknown}. 
Figure~\ref{fig:dataset-histogram} shows the histogram of the label coverage among the processed
documents defined as the percentage of zones labeled with a specific class other than {\it unknown}.
There are many documents (43\%) having more than 90\% of zones labelled, and only those documents
were selected for further processing steps.

We also wanted to be sure that the layout distributions in the entire processed set and the filtered
subset are similar. The layout distribution in a certain document set can be approximated by the
publisher or journal distribution. If poor quality metadata was associated with particular 
publishers or journals, choosing only highly covered documents could result in eliminating
particular layouts entirely, which was to be avoided. We calculated the similarity of publisher
distributions of two sets using the following formula: 

\[ \sm(A,B) = \sum_{p \in P} \min(\dd_A(p), \dd_B(p)) \]

where \( P \) is the set of all publishers in the dataset and \( \dd_A(p) \) and \( \dd_B(p) \) are 
the percentage share of a given publisher in sets \( A \) and \( B \), respectively. The formula
yields 1.0 for identical distributions, and 0.0 in the case of two sets, which do not share any
publishers. The same formula can be used to calculate the similarity with respect to journal
distribution. In our case the similarity of publisher distributions of the originally processed set
and the subset of documents with at least 90\% labelled zones is 0.78, and the similarity of journal
distributions 0.70, thus the distributions are indeed similar.

After filtering the files we randomly chose a sample of 50 documents, which were subjected to a 
manual inspection done by a human expert. The sample was big enough to show common problems 
occurring in the dataset, and small enough to be manually analyzed within a reasonable time. The
inspection revealed a number of repeated errors made by the automatic annotation process. Based on
the results we developed a few simple heuristic-based rules, which would reduce the error rate in
the final dataset. Some examples of the rules are:
\begin{itemize}
\item In some cases the zone labelled as {\it title} contained not only the title as such, but also 
a list of document's authors. We decided to introduce a new label {\it title\_author} for such
cases. We detect those zones based on the similarity measures and assign the new label to them.
\item Pages numbers included in NLM files are usually page ranges from the entire journal volume or
conference proceedings book. In PDF file, on the other hand, pages are sometimes simply numbered
starting from 1. In such cases the zones containing page numbers could not be correctly labelled 
using text content matching. We detect those zones based on their text content and the distance to 
the top or bottom of the page and label them as {\it page\_number}.
\item Figures' captions were often mislabelled as {\it body\_\-con\-tent}, especially when the 
caption contains very similar text to one of the paragraphs. We used a regexp to detect such zones 
and label them as {\it figure}.
\item Similarly, tables' captions mislabelled as {\it body\_content} are recognized by a regexp and 
labelled as {\it table}.
\item A table placed in the document often contains many small zones, especially if every cell forms
a separate zone. The text content of such a small zone can often be found in document's paragraphs 
as well, therefore those zones are sometimes mislabelled as {\it body\_content}. To correct this, we 
detect small zones lying in the close neighbourhood of table zones and label them as {\it table}.
\item Due to the lack of data in NLM files, zones containing information related to document's 
editors, copyright, acknowledgments or conflict statement are sometimes not labelled or mislabelled.
Since those zones are relatively easy to detect based on section titles and characteristic terms, we
use regexps to find them and label them correctly.
\item Zones that occur on every page or every odd/even page and are placed close to the top or 
bottom of the page were often not labelled, in such cases we assigned {\it bib\_info} label to them.
\item The inspection also revealed segmentation problems in a small fraction of pages. The most 
common issue were incorrectly constructed zones, for example when the segmentation process put every
line in a separate zone. Those errors were also corrected automatically by joining the zones that
are close vertically and have the same label.
\end{itemize}

\subsubsection{GROTOAP and GROTOAP2 evaluation}
Every ground truth file in GROTOAP dataset was corrected by a human expert, and thus we can assume 
the quality of GROTOAP is close to perfect. The process of creating GROTOAP2, however, did not 
include manual inspection of every document by a human expert, which allowed us to create a large 
dataset, but also caused the following problems:
\begin{itemize}
\item segmentation errors resulting in incorrectly recognized zones and lines and their bounding
boxes,
\item labeling errors resulting in incorrect zone labels in the dataset.
\end{itemize}

We evaluated GROTOAP2 dataset in order to estimate how accurate the labelling in the ground truth 
files is. We did not perform the evaluation of the segmentation quality. Two kinds of evaluation 
were performed:
\begin{itemize}
\item direct evaluation involved the manual inspection of a small subset of ground truth files from
GROTOAP2,
\item indirect evaluation included comparing the performance of our metadata extraction algorithm
trained on GROTOAP and GROTOAP2 datasets.
\end{itemize}

For the direct manual evaluation we chose a random sample of 50 documents (disjoint with the sample
used before to construct the rules). We evaluated two document sets: files obtained before applying
the rules and the same documents from the final dataset. The groups contain 6,228 and 5,813 zones in
total, respectively (the difference is related to the zone merging step which reduces the overall
number of zones). In both groups the errors were corrected by a human expert, and the original files
were then compared to the corrected ones, which gave the precision and recall values of the
annotation process for each zone label for two stages of the process. The overall accuracy of the
annotation process increased from 0.78 to 0.93 after applying heuristic rules. More details about 
the results of the evaluation can be found in Table~\ref{tab:grotoap2-evaluation}.

\begin{table}
\begin{center}
\begin{tabular}{| l | c | c | c | c | c | c |}
	\hline
	 & \multicolumn{3}{c|}{without rules} & \multicolumn{3}{c|}{with rules}\\ 
	 \cline{2-7}
	 & precision & recall & F-score & precision & recall & F-score \\ \hline\hline
	{\it abstract} & 0.93 & 0.96 & 0.94 & 0.98 & 0.98 & 0.98 \\
	{\it acknowl.} & 0.98 & 0.67 & 0.80 & 1.0 & 0.90 & 0.95 \\ 
	{\it affiliation} & 0.77 & 0.90 & 0.83 & 0.95 & 0.95 & 0.95 \\
	{\it author} & 0.85 & 0.95 & 0.90 & 1.0 & 0.98 & 0.99 \\
	{\it bib\_info} & 0.95 & 0.45 & 0.62 & 0.96 & 0.94 & 0.95 \\
	{\it body} & 0.65 & 0.98 & 0.79 & 0.88 & 0.99 & 0.93 \\
	{\it conflict\_st.} & 0.63 & 0.24 & 0.35 & 0.82 & 0.89 & 0.85 \\ 
	{\it copyright} & 0.71 & 0.94 & 0.81 & 0.93 & 0.78 & 0.85 \\
	{\it corresp.} & 1.0 & 0.72 & 0.84 & 1.0 & 0.97 & 0.99 \\
	{\it dates} & 0.28 & 1.0 & 0.44 & 0.94 & 1.0 & 0.97 \\
	{\it editor} & - & 0 & - & 1.0 & 1.0 & 1.0 \\ 
	{\it equation} & - & - & - & - & - & - \\ 
	{\it figure} & 0.99 & 0.36 & 0.53 & 0.99 & 0.46 & 0.63 \\ 
	{\it glossary} & 1.0 & 1.0 & 1.0 & 1.0 & 1.0 & 1.0 \\
	{\it keywords} & 0.94 & 0.94 & 0.94 & 1.0 & 0.94 & 0.97 \\
	{\it page\_nr.} & 0.99 & 0.53 & 0.69 & 0.98 & 0.97 & 0.98 \\
	{\it references} & 0.91 & 0.95 & 0.93 & 0.99 & 0.95 & 0.97 \\
	{\it table} & 0.98 & 0.83 & 0.90 & 0.98 & 0.96 & 0.97 \\ 
	{\it title} & 0.51 & 1.0 & 0.67 & 1.0 & 1.0 & 1.0 \\
	{\it title\_auth.} & - & 0 & - & 1.0 & 1.0 & 1.0 \\
	{\it type} & 0.76 & 0.46 & 0.57 & 0.89 & 0.47 & 0.62 \\
	{\it unknown} & 0.22 & 0.46 & 0.30 & 0.62 & 0.94 & 0.75 \\\hline
	average & 0.79 & 0.68 & 0.73 & 0.95 & 0.91 & 0.92 \\
	\hline
\end{tabular}
\end{center}
\caption[The results of manual evaluation of GROTOAP2]{The results of manual evaluation of GROTOAP2.
The table shows the precision, recall and F-score values for every zone label for the annotation
process without and with heuristic-based rules. The correct labels for zones were provided by a
human expert. No values appear for labels that were not present in the automatically annotated
dataset.}
\label{tab:grotoap2-evaluation}
\end{table}

For the indirect evaluation we compared the overall performance of our metadata extraction algorithm
trained on the entire GROTOAP dataset and randomly chosen 1000 documents from GROTOAP2 dataset in 
two versions: before applying correction rules and after. The evaluation was done with the use a
PMC-based dataset (more details about the evaluation methodology can be found in 
Section~\ref{sec:extr-eval}). The system achieved the average F-score (calculated as an average over
all metadata categories) 62.41\% when trained on GROTOAP, 75.38\% when trained on GROTOAP2 before
applying rules and 79.34\% when trained on the final version of GROTOAP2. The detailed results are
shown in Table~\ref{tab:grotoap2-training}.

\begin{table}
\renewcommand{\tabcolsep}{3pt}
\renewcommand{\arraystretch}{1.2}
\begin{center}
\begin{tabular}{| l | c | c | c |}
	\hline
	 & \multirow{3}{*}{GROTOAP} & \multicolumn{2}{c|}{GROTOAP2}\\ 
	 \cline{3-4}
	 & & without rules & with rules\\\hline
	Precision & 77.13\% & 81.88\% & 82.22\% \\
	Recall & 55.99\% & 70.94\% & 76.96\% \\
	F-score & 62.41\% & 75.38\% & 79.34\% \\ \hline
\end{tabular}
\end{center}
\caption[The results of the indirect evaluation of GROTOAP2]{The comparison of the performance of
the metadata extraction algorithm trained on GROTOAP, GROTOAP2 before applying improvement rules and
final GROTOAP2. The table shows the mean precision, recall and F-score values calculated as an
average of the values for individual metadata classes.}
\label{tab:grotoap2-training}
\end{table}

GROTOAP and GROTOAP2 datasets are similar in the structure, they both contain the documents in the 
form of hierarchical geometric model serialized using TrueViz format. However, due to the different
nature of their creation procedures resulting in different characteristics, we use them for 
different purposes.

In GROTOAP every document was inspected and corrected by a human expert, and thus we can assume the
structure and labelling in the dataset are close to 100\% correct. GROTOAP was used to evaluate the
performance of the page segmenter, but the dataset was too small and not diverse enough to be useful 
for the zone classifiers.

In GROTOAP2 we omit the manual correction of every file, which resulted in a large and diverse 
dataset, but with the zone labelling only 93\% accurate. Since the page segmenter was directly used
for creating of the dataset, GROTOAP2 could not be used for its evaluation. The zone classifiers, on
the other hand, were not involved in the process of assigning labels, and thus GROTOAP2 could be 
used for the experiments with zone classification: feature selection and SVM parameters adjustment,
final zone classifiers evaluation and training.

\subsection{Metadata Sets}
Metadata sets are used to assess the quality of the entire extraction algorithm in all its aspects:
basic metadata extraction, bibliography extraction and structured content extraction. Metadata sets
contain documents in the form of pairs: a source PDF document and the corresponding ground truth
metadata in NLM JATS format. Metadata sets were built using the resources from PubMed Central Open
Access Subset and Elsevier.

The main dataset we used to assess the performance of the extraction algorithm and to compare it 
with similar solutions is a subset of PMC containing 1,943 documents from 1,943 different journals 
compiled by Alexandru Constantin\footnote{http://pdfx.cs.man.ac.uk/serve/PMC\_sample\_1943.zip}.

We also used Elsevier resources to compile an additional metadata set of 2,508 documents. The 
dataset was created by randomly selecting a group of articles and filtering those with a references
section present in the metadata record. Elsevier dataset was not used to assess the performance of
the section hierarchy extraction, since Elsevier's ground truth metadata records we had access to 
lack this part.

The content classification models in our algorithm were trained on GROTOAP2, which was built from
documents from PMC. Even though we made sure the training, validation and testing subsets were 
pairwise disjoint, our algorithm might have a slight advantage on the PMC dataset over other similar
solutions, depending on what documents and layouts their default versions were trained.

\subsection{Citation Set}
Citation set~\cite{grotoap-citations} contains parsed bibliographic references and is used to 
evaluate and train the reference parser. Citation set was built using the data from 
Cora-ref~\cite{McCallumNRS00} and PMC resources.

Cora-ref already contains parsed references. Unfortunately, due to some differences in the labels
used, labels mapping had to be performed. Labels from original datasets were mapped in the following
way: {\it title} and {\it year} remained the same; {\it journal}, {\it booktitle}, {\it tech} and 
{\it type} were mapped to {\it source}; {\it date} was mapped to {\it year}. Labels {\it author} and
{\it pages} were split respectively into {\it givenname} and {\it surname}, {\it page\_first} and
{\it page\_last} using regular expressions. All remaining tokens were labelled as {\it text}.

The ground truth files from PMC also contain parsed references. Unfortunately, in most cases they do
not preserve the entire reference strings from the original PDF file, separators and punctuation are
often omitted. For this reason the reference set was built using a similar technique as in the case
of GROTOAP2. We extracted reference strings from PDF files using our extraction algorithm and 
labelled them using annotated reference data from NLM files. Only the references with high labelling 
coverage (the percentage of the tokens labelled with a specific label) were selected for the final
set.

The final citation dataset contains parsed citation from both sources combined into a single dataset 
of 6,858 parsed citations in total.

\subsection{Affiliation Set}
Affiliation set~\cite{grotoap-affiliations} contains parsed affiliations and was used to evaluate 
and train the affiliation parser. The dataset is based on PMC resources.

The ground truth files in PMC contain parsed affiliations, but the quality of the labelling varies a
lot from file to file, and the entire raw affiliation strings from the original article are rarely
preserved. To overcome these issues, once again we used a technique similar as in the case of 
GROTOAP2 and the citation dataset. We built the affiliation dataset automatically by matching
labelled metadata from the ground truth files against affiliation strings extracted from PDFs by the
extraction algorithm. We also filtered out the affiliations poorly covered by specific labels, which
was typically caused by poor quality ground truth labelling or extraction errors.

The final affiliation dataset used for parser evaluation contains 8,267 parsed affiliations.

\section{Page Segmentation}
\label{sec:segm}
Page segmentation was evaluated using the entire GROTOAP dataset containing 113 documents. During 
the evaluation the ground truth geometric hierarchical structure of the test document was compared
to the corresponding structure built from scratch by the page segmenter.

The goal of the evaluation was to assess how well on average the algorithm reconstructs the words,
lines and zones of a document. Let $D^G$ be the ground truth document from GROTOAP dataset and $D^T$
the corresponding document with words, lines and zones rebuilt from the characters by the
segmentation algorithm. For every pair $(D^G, D^T)$ we calculated the following scores:
\begin{itemize}
\item the percentage of correctly constructed words: 

\[ \score_{D,W} = \frac{|CW_{D}|}{|W_{D^G}|} \]

where $CW_{D}$ is the set of words correctly built by the segmenter and $W_{D^G}$ is the set of 
words in the ground truth document;
\item the percentage of correctly constructed lines: 

\[ \score_{D,L} = \frac{|CL_{D}|}{|L_{D^G}|} \]

where $CL_{D}$ is the set of lines correctly built by the segmenter and $L_{D^G}$ is the set of 
lines in the ground truth document;
\item the percentage of correctly constructed zones:

\[ \score_{D,Z} = \frac{|CZ_{D}|}{|Z_{D^G}|} \]

where $CZ_{D}$ is the set of zones correctly built by the segmenter and $Z_{D^G}$ is the set of 
zones in the ground truth document.
\end{itemize}

A word is considered correctly constructed if the corresponding word in the ground truth document
contains the same set of characters (the order of the elements is not taken into account):

\[ CW_{D} = W_{D^T} \cap W_{D^G} = \{w_t \in W_{D^T} \enskip | \enskip \exists_{w_g \in W_{D^G}}
\enskip \prt^{-1}(w_t) = \prt^{-1}(w_g)\} \]

where $par$ is a function mapping elements to their parents, in particular characters to their 
words.

Similarly, a line is considered correctly constructed if the corresponding line in the ground truth
document contains the same set of characters (the inner division into words or the order of elements 
are not taken into account):

\[ CL_{D} = \{l_t \in L_{D^T} \enskip | \enskip \exists_{l_g \in L_{D^G}} \enskip \anc_2^{-1}(l_t) =
\anc_2^{-1}(l_g)\} \]

where $\anc_2 = \prt \circ \prt$ is a function mapping elements to the parents of their parents, in
particular characters to their lines.

Finally, a zone is considered correctly constructed if the corresponding zone in the ground truth
document contains the same set of characters (the inner division into words and lines or the order 
of elements are not taken into account):

\[ CZ_{D} = \{z_t \in Z_{D^T} \enskip | \enskip \exists_{z_g \in Z_{D^G}} \enskip \anc_3^{-1}(z_t) =
\anc_3^{-1}(z_g)\} \]

where $\anc_3 = \prt \circ \prt \circ \prt$ is a function mapping elements to the parents of the
parents of their parents, in particular characters to their zones.

The overall scores for the entire test set were obtained by calculating the average values over all 
tested documents.

\begin{figure}
  \centering
  \includegraphics[width=0.75\textwidth]{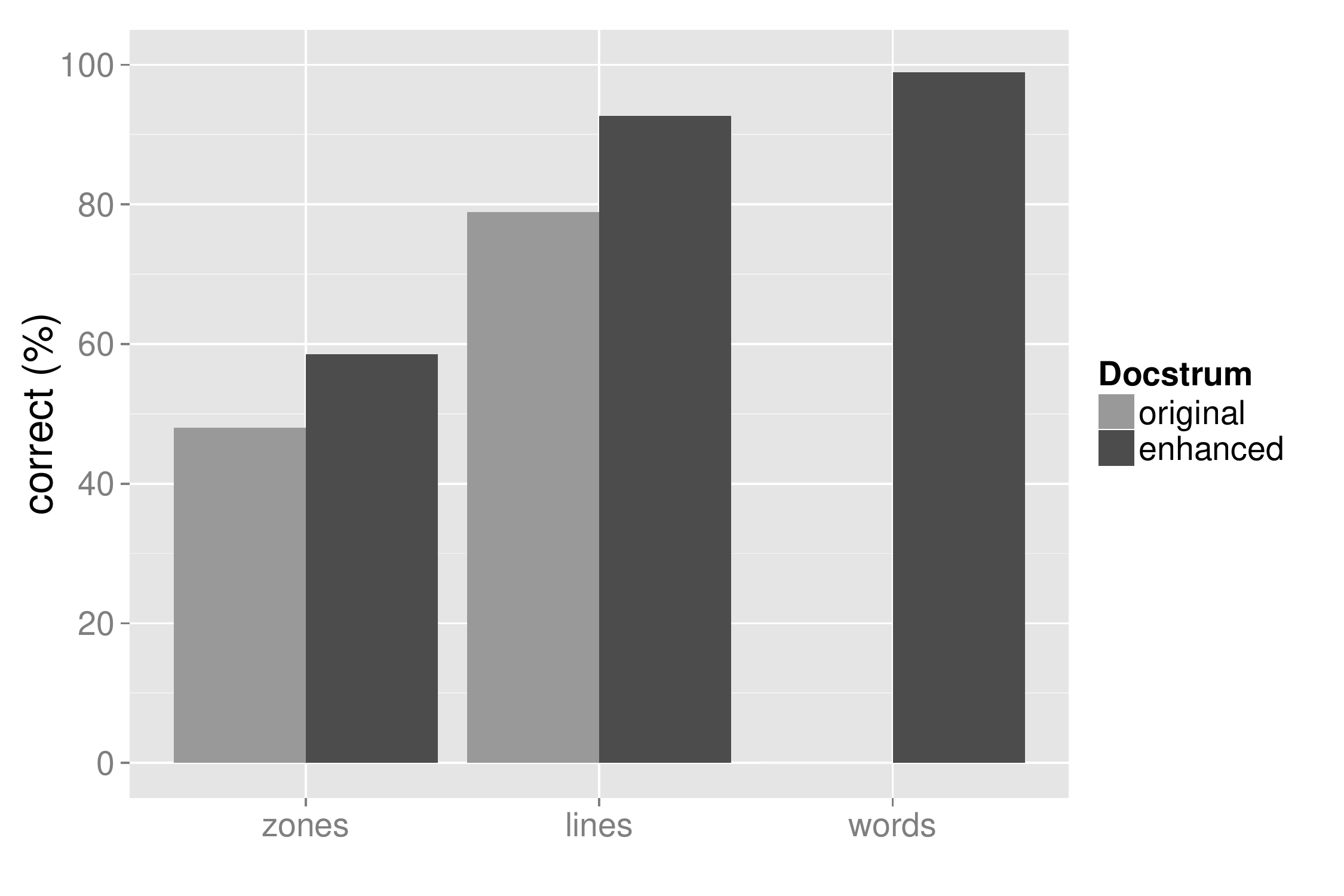}
  \caption[The results of page segmentation evaluation]{The results of page segmentation evaluation.
  The plot shows the accuracy of extracting zones, lines and words for the original Docstrum
  algorithm and Docstrum enhanced with our modifications. Since the original algorithm does not
  determine words, only the score for the enhanced version is given in this case.}
  \label{fig:segm-all}
\end{figure}

Using this methodology we performed the evaluation of two versions of the segmentation algorithm: 
the original Docstrum (only the scores for lines and zones were calculated, as the original
algorithm does not determine words) and the algorithm enhanced with the modifications listed in 
Section~\ref{sec:segm}. The results are shown in Figure~\ref{fig:segm-all}. The modifications 
resulted in increased scores in case of both lines and zones.

There are two main document regions contributing to the low scores in the case of zones: metadata
regions and table/figure regions. When metadata fragments play different roles in the document, but
are displayed close to each other (for example title and authors, authors and affiliations), in 
GROTOAP they are split into separate zones with different labels. The page segmenter, on the other 
hand, analyses only the geometric traits and often will merge such fragments into one zone. What is
more, as a result of an arbitrary decision in GROTOAP every table and figure is placed in one 
separate zone. From the segmentation point of view, however, such areas usually are sparse and 
contain more geometrically separated zones, which is usually the segmenter's output. Since the 
typical document contains much more table/figure zones than metadata zones, they are the main cause 
of the errors.

\begin{figure}
  \centering
  \includegraphics[width=0.75\textwidth]{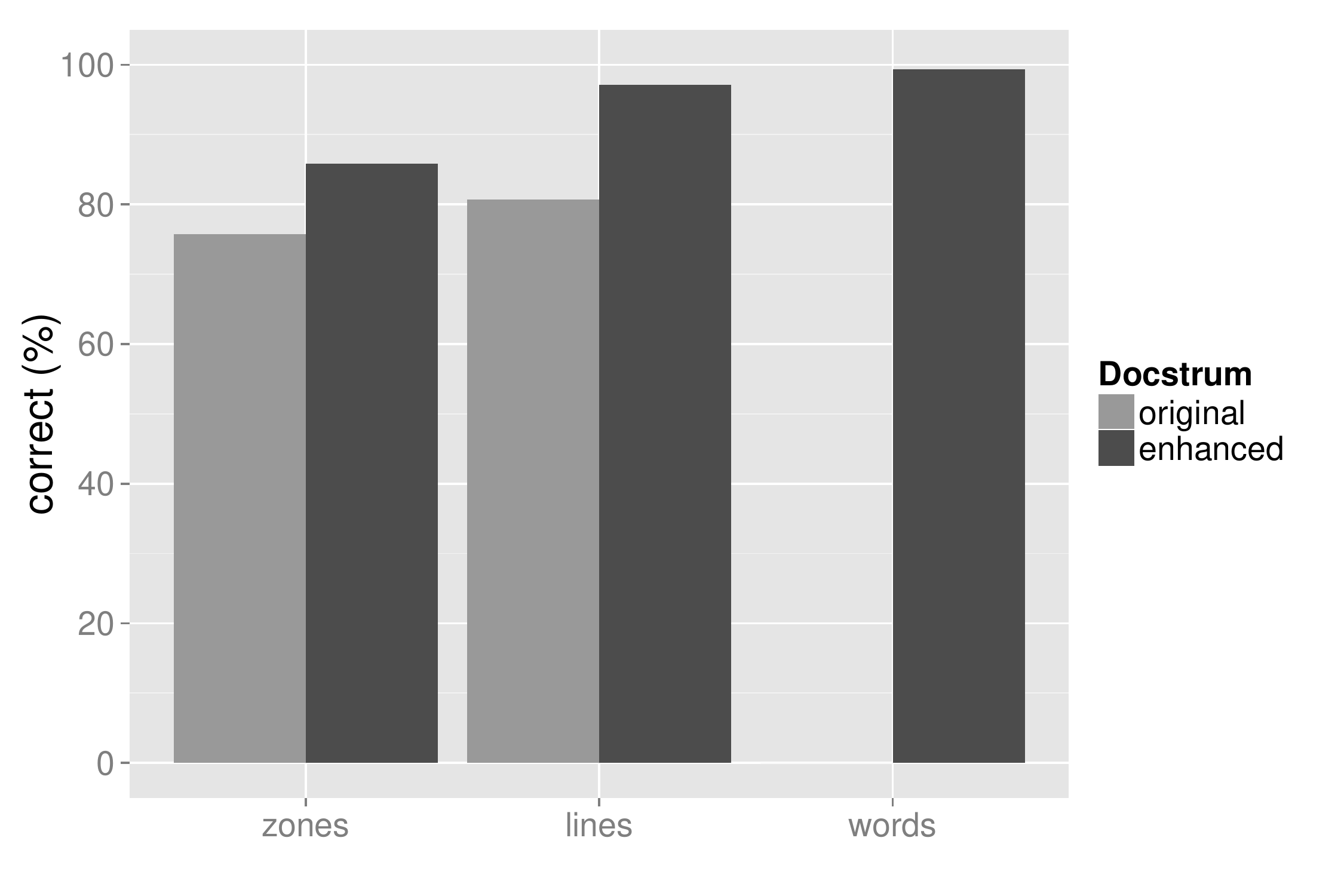}
  \caption[The results of page segmentation evaluation with filtered zones]{The results of page
  segmentation evaluation. The plot shows the accuracy of extracting zones, lines and words for the
  original Docstrum algorithm and Docstrum enhanced with our modifications. The scores are
  calculated for the subset of the documents' zones, with tables and figures regions filtered out.
  Since the original algorithm does not determine words, only the score for the enhanced version is
  given in this case.}
  \label{fig:segm-filtered}
\end{figure}

To illustrate this, we recalculated the scores for the subsets of the documents with the tables and
figures regions filtered out. The results are shown in Figure~\ref{fig:segm-filtered}. Filtering out
the tables/figures regions resulted in all scores increase, the largest increase appears in the case
of the zones.

\section{Content Classification}
\label{sec:cont-class}
The following experiments were performed in the context of content classification: selecting the 
best features for the classifiers (Section~\ref{sec:cont-class-features}), finding the best
classification parameters (Section~\ref{sec:cont-class-pars}) and the evaluation of the classifiers 
(Section~\ref{sec:cont-class-class}). The experiments were performed for every SVM classifier used 
in the algorithm: category classifier (Section~\ref{sec:classification}), metadata classifier
(Section~\ref{sec:metadata-classification}) and body classifier (Section~\ref{sec:text-filtering}).

\subsection{Feature Selection}
\label{sec:cont-class-features}
Feature selection was performed with the use of the validation set, which is a subset of GROTOAP2
containing 100 documents with 14,000 labelled zones in total. For each classifier a total of 103
features were analyzed.

In general feature selection is based on the analysis of the correlations between the features and
between features and expected zone labels. For simplicity we treat all the features as numerical
variables; the values of binary features are decoded as $0$ or $1$. The labels, on the other hand,
are an unordered categorical variable.

Let $L$ be a set of zone labels of a given classifier, $n$ the number of the observations (zones) in
the validation dataset and $k = 103$ the initial number of analysed features. For $i$th feature, 
where $0 \leq i < k$, we can define $f_i \in R^n$, a vector of the values of the $i$th feature for
subsequent observations. Let also $l \in L^n$ be the corresponding vector of zone labels.

In the first step we removed redundant features, highly correlated with other features. For each 
pair of feature vectors we calculated the Pearson correlation score and identified all the pairs
$f_i, f_j \in R^n$, such that

\[ |\cor(f_i, f_j)| > 0.9 \]

Next, for every feature from highly correlated pairs we calculated the mean absolute correlation:

\[ \cor_m(f_i) = \frac{1}{k}\sum_{j=0}^{k-1} \cor(f_i, f_j) \]

and from each highly correlated pair the feature with higher $\cor_m$ was eliminated. Let's denote
the number of remaining features as $k'$.

After eliminating features using correlations between them, we analysed the features using their 
associations with the expected zone labels vector $l$. To calculate the correlation between a single
feature vector $f_i$ (numeric) and label vector $l$ (unordered categorical) we employed Goodman and
Kruskal's $\tau$ (tau) measure~\cite{Goodman63}.

In order to calculate $\tau$ measure, we treat both feature vector $f_i$ and label vector $l$ as 
random variables. Let's denote as $\{\pi_{st}\}$ the joint distribution of $f_i$ and $l$; $\pi_{st}$
is the probability that an observation has a label $t$ and the value of its $i$th feature is $s$. 
Let's also denote as $\{\pi_t\}$ the marginal distribution of labels, and $\{\pi_s\}$ the marginal
distribution of all observed values of the feature $f_i$. Goodman and Kruskal's $\tau$ measure is
calculated as

\[ \tau(f_i, l) = \frac{V(l) - E(V(l|f_i))}{V(l)} \]

where
\begin{itemize}
\item $V(l)$ is a measure of variation for the marginal distribution $\{\pi_t\}$ calculated as 

\[ V(l) = 1 - \sum \pi_t^2 \]

\item $V(l|f_i)$ is a measure of variation for the distribution of labels for a fixed value of $i$th
feature calculated in the same way
\item $E(V(l|f_i))$ is the expected value of the conditional variation $V(l|f_i)$ calculated as

\[ E(V(l|f_i)) = \sum_s \pi_s V(l|s) \]

\end{itemize}

Let $f_0, f_1, ... f_{k'-1}$ be the sequence of the feature vectors ordered by non-decreasing $\tau$ 
measure, that is

$$ \tau(f_0, l) \leq \tau(f_1, l) \leq ... \leq \tau(f_{k'-1}, l) $$

The features were then added to the classifier one by one, starting from the best one (the mostly
correlated with the labels vector, $f_{k'-1}$), and at the end the classifier contained the entire
feature set. At each step we performed a 5-fold cross-validation on the validation dataset and 
calculated the overall F-score as an average over the scores of individual labels. We used feature
scaling and classes weights based on the number of their training samples to set larger penalty for
less represented classes. SVM classifier was trained using RBF kernel.

For $i,j \in L$ let's denote as $C(i,j)$ the number of observations with true label $i$ classified
as $j$ by the classifier. We calculate the overall F-score as

\[ F = \frac{1}{|L|} \sum_{i \in L} 2 (P(i)^{-1}+R(i)^{-1})^{-1} \]

where
\begin{itemize}
\item $P(i)$ is the precision for a specific label $i$ calculated as

\[ P(i) = \frac{C(i,i)}{\sum_{j \in L} C(j,i)} \]

\item $R(i)$ is the recall for a specific label $i$ calculated as

\[ R(i) = \frac{C(i,i)}{\sum_{j \in L} C(i,j)} \]
\end{itemize}

For completeness, we also repeated the same procedure with reversed order of the features, starting
with less useful features.

Using the results, we eliminated a number of the least useful features $f_0, f_1, ... f_{t}$, such 
that the performance of the classifier with the remaining features was similar to the performance of
the classifier using the entire feature set.

\begin{figure}
  \centering
  \includegraphics[width=0.7\textwidth]{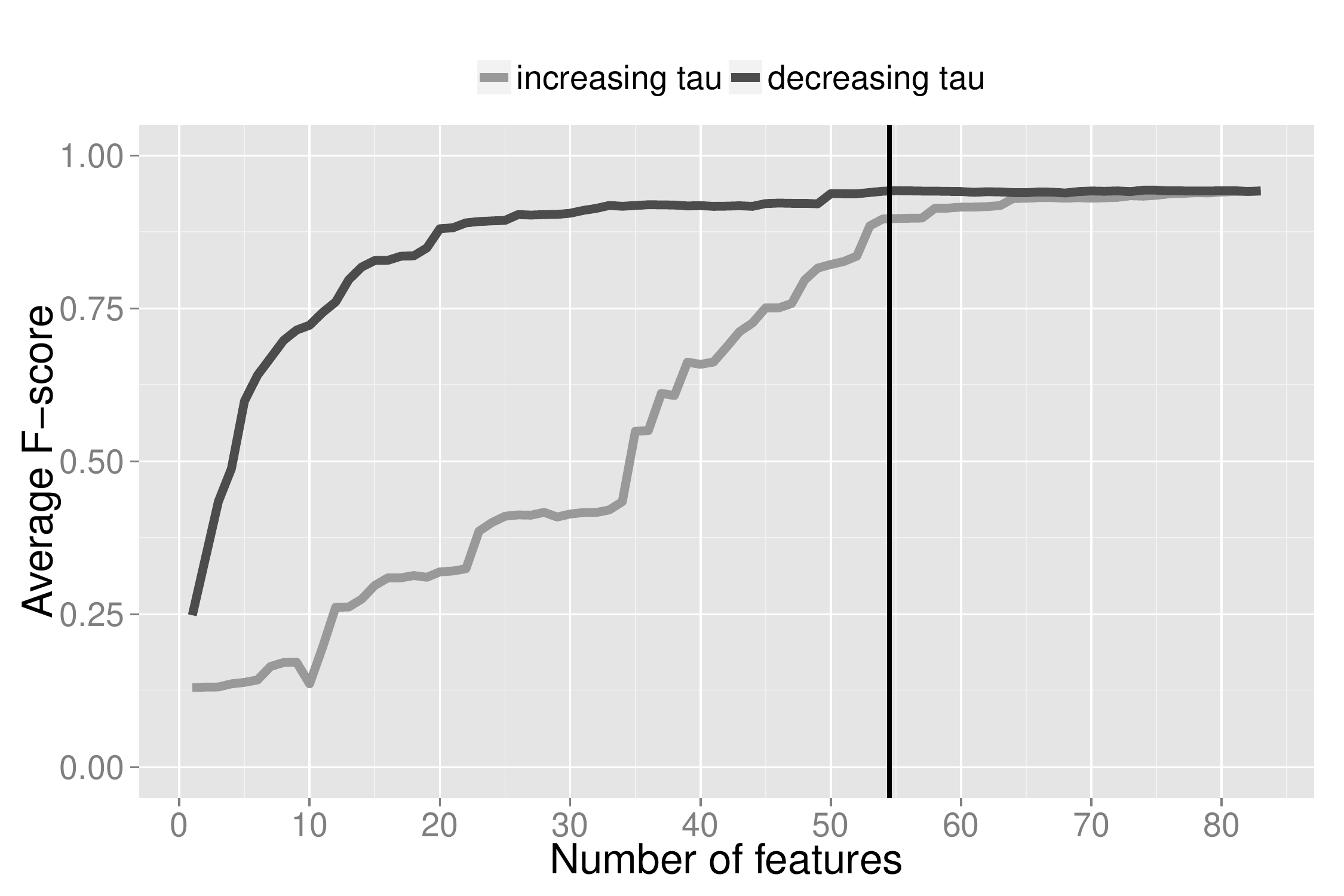}
  \caption[Average F-score for category classifier for various numbers of features]{Average F-score
  for category classifier for 5-fold cross-validation for various numbers of features. Darker line
  shows the change in F-score, while adding features from the most to the least useful one, and the
  lighter line shows the increase with the reversed order. The vertical line marks the feature set
  chosen for the final classifier.}
  \label{fig:init-feat-sel}
\end{figure}

For the category classifier the entire zone set from the validation set was used, and the specific 
labels were mapped to their general categories: {\it metadata}, {\it body}, {\it references} and 
{\it other}. The set contains 14,000 zones represented by features and zone labels. 20 features were 
removed based on between-feature correlation analysis. The remaining 83 features were analyzed using
the $\tau$ measure. The results are shown in Figure~\ref{fig:init-feat-sel}. The final feature set 
for this classifier contains 54 features with the best $\tau$ measure.

\begin{figure}
  \centering
  \includegraphics[width=0.7\textwidth]{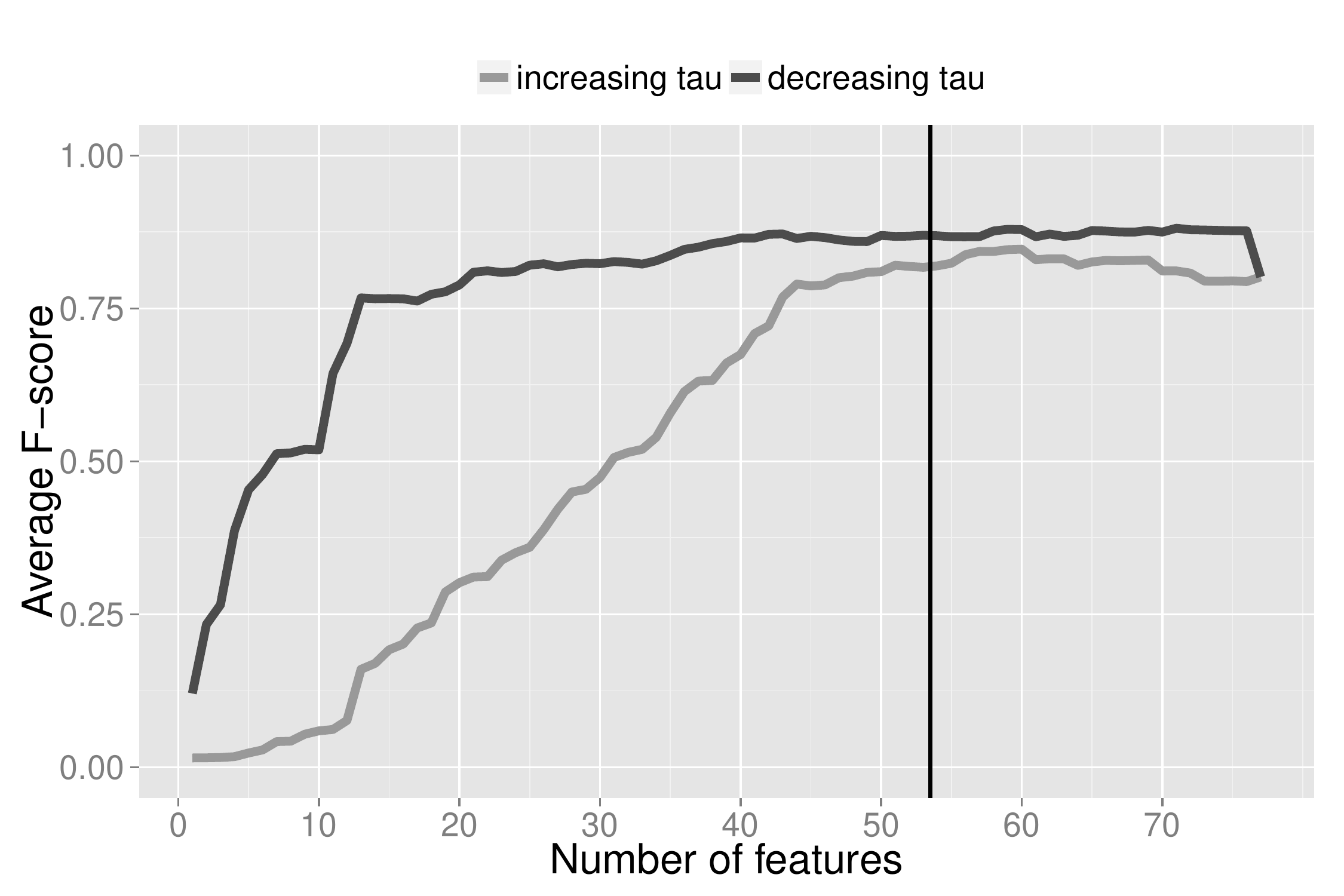}
  \caption[Average F-score for metadata classifier for various numbers of features]{Average F-score
  for metadata classifier for 5-fold cross-validation for various numbers of features. Darker line
  shows the change in F-score, while adding features from the most to the least useful one, and the
  lighter line shows the increase with the reversed order. The vertical line marks the feature set
  chosen for the final classifier.}
  \label{fig:meta-feat-sel}
\end{figure}

For analyzing the metadata classifier we used only the zones of the category {\it metadata}, which 
left us with 2,743 zones represented by features and zone labels. At the beginning 4 features were
eliminated because there was no variation in their values in the metadata zone subset. 22 features
were removed based on between-feature correlation analysis. The remaining 77 features were analyzed
using the $\tau$ measure. The results are shown in Figure~\ref{fig:meta-feat-sel}. The final feature
set for the metadata classifier contains 53 features with the best $\tau$ measure.

\begin{figure}
  \centering
  \includegraphics[width=0.7\textwidth]{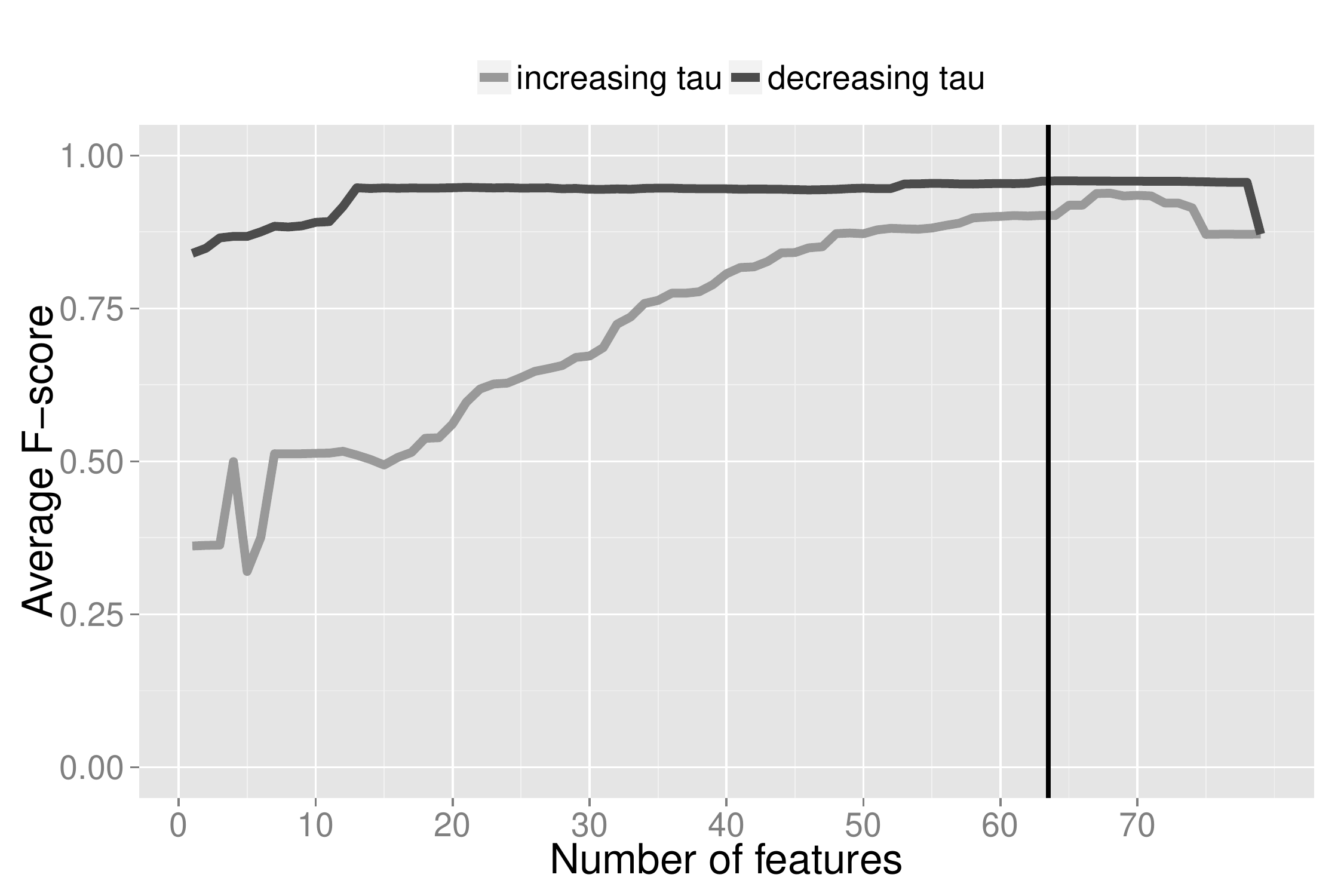}
  \caption[Average F-score for body classifier for various numbers of features]{Average F-score for
  body classifier for 5-fold cross-validation for various numbers of features. Darker line shows the
  change in F-score, while adding features from the most to the least useful one, and the lighter
  line shows the increase with the reversed order. The vertical line marks the feature set chosen
  for the final classifier.}
  \label{fig:body-feat-sel}
\end{figure}

Similarly, for analyzing the body classifier we used only the zones of the category {\it body}, 
which left us with 9,394 zones represented by features and zone labels. At the beginning 3 features
were eliminated because there was no variation in their values in the zone subset. 21 features were 
removed based on between-feature correlation analysis. The remaining 79 features were analyzed using
the $\tau$ measure. The results are shown in Figure~\ref{fig:body-feat-sel}. The final feature set
for the body classifier contains 63 features with the best $\tau$ measure.

34 features are common for all classifiers, although they often differ a lot in their $\tau$
correlation measure. The features present in the case of metadata classifier and missing in other 
classifiers are mostly related to the presence of the keywords characteristic for zones like
affiliation, keywords, addresses, license, etc. The features present in the case of category 
classifier and missing in others are related to the page numbers, references regions and the last
pages of the document in general. The features present in the case of body classifier and missing in 
others are related to the keywords and characters characteristic for zones like figure or table
captions or equations.

\subsection{Parameters Adjustment}
\label{sec:cont-class-pars}
The best SVM parameters were also found using the same validation dataset containing 14,000 zones in
the case of category classifier, 2,743 zones in the case of metadata classifier and 9,394 zones in 
the case of body classifier. For performance reasons, the dataset used for category classifier was
reduced to a smaller set of 10,478 zones randomly chosen from the original set. As a result of the
feature selection procedure, each classifier used a different set of features. We performed a grid
search over all combinations of the parameter values within some ranges and chose the parameters 
which resulted in the highest scores obtained during the cross-validation.

First, the feature vectors were scaled linearly to interval $[0,1]$ according to the bounds found in
the learning samples. In order to find the best parameters for the classifiers we performed a 
grid-search over the parameter space for four different kernels (linear, polynomial, radial-basis
and sigmoid). For every combination of the kernel and its parameters we performed a 5-fold 
cross-validation. We also used classes weights based on the number of their training samples to set
larger penalty for less represented classes. Finally, we chose those parameters, for which we
obtained the highest mean F-score calculated as an average over the scores for individual classes
(the same way as in the case of feature selection).

We testes various ranges of the following parameters:
\begin{itemize}
\item $C = \{2^i | i \in [-5,15]\}$ --- a set of possible values of the penalty parameter,
\item $D = \{2,3,4\}$ --- a set of degrees for polynomial kernels,
\item $\Gamma = \{2^i | i \in [-15,3]\}$ --- a set of possible values of the kernel coefficient
$\gamma$,
\item $R = \{-100, -10, -1, 0, 1, 10, 100\}$ --- a set of possible values of the kernel coefficient
$r$,
\end{itemize}

The following parameter spaces were used:
\begin{itemize}
\item linear kernel: 1-dimensional space $<C>$,
\item polynomial kernel: 4-dimensional space $<D, \Gamma, R, C>$,
\item RBF kernel: 2-dimensional space $<\Gamma, C>$,
\item sigmoid kernel: 3-dimensional space $<\Gamma, R, C>$.
\end{itemize}

Chosen parameters for all the classifiers are shown in Table~\ref{tab:svm-parameters}. The detailed
results, including the best mean F-score values for all classifiers and all kernel function types 
obtained during 5-fold cross-validation, as well as related values of the parameters, can be found
in appendix~\ref{chap:app-results} in Section~\ref{sec:app-params}.

\begin{table}
\renewcommand{\arraystretch}{1.2}
\renewcommand{\tabcolsep}{5pt}
\centering
  \begin{tabular}{ | c | c | c | }
\hline
    classifier & kernel & parameters \\ \hline \hline
    category & RBF & $c=2^5$, $\gamma=2^{-2} $\\ \hline
    metadata & polynomial & $d=3$, $c=2^{-4}$, $\gamma=2^0$, $r=0$\\ \hline
    body & polynomial & $d=4$, $c=2^3$, $\gamma=2^{-3}$, $r=1$\\ \hline
\end{tabular}
\caption[Final SVM kernels and parameters values for SVM classifiers]{The chosen SVM kernels and
parameters values for all the classifiers.}
 \label{tab:svm-parameters}
\end{table}

\subsection{Zone Classification}
\label{sec:cont-class-class}
The final versions of the zone classifiers (with selected features and the best SVM parameters 
found) were evaluated by a 5-fold cross-validation using zone test set, which is a subset of
GROTOAP2 (disjoint with the zone validation set). Zone test set consists of 2,551 documents 
containing 355,779 zones in total, 68,557 of which are metadata zones and 235,126 of which are body 
zones.

During the cross-validation each zone from the test set was classified and tested once. Based on 
these pairs (ground truth label, classifier's label) we constructed the confusion matrix for the 
classification and calculated the precision, recall and F-score measures for individual zone labels.

The confusion matrix is indexed by the set of labels of a given classifier. The value in the row $i$ 
and column $j$ is equal to the total number of zones with true label $i$ labelled as $j$ by the
classifier during the cross-validation. The values on the matrix diagonal represent the number of
correctly classified zones of respective types and the sum of all cells equals to the size of the 
zone test set. The precision, recall and F-score values for individual labels are calculated in the 
same way as before.

The exact confusion matrices for all the zone classifiers can be found in 
appendix~\ref{chap:app-results} in Section~\ref{sec:app-params}. In this chapter we present the
visualizations of the matrices, generated in the following way: the values in the original matrix 
were scaled relatively to the sum of the rows (the number of zones of respective classes in the 
ground truth test set) and logarithmic function was applied to the resulting fractions. As a result,
the value in the cell in row $i$ and column $j$ corresponds to the fraction of the zones with true
label $i$ that were classified by the classifier as $j$.

\begin{figure}
  \centering
  \includegraphics[width=0.7\textwidth]{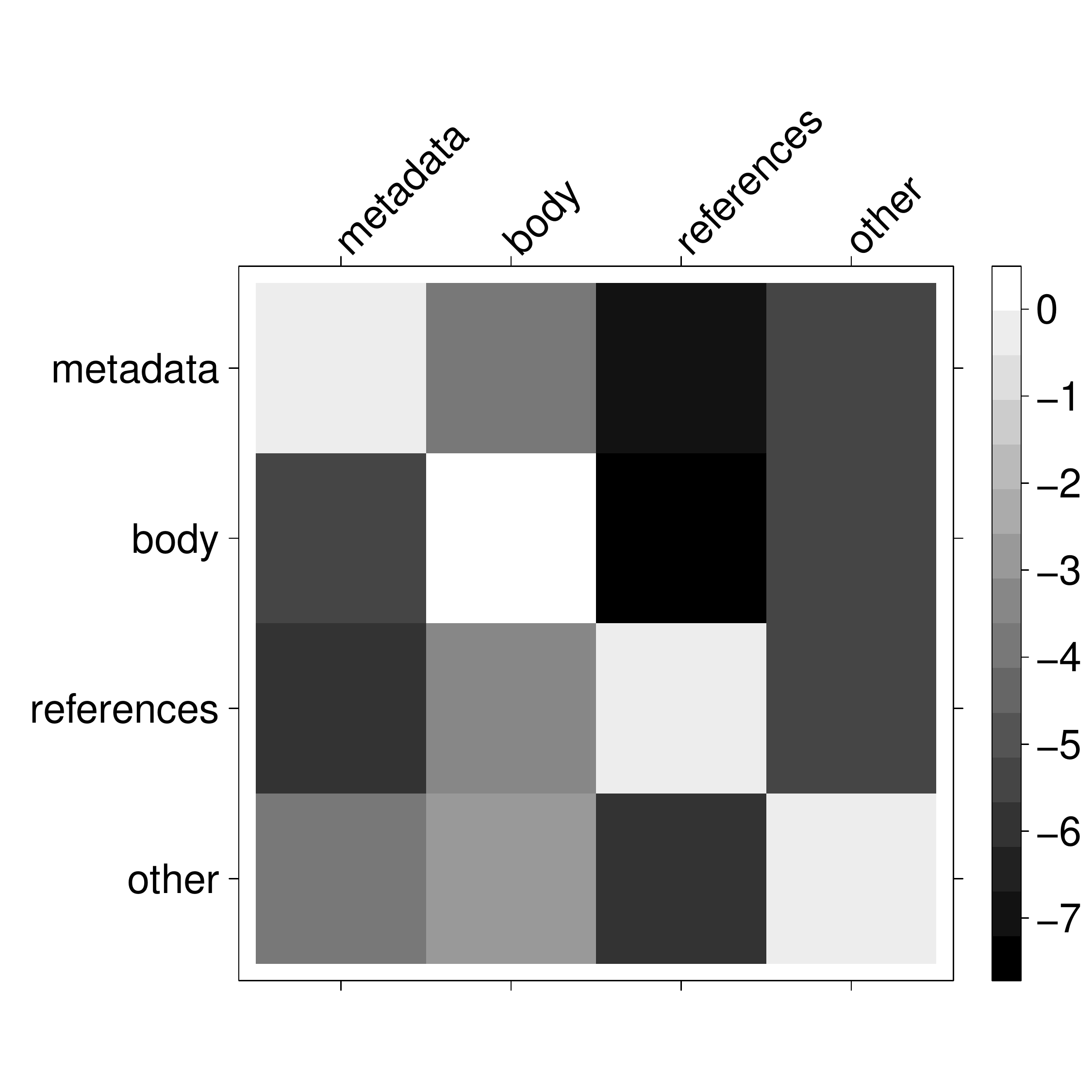}
  \caption[The visualization of the confusion matrix of category classification evaluation]{The
  visualization of the confusion matrix of the category classification. The value in the cell in row
  $i$ and column $j$ corresponds to the fraction of the zones with true label $i$ that were
  classified by the classifier as $j$. The lighter the cell color, the higher the fraction and the
  more confusion between the two classes.}
  \label{fig:hm-init}
\end{figure}

Table~\ref{tab:res-init} shows the precision, recall and F-score values for individual classes for 
category classification. Figure~\ref{fig:hm-init} presents the visualization of the confusion matrix
for this classifier.

\begin{table}
\renewcommand{\arraystretch}{1.3}
\renewcommand{\tabcolsep}{2pt}
\centering
  \begin{tabular}{ | c || r | r | r |}
  	\hline 
  	& precision (\%) & recall (\%) & F-score (\%) \\ \hline \hline
  
  	{\it metadata} & 97.03 & 96.88 & 96.96 \\ \hline
  	{\it body} & 98.12 & 98.98 & 98.55 \\ \hline
    {\it references} & 98.15 & 95.63 & 96.88 \\ \hline
    {\it other} & 96.26 & 91.96 & 94.06 \\ \hline \hline
    average & 97.39 & 95.86 & 96.61 \\ \hline

     \end{tabular}
\caption[The results of category classification evaluation]{Precision, recall and F-score values for
individual classes for category classification.}
\label{tab:res-init}
\end{table}

The results of the category classifier are very good. The lowest scores were achieved in the case of
the label {\it other}, which is a special label for all the zones that do not fit into the three 
main categories, and thus is the least consistent in the feature values.

\begin{figure}
  \centering
  \includegraphics[width=1\textwidth]{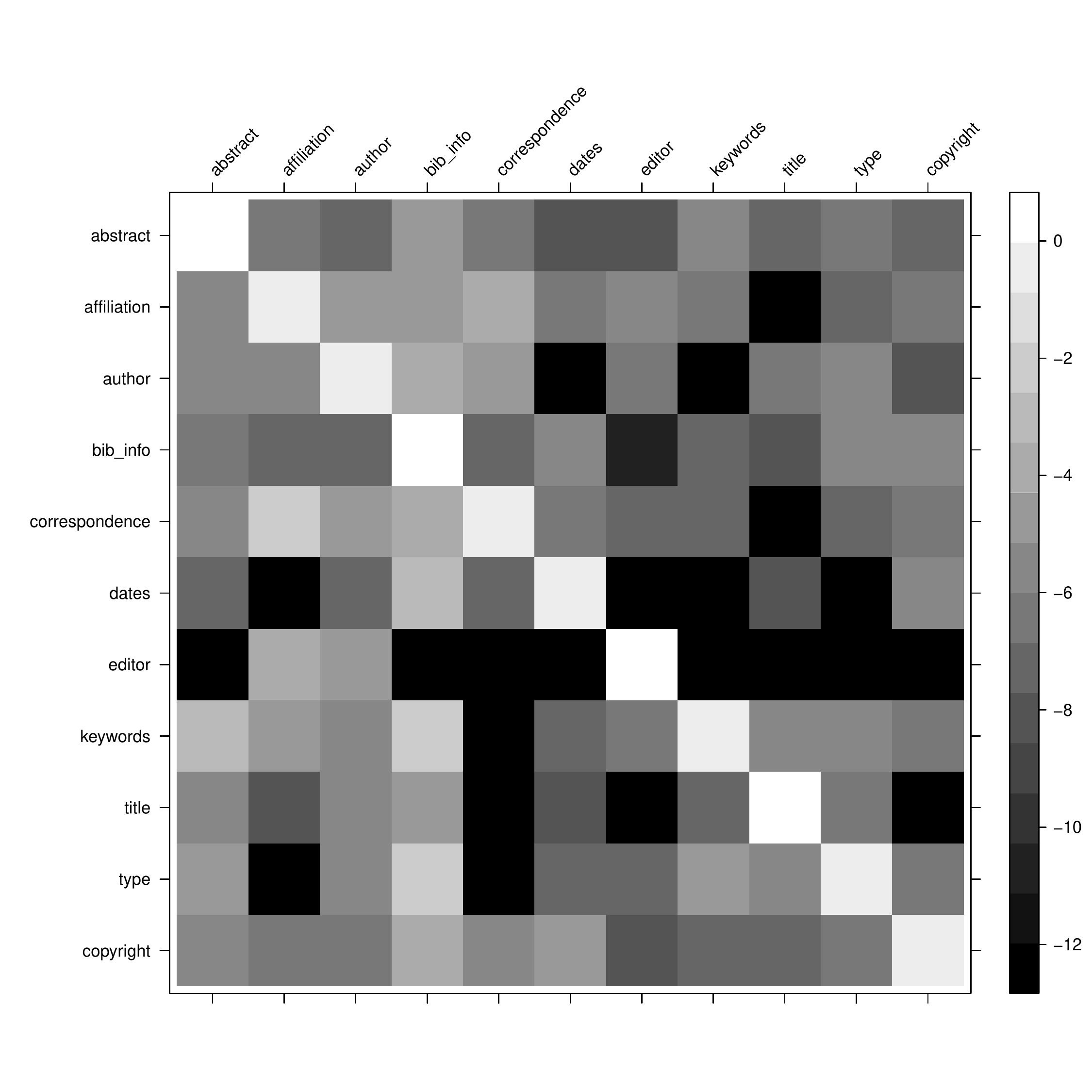}
  \caption[The visualization of the confusion matrix of metadata classification evaluation]{The
  visualization of the confusion matrix of metadata classification. The value in the cell in row $i$
  and column $j$ corresponds to the fraction of the zones with true label $i$ that were classified
  by the classifier as $j$. The lighter the cell color, the higher the fraction and the more
  confusion between the two classes.}
  \label{fig:hm-meta}
\end{figure}

Table~\ref{tab:res-meta} shows the precision, recall and F-score values for individual classes for 
metadata classification. Figure~\ref{fig:hm-meta} presents the visualization of the confusion matrix
for this classifier.

\begin{table}
\renewcommand{\arraystretch}{1.3}
\renewcommand{\tabcolsep}{2pt}
\centering
  \begin{tabular}{ | c || r | r | r | }
	\hline
	& precision (\%) & recall (\%) & F-score (\%) \\ \hline \hline

	{\it abstract} & 97.19 & 98.00 & 97.59\\\hline
	{\it affiliation} & 94.40 & 94.38 & 94.39\\\hline
	{\it author} & 96.13 & 96.41 & 96.27\\\hline
	{\it bib\_info} & 98.11 & 98.68 & 98.40\\\hline
	{\it corresp.} & 91.38 & 88.32 & 89.82\\\hline
	{\it dates} & 94.75 & 93.20 & 93.97 \\\hline
	{\it editor} & 95.67 & 97.48 & 96.57 \\\hline
	{\it keywords} & 92.39 & 79.12 & 85.24 \\\hline
	{\it title} & 98.51 & 98.14 & 98.33 \\\hline
	{\it type} & 89.42 & 87.14 & 88.27\\\hline
	{\it copyright} & 95.41 & 95.15 & 95.28\\\hline \hline
	average & 94.85 & 93.27 & 94.01 \\ \hline
		
\end{tabular}
\caption[The results of metadata classification evaluation]{Precision, recall and F-score values for
individual classes for metadata classification.}
\label{tab:res-meta}
\end{table}

The results of the metadata classifier are also good. {\it bib\_info} zones seem to be often 
confused with other zones, but the precision and recall for this class is one of the highest. This
is caused by the fact that there are a few times more {\it bib\_info} zones in the test set than any
other zone class. The classes pairs often confused with each other are: {\it affiliation} and {\it 
correspondence}, which is related to the fact that affiliation strings sometimes contain postal or
email addresses, and {\it abstract} and {\it keywords}, which is a results of the fact that these 
regions are often placed close to each other, sometimes even in the same geometric zone, and are 
often printed using the same font and formatting style.

\begin{figure}
  \centering
  \includegraphics[width=0.5\textwidth]{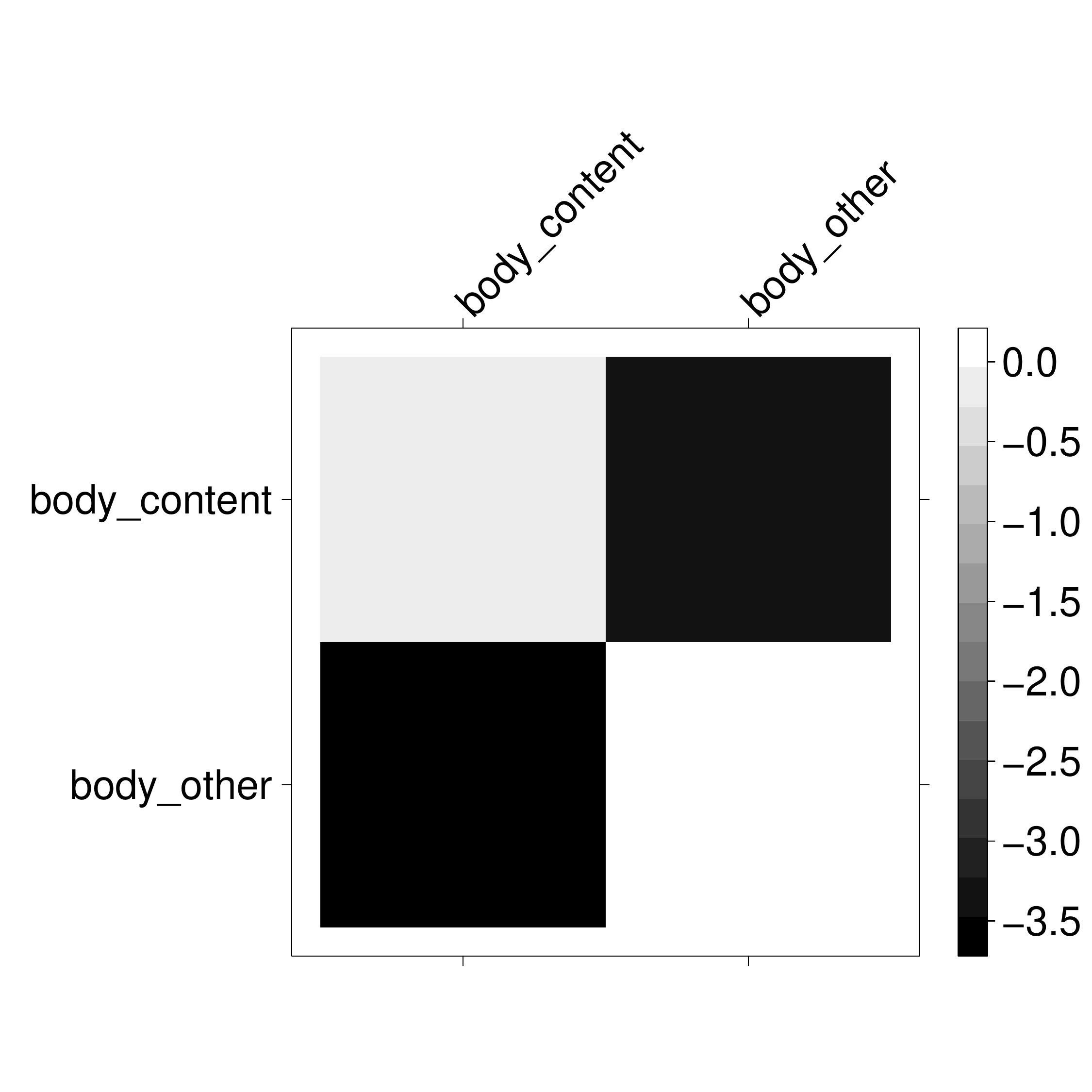}
  \caption[The visualization of the confusion matrix of body classification evaluation]{The
  visualization of the confusion matrix of body classification. The value in the cell in row $i$ and
  column $j$ corresponds to the fraction of the zones with true label $i$ that were classified by
  the classifier as $j$. The lighter the cell color, the higher the fraction and the more confusion
  between the two classes.}
  \label{fig:hm-body}
\end{figure}

Table~\ref{tab:res-body} shows the precision, recall and F-score values for individual classes for 
body classification. Figure~\ref{fig:hm-body} shows the visualization of the confusion matrix for 
this classifier.

\begin{table}
\renewcommand{\arraystretch}{1.3}
\renewcommand{\tabcolsep}{2pt}
\centering
  \begin{tabular}{ | c || r | r | r |}
  	\hline 
  	& precision (\%) & recall (\%) & F-score (\%) \\ \hline \hline
  
  	{\it body\_content} & 96.84 & 96.47 & 96.65\\ \hline
    {\it body\_other} & 96.55 & 97.91 & 96.73 \\ \hline\hline
    average & 96.70 & 96.69 & 96.69 \\ \hline
   
     \end{tabular}
\caption[The results of body classification evaluation]{Precision, recall and F-score values for
individual classes for body classification.}
\label{tab:res-body}
\end{table}

\section{Affiliation Parsing}
\label{sec:aff-parsing}
Affiliation parser was evaluated by a 5-fold cross-validation with the use of the entire set of 
8,267 parsed affiliations from the affiliation dataset.

The affiliation set was randomly divided into 5 disjoint subsets and each subset was processed by 
the affiliation parser trained on the labelled tokens from the remaining 4/5 of the affiliations.
The parsing results in each fold were then compared to the gold standard.

\begin{figure}
  \centering
  \includegraphics[width=0.5\textwidth]{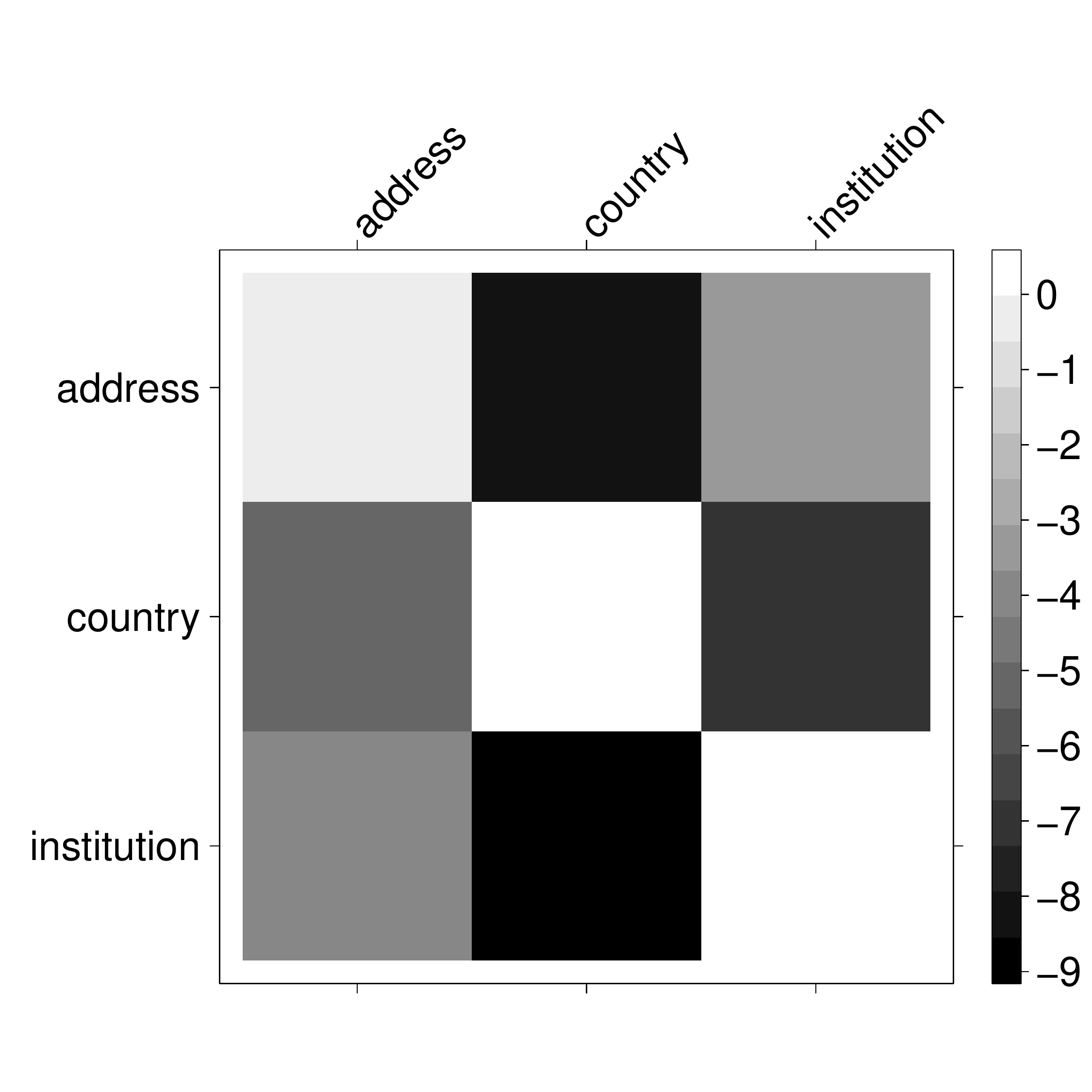}
  \caption[The visualization of the confusion matrix of affiliation token classification]{The
  visualization of the confusion matrix of affiliation token classification. The value in the cell
  in row $i$ and column $j$ corresponds to the fraction of the tokens with true label $i$ that were
  classified by the parser as $j$. The lighter the cell color, the higher the fraction and the more
  confusion between the two classes.}
  \label{fig:hm-affs}
\end{figure}

First, we compared the token labelling obtained from the parser to the labelling in the test set. 
The confusion matrix for token classification was constructed in the same way as confusion matrices
for zone classifiers. The exact values can be found in appendix~\ref{chap:app-results} in 
Section~\ref{sec:app-params}. The precision and recall values for the individual labels were also
calculated the same way and are presented in Table~\ref{tab:res-affs}. Figure~\ref{fig:hm-affs} 
shows the visualization of the confusion matrix for this classifier, generated the same way as
before.

\begin{table}
\renewcommand{\arraystretch}{1.2}
\renewcommand{\tabcolsep}{5pt}
\centering
  \begin{tabular}{ | c || r | r | r |}
  	\hline 
  	& precision (\%) & recall (\%) & F-score (\%) \\ \hline \hline
  
  	{\it address} & 96.74 & 97.25 & 97.00 \\ \hline
  	{\it country} & 99.63 & 99.22 & 99.42 \\\hline
    {\it institution} & 98.66 & 98.45 & 98.55\\ \hline \hline
    average & 98.34 & 98.31 & 98.32 \\ \hline
   
     \end{tabular}
     \vspace{5px}
\caption[The results of affiliation token classification evaluation]{Precision, recall and F-score
values for individual classes for affiliation token classification.}
\label{tab:res-affs}
\end{table}

In addition to evaluating individual token classification, we also checked for how many affiliations
the entire fragments (institution, address or country) were labelled correctly. We considered a
fragment labelled correctly if it was equal to the gold standard data. The following results were
obtained:
\begin{itemize}
\item institution was correctly recognized in 92.39\% of affiliations,
\item address was correctly recognized in 92.12\% of affiliations,
\item country was correctly recognized in 99.44\% of affiliations,
\item 92.05\% of affiliations were entirely correctly parsed.
\end{itemize}

As can be clearly seen from the results, the country is relatively easy to locate in the affiliation 
string, as it usually uses a very limited number of words from an easy to compile dictionary. The
institution and address are more difficult and more often confused with each other.

\section{Reference Parsing}
\label{sec:ref-parsing}
Bibliographic reference parser was evaluated with the use of a 5-fold cross-validation on the subset
of the citation test set containing 2,000 parsed citations.

The test set was randomly divided into 5 disjoint subsets and each subset was processed by the 
parser trained on the labelled tokens from the remaining 4/5 of the test set. After parsing the
citation strings we compared the token labelling to the gold standard labelling from the dataset.

\begin{figure}
  \centering
  \includegraphics[width=1\textwidth]{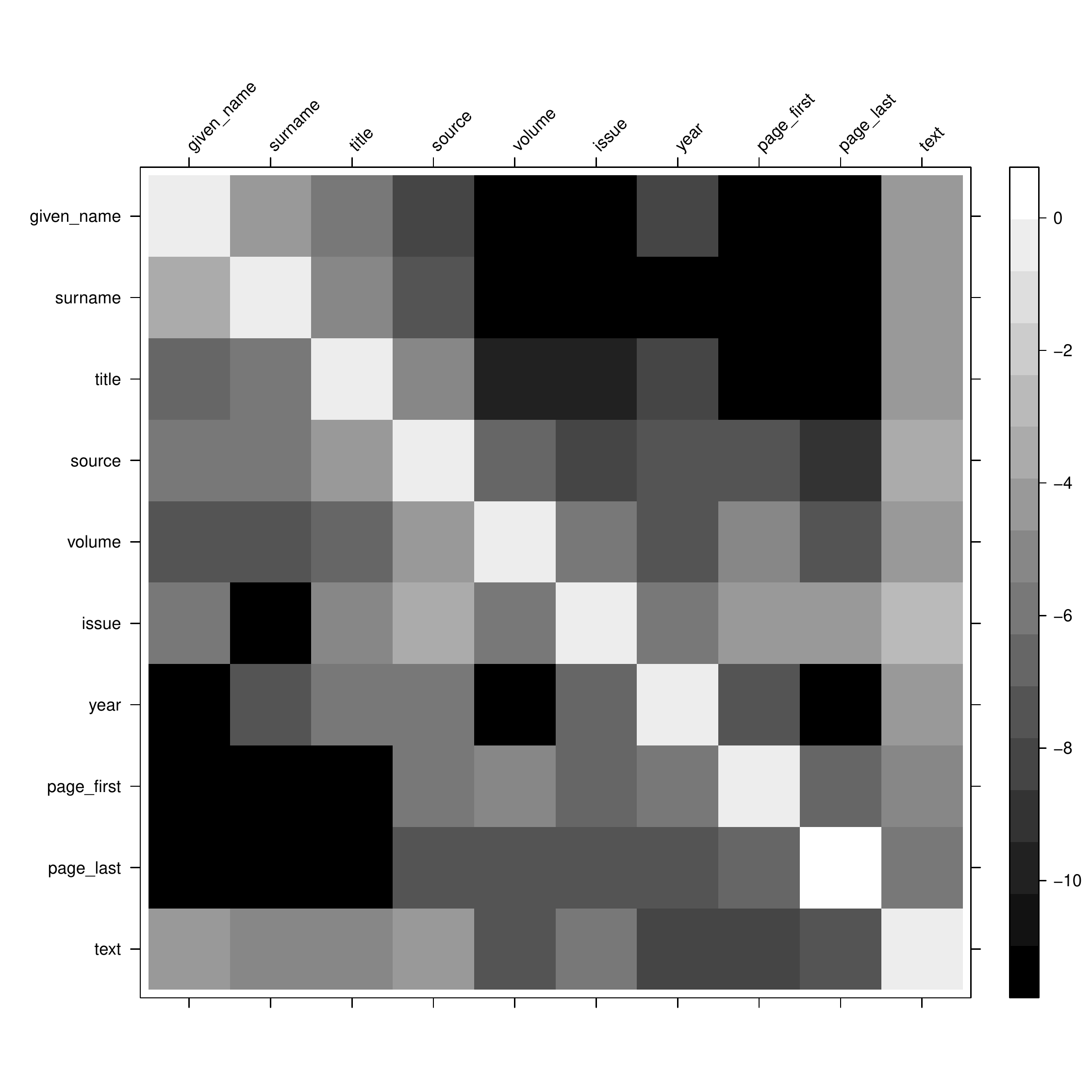}
  \caption[The visualization of the confusion matrix of citation token classification evaluation]
  {The visualization of the confusion matrix of citation token classification. The value in the cell
  in row $i$ and column $j$ corresponds to the fraction of the tokens with true label $i$ that were
  classified by the classifier as $j$. The lighter the cell color, the higher the fraction and the
  more confusion between the two classes.}
  \label{fig:hm-refs}
\end{figure}

The confusion matrix for token classification was constructed in the same way as confusion matrices 
for zone classification and can be found in appendix~\ref{chap:app-results} in 
Section~\ref{sec:app-params}. The precision and recall values for the individual labels were also
calculated the same way and are presented in Table~\ref{tab:res-refs}. Figure~\ref{fig:hm-refs} 
shows the visualization of the confusion matrix for this classifier.

\begin{table}
\renewcommand{\arraystretch}{1.3}
\renewcommand{\tabcolsep}{3pt}
\centering
  \begin{tabular}{ | c || r | r | r | }
	\hline
	& precision (\%) & recall (\%) & F-score (\%) \\ \hline \hline

	{\it given\_name} & 94.66 & 96.12 & 95.38\\\hline
	{\it surname} & 94.05 & 95.49 & 94.76\\\hline
	{\it title} & 97.93 & 97.76 & 97.85\\\hline
	{\it source} & 93.41 & 93.93	& 93.67\\\hline
	{\it volume} & 96.53 & 95.73 & 96.13\\\hline
	{\it issue} & 78.62 & 86.81 & 82.51\\\hline
	{\it year} & 98.53 & 97.64 & 98.08\\\hline
	{\it page\_first} & 98.02 & 97.58 & 97.80\\\hline
	{\it page\_last} & 98.16 & 99.38 & 98.77\\\hline
	{\it text} & 96.75 & 95.77 & 96.26\\\hline\hline
	average & 94.66 & 95.62 & 95.12 \\\hline
\end{tabular}
\caption[The results of citation token classification evaluation]{Precision, recall and F-score
values for individual classes for citation token classification.}
\label{tab:res-refs}
\end{table}

The worst precision and recall were achieved in the case of the label {\it issue}, which is the 
least common label in the test set. The label {\it text} seems to be the most confused with the
other labels, which is caused by the fact that it is the most common label, and also some regions in
the automatically constructed test set might be mislabelled as {\it text}. The mostly confused label 
pairs are: {\it givenname} and {\it surname} (these tokens are almost always close to each other) 
and {\it title} and {\it source} (which are also close to each other, and sometimes it is not a
trivial task to find the border between the two).

We also evaluated the precision and recall of entire task of extracting metadata from citation 
strings. A citation consists of a raw string and a set of substrings labelled with metadata classes.
Let $R^G$ be a parsed citation from the test set and $R^T$ --- the corresponding citation processed
by the parser during the evaluation. Let also $C_{R^G}$ be the set of substrings of the class $C$ in 
the citation $R^G$, and similarly $C_{R^T}$ --- the set of substrings of the class $C$ in the 
citation $R^T$.

We calculate the precision, recall and F-score for citation $R$ and class $C$ as:

\[ P(R,C) = \frac{|C_{R^G} \cap C_{R^T}|}{|C_{R^T}|} \]

\[ R(R,C) = \frac{|C_{R^G} \cap C_{R^T}|}{|C_{R^G}|} \]

\[ F(R,C) = 2 \cdot (P(R,C)^{-1} + R(R,C)^{-1})^{-1} \]

\begin{figure}
  \centering
  \includegraphics[width=0.7\textwidth]{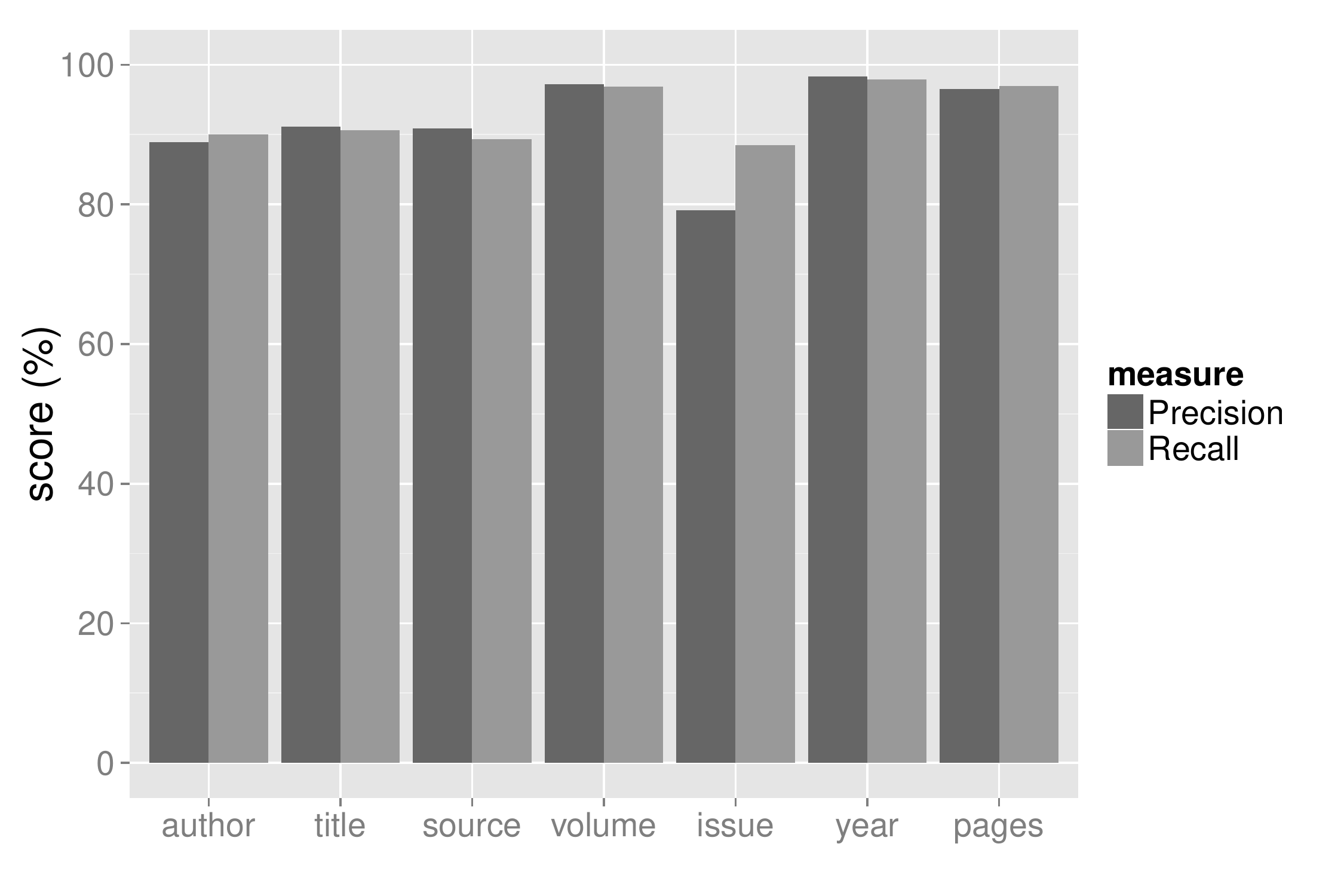}
  \caption[The results of bibliographic reference parser evaluation]{Bibliographic reference parser
  evaluation. The figure shows the average precision and recall values for extracting reference
  fragments belonging to individual metadata classes. A given fragment is considered correctly
  extracted, if it is identical to the ground truth data.}
  \label{fig:eval-refs}
\end{figure}

The average precision, recall and F-score are calculated as mean values over all citations in the
dataset. Figure~\ref{fig:eval-refs} shows the average precision and recall values for individual
metadata classes.

The highest scores were obtained in the case of the labels {\it volume}, {\it year} and {\it pages}.
This is related to the fact, that these fragments are comparatively easy to locate, because they are 
numerical and usually formatted in the same way. The lowest scores were achieved in the case of {\it 
issue}, which is most likely caused by the fact that the test set contains only few cases of 
citations with the issue tagged, as this information is often missing in the reference string.

\section{Extraction Evaluation}
\label{sec:extr-eval}
This sections reports the methodology and results of the evaluation of the entire extraction process
in all its aspects: metadata extraction, bibliography extraction and structured body extraction. The
results are also compared with the results achieved by similar tools on the same datasets.

Two datasets are used for these experiments: a subset of PMC containing 1,943 documents and a subset 
of Elsevier dataset with 2,508 documents in total. The PMC dataset is disjoint with both the 
validation and training sets used before.

We used the test sets to assess the quality of our algorithm and to compare it to other, similar 
systems. The following systems were evaluated: GROBID\footnote{https://github.com/kermitt2/grobid},
PDFX\footnote{http://pdfx.cs.man.ac.uk/}, ParsCit\footnote{http://aye.comp.nus.edu.sg/parsCit/} and
pdf-extract\footnote{https://github.com/CrossRef/pdfextract}. Each system extracts a slightly 
different range of metadata information. Table~\ref{tab:extr-systems} summarizes the scope of the
information extracted by various metadata extraction approaches.

\begin{table}
\renewcommand{\arraystretch}{1.3}
\renewcommand{\tabcolsep}{2pt}
\centering
  \begin{tabular}{| c | C{2.5cm} | C{2.3cm} | C{2.3cm} | C{2.3cm} | C{2.3cm} |}
  \hline
       & our algorithm & PDFX & GROBID & ParsCit & pdf-extract \\ \hline \hline

\textit{title} & \cmark & \cmark & \cmark & \cmark & \cmark\\
\textit{author} & \cmark & \xmark & \cmark & \cmark & \xmark\\
\textit{affiliation} & \cmark & \xmark & \cmark & \cmark & \xmark\\
\textit{affiliation metadata} & \cmark & \xmark & \cmark & \xmark & \xmark\\
\textit{author--affiliation} & \cmark & \xmark & \cmark & \xmark & \xmark\\
\textit{email address} & \cmark & \cmark & \cmark & \cmark & \xmark\\
\textit{author--email} & \cmark & \xmark & \cmark & \xmark & \xmark\\
\textit{abstract} & \cmark & \cmark & \cmark & \cmark & \xmark\\
\textit{keywords} & \cmark & \xmark & \cmark & \xmark & \xmark\\
\textit{journal} & \cmark & \xmark & \cmark & \xmark & \xmark\\
\textit{volume} & \cmark & \xmark & \cmark & \xmark & \xmark\\
\textit{issue} & \cmark & \xmark & \cmark &\xmark & \xmark\\
\textit{pages range} & \cmark & \xmark & \cmark &\xmark & \xmark\\
\textit{year} & \cmark & \xmark & \cmark & \xmark & \xmark\\
\textit{DOI} & \cmark & \cmark & \cmark & \xmark & \xmark\\
\textit{reference} & \cmark & \cmark & \cmark & \cmark & \cmark\\
\textit{reference metadata} & \cmark & \xmark & \cmark & \cmark & \xmark\\
\textit{section headers} & \cmark & \cmark & \cmark & \cmark & \xmark\\
\textit{section hierarchy} & \cmark & \cmark & \xmark & \cmark & \xmark\\
\hline
\end{tabular}
\caption[The scope of the information extracted by various metadata extraction systems]{The
comparison of the scope of the information extracted by various metadata extraction systems. The
table shows simple metadata types (eg. {\it title}, {\it author}, {\it abstract} or {\it 
bibliographic references}), relations between them ({\it author --- affiliation}, {\it author ---
email address}), and also metadata in the structured form ({\it references} and {\it affiliations}
along with their metadata).}
\label{tab:extr-systems}
\end{table}

\subsection{Evaluation Methodology}
The PDF files from each test set were processed by each system, including our algorithm. The 
extraction results were then compared to the gold standard XML data from the test sets, resulting in 
the scores for each metadata category. The only exception was ParsCit, which is not able to directly
process PDF files. In this case we generated text versions of the input files using {\tt pdftotext}
tool, which were then used as the input for ParsCit.

None of the systems processed successfully all the test files. The analysis resulted in an error in
some cases. We also set up a timeout of 20 minutes for processing one files, which was exceeded in a
number of cases. Table~\ref{tab:extr-systems-errors} shows the number and percentage of successfully
processed files for each system.

\begin{table}
\renewcommand{\arraystretch}{1.3}
\renewcommand{\tabcolsep}{3pt}
\centering
  \begin{tabular}{ | c | C{4cm} | C{4cm} | }
  \hline
      system & \textbf{PMC} & \textbf{Elsevier} \\
      \hline \hline

our algorithm & 1940 (99.8\%) & 2508 (100\%)\\\hline
GROBID & 1941 (99.9\%) & 2506 (99.9\%) \\\hline
PDFX & 1907 (98.1\%) & 2439 (97.2\%) \\\hline
ParsCit & 1924 (99.0\%) & 2505 (99.9\%) \\\hline
pdf-extract & 1912 (98.4\%) & 2352 (93.8\%) \\\hline

\end{tabular}
\caption[The number and percentage of successfully processed files for each system]{The number and
percentage of successfully processed files from each dataset for each system.}
\label{tab:extr-systems-errors}
\end{table}

We evaluated how well the systems extract the following metadata categories:
\begin{itemize}
\item title (a single string),
\item authors (a list of authors' full names),
\item affiliations (understood as raw strings, the affiliations' metadata was not evaluated in these
experiments),
\item relations author--affiliation (a list of pairs author full name -- affiliation raw string),
\item email addresses (a list of strings), 
\item relations author--email (a list of pairs author full name -- email string),
\item abstract (a single string),
\item keywords (a list of strings),
\item journal (a single string),
\item volume (a single string),
\item issue (a single string),
\item pages range (a single relation first page -- last page),
\item year (a single string), 
\item DOI identifier (a single string),
\item bibliographic references (understood as raw strings, the references' metadata was not 
evaluated in these experiments),
\item section headers (a list of strings),
\item section headers hierarchy (a list of pairs section level -- section header).
\end{itemize}

Each metadata category was evaluated separately. For each category we report the overall performance 
scores (precision, recall and F-score) on each dataset for each system that is able to extract the
given category. We also performed statistical analysis to find out which differences in performance
are statistically significant.

In general we deal with two types of metadata categories: those that appear at most once per 
document ("single" types: title, abstract, journal, volume, issue, pages range, year and DOI) and
those present as lists ("list" types: authors, affiliations, email addresses, all relations,
keywords, bibliographic references and section headers). Once again ParsCit is an exception to this
rule. Since the system returns labelled document text instead of outputting the extraction results
in the form of a structured metadata record, all metadata categories may appear more than once per
document, along with the confidence scores. In the case of single metadata types we decided to
choose for testing the information with the highest confidence returned by ParsCit.

\subsubsection{Performance Scores}
We calculated the overall performance scores (precision, recall and F-score) for each combination of
extraction system, dataset and metadata category, provided that a given system is able to extract 
the information of a given category.

The overall performance scores are calculated based on the scores for the individual files. Let 
$D^G$ be the ground truth document from the test set and $D^T$ --- the corresponding metadata
document created by the evaluated system. For a metadata class $C$ we denote as $C_{D^G}$ the list
of metadata instances of class $C$ associated with the document $D^G$, and similarly, $C_{D^T}$ is
the list of metadata instances of class $C$ associated with the document $D^T$. Both lists can be
empty, if a certain metadata category is missing in the record. In the case of "single" metadata
types the lists contain at most one element.

Precision, recall and F-score for a document $D$ and metadata category $C$ are then calculated in   
the following way:

\[ P(D,C) =
  \begin{cases}
    null       & \quad \text{if } C_{D^T} = \emptyset \\
    \frac{|C_{D^G} \cap C_{D^T}|}{|C_{D^T}|}  & \quad \text{otherwise}\\
  \end{cases}
\]

\[ R(D,C) =
  \begin{cases}
    null       & \quad \text{if } C_{D^G} = \emptyset \\
    \frac{|C_{D^G} \cap C_{D^T}|}{|C_{D^G}|}  & \quad \text{otherwise}\\
  \end{cases}
\]

\[ F(D,C) =
  \begin{cases}
    null       & \quad \text{if } P(D,C) = \text{null and } R(D,C) = \text{null} \\
    0       	& \quad \text{if } P(D,C) = \text{null or } R(D,C) = \text{null} \\
    0       	& \quad \text{if } P(D,C) \cdot R(D,C) = 0 \\
    2(P(C)^{-1}+R(C)^{-1})^{-1}  & \quad \text{otherwise}\\
  \end{cases}
\]

The intersection $C_{D^G} \cap C_{D^T}$ is a set of elements present in both lists, with respect to
the comparison method, which differs between metadata categories.

The overall precision, recall and F-score for the entire test set are computed as averages over all
non-null values. As a result, in general the overall F-score is not equal to the harmonic mean of
the overall precision and recall scores.

\subsubsection{Statistical Analysis}
For each metadata category and each dataset we compared the performances of every pair of relevant 
systems in order to test the following null hypothesis: the performances of the two systems are on
average the same. In all tests we compare two lists of F-scores obtained for individual documents,
with all records containing null values removed prior to the test. The lists contained only the
results for the documents successfully processed by the two systems of interest.

In the case of single metadata categories, the F-score for a given document is always equal to 1
(the metadata information was extracted correctly) or 0 (the metadata information was extracted 
incorrectly). Since we were interested in a difference between two paired proportions, we used
McNemar's test~\cite{McNemar47}.

In the case of list metadata categories, we were interested in analysing paired differences between 
F-scores for individual documents and we used Wilcoxson signed-rank test~\cite{Wilcoxon45}, a paired
difference test in which we do not have to assume normal distribution in the population, in contrast
to more popular paired Student's t-test.

To make sure we do not obtain significant results simply by chance, as a result of performing many
statistical tests, we used the Bonferroni correction~\cite{Dunn59} and compared the p-values to the 
significance level of 5\% divided by the number of tests performed within each category.

In every test we compute the p-value and compare it to the significance level adjusted accordingly
to the number of executed tests. Based on this we decide whether we accept the null hypothesis or 
not, and thus whether the difference in performance between the two tested approaches is 
statistically significant.

\subsubsection{Comparison Details}
The metadata information extracted from the documents are compared in various manners, depending on
the metadata category.

The titles and abstracts are compared in a way that takes into account some minor differences 
related to encoding, character case, the presence of spaces, the representation of accents, etc. The
strings are considered equal if at least one of the following two conditions is met: if normalized
strings (with all non-alphanumerical characters removed) are identical, or if the similarity
calculated using the Smith-Watermann sequence alignment algorithm exceeds a predefined threshold. To
calculate the Smith-Watermann similarity we use the following formula:

\[ \sm(s_1,s_2) = \frac{2 \cdot \sw(s_1,s_2)}{\lnt(s_1) + \lnt(s_2)}\]

where $s_1$ and $s_2$ are the sequences of tokens of the compared strings, $\lnt(s)$ is the size of
the sequence $s$ and $\sw(s_1,s_2)$ is the Smith-Watermann distance of the two sequences $s_1$ and
$s_2$.

Two keyword strings are considered equal if they are identical after a basic (default in all our
experiments) normalization, which includes converting to lower case an trimming. Since ParsCit 
system only marks the entire keywords section, but does not extract the list of individual keywords,
it was excluded from the keywords evaluation.

The author names are considered equal if their normalized forms (with all non-letters removed) are
identical. This allows to take into account some minor differences like the presence or absence of
spaces or dots following the initials. However, in case of the problems with the encoding or when
one name is given in the full form and the other using initials, the names will not match.

The affiliations are tokenized and compared using cosine distance with a threshold. Cosine distance 
allows not to take into account the order of words, which helps in the case of systems which do not 
preserve the original affiliation string and the order of its tokens, but rather output only the
structured affiliation metadata. In the case of extracting the affiliations only the raw affiliation
strings are evaluated, the affiliation metadata is omitted.

Emails are compared after normalization, which includes removing all characters that are not
alphanumerical or '@'. We also remove prefixes like "E-mail:" and similar from the email strings, as
some systems, in particular GROBID and ParsCit, leave it in the output.

A relation (author-email or author-affiliation) is considered correct if both elements match their 
respective elements in compared pair. Individual elements are compared in the same way as described
above.

As journal name is often abbreviated in the input PDF document and given in the full form in the 
ground truth metadata record, we consider it extracted correctly if its normalized version (with all 
non-letters removed) is a non-empty subsequence of the ground truth journal name.

The pages range is correct if both first and the last page number is identical to the ground truth 
data. Similarly, the volume, issue, year and DOI are correct only if identical to the ground truth 
data.

The references are compared similarly as affiliations: using a cosine distance with a threshold. As
a result we do not take into account the order of the tokens, and we can reliably evaluate the
systems which return the references in a highly structured form without preserving the original
references strings.

The section headers are compared similarly as the title and abstract: with a use of Smith-Watermann 
similarity with a threshold. In addition to comparing the lists of section headers, we also assess
the performance of extracting the section hierarchy by comparing the relations of the form section
level -- section title. The levels are equal only if identical, and section titles are compared as 
before.

\subsection{Evaluation Results}
This section contains the summary of the results of the evaluation described above. The detailed
precision, recall and F-score values, as well as the p-values obtained in all statistical tests can
be found in the appendix~\ref{chap:app-results}.

\subsubsection{Document Metadata}
\label{sec:basic-metadata}
In this section we report the evaluation results related to the document metadata. We divided the
metadata categories into three groups: basic document metadata (the title, abstract and keywords),
authorship metadata (author names, affiliations, relations author--affiliation, email addresses,
relations author--email) and bibliographic metadata (journal name, volume, issue, pages range, year
of publication and DOI identifier).

\begin{figure}
  \centering
  \includegraphics[width=0.8\textwidth]{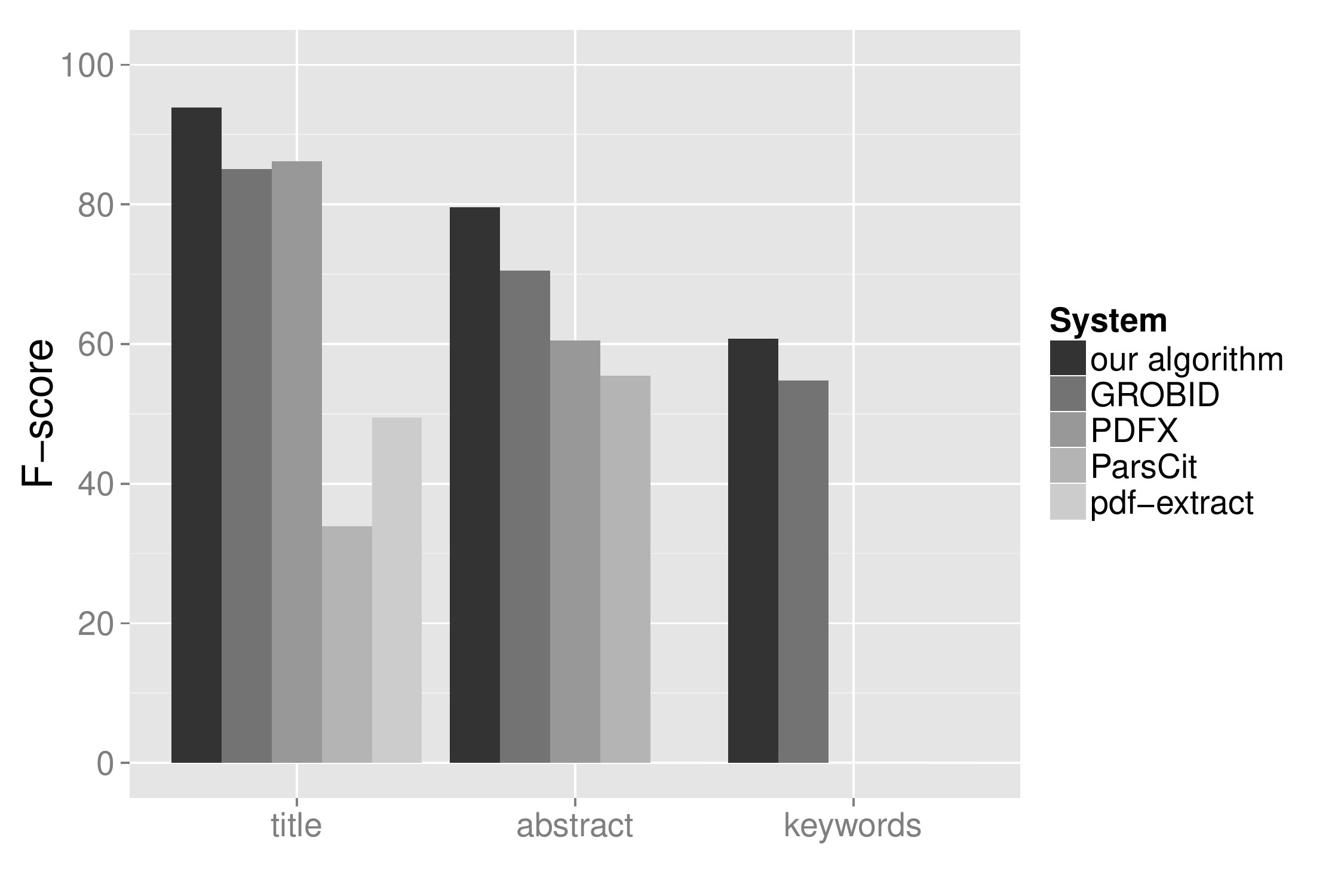}
  \caption[The results of the evaluation of extracting basic metadata on PMC]{The results of the
  evaluation of five different extraction systems with respect to the basic document metadata. The
  figure shows the average F-score over all documents from PMC test set.}
  \label{fig:eval-basic-pmc}
\end{figure}

Figure~\ref{fig:eval-basic-pmc} shows the average F-score for the basic metadata categories for all
tested systems on the PMC test set. In this set our algorithm achieved the best scores in all 
categories. 

\begin{figure}
  \centering
  \includegraphics[width=0.8\textwidth]{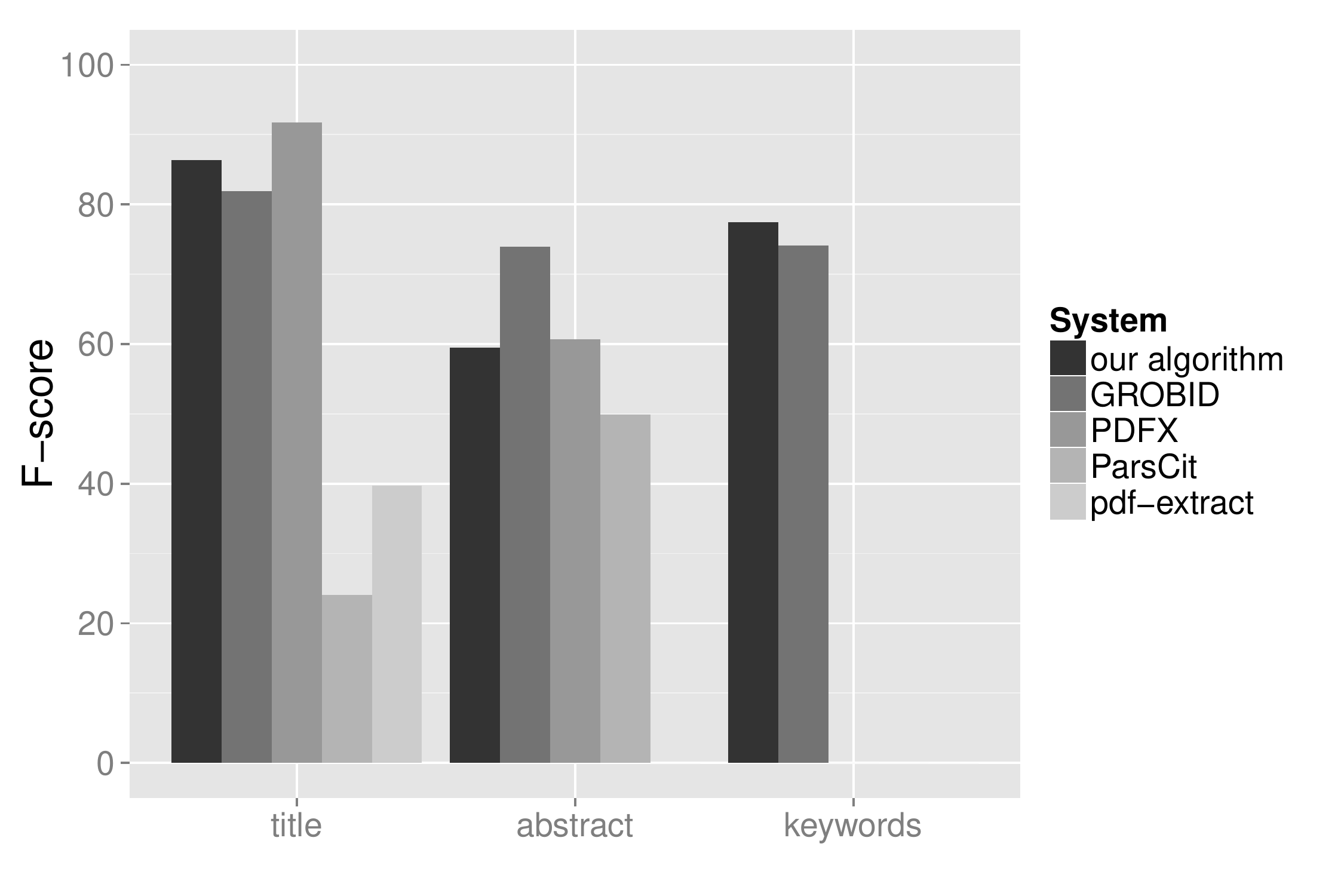}
  \caption[The results of the evaluation of extracting basic metadata on Elsevier]{The results of
  the evaluation of five different extraction systems with respect to the basic document metadata.
  The figure shows the average F-score over all document from Elsevier test set.}
  \label{fig:eval-basic-els}
\end{figure}

Figure~\ref{fig:eval-basic-els} shows the average F-score for the basic metadata categories for all
tested systems on the Elsevier test set. In this set every category has a different winner, and 
there are small differences between our algorithm, GROBID and PDFX in the case of title, and between
our algorithm and GROBID in the case of keywords.

\begin{figure}
  \centering
  \includegraphics[width=0.8\textwidth]{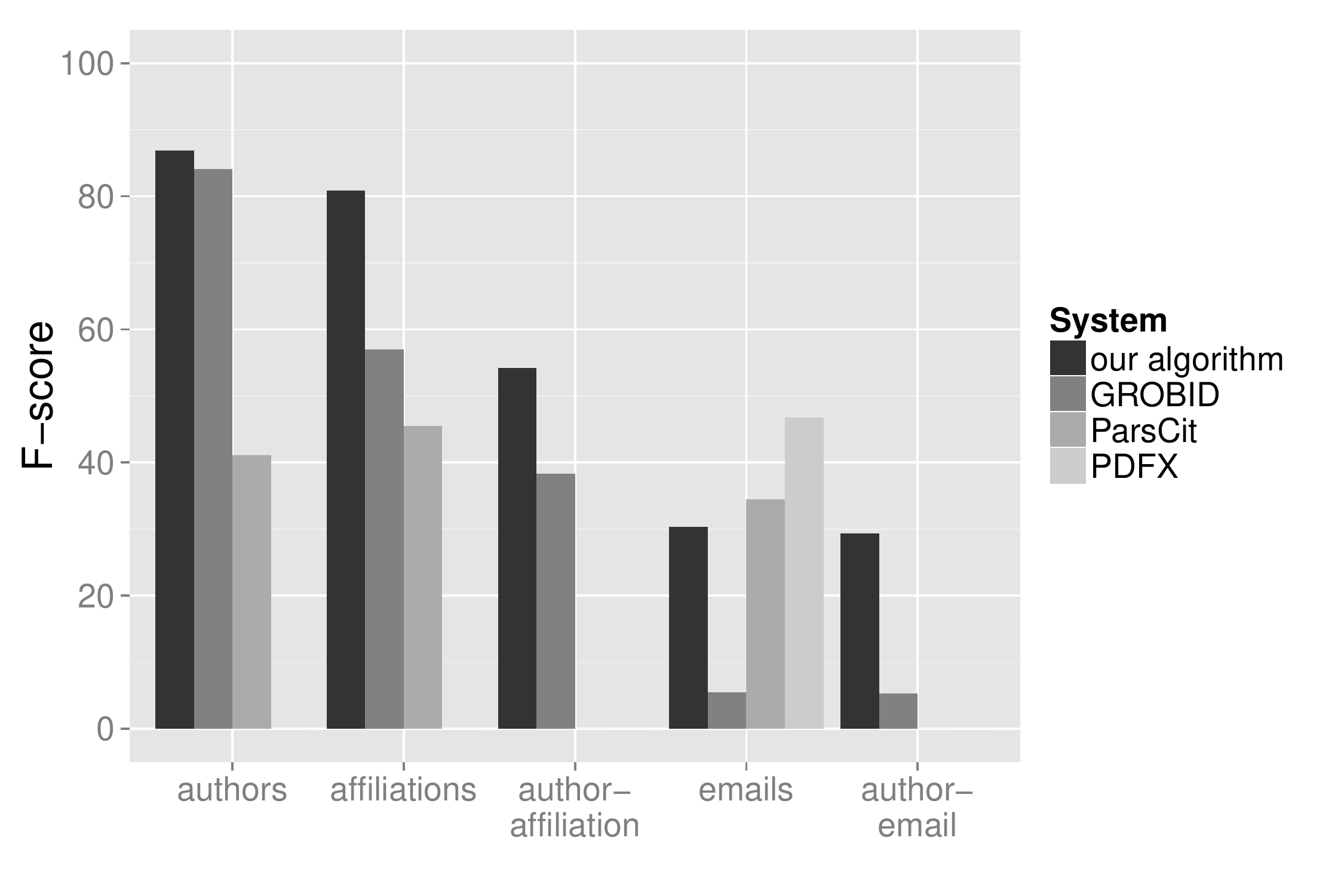}
  \caption[The results of the evaluation of extracting authorship metadata on PMC]{The results of
  the evaluation of four different extraction systems with respect to the authorship-related
  metadata. The figure shows the average F-score over all documents in PMC test set.}
  \label{fig:eval-auth-pmc}
\end{figure}

Figure~\ref{fig:eval-auth-pmc} shows the average F-score for the authorship metadata categories for
all tested systems in the PMC test set. Our algorithm achieved the best scores in all categories
except for the email address, where it was outperformed by PDFX. In the case of authors there is 
only a small difference between our solution and GROBID.

\begin{figure}
  \centering
  \includegraphics[width=0.8\textwidth]{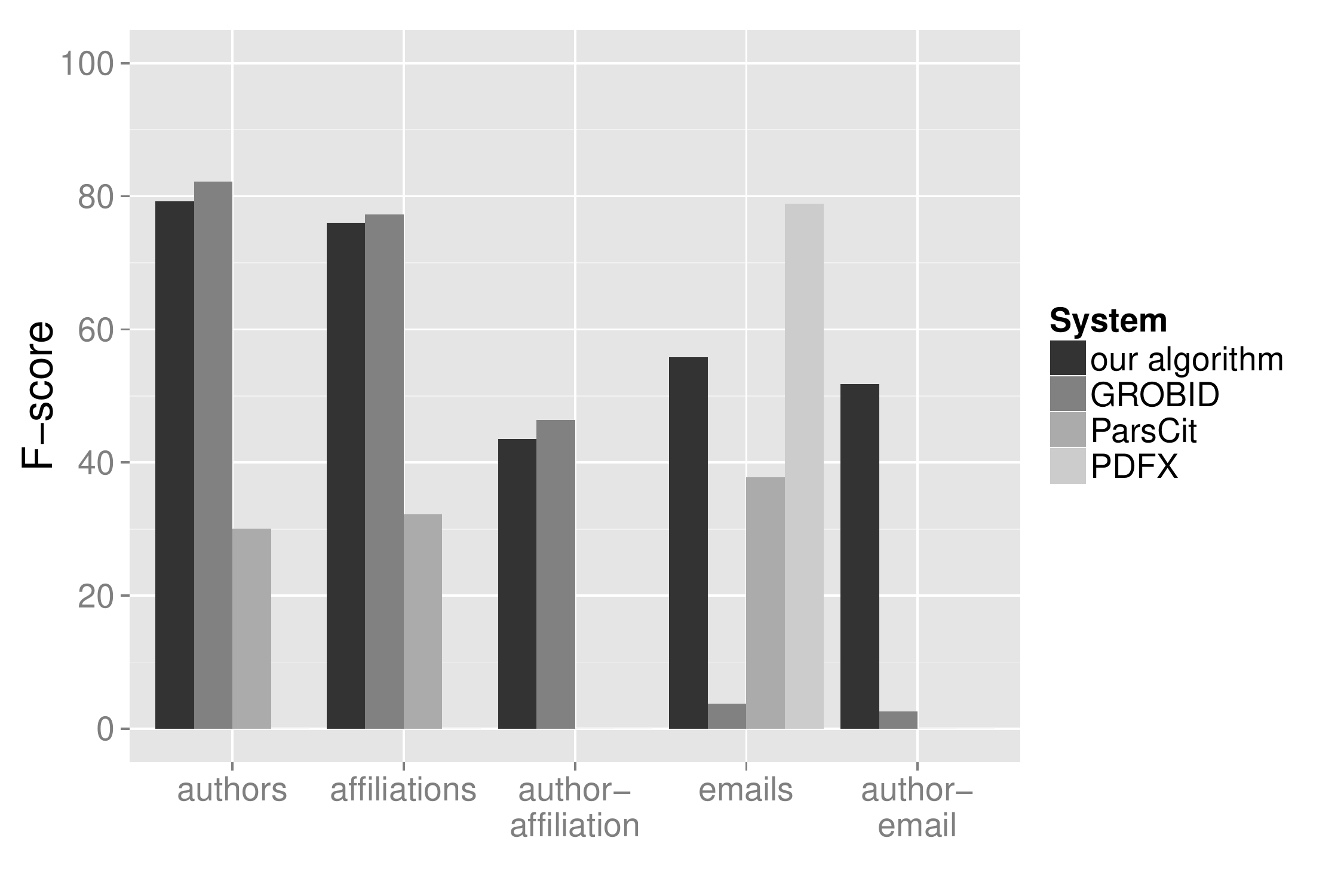}
  \caption[The results of the evaluation of extracting authorship metadata on Elsevier]{The results
  of the evaluation of four different extraction systems with respect to the authorship-related
  metadata. The figure shows the average F-score over all documents from Elsevier test set.}
  \label{fig:eval-auth-els}
\end{figure}

Figure~\ref{fig:eval-auth-els} shows the average F-score for the authorship metadata categories for
all tested systems for the Elsevier-based test set. Once again PDFX proved to be the best in 
extracting email addresses. In the first three categories GROBID achieved the best results and our
algorithm was the second best, with small differences between the two approaches. Our algorithm
performed the best in extracting author-email relations.

\begin{figure}
  \centering
  \includegraphics[width=0.8\textwidth]{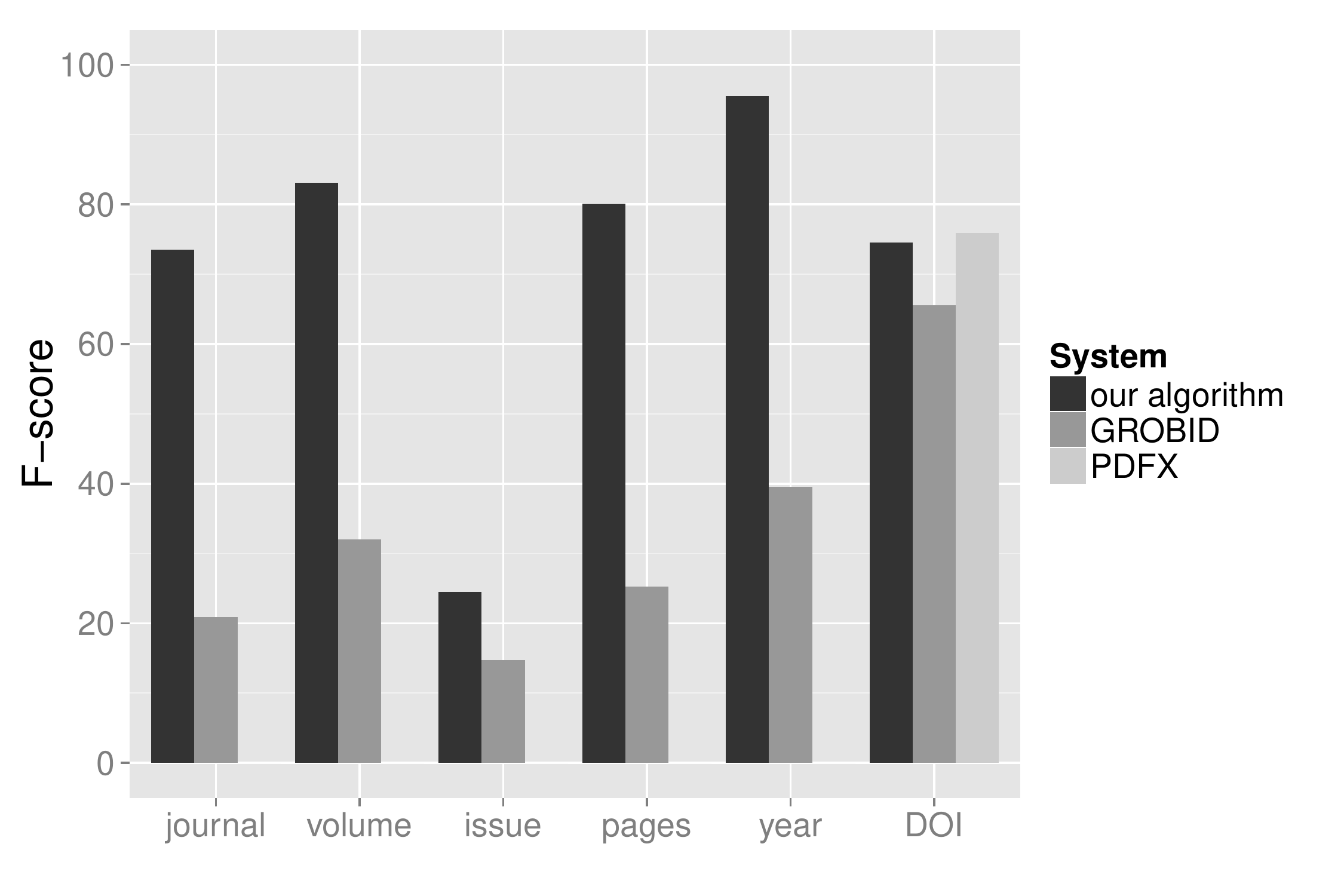}
  \caption[The results of the evaluation of extracting bibliographic metadata on PMC]{The results of
  the evaluation of three different extraction systems with respect to the bibliographic metadata.
  The figure shows the average F-score over all documents from PMC test set.}
  \label{fig:eval-journ-pmc}
\end{figure}

Figure~\ref{fig:eval-journ-pmc} shows the average F-score for the bibliographic metadata categories
for all tested systems for the PMC-based test set. Our algorithm achieved the best scores in all
categories except for DOI, where the differences in F-score for the three systems are relatively 
small.

\begin{figure}
  \centering
  \includegraphics[width=0.8\textwidth]{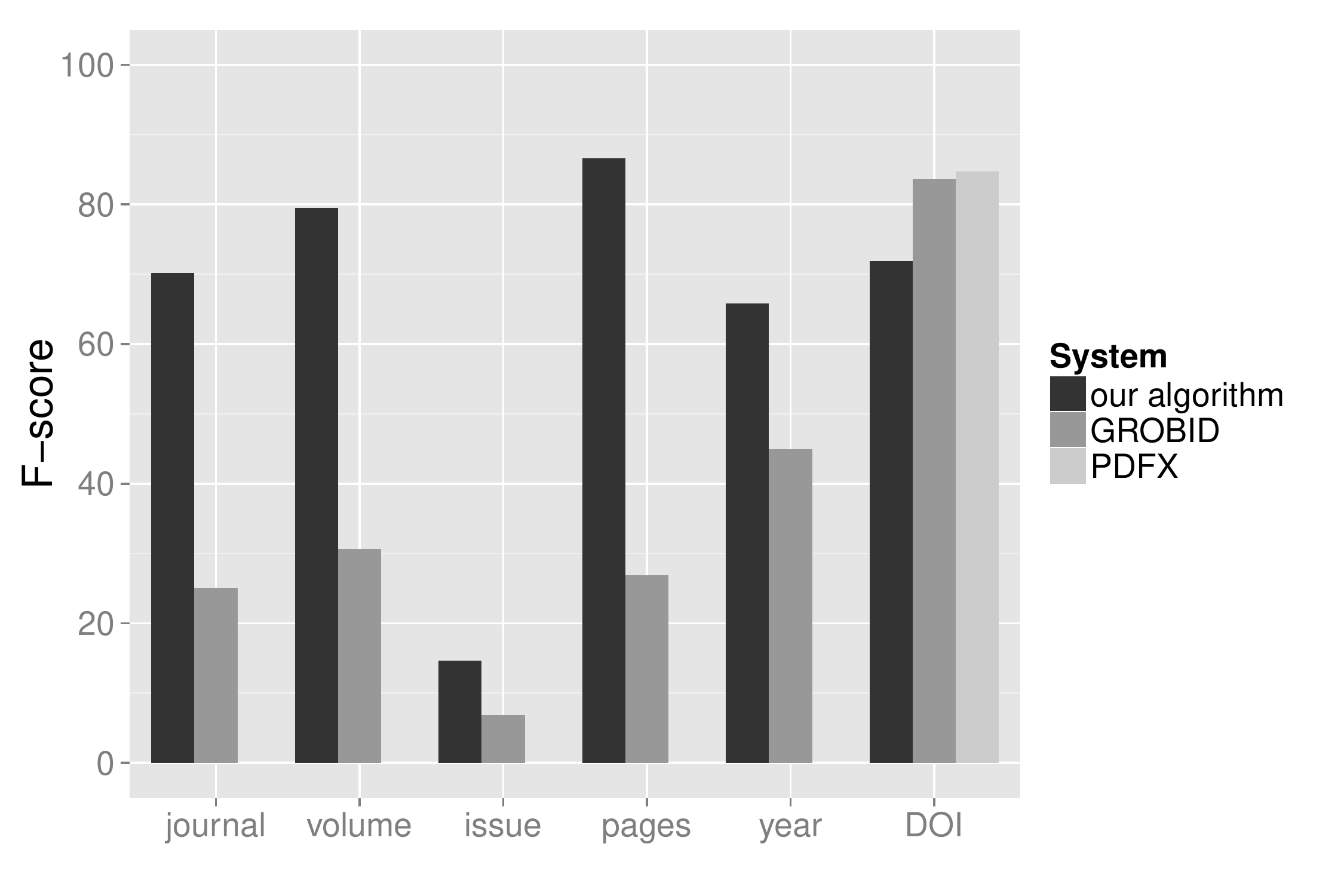}
  \caption[The results of the evaluation of extracting bibliographic metadata on Elsevier]{The
  results of the evaluation of three different extraction systems with respect to the bibliographic
  metadata. The figure shows the average F-score over all document from Elsevier test set.}
  \label{fig:eval-journ-els}
\end{figure}

Figure~\ref{fig:eval-journ-els} shows the average F-score for the bibliographic metadata categories
for all tested systems for the Elsevier-based test set. Similarly as before, our algorithm achieved
the best results in all categories except for DOI. In the case of DOI our algorithm performed the
worst, and the differences between the scores achieved by PDFX and GROBID are relatively small.

\subsubsection{Bibliographic References}
In this section we report the evaluation results with respect to extracting references from the 
bibliography section.

\begin{figure}
  \centering
  \includegraphics[width=0.8\textwidth]{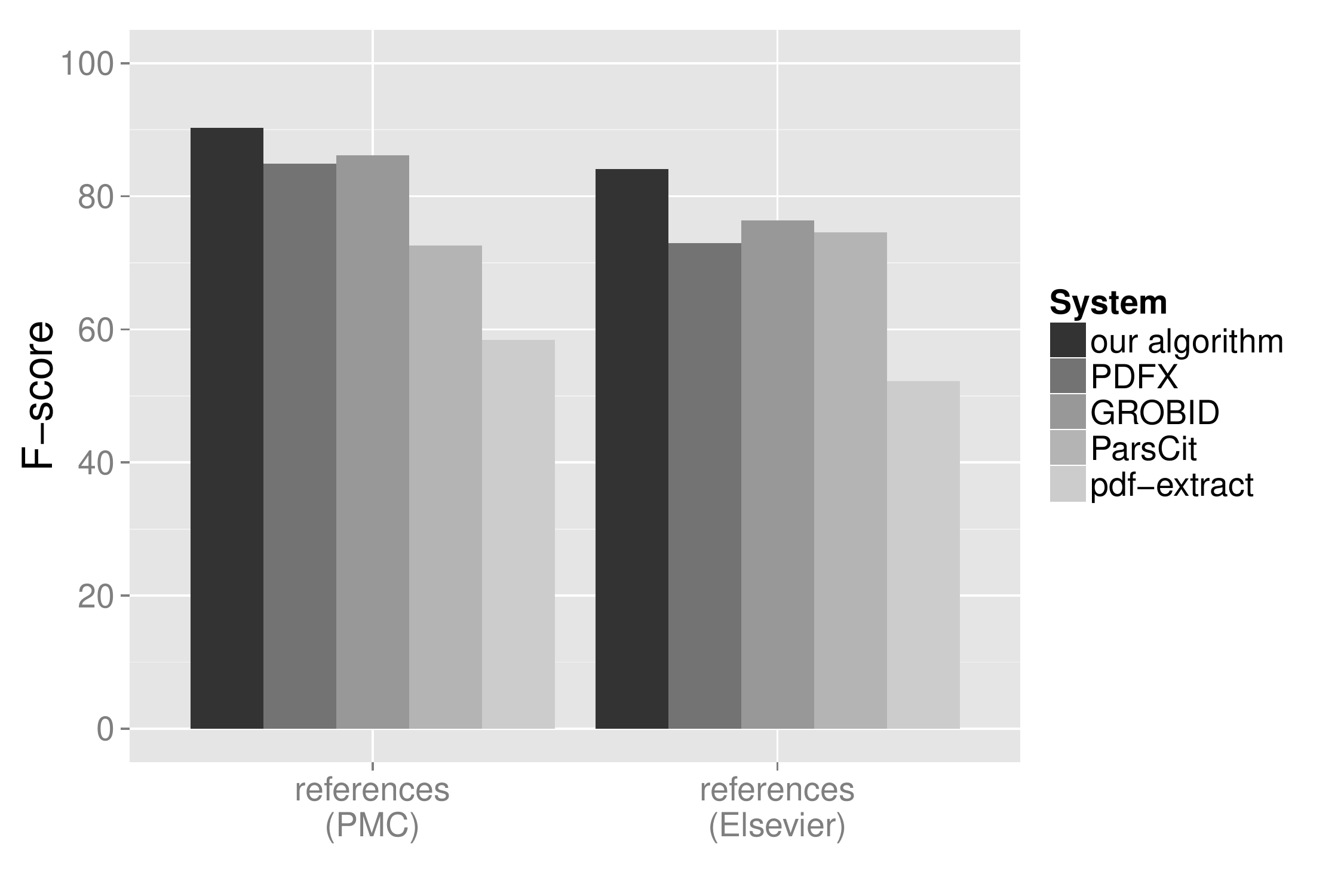}
  \caption[The results of the evaluation of extracting bibliographic references]{The results of the
  evaluation of various extraction systems with respect to bibliographic references extraction. The
  figure shows the average F-score over all documents from both PMC and Elsevier test sets.}
  \label{fig:eval-extr-refs}
\end{figure}

The results are shown in Figure~\ref{fig:eval-extr-refs}. In both datasets our algorithm performed 
the best, although the differences between the algorithms are not large.

\subsubsection{Section Hierarchy}
This section contains the results of the evaluation of extracting the hierarchy of section headers.

\begin{figure}
  \centering
  \includegraphics[width=0.8\textwidth]{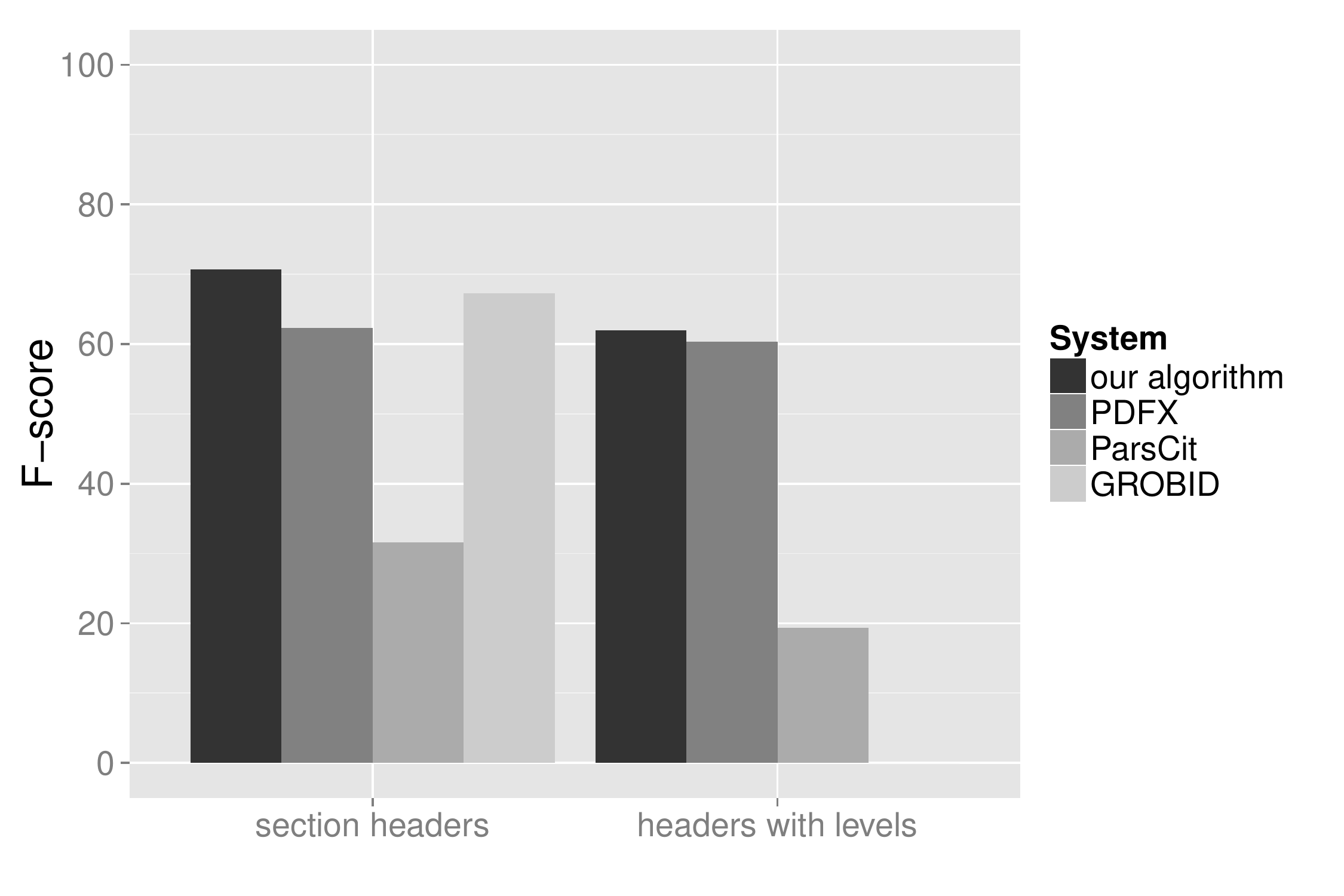}
  \caption[The results of the evaluation of extracting document headers on PMC]{The results of the
  evaluation of various extraction systems with respect to section headers extraction. The figure
  shows the average F-score over all documents from PMC test set for two tasks: extracting the
  header titles and extracting the header hierarchy.}
  \label{fig:eval-headers-pmc}
\end{figure}

The results are shown in Figure~\ref{fig:eval-headers-pmc}. We only used PMC test set for this 
experiment, as Elsevier data does not contain any information related to the section hierarchy. In
both tasks (extracting the section headers and extracting their hierarchy) our algorithm performed 
the best, although the differences between algorithms are not large.

\subsubsection{Summary}
\label{sec:eval-summary}
We evaluated five different extraction system (our algorithm, GROBID, PDFX, ParsCit and pdf-extract)
on two test sets (PMC and Elsevier). The systems processed the input PDF files and the resulting 
metadata records were compared to the ground truth metadata records in order to assess how well
various metadata categories are extracted. The evaluation comprised seventeen extraction tasks:
extracting title of the document, abstract, keywords, authors' full names, authors' affiliation
strings, relations author--affiliation, authors' email addresses, relations author--email, journal 
name, volume and issue numbers, pages range, year of publication, DOI identifier, reference strings
placed in the document, the titles of the sections and relations section level--section title.

Table~\ref{tab:summary-winners} presents the winner for each test set and each metadata category. In
some cases the difference between the winner and other systems were small, but in almost all 
statistical tests we performed the differences between the winner and other systems proved to be
statistically significant, the only exception being GROBID's win in extracting affiliations on
Elsevier dataset. More detailed results related to the p-values obtained can be found in the
appendix~\ref{chap:app-results}.

\begin{table}
\renewcommand{\arraystretch}{1.3}
\renewcommand{\tabcolsep}{3pt}
\centering
\begin{tabular}{ | c || c | c | }
	\hline
  	& PMC & Elsevier	\\	\hline \hline

	{\it title} 		& {\bf our algorithm} & PDFX \\\hline
	{\it abstract} 	& {\bf our algorithm} & GROBID \\\hline
	{\it keywords} 	& {\bf our algorithm} & {\bf our algorithm} \\\hline
	{\it authors} 		& {\bf our algorithm} & GROBID \\\hline
	{\it affiliations} 	& {\bf our algorithm} & GROBID* \\\hline
	{\it author--affiliation} 		& {\bf our algorithm} & GROBID \\\hline
	{\it email addresses} 			& PDFX & PDFX \\\hline
	{\it author--email} 		& {\bf our algorithm} & {\bf our algorithm} \\\hline
	{\it journal name}	& {\bf our algorithm} & {\bf our algorithm} \\\hline
	{\it volume} 	& {\bf our algorithm} & {\bf our algorithm} \\\hline
	{\it issue} 		& {\bf our algorithm} & {\bf our algorithm} \\\hline
	{\it pages range} 		& {\bf our algorithm} & {\bf our algorithm} \\\hline
	{\it year} 		& {\bf our algorithm} & {\bf our algorithm} \\\hline
	{\it DOI} 		& PDFX & PDFX \\\hline
	{\it references} 		& {\bf our algorithm} & {\bf our algorithm} \\\hline
	{\it section titles} 		& {\bf our algorithm} & --- \\\hline
	{\it level--section title} 		& {\bf our algorithm} & ---\\\hline
\end{tabular}
\caption[The winners of extracting each metadata category in each test sets]{The winner of
extracting each metadata category in each test sets. In almost every case the difference between the
winner and other approaches was statistically significant. The only exception is extracting
affiliations on Elsevier dataset won by GROBID (marked with a star), where the difference between
GROBID and our algorithm was not statistically significant.}
\label{tab:summary-winners}
\end{table}

In the case of PMC dataset our algorithm was the winner in all categories except for email addresses 
and DOI. In the case of Elsevier the results are worse, our algorithm was the winner in eight out of
fifteen categories. PDFX achieved the best scores in the case of email addresses and DOI identifier
in both test sets. GROBID performed the best in extracting abstract, authors and affiliations in
Elsevier test set. ParsCit and pdf-extract systems did not win in any category.
Table~\ref{tab:summary} shows for every system the number of categories the system was the best in,
either in both datasets or one of them.

\begin{table}
\renewcommand{\arraystretch}{1.3}
\renewcommand{\tabcolsep}{3pt}
\centering
\begin{tabular}{ | c || c | c | }
	\hline
	& \multicolumn{2}{c|}{number of wins in} \\
  	& both test sets & one test set
  	\\	\hline \hline

	{\it our algorithm} 		& 8 & 7 \\\hline
	{\it GROBID} 			& 0 & 4 \\\hline
	{\it PDFX} 				& 2 & 1 \\\hline
	{\it ParsCit} 			& 0 & 0 \\\hline
	{\it pdf-extract} 		& 0 & 0 \\\hline
\end{tabular}
\caption[The summary of the metadata extraction systems comparison]{The summary of the systems
comparison. The table shows the number of categories won by every extraction system in (1) both test
sets, and (2) one of the test sets. In 8 categories our algorithm was the best in both test sets,
and in 7 categories --- in one test set.}
\label{tab:summary}
\end{table}

\subsection{Error Analysis}
The extraction errors made by our algorithm can be divided into two groups: metadata was not
extracted or the extracted information is incorrect. The majority of errors happen in the following 
situations:
\begin{itemize}
\item When two (or more) zones with different roles in the document are placed close to each other, 
they are often merged together by the segmenter. In this case the classification is more difficult
and by design only one label is assigned to such a hybrid zone. A potential solution would be to
introduce additional labels for pairs of labels that often appear close to each other, for example
{\it title\_author} or {\it author\_affiliation}, and split the content of such zones later in the
workflow.
\item The segmenter introduces other errors as well, such as incorrectly attaching an upper index to
the line above the current line, or merging text written in two columns. These errors can be 
corrected by further improvement of the page segmenter.
\item Zone classification errors are also responsible for a lot of extraction errors. These errors
can be improved by adding training instances to the training set and improving the labelling 
accuracy in GROTOAP2.
\item Sometimes the metadata, usually keywords, volume, issue or pages, is not explicitly given in
the input PDF file. Since our algorithm analyses the PDF file only, such information cannot be 
extracted. This is in fact not an extraction error. Unfortunately, since ground truth NLM data in
PMC usually contains such information, whether it is written in the PDF or not, these situations 
also contribute to the overall error rates (equally for all evaluated systems).
\end{itemize}

The most common extraction errors include:
\begin{itemize}
\item Title merged with other parts of the document, when title zone is placed close to another 
region.
\item Title not recognized, for example when it appears on the second page of the PDF file.
\item Title zone split by the segmenter into a few zones, and only a subset of them is correctly 
classified.
\item Authors zone not labelled, in that case the authors are missing.
\item Authors zone merged with other fragments, such as affiliations or research group name, in such
cases additional fragments appear in the authors list.
\item Affiliation zone not properly recognized by the classifier, for example when it is not
visually separated from other zones, or placed at the end of the document. Affiliations are missing
in that case.
\item The entire abstract or a part of it recognized as {\it body} by the classifier, as a result
the abstract or a part of it is missing.
\item The first {\it body} paragraph recognized incorrectly as {\it abstract}, as a result the
extracted abstract contains a fragment of the document's proper text.
\item Bibliographic information missing from a PDF file or not recognized by the classifiers, as a
result journal name, volume, issue and/or pages range are not extracted.
\item Keywords missing because the zone was not recognized or not included in the PDF file.
\item A few of the references zones classified as {\it body}, in such cases some or all of the
references are missing.
\item When header titles are too similar to paragraph lines, the algorithm might not recognize them
properly, and as a result they will be missing in the section hierarchy.
\item If some parts of the document's body, such as table/figure fragments, are not filtered out
during the text content filtering, they are sometimes treated incorrectly as header lines.
\end{itemize}

\section{Time Performance}
\label{sec:time}
In addition to evaluating the performance of the algorithm, we also measured the time required to
process a scientific article. For these experiments we used a sample of 200 articles from PMC, 
selected randomly from the subset successfully process by all the tested algorithms.

\begin{figure}
  \centering
  \includegraphics[width=0.8\textwidth]{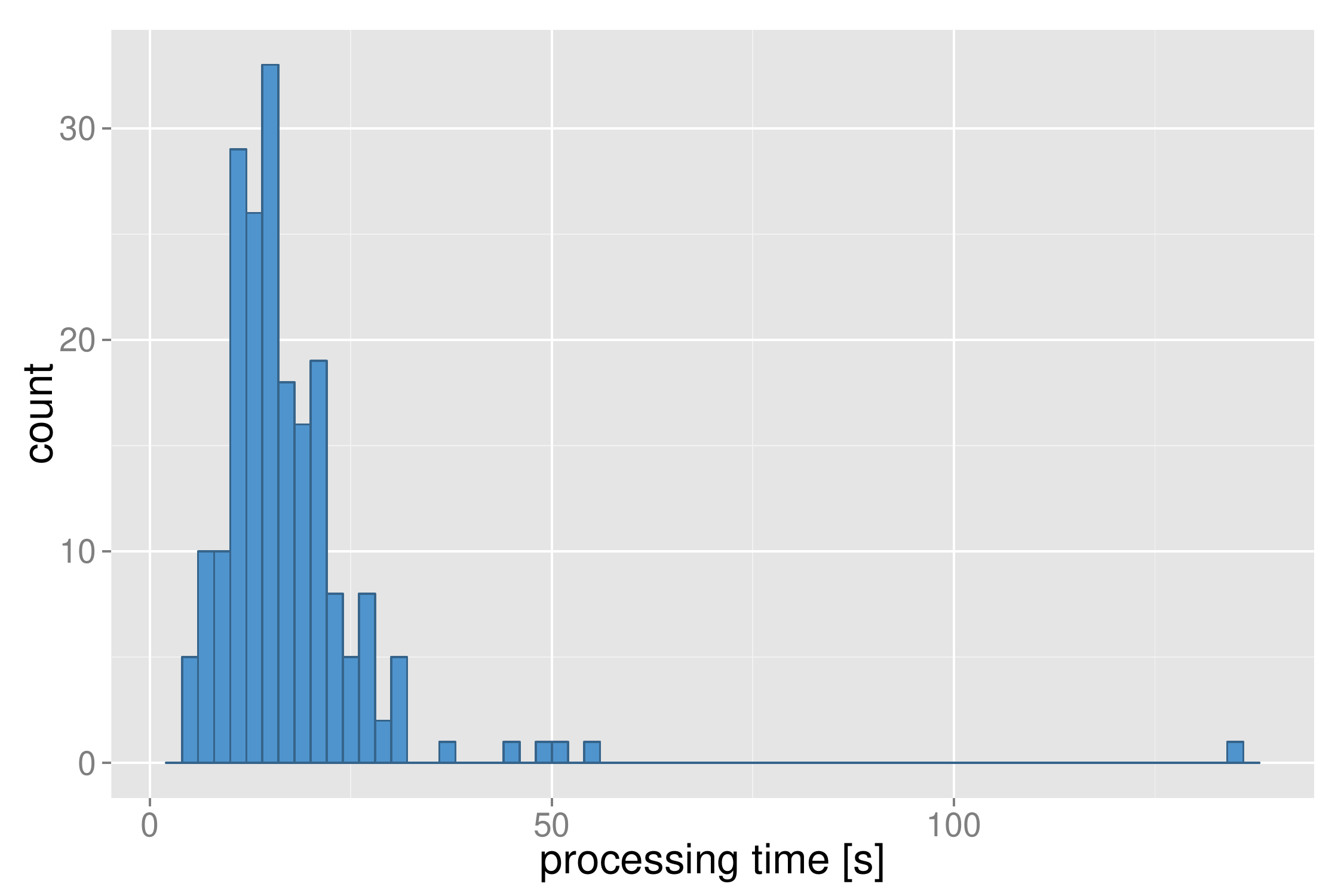}
  \caption[The histogram of the extraction algorithm's processing time]{The histogram of our
  algorithm's processing time for a random subset of 200 documents from PMC.}
  \label{fig:time-histogram}
\end{figure}

Figure~\ref{fig:time-histogram} shows the histogram of our algorithm's processing time for this
subset. The mean processing time was 16.74 seconds.

\begin{figure}
  \centering
  \includegraphics[width=0.8\textwidth]{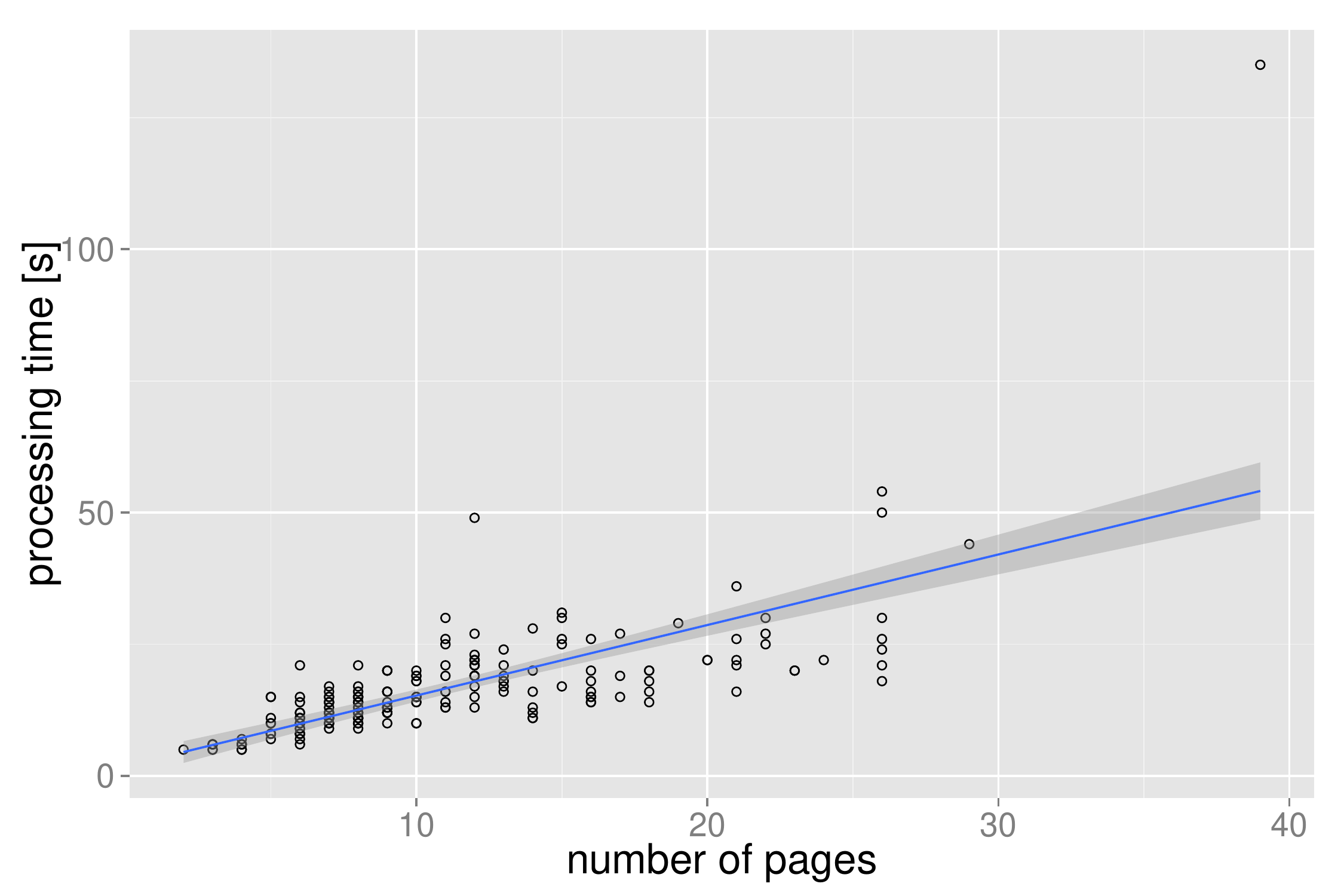}
  \caption[Processing time as a function of the number of pages of a document]{The processing time
  as a function of the number of pages of a document.}
  \label{fig:time-pages}
\end{figure}

In our algorithm the processing time of a document depends linearly on its number of pages. The
Pearson correlation between these variables on the tested subset was 0.70. 
Figure~\ref{fig:time-pages} shows the processing time as a function of the number of the document's
pages.

\begin{figure}
  \centering
  \includegraphics[width=0.8\textwidth]{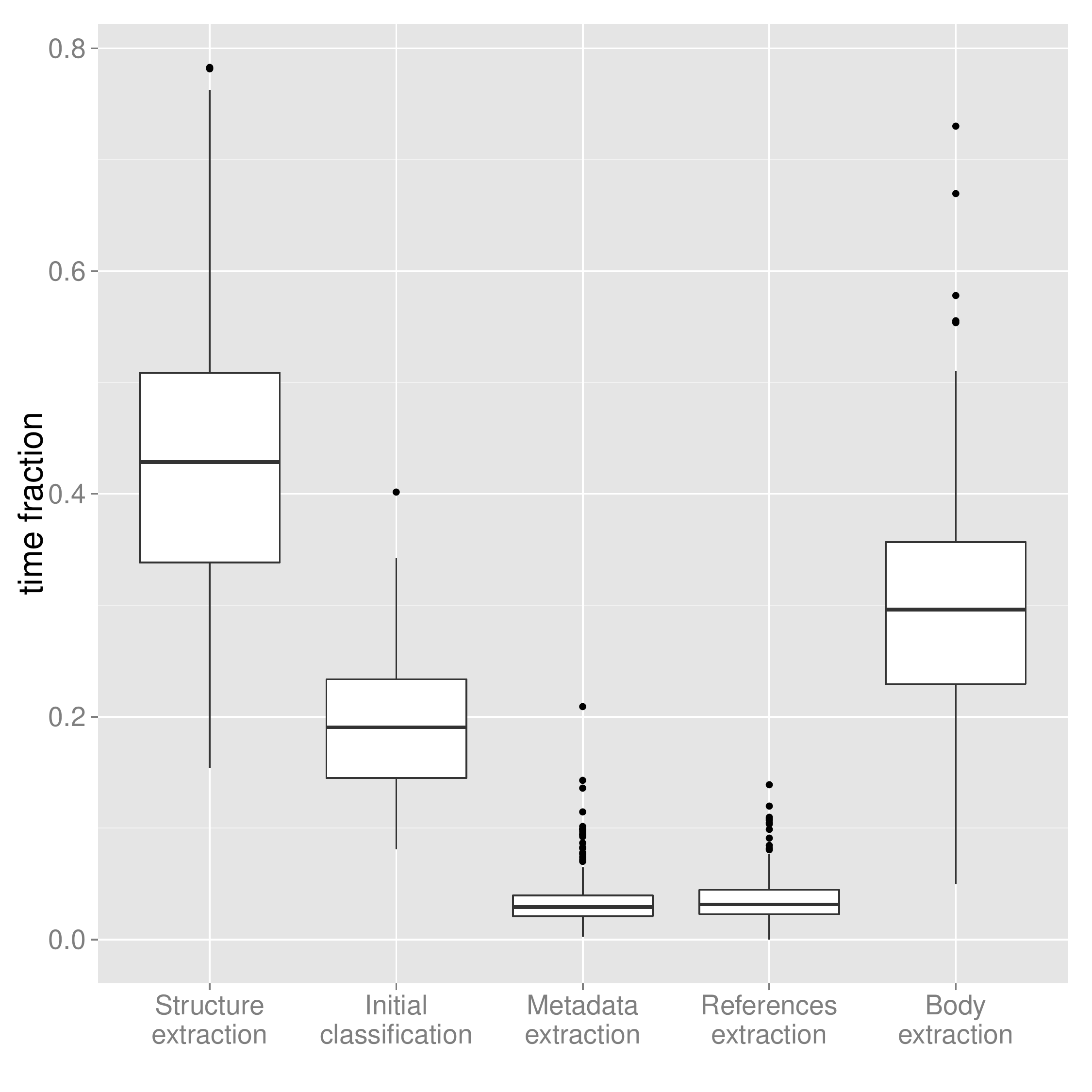}
  \caption[The boxplots of the percentage of time spent on each phase of the extraction]{The
  boxplots of the percentage of time spent on each phase. Character extraction, category
  classification and body extraction are the most time-consuming phases of the algorithm.}
  \label{fig:time-phase}
\end{figure}

The most time-consuming phases of the algorithm are: structure extraction (which inspects the entire 
input document analyzing the individual characters and their mutual positions), body extraction 
(which processes a large, middle region of the publication) and category classification (which
processes all the zones in the document sequentially). Figure~\ref{fig:time-phase} shows the 
boxplots of the percentage of time spent on each phase.

\begin{figure}
  \centering
  \includegraphics[width=0.8\textwidth]{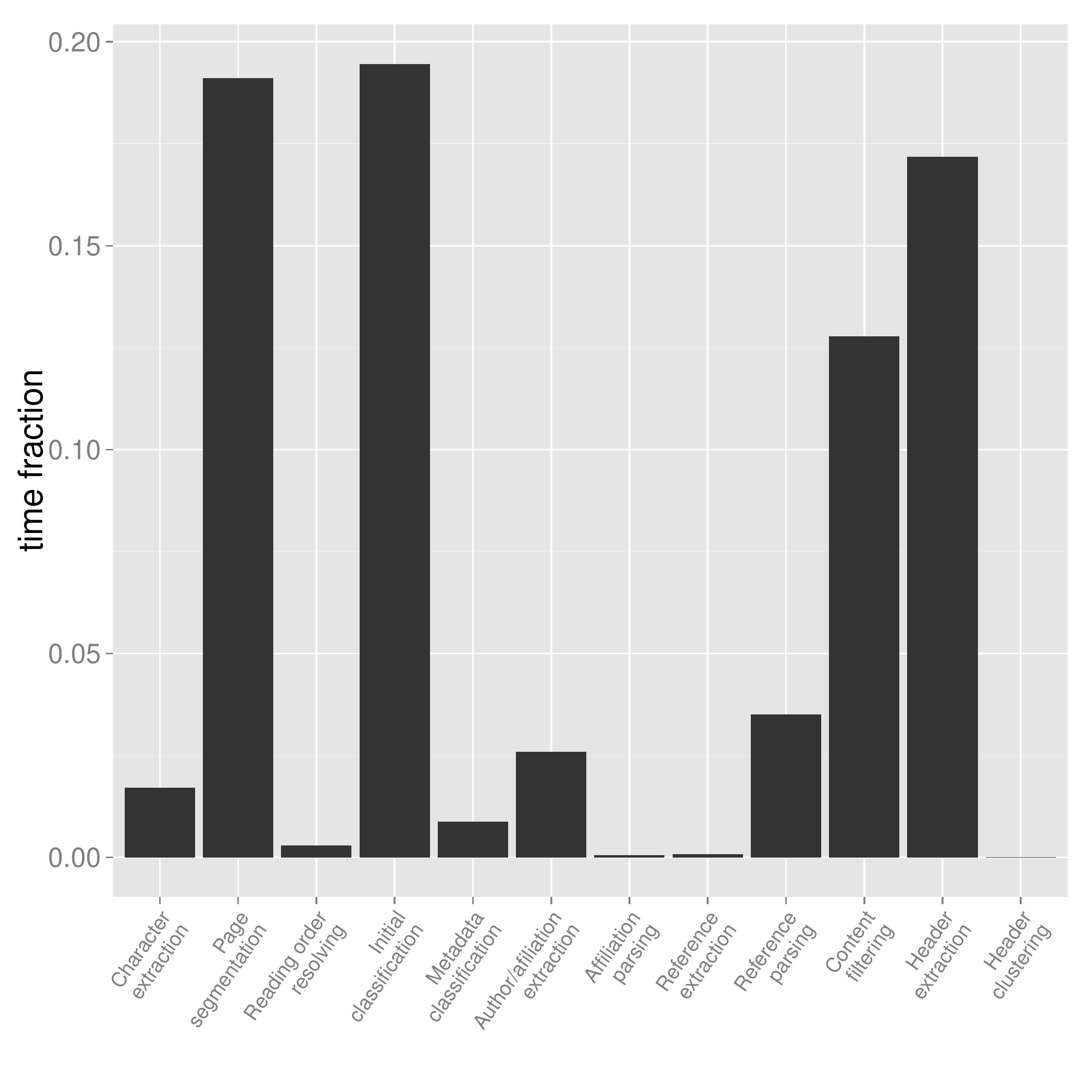}
  \caption[The percentage of time spent on each algorithm step]{The percentage of time spent on each
  algorithm step (the cleaning steps are omitted).}
  \label{fig:time-steps}
\end{figure}

The most time consuming steps of the algorithm are the following: category classification (which 
processes all the zones in the document sequentially), page segmentation (which processes all the 
pages and depends quadratically on the number of the characters on the page), header extraction 
(which iterates over all lines of the text of the document) and content filtering (which classifies
all the zones of the text of the document). Figure~\ref{fig:time-steps} shows the percentage of time
spent on each algorithm step.

\begin{figure}
  \centering
  \includegraphics[width=0.8\textwidth]{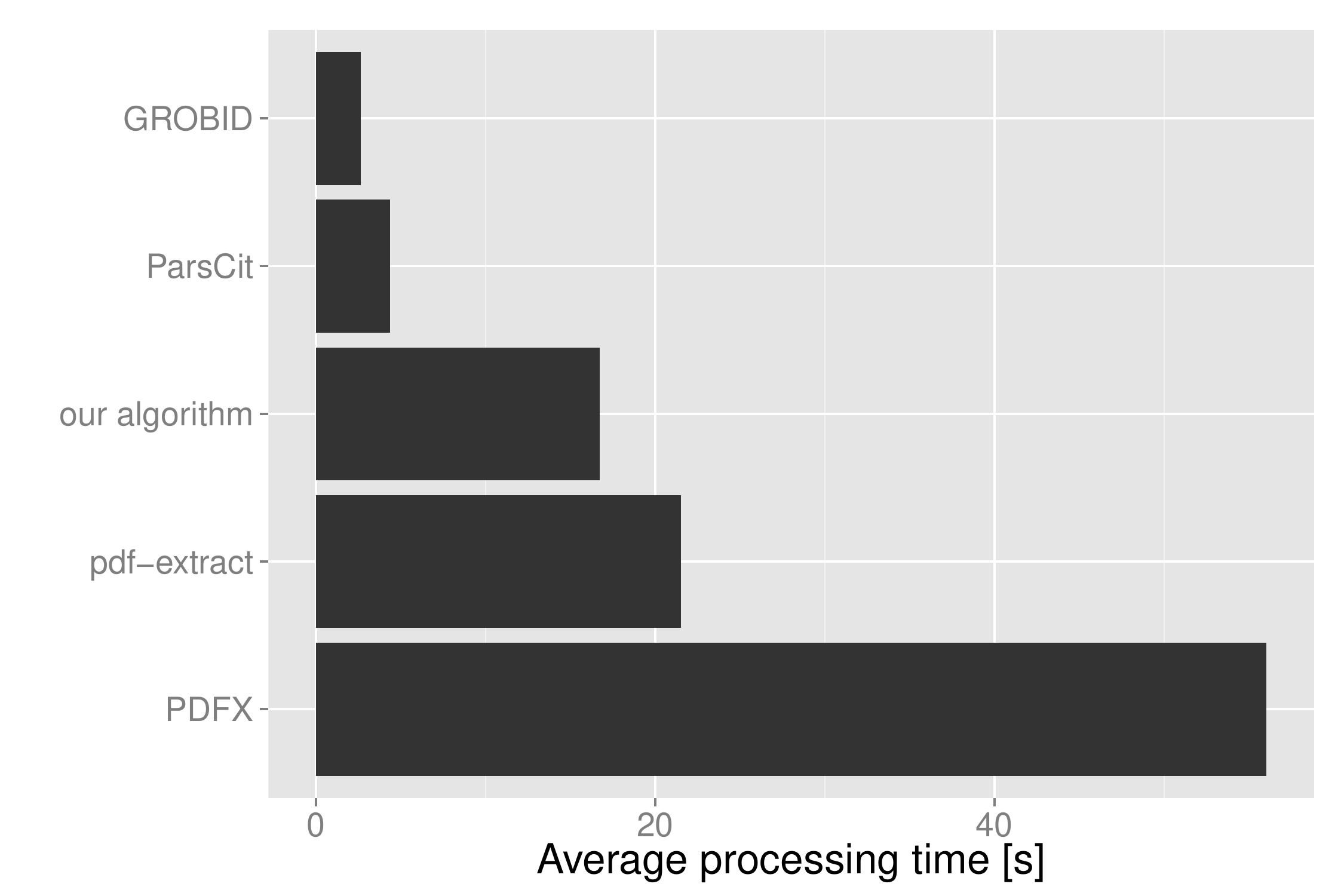}
  \caption[The comparison of the average processing time of different extraction algorithms]{The
  comparison of the average processing time of all tested extraction algorithms on a random subset
  of 200 documents from PMC.}
  \label{fig:time-comparison}
\end{figure}

There are also large differences in processing time between all five tested tools. 
Figure~\ref{fig:time-comparison} compares the average processing time of all tested algorithms.
GROBID and ParsCit proved to be the fastest. The slowest is PDFX, which is a result of the system
being available only as a web service and thus suffering from network transmission overhead.

\clearpage
\thispagestyle{empty}
\cleardoublepage

\chapter{Summary and Future Work}
\label{chap:summary}
In this chapter we conclude the thesis by summarizing the research, listing the most important
achievements of our work and discussing the potential improvements and extensions for the future.

\section{Summary}
The background of the research described in the thesis is closely related to the problems of 
scholarly communication in the digital era. With huge and still growing volume of scientific
literature, keeping track with the latest scientific achievements becomes a major challenge for the
researchers. Scientific information overload problem contributes to slowing down the scholarly
communication and knowledge propagation across the academia.

The main objective of our work was to tackle the problem of scientific information overload. This 
goal is accomplished by equipping digital libraries and research infrastructures with means allowing
them to support the process of consuming the growing volume of scientific literature by interested
readers. This in turn will boost the communication among the scientists and facilitate the knowledge 
propagation in the scientific world. 

We proposed an automatic, novel method for extracting machine-readable, structured metadata from
unstructured scientific publications in born-digital PDF format. Our approach can be used within a
digital library whenever it has to deal with resources with missing, erroneous or fragmentary
document metadata. The algorithm can be used both for assisting the users in the process of
providing the rich metadata for the documents they submit and also automatically processing large 
volumes of already existing resources.

Based on the metadata extraction results, a digital library can provide reliable services such as
intelligent search tools, proposing similar and related documents, building and visualizing
interactive citation and author networks, providing various citations-based statistics, and so on.
This in turn enables the users to effectively explore the map of science, quickly get familiar with
the current state of the art of a given problem and reduce the volume of articles to read by
retrieving only the most relevant and interesting positions.

\section{Conclusions}
We designed and implemented an accurate automatic algorithm for extracting rich metadata directly
from a scientific publication in PDF format. Proposed algorithm takes a single publication on the
input, performs a thorough analysis of the document and outputs a structured machine-readable
metadata record containing:
\begin{itemize}
\item a rich set of document's metadata, such as title, abstract, keywords, authors' full names, 
their affiliations and email addresses, journal name, volume, issue, pages range, year of 
publication, etc.,
\item a list of references to other documents given in the article in a structured form along with
their metadata,
\item structured full text with sections and subsections hierarchy.
\end{itemize}

Designed as a universal solution, the algorithm is able to handle a vast variety of scientific 
publications reasonably well, instead of being perfect in processing a limited number of document
layouts only. We achieved this by employing supervised and unsupervised machine-learning algorithms
trained on large diverse datasets, which also resulted in increased maintainability of the system,
as well as its ability to adapt to new, previously unseen document layouts.

Since our main objective was to provide a useful, accurate solution to a practical problem, machine
learning-based solutions are accompanied with a number of heuristics. This approach proved to 
achieve good results in practice, although perhaps lacks the simplicity and elegance of algorithms
based purely on machine learning.

The evaluation we conducted showed good performance of the proposed metadata extraction algorithm.
The comparison to other similar systems also proved our algorithm performs better than competition
for most metadata types, winning 23 out of 32 evaluation tasks.

The proposed extraction algorithm is based to a great extent on a well-known supervised and
unsupervised machine-learning techniques accompanied with heuristics. Nevertheless, the research
contains a number of innovatory ideas and extensions:
\begin{enumerate}
\item One of the key contributions is the architecture of the extraction workflow and the 
decomposition of the entire extraction problem into smaller tasks.
\item We enhanced the Docstrum-based implementation of page segmentation with a few modifications
listed in Section~\ref{sec:segm}, resulting in increased algorithm accuracy.
\item We developed a large set of 103 zone features capturing all aspects of the zone content and
appearance, which allows to classify zones with high accuracy (Sections~\ref{sec:classification},
\ref{sec:metadata-classification} and~\ref{sec:text-filtering}).
\item We developed a set of features for affiliation tokens allowing to parse affiliations with high
accuracy (Section~\ref{sec:aff-parsing}).
\item We proposed a clustering-based algorithm for extracting reference strings 
(Section~\ref{sec:refs-extr}).
\item We developed a set of features for reference tokens allowing to parse references with high
accuracy (Section~\ref{sec:refs-parsing}).
\item We proposed an algorithm based on normal scores for selecting section header lines from the
text content of the document (Section~\ref{sec:toc-headers}).
\item We proposed a simple clustering approach for building the section hierarchy 
(Section~\ref{sec:toc-hierarchy}).
\item We proposed a scalable, efficient method for constructing gold standard publication datasets 
(Section~\ref{sec:str-sets}).
\end{enumerate}

\section{Outlook}
There are several areas where the extraction algorithm can still be improved to extract higher 
quality metadata or extended to capture more information hidden in unstructured documents.

The algorithm currently processes only born-digital document, in which the text is present as PDF
stream rather than the images of scanned pages. The documents containing scanned pages are not
properly processed, which concerns older resources in particular. We plan to improve this by
extending the workflow and adding an optical character recognition step to the algorithm.

Some algorithm steps, such as parsing authors' zones, associating authors with affiliations, 
detecting header titles, are currently performed using heuristics. In the future we plan to
experiment with machine learning-based approaches as well, which might make the algorithm generalize
better and increase its flexibility.

We would also like to experiment with an unsupervised approach to assist the citation parsing. This
would be based on an observation that the citations usually share the same format within a document,
which could provide useful information in the cases of citations more problematic to parse the
default way.

Currently the algorithm ignores certain document's regions, in particular tables, which contain
useful information present in the documents in an unstructured form. In the future we plan to extend
the workflow so that it is able to extract tables from the documents and present their content in a
machine-readable, structured form.

Apart from the tables, the algorithm currently ignores other useful information that might be 
present in the analyzed document, such as document's categories listed in the first page, the
funding information or the roles of the authors described in a separate section. In the future we
plan to successfully extend the range of information the algorithm is able to extract.

Last but not least, in documents of various domains there are a lot of useful and important
information present in the text of the document, but not expressed directly, such as the reasons the 
document cites other documents, the methods used in the paper, the problem the paper addresses, or
the experiment results. In the future we would like to experiment with machine learning and natural
language processing techniques in order to acquire a deeper understanding of the text of the input
scientific publication.

\clearpage
\thispagestyle{empty}
\cleardoublepage

\appendix
\chapter{Detailed Evaluation Results}
\label{chap:app-results}
The appendix contains detailed results of the performed experiments (the best SVM parameters and
average F-scores for all kernels and all zone classifiers, the confusion matrices for the evaluation
of all zone and token classifiers), the systems evaluation and comparison results (the precision,
recall and F-scores for all tested systems, all test sets and all metadata categories, as well as
the p-values for all statistical tests).

\section{Content Classification Parameters}
\label{sec:app-params}
The best SVM parameters for the zone classifiers were chosen by a grid-search with the use of the
zone validation dataset. For various SVM kernels and parameters combinations we performed a 5-fold
cross-validation on the validation dataset and calculated the overall F-score as an average over 
F-scores for individual class labels. For the final classifiers the kernel and parameters that gave
the best score were chosen.

Tables~\ref{tab:svm-parameters-init}-\ref{tab:svm-parameters-body} present the best SVM parameters
for all the SVM kernels for all zone classifiers (category, metadata and body classifier) as well as
the F-scores obtained during the parameter searching.

\begin{table}
\small
\renewcommand{\arraystretch}{1.2}
\renewcommand{\tabcolsep}{5pt}
\centering
  \begin{tabular}{ | c | C{2cm} | C{2cm} | C{2cm} | C{2cm} | }
\hline
    \multicolumn{5}{|c|}{Category classification} \\
    \hline\hline
    kernel & linear & polynomial & {\bf RBF} & sigmoid \\ \hline
$log_2(c)$ 			& 10 & -2 & {\bf 5} & 10 \\ \hline
$d$ 			 		& - & 3 & - & - \\ \hline
$log_2(\gamma)$  	& - & 0 & {\bf -2} & -4 \\ \hline
$r$				 	& - & 1 & - & -1 \\ \hline
mean F-score (\%) 		& 91.55 & 94.90 & {\bf 94.99} & 94.50 \\ \hline
\end{tabular}
\caption[The results of SVM parameters searching for category classification]{The results of SVM
parameters searching for category classification. The table shows the best mean F-score values for
all kernel function types obtained during 5-fold cross-validation, as well as related values of the
parameters. The best parameters are bolded.} 
 \label{tab:svm-parameters-init}
\end{table}

\begin{table}
\small
\renewcommand{\arraystretch}{1.2}
\renewcommand{\tabcolsep}{5pt}
\centering
  \begin{tabular}{ | c | C{2cm} | C{2cm} | C{2cm} | C{2cm} | }
\hline
    \multicolumn{5}{|c|}{Metadata classification} \\
    \hline\hline
    kernel & linear & {\bf polynomial} & RBF & sigmoid \\ \hline
$log_2(c)$ 			& 5 & {\bf -4} & 5 & 8 \\ \hline
$d$ 			 		& - & {\bf 3} & - & - \\ \hline
$log_2(\gamma)$  	& - & {\bf 0} & -3 & -4 \\ \hline
$r$				 	& - & {\bf 0} & - & -1 \\ \hline
mean F-score (\%) 		& 86.02 & {\bf 88.61} & 88.01 & 87.80 \\ \hline
\end{tabular}
\caption[The results of SVM parameters searching for metadata classification]{The results of SVM
parameters searching for metadata classification. The table shows the best mean F-score values for
all kernel function types obtained during 5-fold cross-validation, as well as related values of the
parameters. The best parameters are bolded.} 
 \label{tab:svm-parameters-meta}
\end{table}

\begin{table}
\small
\renewcommand{\arraystretch}{1.2}
\renewcommand{\tabcolsep}{5pt}
\centering
  \begin{tabular}{ | c | C{2cm} | C{2cm} | C{2cm} | C{2cm} | }
\hline
    \multicolumn{5}{|c|}{Body classification} \\
    \hline\hline
    kernel & linear & {\bf polynomial} & RBF & sigmoid \\ \hline
$log_2(c)$ 			& 7 & {\bf 3} & 4 & 14 \\ \hline
$d$ 			 		& - & {\bf 4} & - & - \\ \hline
$log_2(\gamma)$  	& - & {\bf -3} & -2 & -7 \\ \hline
$r$				 	& - & {\bf 1} & - & -1 \\ \hline
mean F-score (\%) 		& 94.49 & {\bf 96.03} & 95.98 & 95.74 \\ \hline
\end{tabular}
\caption[The results of SVM parameters searching for body classification]{The results of SVM
parameters searching for body classification. The table shows the mean F-score values for all kernel
function types obtained during 5-fold cross-validation, as well as related values of the parameters.
The best parameters are bolded.} 
 \label{tab:svm-parameters-body}
\end{table}

\section{Content Classification Evaluation}
\label{sec:app-class}
All the classifiers used in the extraction algorithm were evaluated by a 5-fold cross-validation on
a respective test set. In every case we generated a confusion matrix and calculated the precision,
recall and F-score for individual class labels.

The confusion matrix $CM$ for a given classifier is indexed by the set of the classifier's labels
$L$. The value $CM(i,j)$, where $i,j \in L$ is equal to the total number of instances with true
label $i$ labelled as $j$ by the classifier during the validation. The values on the matrix diagonal
represent the number of correctly classified instances of respective types and the sum of all cells
equals to the size of the test set.

The precision for a given class label $l \in L$ is calculated as the fraction of the number of
instances labelled as $l$ by the classifier that were correctly labelled. Similarly, the recall for
the label $l$ is calculated as the fraction of the number of instances with true label $l$ that were
correctly labelled.

Tables~\ref{tab:matrix-init}-~\ref{tab:matrix-body} present the confusion matrices, as well as
precision, recall and F-score values for individual class labels obtained from the evaluation of the
zone classifiers (category, metadata and body classifier). 

\begin{table}
\small
\renewcommand{\arraystretch}{1.3}
\renewcommand{\tabcolsep}{2pt}
\centering
  \begin{tabular}{ | c || r | r | r | r || r | r |}
  \hline
  \multicolumn{7}{|c|}{Category classification} \\
  	\hline \hline
  	& \multicolumn{1}{c|}{\it metadata} & \multicolumn{1}{c|}{\it body} & 
  	\multicolumn{1}{c|}{\it references} & \multicolumn{1}{c||}{\it other} & 
  	\multicolumn{1}{c|}{precision (\%)} & \multicolumn{1}{c|}{recall (\%)} \\ \hline \hline
  
  	{\it metadata} & {\bf 66,421} & 1,819 & 76 & 241 & 97.03 & 96.88 \\ \hline
  	{\it body} & 1,324 & {\bf 232,739} & 173 & 890 & 98.12 & 98.98 \\ \hline
    {\it references} & 39 & 692 & {\bf 17,605} & 73 & 98.15 & 95.63 \\ \hline
    {\it other} & 668 & 1,960 & 82 & {\bf 30,977} & 96.26 & 91.96 \\ \hline
   
     \end{tabular}
\caption[The confusion matrix of category classification evaluation]{Confusion matrix for category
classification for 5-fold cross-validation. Rows and columns represent the desired and obtained
classification result, respectively. Bold values on the main matrix diagonal represent correctly
classified zones of respective classes.}
\label{tab:matrix-init}
\end{table}

\begin{table}
\small
\renewcommand{\arraystretch}{1.3}
\renewcommand{\tabcolsep}{2pt}
\centering
  \begin{tabular}{ | c || r | r | r | r | r | r | r | r | r | r | r || r | r | }
  \hline
  \multicolumn{14}{|c|}{Metadata classification} \\
	\hline\hline
  	& \multicolumn{1}{c|}{\rotatebox{90}{\it abstract}} 
  	& \multicolumn{1}{c|}{\rotatebox{90}{\it affiliation}} 
  	& \multicolumn{1}{c|}{\rotatebox{90}{\it author}} 
  	& \multicolumn{1}{c|}{\rotatebox{90}{\it bib\_info}} 
  	& \multicolumn{1}{c|}{\rotatebox{90}{\it correspondence\hspace{5px}}} 
  	& \multicolumn{1}{c|}{\rotatebox{90}{\it dates}} 
  	& \multicolumn{1}{c|}{\rotatebox{90}{\it editor}} 	
  	& \multicolumn{1}{c|}{\rotatebox{90}{\it keywords}} 
  	& \multicolumn{1}{c|}{\rotatebox{90}{\it title}} 
  	& \multicolumn{1}{c|}{\rotatebox{90}{\it type}} 
  	& \multicolumn{1}{c||}{\rotatebox{90}{\it copyright}} 
  	& \multicolumn{1}{c|}{\rotatebox{90}{precision (\%)}} 
  	& \multicolumn{1}{c|}{\rotatebox{90}{recall (\%)}} \\	\hline \hline

	{\it abstract} & {\bf 6,858} & 8 & 6 & 68 & 10 & 3 & 2 & 21 & 5 & 11 & 6 & 97.19 & 98.00
	\\\hline
	{\it affiliation} & 21 & {\bf 3,474} & 24 & 45 & 88 & 6 & 10 & 4 & 0 & 3 & 6 & 94.40 & 94.38
	\\\hline
	{\it author} & 8 & 16 & {\bf 2,682} & 40 & 17 & 0 & 3 & 0 & 5 & 10 & 1 & 96.13 & 96.41 \\\hline
	{\it bib\_info} & 83 & 24 & 30 & {\bf 40,964} & 23 & 110 & 1 & 27 & 16 & 127 & 105 & 98.11 & 98.68
	\\\hline
	{\it corresp.} & 5 & 135 & 16 & 45 & {\bf 1,580} & 2 & 1 & 1 & 0 & 1 & 3 & 91.38 & 88.32
	\\\hline
	{\it dates} & 3 & 0 & 2 & 183 & 2 & {\bf 2,796} & 0 & 0 & 1 & 0 & 13 & 94.75 & 93.20 \\\hline
	{\it editor} & 0 & 7 & 5 & 0 & 0 & 0 & {\bf 464} & 0 & 0 & 0 & 0 & 95.67 & 97.48 \\\hline
	{\it keywords} & 38 & 10 & 4 & 154 & 0 & 1 & 2 & {\bf 826} & 4 & 3 & 2 & 92.39 & 79.12 \\\hline
	{\it title} & 10 & 1 & 10 & 20 & 0 & 1 & 0 & 2 & {\bf 2,584} & 5 & 0 & 98.51 & 98.14 \\\hline
	{\it type} & 13 & 0 & 4 & 169 & 0 & 1 & 1 & 11 & 5 & {\bf 1,403} & 3 & 89.42 & 87.14 \\\hline
	{\it copyright} & 17 & 5 & 7 & 66 & 9 & 31 & 1 & 2 & 3 & 6 & {\bf 2,887} & 95.41 & 95.15 \\\hline
\end{tabular}
\caption[The confusion matrix of metadata classification evaluation]{Confusion matrix for metadata
classification for 5-fold cross-validation. Rows and columns represent the desired and obtained
classification result, respectively. Bold values on the main matrix diagonal represent correctly
classified zones of respective classes.}
\label{tab:matrix-meta}
\end{table}

\begin{table}
\small
\renewcommand{\arraystretch}{1.3}
\renewcommand{\tabcolsep}{2pt}
\centering
  \begin{tabular}{ | c || r | r || r | r |}
  \hline
  \multicolumn{5}{|c|}{Body classification} \\
  	\hline \hline
  	& \multicolumn{1}{c|}{\it body\_content} & \multicolumn{1}{c||}{\it  body\_other} & 
 	\multicolumn{1}{c|}{precision (\%)} & \multicolumn{1}{c|}{recall (\%)} \\ \hline \hline
  
  	{\it body\_content} & {\bf 112,315} & 4,112 & 96.84 & 96.47 \\ \hline
    {\it body\_other} & 3,662 & {\bf 115,037} & 96.55 & 97.91 \\ \hline
   
     \end{tabular}
\caption[The confusion matrix of body content classification evaluation]{Confusion matrix for body
content classification for 5-fold cross-validation. Rows and columns represent the desired and
obtained classification result, respectively. Bold values on the main matrix diagonal represent
correctly classified zones of respective classes.}
\label{tab:matrix-body}
\end{table}

Tables~\ref{tab:matrix-pars}~\ref{tab:matrix-refs} presents the confusion matrices, as well as the
precision, recall and F-score values for individual class labels for affiliation and citation token
classification, respectively.

\begin{table}
\renewcommand{\arraystretch}{1.2}
\renewcommand{\tabcolsep}{5pt}
\centering
  \begin{tabular}{ | c || r | r | r || r | r |}
  \hline
  \multicolumn{6}{|c|}{Affiliation token classification} \\
  	\hline \hline
  	& \multicolumn{1}{c|}{\it address} & \multicolumn{1}{c|}{\it country} 
  	& \multicolumn{1}{c||}{\it institution} & \multicolumn{1}{c|}{precision (\%)} 
  	& \multicolumn{1}{c|}{recall (\%)} \\ \hline \hline
  
  	{\it address} & {\bf 44,463} & 12 & 1243 & 96.74 & 97.25 \\ \hline
  	{\it country} & 55 & {\bf 8,102} & 9 & 99.63 & 99.22 \\ \hline
    {\it institution} & 1442 & 18 & {\bf 92,449} & 98.66 & 98.45 \\ \hline
   
     \end{tabular}
     \vspace{5px}
\caption[The confusion matrix of affiliation token classification evaluation]{Confusion matrix for
affiliation token classification for 5-fold cross-validation. Rows and columns represent the desired
and obtained classification result, respectively. Bold values on the main matrix diagonal represent
correctly classified tokens of respective classes.}
\label{tab:matrix-pars}
\end{table}

\begin{table}
\footnotesize
\renewcommand{\arraystretch}{1.3}
\renewcommand{\tabcolsep}{3pt}
\centering
  \begin{tabular}{ | c || r | r | r | r | r | r | r | r | r | r || r | r | }
  \hline
  \multicolumn{13}{|c|}{Citation token classification} \\
	\hline\hline
  	& \multicolumn{1}{c|}{\rotatebox{90}{\it given\_name}} 
  	& \multicolumn{1}{c|}{\rotatebox{90}{\it surname}} 
  	& \multicolumn{1}{c|}{\rotatebox{90}{\it title}} 
  	& \multicolumn{1}{c|}{\rotatebox{90}{\it source}} 
  	& \multicolumn{1}{c|}{\rotatebox{90}{\it volume}} 
  	& \multicolumn{1}{c|}{\rotatebox{90}{\it issue}} 
  	& \multicolumn{1}{c|}{\rotatebox{90}{\it year}} 
  	& \multicolumn{1}{c|}{\rotatebox{90}{\it page\_first}} 
  	& \multicolumn{1}{c|}{\rotatebox{90}{\it page\_last}}
  	& \multicolumn{1}{c||}{\rotatebox{90}{\it text}} 
  	& \multicolumn{1}{c|}{\rotatebox{90}{precision (\%)}} 
  	& \multicolumn{1}{c|}{\rotatebox{90}{recall (\%)}} \\	\hline \hline

	{\it given\_name} & {\bf 9,529} & 172 & 20 & 3 & 0 & 0 & 2 & 0 & 0 & 188 & 94.66 & 96.12 \\\hline
	{\it surname} & 163 & {\bf 6,858} & 30 & 6 & 0 & 0 & 0 & 0 & 0 & 125 & 94.05 & 95.49 \\\hline
	{\it title} & 23 & 55 & {\bf 20,761} & 168 & 1 & 1 & 5 & 0 & 0 & 222 & 97.93 & 97.76 \\\hline
	{\it source} & 21 & 30 & 165 & {\bf 8,544} & 12 & 2 & 4 & 6 & 1 & 311 & 93.41 & 93.93	\\\hline
	{\it volume} & 1 & 1 & 2 & 21 & {\bf 1,279} & 3 & 1 & 7 & 1 & 20 & 96.53 & 95.73 \\\hline
	{\it issue} & 1 & 0 & 2 & 7 & 1 & {\bf 250} & 1 & 4 & 5 & 17 & 78.62 & 86.81 \\\hline
	{\it year} & 0 & 1 & 8 & 7 & 0 & 3 & {\bf 1,945} & 1 & 0 & 27 & 98.53 & 97.64 \\\hline
	{\it page\_first} & 0 & 0 & 0 & 3 & 14 & 2 & 4 & {\bf 1,533} & 2 & 13 & 98.02 & 97.58 \\\hline
	{\it page\_last} & 0 & 0 & 0 & 1 & 1 & 1 & 1 & 2 & {\bf 1,440} & 3 & 98.16 & 99.38 \\\hline
	{\it text} & 329 & 175 & 212 & 387 & 17 & 56 & 11 & 11 & 18 & {\bf 27,547} & 96.75 & 95.77 \\\hline
\end{tabular}
\caption[The confusion matrix of citation token classification evaluation]{Confusion matrix for
citation token classification for 5-fold cross-validation. Rows and columns represent the desired
and obtained classification result, respectively. Bold values on the main matrix diagonal represent
correctly classified tokens of respective classes.}
\label{tab:matrix-refs}
\end{table}

\section{Performance Scores}
We performed the evaluation of five different extraction systems (our algorithm, GROBID, PDFX,
ParsCit and Pdf-extract) with respect to extracting various metadata types on two different test 
sets (PMC and Elsevier).

Tables~\ref{tab:eval-full-basic}-\ref{tab:eval-full-headers} show the precision, recall and F-scores
for all metadata categories for the tested systems.

\begin{table}
\small
\renewcommand{\arraystretch}{1.2}
\renewcommand{\tabcolsep}{3pt}
\centering
  \begin{tabular}{| c | C{2.3cm} | C{2.3cm} | C{2.3cm} | C{2.3cm} | C{2.3cm} |}
  \hline
  \multicolumn{6}{|c|}{{\bf Basic Metadata}} \\ \hline\hline
     & our algorithm & PDFX & GROBID & ParsCit & Pdf-extract \\ \hline \hline
\multicolumn{6}{|c|}{PMC} \\ \hline\hline
\multirow{3}{*}{\it title} 
 & {\bf 95.84} & 86.25 & 89.88 & 38.31 & 49.48 \\
 & {\bf 93.87} & 86.16 & 85.11 & 33.89 & 49.48 \\
 & {\bf 93.87} & 86.16 & 85.11 & 33.89 & 49.48 \\ \hline
\multirow{3}{*}{\it abstract} 
 & {\bf 83.19} & 67.63 & 77.61 & 56.42 & \\
 & {\bf 80.40} & 60.85 & 70.82 & 56.18 & --- \\
 & {\bf 79.61} & 60.53 & 70.53 & 55.45 & \\ \hline
\multirow{3}{*}{\it keywords} 
 & {\bf 91.61} & & 87.34 & & \\
 & {\bf 60.42} & --- & 54.79 & --- & --- \\
 & {\bf 60.80} & & 54.80 & & \\ \hline\hline
\multicolumn{6}{|c|}{Elsevier} \\ \hline\hline
\multirow{3}{*}{\it title} 
 & 92.04 & {\bf 93.68} & 88.36 & 31.32 & 39.75 \\
 & 86.42 & {\bf 91.91} & 81.93 & 24.12 & 39.82 \\
 & 86.35 & {\bf 91.76} & 81.86 & 24.08 & 39.75\\ \hline
\multirow{3}{*}{\it abstract} 
 & 64.40 & 66.68 & {\bf 84.35} & 50.39 &  \\
 & 64.98 & 62.83 & {\bf 75.94} & 55.72 & --- \\
 & 59.46 & 60.70 & {\bf 73.98} & 49.92 & \\ \hline
\multirow{3}{*}{\it keywords} 
 & 89.58 & & {\bf 91.53} & & \\
 & {\bf 78.58} & --- & 74.75 & --- & --- \\
 & {\bf 77.44} & & 74.09 &  & \\ \hline
\end{tabular}
\caption[The results of the comparison evaluation of basic metadata extraction]{The results of the
evaluation of five different extraction systems with respect to the basic document metadata. Each
cell shows the average precision, recall and F-score. In every category the best score is bolded.}
\label{tab:eval-full-basic}
\end{table}

\begin{table}
\small
\renewcommand{\arraystretch}{1.1}
\renewcommand{\tabcolsep}{3pt}
\centering
  \begin{tabular}{| c | C{2.3cm} | C{2.3cm} | C{2.3cm} | C{2.3cm} |}
  \hline
  \multicolumn{5}{|c|}{{\bf Authorship Metadata}} \\ \hline\hline
     & our algorithm & PDFX & GROBID & ParsCit \\ \hline \hline
     
\multicolumn{5}{|c|}{PMC} \\ \hline\hline

\multirow{3}{*}{\it authors} 
 & {\bf 89.98} &  & 86.45 & 52.31 \\
 & 88.09 & --- & {\bf 87.75} & 44.23 \\
 & {\bf 86.89} & & 84.09 & 41.09\\\hline

\multirow{3}{*}{\it affiliations} 
 & 83.97 & & {\bf 84.04} & 72.34 \\
 & {\bf 84.41} & --- & 61.50 & 44.44 \\
 & {\bf 80.84} & & 56.96 & 45.49 \\ \hline

\multirow{3}{*}{\it author-affiliation} 
 & 55.86 &  & {\bf 62.75} & \\
 & {\bf 71.71} & --- & 45.70 & --- \\
 & {\bf 54.23} & & 38.28 & \\ \hline

\multirow{3}{*}{\it emails} 
 & 52.11 & {\bf 54.55} & 41.48 & 48.27 \\
 & 43.58 & {\bf 82.22} & 5.61 & 55.65 \\
 & 30.35 & {\bf 46.73} & 5.44 & 34.42 \\ \hline
 
\multirow{3}{*}{\it author-email} 
 & {\bf 50.63} & & 46.63 & \\
 & {\bf 42.12} & --- & 5.48 & --- \\
 & {\bf 29.36} & & 5.32 & \\ \hline \hline
 
\multicolumn{5}{|c|}{Elsevier} \\ \hline\hline

\multirow{3}{*}{\it authors} 
 & {\bf 85.27} & & 88.63 & 41.62 \\
 & 81.43 & --- & {\bf 84.66} & 32.17 \\
 & 79.28 & & {\bf 82.24} & 30.04 \\ \hline

\multirow{3}{*}{\it affiliations} 
 & 81.67 & & {\bf 85.58} & 66.21 \\
 & 80.36 & --- & {\bf 81.83} & 30.82 \\
 & 76.05 & & {\bf 77.25} & 32.17 \\ \hline
 
\multirow{3}{*}{\it author-affiliation} 
 & 46.71 &  & {\bf 50.39} & \\
 & 70.34 & --- & {\bf 72.99} & --- \\
 & 43.51 & & {\bf 46.38} & \\ \hline

\multirow{3}{*}{\it emails} 
 & {\bf 84.66} & 79.79 & 80.43 & 65.81 \\
 & 61.26 & {\bf 97.41} & 3.74 & 48.53 \\
 & 55.85 & {\bf 78.92} & 3.79 & 37.76 \\ \hline

\multirow{3}{*}{\it author-email} 
 & {\bf 78.52} & & 76.09 & \\
 & {\bf 56.74} & --- & 3.55 & --- \\
 & {\bf 51.76} & & 3.59 & \\ \hline

\end{tabular}
\caption[The results of the comparison evaluation of authorship metadata extraction]{The results of
comparing the performance of various metadata extraction systems with respect to the 
authorship-related metadata. In every cell the precision, recall and F-score values are shown. The
best results in every category are bolded.}
\label{tab:eval-full-auth}
\end{table}

\begin{table}
\scriptsize
\renewcommand{\tabcolsep}{3pt}
\renewcommand{\arraystretch}{1.2}
\centering
  \begin{tabular}{| c | C{1.8cm} | C{1.8cm} | C{1.8cm} |}
  \hline
  \multicolumn{4}{|c|}{{\bf Bibliographic Metadata}} \\ \hline\hline
     & our algorithm & PDFX & GROBID \\ \hline \hline
     
\multicolumn{4}{|c|}{PMC} \\ \hline\hline

\multirow{3}{*}{\it journal} 
 & {\bf 79.80} &  &  74.09 \\
 & {\bf 73.51} & --- & 20.92 \\
 & {\bf 73.51} & & 20.92 \\ \hline

\multirow{3}{*}{\it volume} 
 & {\bf 93.13} &  & 32.07 \\
 & {\bf 83.14} & --- & 32.01 \\
 & {\bf 83.14} & & 31.99 \\ \hline

\multirow{3}{*}{\it issue} 
 & {\bf 52.65} &  & 21.17 \\
 & {\bf 28.02} & --- & 17.97 \\
 & {\bf 24.53} & & 14.75 \\ \hline

\multirow{3}{*}{\it pages} 
 & {\bf 87.10} & & 25.67 \\
 & {\bf 80.46} & --- & 27.27 \\
 & {\bf 80.14} & & 25.26 \\ \hline
 
\multirow{3}{*}{\it year} 
 & {\bf 96.86} &  & 93.20\\
 & {\bf 95.52} & --- & 39.52 \\
 & {\bf 95.52} & & 39.52 \\ \hline 

\multirow{3}{*}{\it DOI} 
 & 98.13 & 75.93 & {\bf 99.29} \\
 & 74.58 & {\bf 77.35} & 65.62 \\
 & 74.54 & {\bf 75.93} & 65.58 \\ \hline \hline
 
\multicolumn{4}{|c|}{Elsevier} \\ \hline\hline

\multirow{3}{*}{\it journal} 
 & 81.63 &  & {\bf 84.41} \\
 & {\bf 70.18} & --- & 25.06 \\
 & {\bf 70.18} & & 25.06 \\ \hline

\multirow{3}{*}{\it volume} 
 & {\bf 87.53} &  & 31.46\\
 & {\bf 79.54} & --- & 30.66\\
 & {\bf 79.51} & & 30.65\\ \hline

\multirow{3}{*}{\it issue} 
 & {\bf 17.99} &  & 11.89 \\
 & {\bf 14.70} & --- & 6.84 \\
 & {\bf 14.68} & & 6.83 \\ \hline
 
\multirow{3}{*}{\it pages} 
 & {\bf 93.18} &  & 27.94\\
 & {\bf 86.64} & --- & 26.87 \\
 & {\bf 86.60} & & 26.86 \\ \hline

\multirow{3}{*}{\it year} 
 & 67.64 &  & {\bf 82.87} \\
 & {\bf 65.83} & --- & 44.97 \\
 & {\bf 65.83} & & 44.97 \\ \hline

\multirow{3}{*}{\it DOI} 
 & 96.21 & 84.75 & {\bf 97.52} \\
 & 72.12 & {\bf 88.90} & 83.81 \\
 & 71.93 & {\bf 84.75} & 83.60 \\ \hline

\end{tabular}
\caption[The results of the comparison evaluation of bibliographic metadata extraction]{The results
of comparing the performance of various metadata extraction systems with respect to the
bibliographic metadata. In every cell the precision, recall and F-score values are shown. The best
results in every category are bolded.}
\label{tab:eval-full-journ}
\end{table}

\begin{table}
\renewcommand{\arraystretch}{1.2}
\small
\renewcommand{\tabcolsep}{3pt}
\centering
  \begin{tabular}{| c | C{2.3cm} | C{2.3cm} | C{2.3cm} | C{2.3cm} | C{2.3cm} |}
  \hline
   \multicolumn{6}{|c|}{{\bf References}} \\ \hline\hline
     & our algorithm & PDFX & GROBID & ParsCit & Pdf-extract \\ \hline \hline

\multicolumn{6}{|c|}{PMC} \\ \hline\hline

\multirow{3}{*}{\it references} 
 & {\bf 96.17} & 89.68 & 86.31 & 81.21 & 80.40 \\
 & {\bf 90.45} & 84.78 & 87.81 & 71.81 & 57.48 \\
 & {\bf 90.29} & 84.93 & 86.11 & 72.64 & 58.44 \\ \hline\hline

\multicolumn{6}{|c|}{Elsevier} \\ \hline\hline
\multirow{3}{*}{\it references} 
 & {\bf 85.98} & 81.91 & 77.92 & 81.71 & 68.58 \\
 & {\bf 84.16} & 72.55 & 78.35 & 74.57 & 51.39 \\
 & {\bf 84.09} & 72.99 & 76.39 & 74.56 & 52.27 \\ \hline

\end{tabular}
\caption[The results of the comparison evaluation of bibliographic references extraction]{The
results of comparing the performance of various metadata extraction systems with respect to the
bibliographic references extraction. In every cell the precision, recall and F-score values are
shown. The best results in every category are bolded.}
\label{tab:eval-full-refs}
\end{table}

\begin{table}
\small
\renewcommand{\arraystretch}{1.2}
\renewcommand{\tabcolsep}{3pt}
\centering
  \begin{tabular}{| c | C{2.3cm} | C{2.3cm} | C{2.3cm} | C{2.3cm} |}
  \hline
  \multicolumn{5}{|c|}{{\bf Section Hierarchy Metadata}} \\ \hline\hline
     & our algorithm & PDFX & GROBID & ParsCit \\ \hline \hline
     
\multicolumn{5}{|c|}{PMC} \\ \hline\hline

\multirow{3}{*}{\it header titles} 
 & {\bf 75.82} & 71.34 & 72.20 & 41.41 \\
 & {\bf 70.95} & 61.29 & 67.28 & 30.36 \\
 & {\bf 70.70} & 62.34 & 67.28 & 31.62 \\\hline

\multirow{3}{*}{\it level-header title} 
 & 66.56 & {\bf 68.96} &  & 24.28 \\
 & {\bf 61.72} & 59.28 & --- & 18.11 \\
 & {\bf 61.93} & 60.35 &  & 19.32 \\\hline

\end{tabular}
\caption[The results of the comparison evaluation of section headers extraction]{The results of
comparing the performance of various metadata extraction systems with respect to the section
hierarchy-related metadata. In every cell the precision, recall and F-score values are shown. The
best results in every category are bolded.}
\label{tab:eval-full-headers}
\end{table}

\section{Statistical Analysis}
We performed statistical tests for both test sets, for every metadata category and every pair of 
systems that are able to extract the category. The aim of a single test is to establish whether the
difference in the systems' performance is statistically significant. This is done by comparing the
obtained p-value to the significance level adjusted using the Bonferroni correction.

Tables~\ref{tab:stat-full-title}-\ref{tab:stat-full-levheaders} show the p-values for all performed
statistical tests.

\begin{table}
\small
\renewcommand{\arraystretch}{1.3}
\renewcommand{\tabcolsep}{3pt}
\centering
\begin{tabular}{ | c || r | r | r | r | r | }
	\hline
	\multicolumn{6}{|c|}{Title} \\ \hline\hline
	& \multicolumn{1}{c|}{\it our algorithm} & \multicolumn{1}{c|}{\it GROBID} 
  	& \multicolumn{1}{c|}{\it PDFX} & \multicolumn{1}{c|}{\it ParsCit}
  	& \multicolumn{1}{c|}{\it Pdf-extract}
  	\\	\hline \hline
	\multicolumn{6}{|c|}{PMC} \\ \hline\hline

	{\it our algorithm} 	& - & {\bf 3.89e-22} & {\bf 9.43e-17} & {\bf 1.12e-241} & {\bf 9.21e-175} \\\hline
	{\it GROBID} 		& {\bf 3.89e-22} & - & 0.35 & {\bf 7.01e-194} & {\bf 1.04e-118} \\\hline
	{\it PDFX} 			& {\bf 9.43e-17} & 0.35 & - & {\bf 9.32e-189} & {\bf 4.65e-135} \\\hline
	{\it ParsCit} 		& {\bf 1.12e-241} & {\bf 7.01e-194} & {\bf 9.32e-189} & - & {\bf 8.05e-28} \\\hline
	{\it Pdf-extract} 	& {\bf 9.21e-175} & {\bf 1.04e-118} & {\bf 4.65e-135} & {\bf 8.05e-28} & - \\\hline\hline

\multicolumn{6}{|c|}{Elsevier} \\ \hline\hline

	{\it our algorithm} 	& - & {\bf 3.30e-08} & {\bf 4.23e-13} & {\bf 3.01e-310} & {\bf 3.40e-231} \\\hline
	{\it GROBID} 		& {\bf 3.30e-08} & - & {\bf 7.39e-32} & {\bf 5.76e-291} & {\bf 7.87e-177} \\\hline
	{\it PDFX} 			& {\bf 4.23e-13} & {\bf 7.39e-32} & - & {\bf 0} & {\bf 2.62e-266} \\\hline
	{\it ParsCit} 		& {\bf 3.01e-310} & {\bf 5.76e-291} & {\bf 0} & - & {\bf 4.81e-37} \\\hline
	{\it Pdf-extract} 	& {\bf 3.40e-231} & {\bf 7.87e-177} & {\bf 2.62e-266} & {\bf 4.81e-37} & - \\\hline
\end{tabular}
\caption[The results of statistical tests of the performance of title extraction]{P-values obtained
in pairwise statistical tests comparing the performance of extracting the title by various systems.
Significance level for these tests was 0.0025. All the statistically significant results are
bolded.}
\label{tab:stat-full-title}
\end{table}

\begin{table}
\small
\renewcommand{\arraystretch}{1.3}
\renewcommand{\tabcolsep}{3pt}
\centering
\begin{tabular}{ | c || r | r | r | r | }
	\hline
	\multicolumn{5}{|c|}{Abstract} \\ \hline\hline
  	& \multicolumn{1}{c|}{\it our algorithm} & \multicolumn{1}{c|}{\it GROBID}
  	& \multicolumn{1}{c|}{\it PDFX} & \multicolumn{1}{c|}{\it ParsCit}
  	\\	\hline \hline

\multicolumn{5}{|c|}{PMC} \\ \hline\hline

	{\it our algorithm} 	& - & {\bf 4.64e-14} & {\bf 3.13e-54} & {\bf 5.04e-67} \\\hline
	{\it GROBID} 		& {\bf 4.64e-14} & - & {\bf 8.48e-12} & {\bf 5.26e-28} \\\hline
	{\it PDFX} 			& {\bf 3.13e-54} & {\bf 8.48e-12} & - & {\bf 0.0028} \\\hline
	{\it ParsCit} 		& {\bf 5.04e-67} & {\bf 5.26e-28} & {\bf 0.0028} & - \\\hline
	
	\multicolumn{5}{|c|}{Elsevier} \\ \hline\hline
	
	{\it our algorithm} 	& - & {\bf 8.95e-22} & 0.042 & {\bf 4.80e-14} \\\hline
	{\it GROBID} 		& {\bf 8.95e-22} & - & {\bf 1.66e-27} & {\bf 1.06e-60} \\\hline
	{\it PDFX} 			& 0.042 & {\bf 1.66e-27} & - & {\bf 7.76e-09} \\\hline
	{\it ParsCit} 		& {\bf 4.80e-14} & {\bf 1.06e-60} & {\bf 7.76e-09} & -  \\\hline
\end{tabular}
\caption[The results of statistical tests of the performance of abstract extraction]{P-values
obtained in pairwise statistical tests comparing the performance of extracting the abstract by
various systems. Significance level for these tests were 0.0042. All the statistically significant
results are bolded.}
\label{tab:stat-full-abstract}
\end{table}

\begin{table}
\small
\renewcommand{\arraystretch}{1.3}
\renewcommand{\tabcolsep}{3pt}
\centering
\begin{tabular}{ | c || r | r | }
	\hline
	\multicolumn{3}{|c|}{Keywords} \\ \hline\hline
  	& \multicolumn{1}{c|}{\it our algorithm} & \multicolumn{1}{c|}{\it GROBID}
  	\\	\hline \hline

\multicolumn{3}{|c|}{PMC} \\ \hline\hline

	{\it our algorithm} 	& - & {\bf 0.0031}  \\\hline
	{\it GROBID} 		& {\bf 0.0031} & - \\\hline
	
	\multicolumn{3}{|c|}{Elsevier} \\ \hline\hline
	
	{\it our algorithm} 	& - & {\bf 4.61e-05} \\\hline
	{\it GROBID} 		& {\bf 4.61e-05} & -  \\\hline
	
\end{tabular}
\caption[The results of statistical tests of the performance of keywords extraction]{P-values
obtained in pairwise statistical tests comparing the performance of extracting the keywords by
various systems. Significance level for these tests were 0.025. All the statistically significant
results are bolded.}
\label{tab:stat-full-keywords}
\end{table}

\begin{table}
\small
\renewcommand{\arraystretch}{1.3}
\renewcommand{\tabcolsep}{3pt}
\centering
\begin{tabular}{ | c || r | r | r | }
	\hline
	\multicolumn{4}{|c|}{Authors} \\ \hline\hline
  	& \multicolumn{1}{c|}{\it our algorithm} & \multicolumn{1}{c|}{\it GROBID}
  	& \multicolumn{1}{c|}{\it ParsCit}
  	\\	\hline \hline

\multicolumn{4}{|c|}{PMC} \\ \hline\hline

	{\it our algorithm} 	& - & {\bf 2.02e-06} & {\bf 3.05e-192} \\\hline
	{\it GROBID} 		& {\bf 2.02e-06} & - & {\bf 6.95e-191} \\\hline
	{\it ParsCit} 		& {\bf 3.05e-192} & {\bf 6.95e-191} & -  \\\hline

\multicolumn{4}{|c|}{Elsevier} \\ \hline\hline

	{\it our algorithm} 	& - & {\bf 1.61e-05} & {\bf 5.43e-244} \\\hline
	{\it GROBID} 		& {\bf 1.61e-05} & - & {\bf 1.20e-268} \\\hline
	{\it ParsCit} 		& {\bf 5.43e-244} & {\bf 1.20e-268} & -  \\\hline

\end{tabular}
\caption[The results of statistical tests of the performance of authors extraction]{P-values
obtained in pairwise statistical tests comparing the performance of extracting the authors by
various systems. Significance level for these tests were 0.0083. All the statistically significant
results are bolded.}
\label{tab:stat-full-authors}
\end{table}

\begin{table}
\small
\renewcommand{\arraystretch}{1.3}
\renewcommand{\tabcolsep}{3pt}
\centering
\begin{tabular}{ | c || r | r | r | }
	\hline
	\multicolumn{4}{|c|}{Affiliations} \\ \hline\hline
  	& \multicolumn{1}{c|}{\it our algorithm} & \multicolumn{1}{c|}{\it GROBID}
  	& \multicolumn{1}{c|}{\it ParsCit}
  	\\	\hline \hline

\multicolumn{4}{|c|}{PMC} \\ \hline\hline

	{\it our algorithm} 	& - & {\bf 3.24e-76} & {\bf 5.57e-144} \\\hline
	{\it GROBID} 		& {\bf 3.24e-76} & - & {\bf 1.34e-17} \\\hline
	{\it ParsCit} 		& {\bf 5.57e-144} & {\bf 1.34e-17} & -  \\\hline

\multicolumn{4}{|c|}{Elsevier} \\ \hline\hline

	{\it our algorithm} 	& - & 0.030 & {\bf 5.41e-231} \\\hline
	{\it GROBID} 		& 0.030 & - & {\bf 4.58e-227} \\\hline
	{\it ParsCit} 		& {\bf 5.41e-231} & {\bf 4.58e-227} & -  \\\hline

\end{tabular}
\caption[The results of statistical tests of the performance of affiliations extraction]{P-values
obtained in pairwise statistical tests comparing the performance of extracting the affiliations by
various systems. Significance level for these tests were 0.0083. All the statistically significant
results are bolded.}
\label{tab:stat-full-affiliations}
\end{table}

\begin{table}
\small
\renewcommand{\arraystretch}{1.3}
\renewcommand{\tabcolsep}{3pt}
\centering
\begin{tabular}{ | c || r | r | }
	\hline
	\multicolumn{3}{|c|}{Author-affiliation} \\ \hline\hline
  	& \multicolumn{1}{c|}{\it our algorithm} & \multicolumn{1}{c|}{\it GROBID}
  	\\	\hline \hline

\multicolumn{3}{|c|}{PMC} \\ \hline\hline

	{\it our algorithm} 	& - & {\bf 1.77e-49}  \\\hline
	{\it GROBID} 		& {\bf 1.77e-49} & - \\\hline
	
	\multicolumn{3}{|c|}{Elsevier} \\ \hline\hline
	
	{\it our algorithm} 	& - & {\bf 0.0042} \\\hline
	{\it GROBID} 		& {\bf 0.0042} & -  \\\hline
	
\end{tabular}
\caption[The results of statistical tests of the performance of relations author-affiliation
extraction]{P-values obtained in pairwise statistical tests comparing the performance of extracting
the relations author-affiliation by various systems. Significance level for these tests were 0.025.
All the statistically significant results are bolded.}
\label{tab:stat-full-autaffs}
\end{table}

\begin{table}
\small
\renewcommand{\arraystretch}{1.3}
\renewcommand{\tabcolsep}{3pt}
\centering
\begin{tabular}{ | c || r | r | r | r | }
	\hline
	\multicolumn{5}{|c|}{Emails} \\ \hline\hline
  	& \multicolumn{1}{c|}{\it our algorithm} & \multicolumn{1}{c|}{\it GROBID}
  	& \multicolumn{1}{c|}{\it PDFX} & \multicolumn{1}{c|}{\it ParsCit}
  	\\	\hline \hline

\multicolumn{5}{|c|}{PMC} \\ \hline\hline

	{\it our algorithm} 	& - & {\bf 1.15e-82} & {\bf 1.53e-88} & 0.011 \\\hline
	{\it GROBID} 		& {\bf 1.15e-82} & - & {\bf 2.80e-175} & {\bf 2.41e-90} \\\hline
	{\it PDFX} 			& {\bf 1.53e-88} & {\bf 2.80e-175} & - & {\bf 2.43e-90} \\\hline
	{\it ParsCit} 		& 0.011 & {\bf 2.41e-90} & {\bf 2.43e-90} & - \\\hline
	
	\multicolumn{5}{|c|}{Elsevier} \\ \hline\hline
	
	{\it our algorithm} 	& - & {\bf 5.71e-218} & {\bf 6.31e-128} & {\bf 8.30e-36} \\\hline
	{\it GROBID} 		& {\bf 5.71e-218} & - & {\bf 0} & {\bf 2.76e-142} \\\hline
	{\it PDFX} 			& {\bf 6.31e-128} & {\bf 0} & - & {\bf 1.87e-205} \\\hline
	{\it ParsCit} 		& {\bf 8.30e-36} & {\bf 2.76e-142} & {\bf 1.87e-205} & - \\\hline
	
\end{tabular}
\caption[The results of statistical tests of the performance of email extraction]{P-values obtained
in pairwise statistical tests comparing the performance of extracting the email addresses by various
systems. Significance level for these tests were 0.0042. All the statistically significant results
are bolded.}
\label{tab:stat-full-email}
\end{table}

\begin{table}
\small
\renewcommand{\arraystretch}{1.3}
\renewcommand{\tabcolsep}{3pt}
\centering
\begin{tabular}{ | c || r | r | }
	\hline
	\multicolumn{3}{|c|}{Author-email} \\ \hline\hline
  	& \multicolumn{1}{c|}{\it our algorithm} & \multicolumn{1}{c|}{\it GROBID}
  	\\	\hline \hline

\multicolumn{3}{|c|}{PMC} \\ \hline\hline

	{\it our algorithm} 	& - & {\bf 6.13e-80}  \\\hline
	{\it GROBID} 		& {\bf 6.13e-80} & - \\\hline
	
	\multicolumn{3}{|c|}{Elsevier} \\ \hline\hline
	
	{\it our algorithm} 	& - & {\bf 1.41e-201} \\\hline
	{\it GROBID} 		& {\bf 1.41e-201} & -  \\\hline
	
\end{tabular}
\caption[The results of statistical tests of the performance of relations author-email extraction]
{P-values obtained in pairwise statistical tests comparing the performance of extracting the 
relations author-email by various systems. Significance level for these tests were 0.025. All the
statistically significant results are bolded.}
\label{tab:stat-full-autemails}
\end{table}

\clearpage

\begin{table}
\small
\renewcommand{\arraystretch}{1.3}
\renewcommand{\tabcolsep}{3pt}
\centering
\begin{tabular}{ | c || r | r | }
	\hline
	\multicolumn{3}{|c|}{Journal} \\ \hline\hline
  	& \multicolumn{1}{c|}{\it our algorithm} & \multicolumn{1}{c|}{\it GROBID}
  	\\	\hline \hline

\multicolumn{3}{|c|}{PMC} \\ \hline\hline

	{\it our algorithm} 	& - & {\bf 4.32e-190}  \\\hline
	{\it GROBID} 		& {\bf 4.32e-190} & - \\\hline
	
	\multicolumn{3}{|c|}{Elsevier} \\ \hline\hline
	
	{\it our algorithm} 	& - & {\bf 3.81e-205} \\\hline
	{\it GROBID} 		& {\bf 3.81e-205} & -  \\\hline
	
\end{tabular}
\caption[The results of statistical tests of the performance of journal extraction]{P-values
obtained in pairwise statistical tests comparing the performance of extracting the journal by
various systems. Significance level for these tests were 0.025. All the statistically significant
results are bolded.}
\label{tab:stat-full-journal}
\end{table}

\begin{table}
\small
\renewcommand{\arraystretch}{1.3}
\renewcommand{\tabcolsep}{3pt}
\centering
\begin{tabular}{ | c || r | r | }
	\hline
	\multicolumn{3}{|c|}{Volume} \\ \hline\hline
  	& \multicolumn{1}{c|}{\it our algorithm} & \multicolumn{1}{c|}{\it GROBID}
  	\\	\hline \hline

\multicolumn{3}{|c|}{PMC} \\ \hline\hline

	{\it our algorithm} 	& - & {\bf 4.45e-177}  \\\hline
	{\it GROBID} 		& {\bf 4.45e-177} & - \\\hline
	
	\multicolumn{3}{|c|}{Elsevier} \\ \hline\hline
	
	{\it our algorithm} 	& - & {\bf 2.73e-222} \\\hline
	{\it GROBID} 		& {\bf 2.73e-222} & -  \\\hline
	
\end{tabular}
\caption[The results of statistical tests of the performance of volume extraction]{P-values obtained
in pairwise statistical tests comparing the performance of extracting the volume by various systems.
Significance level for these tests were 0.025. All the statistically significant results are
bolded.}
\label{tab:stat-full-volume}
\end{table}

\begin{table}
\small
\renewcommand{\arraystretch}{1.3}
\renewcommand{\tabcolsep}{3pt}
\centering
\begin{tabular}{ | c || r | r | }
	\hline
	\multicolumn{3}{|c|}{Issue} \\ \hline\hline
  	& \multicolumn{1}{c|}{\it our algorithm} & \multicolumn{1}{c|}{\it GROBID}
  	\\	\hline \hline

\multicolumn{3}{|c|}{PMC} \\ \hline\hline

	{\it our algorithm} 	& - & {\bf 8.01e-15} \\\hline
	{\it GROBID} 		& {\bf 8.01e-15} & - \\\hline
	
	\multicolumn{3}{|c|}{Elsevier} \\ \hline\hline
	
	{\it our algorithm} 	& - & {\bf 2.57e-26} \\\hline
	{\it GROBID} 		& {\bf 2.57e-26} & -  \\\hline
	
\end{tabular}
\caption[The results of statistical tests of the performance of issue extraction]{P-values obtained
in pairwise statistical tests comparing the performance of extracting the issue by various systems.
Significance level for these tests were 0.025. All the statistically significant results are 
bolded.}
\label{tab:stat-full-issue}
\end{table}

\begin{table}
\small
\renewcommand{\arraystretch}{1.3}
\renewcommand{\tabcolsep}{3pt}
\centering
\begin{tabular}{ | c || r | r | }
	\hline
	\multicolumn{3}{|c|}{Pages} \\ \hline\hline
  	& \multicolumn{1}{c|}{\it our algorithm} & \multicolumn{1}{c|}{\it GROBID}
  	\\	\hline \hline

\multicolumn{3}{|c|}{PMC} \\ \hline\hline

	{\it our algorithm} 	& - & {\bf 5.52e-194}  \\\hline
	{\it GROBID} 		& {\bf 5.52e-194} & - \\\hline
	
	\multicolumn{3}{|c|}{Elsevier} \\ \hline\hline
	
	{\it our algorithm} 	& - & {\bf 2.47e-315} \\\hline
	{\it GROBID} 		& {\bf 2.47e-315} & -  \\\hline
	
\end{tabular}
\caption[The results of statistical tests of the performance of pages range extraction]{P-values
obtained in pairwise statistical tests comparing the performance of extracting the pages by various
systems. Significance level for these tests were 0.025. All the statistically significant results
are bolded.}
\label{tab:stat-full-pages}
\end{table}

\begin{table}
\small
\renewcommand{\arraystretch}{1.3}
\renewcommand{\tabcolsep}{3pt}
\centering
\begin{tabular}{ | c || r | r | }
	\hline
	\multicolumn{3}{|c|}{Year} \\ \hline\hline
  	& \multicolumn{1}{c|}{\it our algorithm} & \multicolumn{1}{c|}{\it GROBID}
  	\\	\hline \hline

\multicolumn{3}{|c|}{PMC} \\ \hline\hline

	{\it our algorithm} 	& - & {\bf 2.44e-228}  \\\hline
	{\it GROBID} 		& {\bf 2.44e-228} & - \\\hline
	
	\multicolumn{3}{|c|}{Elsevier} \\ \hline\hline
	
	{\it our algorithm} 	& - & {\bf 2.13e-47} \\\hline
	{\it GROBID} 		& {\bf 2.13e-47} & -  \\\hline
	
\end{tabular}
\caption[The results of statistical tests of the performance of year extraction]{P-values obtained
in pairwise statistical tests comparing the performance of extracting the year by various systems.
Significance level for these tests were 0.025. All the statistically significant results are
bolded.}
\label{tab:stat-full-year}
\end{table}

\begin{table}
\small
\renewcommand{\arraystretch}{1.3}
\renewcommand{\tabcolsep}{3pt}
\centering
\begin{tabular}{ | c || r | r | r | }
	\hline
	\multicolumn{4}{|c|}{DOI} \\ \hline\hline
  	& \multicolumn{1}{c|}{\it our algorithm} & \multicolumn{1}{c|}{\it GROBID}
  	& \multicolumn{1}{c|}{\it PDFX}
  	\\	\hline \hline

\multicolumn{4}{|c|}{PMC} \\ \hline\hline

	{\it our algorithm} 	& - & {\bf 1.79e-18} & {\bf 3.40e-07} \\\hline
	{\it GROBID} 		& {\bf 1.79e-18} & - & {\bf 8.30e-36} \\\hline
	{\it PDFX} 			& {\bf 3.40e-07} & {\bf 8.30e-36} & -  \\\hline

\multicolumn{4}{|c|}{Elsevier} \\ \hline\hline

	{\it our algorithm} 	& - & {\bf 6.84e-52} & {\bf 3.61e-65} \\\hline
	{\it GROBID} 		& {\bf 6.84e-52} & - & {\bf 2.45e-13} \\\hline
	{\it PDFX} 			& {\bf 3.61e-65} & {\bf 2.45e-13} & -  \\\hline

\end{tabular}
\caption[The results of statistical tests of the performance of DOI extraction]{P-values obtained in
pairwise statistical tests comparing the performance of extracting the DOI identifier by various
systems. Significance level for these tests were 0.0083. All the statistically significant results
are bolded.}
\label{tab:stat-full-doi}
\end{table}

\begin{table}
\small
\renewcommand{\arraystretch}{1.3}
\renewcommand{\tabcolsep}{3pt}
\centering
\begin{tabular}{ | c || r | r | r | r | r | }
	\hline
	\multicolumn{6}{|c|}{References} \\ \hline\hline
	& \multicolumn{1}{c|}{\it our algorithm} & \multicolumn{1}{c|}{\it GROBID} 
  	& \multicolumn{1}{c|}{\it PDFX} & \multicolumn{1}{c|}{\it ParsCit}
  	& \multicolumn{1}{c|}{\it Pdf-extract}
  	\\	\hline \hline
	\multicolumn{6}{|c|}{PMC} \\ \hline\hline

	{\it our algorithm} 	& - & {\bf 1.79e-118} & {\bf 1.06e-14} & {\bf 3.13e-160} & {\bf 1.60e-169} \\\hline
	{\it GROBID} 		& {\bf 1.79e-118} & - & {\bf 1.37e-44} & {\bf 3.15e-56} & {\bf 5.60e-73} \\\hline
	{\it PDFX} 			& {\bf 1.06e-14} & {\bf 1.37e-44} & - & {\bf 7.04e-78} & {\bf 4.04e-120} \\\hline
	{\it ParsCit} 		& {\bf 3.13e-160} & {\bf 3.15e-56} & {\bf 7.04e-78} & - & {\bf 5.07e-19} \\\hline
	{\it Pdf-extract} 	& {\bf 1.60e-169} & {\bf 5.60e-73} & {\bf 4.04e-120} & {\bf 5.07e-19} & - \\\hline\hline

\multicolumn{6}{|c|}{Elsevier} \\ \hline\hline

	{\it our algorithm} 	& - & {\bf 1.61e-138} & {\bf 1.52e-77} & {\bf 5.00e-78} & {\bf 2.33e-198} \\\hline
	{\it GROBID} 		& {\bf 1.61e-138} & - & 0.0037 & 0.058 & {\bf 4.88e-95} \\\hline
	{\it PDFX} 			& {\bf 1.52e-77} & 0.0037 & - & 0.18 & {\bf 5.01e-92} \\\hline
	{\it ParsCit} 		& {\bf 5.00e-78} & 0.058 & 0.18 & - & {\bf 7.38e-92} \\\hline
	{\it Pdf-extract} 	& {\bf 2.33e-198} & {\bf 4.88e-95} & {\bf 5.01e-92} & {\bf 7.38e-92} & - \\\hline

\end{tabular}
\caption[The results of statistical tests of the performance of references extraction]{P-values
obtained in pairwise statistical tests comparing the performance of extracting the references by
various systems. Significance level for these tests were 0.0025. All the statistically significant
results are bolded.}
\label{tab:stat-full-references}
\end{table}

\begin{table}
\small
\renewcommand{\arraystretch}{1.3}
\renewcommand{\tabcolsep}{3pt}
\centering
\begin{tabular}{ | c || r | r | r | r | }
	\hline
	\multicolumn{5}{|c|}{Headers} \\ \hline\hline
  	& \multicolumn{1}{c|}{\it our algorithm} & \multicolumn{1}{c|}{\it GROBID}
  	& \multicolumn{1}{c|}{\it PDFX} & \multicolumn{1}{c|}{\it ParsCit}
  	\\	\hline \hline

\multicolumn{5}{|c|}{PMC} \\ \hline\hline

	{\it our algorithm} 	& - & {\bf 1.03-18} & {\bf 2.33e-29} & {\bf 2.58e-239} \\\hline
	{\it GROBID} 		& {\bf 1.03-18} & - & {\bf 3.69e-11} & {\bf 1.11e-267} \\\hline
	{\it PDFX} 			& {\bf 2.33e-29} & {\bf 3.69e-11} & - & {\bf 2.57e-218} \\\hline
	{\it ParsCit} 		& {\bf 2.58e-239} & {\bf 1.11e-267} & {\bf 2.57e-218} & - \\\hline
	
\end{tabular}
\caption[The results of statistical tests of the performance of section titles extraction]{P-values
obtained in pairwise statistical tests comparing the performance of extracting the section titles by
various systems. Significance level for these tests were 0.0042. All the statistically significant
results are bolded.}
\label{tab:stat-full-headers}
\end{table}

\begin{table}
\small
\renewcommand{\arraystretch}{1.3}
\renewcommand{\tabcolsep}{3pt}
\centering
\begin{tabular}{ | c || r | r | r | }
	\hline
	\multicolumn{4}{|c|}{Level-header} \\ \hline\hline
  	& \multicolumn{1}{c|}{\it our algorithm} & \multicolumn{1}{c|}{\it PDFX}
  	& \multicolumn{1}{c|}{\it ParsCit}
  	\\	\hline \hline

\multicolumn{4}{|c|}{PMC} \\ \hline\hline

	{\it our algorithm} 	& - & {\bf 0.0030} & {\bf 3.00e-199} \\\hline
	{\it PDFX} 			& {\bf 0.0030} & - & {\bf 1.37e-255} \\\hline
	{\it ParsCit} 		& {\bf 3.00e-199} & {\bf 1.37e-255} & -  \\\hline

\end{tabular}
\caption[The results of statistical tests of the performance of relations level-header extraction]
{P-values obtained in pairwise statistical tests comparing the performance of extracting the 
relations level-header by various systems. Significance level for these tests were 0.0083. All the 
statistically significant results are bolded.}
\label{tab:stat-full-levheaders}
\end{table}

\clearpage
\thispagestyle{empty}
\cleardoublepage
\chapter{System Architecture}
\label{chap:app-arch}
This chapter provides all the details related to the implementation of the metadata extraction 
algorithm, which is available as an open-source CERMINE system (Content ExtRactor and MINEr).

CERMINE provides the entire functionality of the described extraction algorithm. The system is
available as a web service\footnote{http://cermine.ceon.pl} and open-source library written in 
Java\footnote{https://github.com/CeON/CERMINE}. CERMINE is licensed under GNU Affero General Public 
License version 3.

\section{System Usage}
\begin{figure}[h]
  \centering
  \includegraphics[width=0.7\textwidth]{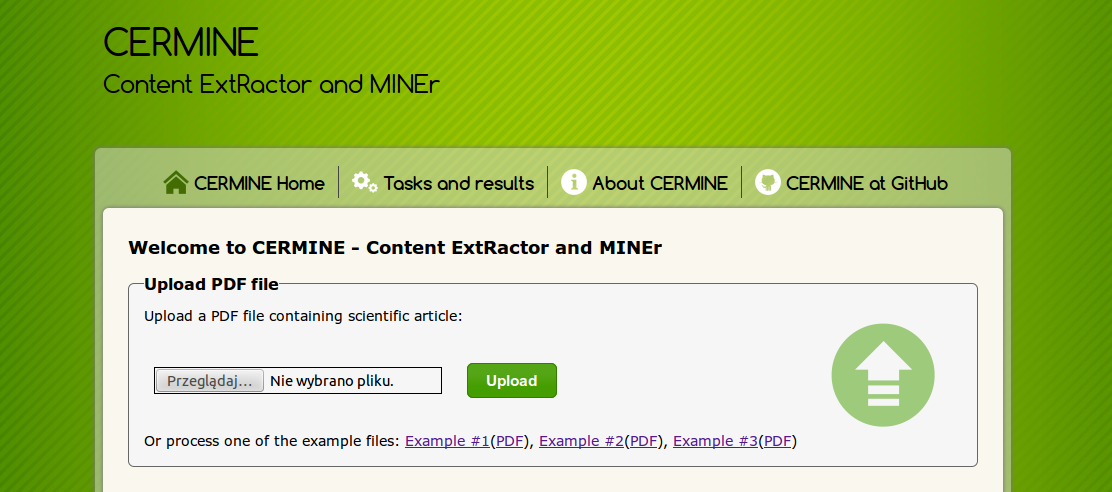}
  \caption[The extraction algorithm demonstrator service]{CERMINE demonstrator service. It allows to
  extract metadata, structured full text and bibliography from a PDF document.}
  \label{fig:web-service}
\end{figure}

The system allows the user to perform the following tasks:
\begin{itemize}
\item Extracting metadata, structured full text and bibliography from a PDF document. The output
format is NLM JATS.
\item Extracting metadata from a citation string. The output format is NLM JATS or BibTeX.
\item Extracting metadata from an affiliation string. The output format in NLM JATS.
\end{itemize}

The demonstrator of the system is available at http://cermine.ceon.pl 
(Figure~\ref{fig:web-service}). Currently it allows to perform only the first task.

There are three ways of using CERMINE system: RESTful services, Java API and executable JAR file.

CERMINE provides three RESTful services, which can be accessed for example using {\tt cURL} tool. 
Listing~\ref{lst:ws} shows how to access the services in order to process entire PDF file, parse a
citation or parse an affiliation.

\begin{lstlisting}[caption=The usage of CERMINE RESTful services.,label=lst:ws]
$ curl -X POST --data-binary @article.pdf \
  --header "Content-Type: application/binary"\
  http://cermine.ceon.pl/extract.do

$ curl -X POST --data "reference=the text of the reference" \
  http://cermine.ceon.pl/parse.do

$ curl -X POST --data "affiliation=the text of the affiliation" \
  http://cermine.ceon.pl/parse.do
\end{lstlisting}

CERMINE can also be used directly in Java projects by adding a dependency and repository to the
project's {\tt pom.xml} file. Listing~\ref{lst:dep} shows the relevant fragment of the {\tt pom.xml}
file and listing~\ref{lst:code} shows example code snippets using CERMINE's Java API.

\begin{lstlisting}[caption=CERMINE's dependency and repository,label=lst:dep]
<dependency>
    <groupId>pl.edu.icm.cermine</groupId>
    <artifactId>cermine-impl</artifactId>
    <version>1.6</version>
</dependency>

<repository>
    <id>icm</id>
    <name>ICM repository</name>
    <url>http://maven.icm.edu.pl/artifactory/repo</url>
</repository>
\end{lstlisting}

\begin{lstlisting}[caption=Example code using CERMINE API.,label=lst:code]
PdfNLMContentExtractor extractor = new PdfNLMContentExtractor();
InputStream inputStream = new FileInputStream("path/to/pdf/file");
Element result = extractor.extractContent(inputStream);

CRFBibReferenceParser parser = CRFBibReferenceParser.getInstance();
BibEntry reference = parser.parseBibReference(referenceText);

CRFAffiliationParser parser = new CRFAffiliationParser();
Element affiliation = parser.parse(affiliationText);
\end{lstlisting}

Lastly, CERMINE can be used as an executable JAR file that can be downloaded from the Maven
repository\footnote{http://maven.ceon.pl/artifactory/simple/kdd-shapshots/pl/edu/icm/cermine/cermine-impl/}. Listing~\ref{lst:jar} shows example commands that can 
be used for processing a PDF file, parsing citations or affiliations.

\begin{lstlisting}[caption=Example commands using executable JAR file,label=lst:jar]
$ java -cp cermine-impl-1.7-SNAPSHOT-jar-with-dependencies.jar \       
  pl.edu.icm.cermine.PdfNLMContentExtractor \
  -path path/to/directory/with/pdfs/or/a/single/pdf

$ java -cp cermine-impl-1.7-SNAPSHOT-jar-with-dependencies.jar \
  pl.edu.icm.cermine.bibref.CRFBibReferenceParser \
  -reference "the text of the reference"

$ java -cp cermine-impl-1.7-SNAPSHOT-jar-with-dependencies.jar \
  pl.edu.icm.cermine.metadata.affiliation.CRFAffiliationParser \
  -affiliation "the text of the affiliation"
\end{lstlisting}

\section{System Components}
The source code of the system and all the components is available on 
GitHub\footnote{https://github.com/CeON/CERMINE}. The project contains the following subprojects:
\begin{itemize}
\item {\tt cermine-impl} --- the entire extraction workflow,
\item {\tt cermine-tools} --- tools that are not part of the main workflow, for example SVM best
parameters searching, models building, datasets creation,
\item {\tt cermine-web} --- the demonstrator and RESTful services.
\end{itemize}

The code structure reflects the decomposition of the workflow into paths and steps. The most 
important packages are:
\begin{itemize}
\item {\tt pl.edu.icm.cermine} --- main extraction classes,
\item {\tt pl.edu.icm.cermine.structure} --- preprocessing, layout analysis and category 
classification,
\item {\tt pl.edu.icm.cermine.metadata} --- basic metadata extraction,
\item {\tt pl.edu.icm.cermine.content} --- structured full text extraction,
\item {\tt pl.edu.icm.cermine.bibref} --- bibliography extraction,
\item {\tt pl.edu.icm.cermine.tools} --- utility classes, related mainly to various machine learning 
algorithms used.
\end{itemize}

CERMINE uses the following external libraries: iText for parsing PDF 
stream\footnote{http://itextpdf.com/}, LibSVM~\cite{ChangL11} for Support Vector Machines
classifiers and GRMM and MALLET~\cite{McCallum02} for linear-chain Conditional Random Fields.

\clearpage
\thispagestyle{empty}
\cleardoublepage

\addcontentsline{toc}{chapter}{Acknowledgements}
\chapter*{Acknowledgements}

First of all, I would like to thank my supervisors Prof. Marek Niezg\'odka and \L{}ukasz Bolikowski 
for the inspiration, advice, motivation and creating a comfortable environment for research.

Next, I would like to thank my colleagues at Interdisciplinary Centre for Mathematical and 
Computational Modelling involved in the development of CERMINE system and GROTOAP datasets: Artur
Czeczko, Jan Lasek, Aleksander Nowi\'{n}ski, Pawe\l{} Szostek, \L{}ukasz Pawe\l{}czak, Krzysztof
Rusek and Bartosz Tarnawski. I would also like to thank the colleagues at ICM, whose enlightened 
advice, fruitful discussions and solid debugging effort were priceless: Piotr Jan Dendek, Mateusz
Fedoryszak, Marek Horst, Jakub Jurkiewicz and Mateusz Kobos.

Last but not least, this thesis would never come into existence without a tremendous amount of
support and motivation I received over many years from my former colleague, friend and mentor,
Sebastian Zagrodzki.

\printbibliography
\end{document}